\documentclass[aps,prd,preprintnumbers,nofootinbib,superscriptaddress,preprint]{revtex4-1}
\usepackage{graphicx}
\usepackage{amssymb}
\usepackage{amsmath}
\usepackage{color}
\usepackage{grffile}
\usepackage{booktabs}
\usepackage{dcolumn}

\graphicspath{{fig/}}

\newcommand{\beq}{\begin{eqnarray}}
\newcommand{\eeq}{\end{eqnarray}}

\def\fsl#1{\setbox0=\hbox{$#1$}
   \dimen0=\wd0
   \setbox1=\hbox{/} \dimen1=\wd1
   \ifdim\dimen0>\dimen1
      \rlap{\hbox to \dimen0{\hfil/\hfil}}
      #1
   \else
      \rlap{\hbox to \dimen1{\hfil$#1$\hfil}}
      /
   \fi}

\begin{document}

\preprint{KEK-CP-348}

\title{Light flavor-singlet scalars and walking signals in $N_f=8$ QCD on the lattice}

\date{\today}

\author{Yasumichi~Aoki}
\affiliation{Institute of Particle and Nuclear Studies,\\
High Energy Accelerator Research Organization (KEK), Tsukuba 305-0801, Japan}
\affiliation{RIKEN BNL Research Center, Brookhaven National Laboratory,\\ Upton, NY, 11973, USA}

\author{Tatsumi~Aoyama}
\affiliation{Kobayashi-Maskawa Institute for the Origin of Particles and the Universe, \\ Nagoya University, Nagoya 464-8602, Japan}

\author{Ed~Bennett}
\affiliation{College of Science, Swansea University, Singleton Park, \\ Swansea, SA2 8PP, UK}

\author{Masafumi~Kurachi}
\affiliation{Institute of Particle and Nuclear Studies,\\
High Energy Accelerator Research Organization (KEK), Tsukuba 305-0801, Japan}

\author{Toshihide~Maskawa}
\affiliation{Kobayashi-Maskawa Institute for the Origin of Particles and the Universe, \\ Nagoya University, Nagoya 464-8602, Japan}

\author{Kohtaroh~Miura}
\affiliation{Centre de Physique Theorique(CPT), Aix-Marseille University,\\ Campus de Luminy, Case 907, 163 Avenue de Luminy, 13288 Marseille cedex 9, France}
\affiliation{Kobayashi-Maskawa Institute for the Origin of Particles and the Universe, \\ Nagoya University, Nagoya 464-8602, Japan}

\author{Kei-ichi~Nagai}
\affiliation{Kobayashi-Maskawa Institute for the Origin of Particles and the Universe, \\ Nagoya University, Nagoya 464-8602, Japan}

\author{Hiroshi~Ohki}
\affiliation{RIKEN BNL Research Center, Brookhaven National Laboratory,\\ Upton, NY, 11973, USA}

\author{Enrico~Rinaldi}
\affiliation{RIKEN BNL Research Center, Brookhaven National Laboratory,\\ Upton, NY, 11973, USA}
\affiliation{Nuclear and Chemical Sciences Division, Lawrence Livermore National Laboratory, Livermore CA 94550, USA}

\author{Akihiro~Shibata}
\affiliation{Computing Research Center, High Energy Accelerator Research Organization (KEK),\\ Tsukuba 305-0801, Japan}

\author{Koichi~Yamawaki}
\affiliation{Kobayashi-Maskawa Institute for the Origin of Particles and the Universe, \\ Nagoya University, Nagoya 464-8602, Japan}

\author{Takeshi~Yamazaki}
\affiliation{Faculty of Pure and Applied Sciences, University of Tsukuba, \\ Tsukuba, Ibaraki 305-8571, Japan}
\affiliation{Center for Computational Sciences, University of Tsukuba, \\ Tsukuba, Ibaraki 305-8577, Japan}

\collaboration{LatKMI Collaboration}
\noaffiliation

\begin{abstract}
Based on the highly improved staggered quark action, we perform lattice simulations of $N_f=8$ QCD and confirm our previous observations, both of a flavor-singlet scalar meson (denoted as $\sigma$) as light as the pion, and of various ``walking signals'' through the low-lying spectra, with higher statistics, smaller fermion masses $m_f$, and larger volumes.
We measure $M_\pi$, $F_\pi$, $M_\rho$, $M_{a_0}$, $M_{a_1}$, $M_{b_1}$, $M_N$, $M_\sigma$, $F_\sigma$, $\langle \overline{\psi} \psi\rangle$ (both directly and through the GMOR relation), and the string tension.
The data are consistent with the spontaneously broken phase of the chiral symmetry, in agreement with the previous results: ratios of the quantities to $M_\pi$ monotonically increase in the smaller $m_f$ region towards the chiral limit similarly to $N_f=4$ QCD, in sharp contrast to $N_f=12$ QCD where the ratios become flattened.
We perform fits to chiral perturbation theory, with the  value of $F_\pi$ found in the chiral limit extrapolation reduced dramatically to roughly 2/3 of the previous result, suggesting the theory is much closer to the conformal window.
In fact,  each quantity obeys the respective hyperscaling relation throughout a more extensive $m_f$ region compared with earlier works.
The hyperscaling relation holds with roughly a universal value of the anomalous dimension, $\gamma_m \simeq 1$, with the notable exception of  $M_\pi$ with $\gamma_m \simeq 0.6$ as in the previous results, which reflects the above growing up of the ratios towards the chiral limit.
This is  a salient feature (``walking signal'') of $N_f=8$, unlike either $N_f=4$ which has no hyperscaling relation at all, or $N_f=12$ QCD which exhibits universal hyperscaling.
The effective $\gamma_m\equiv\gamma_m(m_f)$ of $M_\pi$ defined for each $m_f$ region has a tendency to grow towards unity near the chiral limit, in conformity with the Nambu-Goldstone boson nature, as opposed to the case of $N_f=12$ QCD where it is almost constant. We further confirm the previous observation of the light $\sigma$ with mass comparable to the pion in the studied $m_f$ region.
In a chiral limit extrapolation of the $\sigma$ mass using the dilaton chiral perturbation theory and also using the simple linear fit, we find the value consistent with the 125 GeV Higgs boson within errors.
Our results suggest that the theory could be a good candidate for walking technicolor model, having anomalous dimension $\gamma_m \simeq 1$ and a light flavor-singlet scalar meson as a technidilaton, which can be identified with the 125 GeV composite Higgs in the $N_f=8$ one-family model.
\end{abstract}

\maketitle

%\tableofcontents

\section{Introduction}
\label{sec:intro}

\subsection{Walking Technicolor and mass deformation}

The Higgs boson, with a mass of 125 GeV, has been discovered.
Its properties are so far consistent with the Standard Model (SM) of particle physics.
However, there remain many unsolved problems within the SM, one of which is {the Higgs boson mass itself as the origin of the electroweak scale}.
This is expected to be solved in an underlying theory beyond the SM (BSM).

One of the candidates for such a BSM theory is walking technicolor, {an approximately scale-invariant and strongly-coupled gauge dynamics. This theory was proposed based on the results of the ladder Schwinger-Dyson (SD) equation. It predicted} a technidilaton, a light Higgs-like particle, as a composite pseudo Nambu-Goldstone (NG) boson of the approximate scale symmetry, as well as a large anomalous dimension $\gamma_m \simeq 1$ to resolve the Flavor-Changing Neutral Current (FCNC) problem~\cite{Yamawaki:1985zg,Bando:1986bg}.\footnote{Similar works for the FCNC problem in the technicolor were also done without a technidilaton or consideration of the anomalous dimension and the scale symmetry~\cite{Holdom:1984sk,Akiba:1985rr,Appelquist:1986an}.}

It has in fact been shown that the technidilaton can be identified with the 125 GeV Higgs~\cite{Matsuzaki:2012xx,Matsuzaki:2012mk}.
{Moreover, in terms of UV completions for the SM Higgs sector, the identification of the Higgs boson with a dilaton is one of the most natural and immediate possibilities. The SM Higgs itself is a pseudo-dilaton near the BPS limit (conformal limit) of the SM Higgs Lagrangian when rewritten, via a polar decomposition, into a scale-invariant non-linear sigma model. The NG-boson nature of the SM Higgs in this context is evident because its mass vanishes in the BPS limit with the quartic coupling $\lambda \rightarrow 0$ and the VEV $v (\ne 0)$ fixed (see~\cite{Yamawaki:2016qux} and references therein).}

Besides the technidilaton as a light composite Higgs, walking technicolor generically predicts new composite states in the TeV region, such as technirhos and technipions---a prediction which will be tested at the LHC.

Such a walking theory has an almost non-running coupling; this may be realized for a large number of massless flavors $N_f (\gg 2)$ of the asymptotically-free SU$(N_c)$ gauge theory, dubbed large $N_f$ QCD~\cite{Appelquist:1996dq,Miransky:1996pd}.
In this theory the two-loop beta function has the Caswell-Banks-Zaks (CBZ) infrared (IR) fixed point~\cite{Caswell:1974gg,Banks:1981nn}  $\alpha_*=\alpha_*(N_f,N_c)$ for large enough $N_f$, before losing asymptotic freedom, such that the coupling is small enough to be perturbative. While the coupling runs asymptotically free in units of $\Lambda_{\rm QCD}$ in the ultraviolet region $\mu>\Lambda_{\rm QCD}$, it is almost non-running in the infrared region $\alpha(\mu) \simeq \alpha_*$ for $0<\mu<\Lambda_{\rm QCD}$, where $\Lambda_{\rm QCD}$  is the  intrinsic scale of the theory, analogous to that of ordinary QCD generated by the trace anomaly, which breaks the scale symmetry explicitly.
The CBZ IR fixed point $\alpha_*=\alpha_*(N_f,N_c)$  exists for $N_f^*<N_f < 11N_c/2$ such that $0=\alpha_*(11N_c/2, N_c) <\alpha_*<  \alpha_*(N_f^*,N_c)=\infty$ ($N_f^* \simeq 8 $ for $N_c=3$).
As $N_f$ decreases from $11N_c/2$, $\alpha_*$ increases to the order of $N_c \alpha_*={\cal O} (1) $ at a certain $N_f (> N_f^*) $,  invalidating the assumption about a perturbative IR fixed point before reaching the lower end $N_f^*$.

 Nevertheless, as far as $\alpha_*= {\cal O} (1/N_c)
 $, the slowly-running coupling would still be present for $0<\mu<\Lambda=\Lambda_{\rm QCD}$, where the nonperturbative dynamics can be described---at least qualitatively---by the ladder SD equation with non-running coupling $\alpha(\mu)  \equiv \alpha =\alpha_*$.
The original explicit calculation~\cite{Yamawaki:1985zg} of the large anomalous dimension $\gamma_m=1$ and the technidilaton was actually done in this framework applied to the strong coupling phase $\alpha>\alpha_{\rm cr}$. This phase is characterized by spontaneous chiral symmetry breaking (S$\chi$SB) together with spontaneous (approximate) scale symmetry breaking due to the chiral condensate responsible for the electroweak symmetry breaking. In contrast, the weak coupling $\alpha<\alpha_{\rm cr}$ phase does not have a chiral condensate (``conformal window'').
In fact, the ladder critical coupling  is $\alpha_{\rm cr}=\pi/(3 C_2)= 2N_c \pi/[3(N_c^2-1)]$ ($=\pi/4$ for $N_c=3$), which suggests that
$\alpha_*>\alpha_{\rm cr}$ is realized for $(N_f^* <) \, N_f< 4 N_c$ ($(8<) N_f<12$ for $N_c=3)$~\cite{Appelquist:1996dq},
although the perturbative estimate of $\alpha_*$  (and hence $N_f^{\rm cr}$ such that $\alpha_*(N_f^{\rm cr},N_c)=\alpha_{\rm cr}$) is quantitatively unreliable for such a large $\alpha_*$: $N_c\alpha_*>N_c\alpha_{\rm cr}= {\cal O}(1)$.

In the conformal window, there exist no bound states $H$ of massless fermions (dubbed ``unparticles''), and bound states are only possible in the presence of an explicit fermion mass $m_f$, in such a way that the physical quantities $M_H$ obey the hyperscaling relation $M_H \sim C^{M_H}  m_f^{1/(1+\gamma)}$, with $\gamma= \gamma_m$ and $C^{M_H}$ a constant depending on the quantity.
{To be more specific, bound states in the weakly coupled Coulomb phase (conformal window) would have mass $M_H \sim 2 m_f^{(R)}\sim m_f^{1/(1+\gamma_m)}$, where the renormalized mass (or ``current quark mass'') $m_f^{(R)} = Z_m^{-1} m_f $ is given by the solution of the SD equation as $m_f^{(R)} \sim \Lambda^{\gamma_m/(1+\gamma_m)} m_f^{1/(1+\gamma_m)}$, with $Z_m^{-1}|_{\mu=m_f^{(R)}}= \left(\Lambda/m_f^{(R)} \right)^{\gamma_m}$ and $\Lambda$ being some UV scale such that $m_f=m_f^{(R)}(\mu=\Lambda)$~\cite{Aoki:2012ve}.}
{\footnote{Hereafter we shall not distinguish between $m_f$ and $m_f^{(R)}$ for the qualitative discussions in the region: $m_f< m_f^{(R)}\ll \Lambda_{\rm QCD}$. See also footnote in section~\ref{sec:summary}.}}

A walking theory is expected to be in the broken phase, slightly outside of the conformal window, and hence bound states already exist, even at $m_f=0$, such that $M_\pi=0$ and $M_{H \ne \pi} > m_D \ne 0 $.
{$m_D (\ll \Lambda_{\rm QCD})$ is the dynamical mass of the fermions and it is customarily given by the spontaneously broken solution of the SD gap equation for the mass function in the full fermion propagator, such that $\Sigma(-p^2=m_D^2)=m_D$ in the chiral limit, where it coincides with the so-called ``constituent quark mass'' $m_F$ (distinct from the ``current quark mass'' (renormalized mass) $m_f^{(R)}$.) 

For $m_f^{(R)} \ll m_D$ and  $m_f^{(R)} \gg m_D$, the solution of the  SD solution takes the form $m_F \simeq m_D + m_f^{(R)}$.}%
Once the chiral condensate is generated, the would-be CBZ IR fixed point is actually washed out by the presence of $m_D (\ll \Lambda_{\rm QCD})$ in such a way that the coupling in the region $\mu<\Lambda_{\rm QCD}$ is now nonperturbatively walking in units of $m_D$ (instead of $\Lambda_{\rm QCD}$ when $\mu>\Lambda_{\rm QCD}$), with $\alpha_{\rm cr} (\simeq \alpha_*)$ acting as an ultraviolet fixed point in the IR region $\mu<\Lambda=\Lambda_{\rm QCD}$  as in the original ladder SD arguments~\cite{Yamawaki:1985zg}.
The approximate scale symmetry would still be present for the wide IR walking region $m_D < \mu<\Lambda_{\rm QCD}$ with the mass anomalous dimension $\gamma_m \simeq 1$ and a light (pseudo) dilaton $\sigma$, with mass $M_\sigma ={\cal O} (m_D)$.
The latter is given by the nonperturbative trace anomaly $\langle \theta^\mu_\mu\rangle \sim - N_f N_c m_D^4$ generated by
 $m_D$ in the chiral limit $m_f=0$, such that $M_\sigma^2  = -d_{\theta} \langle  \theta_\mu^\mu \rangle/F_\sigma^2= {\cal O} (m_D^2)$, from the Partially Conserved Dilatation Current (PCDC) relation, with $d_\theta =4$ and the dilaton decay constant $F_\sigma$ as given by $F_\sigma^2 \sim N_f N_c m_D^2$~\cite{Matsuzaki:2015sya}.

\begin{figure}[!tbp]
\includegraphics[scale=0.45]{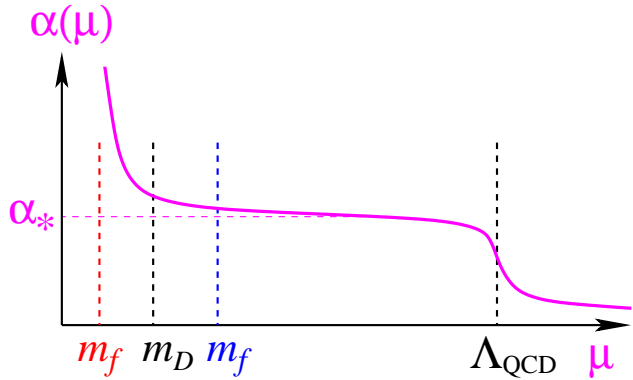}
\caption{
A schematic picture of  the gauge coupling of massless large-$N_f$ QCD as a walking gauge theory in the S$\chi$SB phase near the conformal window. $m_D$ is the dynamical mass of the fermions generated by the S$\chi$SB. The effects of the bare mass of the fermion $m_f$ would be qualitatively different depending on the cases: Case 1: $m_f \ll m_D$ (the red dotted line), which is well described by ChPT, and Case 2: $m_f \gg m_D$ (blue dotted line), which is well described by the hyperscaling, with a possible non-universal exponent for $M_\pi$.
}
\label{fig:walking}
\end{figure}

In the presence of $m_f \ne 0$, the walking theories may be characterized by  $0< m_D, m_f  \ll \Lambda_{\rm QCD}$ as in Fig.~\ref{fig:walking} which was illustrated in our previous paper~\cite{Aoki:2013xza}.
This is not fulfilled in ordinary QCD with $m_D={\cal O} (\Lambda_{\rm QCD})$.
The bound states in theories with a coupling behaving as in Fig.~\ref{fig:walking} are expected to produce ``walking signals'' based on the following two mass regimes:
\begin{enumerate}

\item $\mathbf {m_D\ll m_f\ll \Lambda_{\rm QCD}}$ The approximate hyperscaling relation for the quantities $M_H$ {\it other than $M_\pi$} holds, $M_H \sim C^{M_H} m_f^{1/(1+\gamma)}+ c$, with the same power $\gamma$ independent of $H$, $\gamma \simeq \gamma_m \simeq 1$, where the S$\chi$SB effects $c={\cal O} (m_D) \ll C^{M_H} m_f^{1/(1+\gamma)}$ are negligible.
The mass of the pion $M_\pi$, as a pseudo NG boson, may have a $m_f$ dependence different than other quantities, as $M_\pi^2 \sim C_\pi m_f + C_\pi^\prime m_f^2   + \cdots$, with $C_\pi m_f ={\cal O}(m_D m_f) \ll m_f^2$.
Potentially large corrections $C_\pi^\prime m_f^2$ to Chiral Perturbation Theory (ChPT), which holds in the general case, are possible.
So, even if $M_\pi$ appears to follow hyperscaling, the validity of it may be restricted to a small region of $m_f$, or $\gamma$ should be different from others.
For example, it may change depending on the region of $m_f$, $\gamma=\gamma(m_f)$, reflecting  $m_f$ corrections to hyperscaling inherent to ChPT.
Thus the hyperscaling for individual quantities---if it is observed at all---is expected to be non-universal.

\item $\mathbf {m_f\ll m_D\ll \Lambda_{\rm QCD}}$ The quantities $M_H$ {\it other than} $M_\pi$ go to a non-zero value in such a way that the hyperscaling relation breaks down or $\gamma \rightarrow \infty$ for $m_f\rightarrow 0$.
On the other hand, $M_\pi^2 \rightarrow 0$ (and $F_\pi\rightarrow \ne 0$) behaves according to ChPT with a chiral log, although the ChPT behavior for the $m_f\ll m_D$ region may appear to mimic hyperscaling with $\gamma=1$, $M_\pi^2 \sim m_f$ (up to the chiral log) without a constant term.

\end{enumerate}

In either $m_f$ region the hyperscaling is expected to be non-universal.
Thus the simultaneous validity of a ChPT fit and non-universal hyperscaling may be regarded as the ``walking signals'' to be contrasted with the theory in the conformal window (universal hyperscaling without a good ChPT fit) and that in deep S$\chi$SB phase such as ordinary QCD (a good ChPT fit and the breakdown of even individual (non-universal) hyperscaling).

\subsection{Motivations for lattice studies of large-$N_f$ QCD}

In search of a candidate theory for walking technicolor based on the signals described above, there have recently been many lattice studies on large-$N_f$ QCD.
See for reviews,~\cite{Kuti:2014epa,DeGrand:2015zxa,Aoki:lat2014,Hasenfratz:lat2015}.\footnote{For earlier studies in other contexts, see Ref.~\cite{Iwasaki:1991mr} }
Among large-$N_f$ QCD with a CBZ IR fixed point for $N_c=3$, particular interest was paid to the cases of $N_f=12$ and $N_f=8$ with staggered fermions, partly because the phase boundary is expected to exist somewhere around  $8 <N_f<12$, as suggested by the ladder SD equation and the two-loop CBZ IR fixed point mentioned above.

In the case of $N_f=12$ QCD on the lattice, we obtained results~\cite{Aoki:2012eq} consistent with the conformal window, in agreement with other groups, except for Ref.~\cite{Fodor:2011tu} (see~\cite{DeGrand:2015zxa,Kuti:2014epa,Aoki:lat2014,Hasenfratz:lat2015}).
If it is the case, the walking theory should be realized for $N_f<12$.
It was argued that $N_f=10$ is also consistent with the conformal window~\cite{Hayakawa:2010yn}.

How about $N_f=8$?
Besides lattice studies to be mentioned below, the $N_f=8$ theory is of particular interest as a candidate for walking technicolor for various phenomenological reasons.
First of all the SU(3) gauge theory with $N_f=8$ and four weak-doublets ($N_{\rm D}=N_f/2=4$) is the one-family technicolor model~\cite{Dimopoulos:1979sp,Farhi:1980xs}.
This is the simplest and most straightforward model building of Extended Technicolor (ETC)~\cite{Dimopoulos:1979es,Eichten:1979ah}, to give mass to the SM fermions by unifying the SM fermions and the technifermions.

Moreover, this same model includes a 125 GeV Higgs as the technidilaton~\cite{Matsuzaki:2012xx,Matsuzaki:2012mk,Matsuzaki:2015sya}:
The chiral breaking scale\footnote{Our $F_\pi$ throughout this paper corresponds to $\sqrt{2} \times 93$ MeV in usual QCD.}
$\Lambda_\chi= 4\pi F_\pi/\sqrt{2N_f}=4 \pi \sqrt{2} v_{\rm EW}/N_f$ with $v_{\rm EW}=246$ GeV is much smaller than a naive scale-up of ordinary QCD with $N_f=2,N_c=3$, $\Lambda_\chi \simeq 2$ TeV, by the kinematical factor $1/N_{\rm D}=2/N_f=1/4$, down to
  $\Lambda_\chi \simeq 500\cdot\sqrt{3/N_c}$ GeV.
 This is already close to 125 GeV, even without reference to the detailed conformal dynamics, and naturally accommodates a technidilaton as light as 125 GeV by further reduction via the PCDC, $M_\sigma ={\cal O}(m_D)={\cal O}(\Lambda_\chi/\sqrt{N_c})$, due to the pseudo NG boson nature of the spontaneously broken scale symmetry, similarly to the pion~\cite{Matsuzaki:2015sya}.
In fact a ladder calculation and a holographic estimate in the one-family walking technicolor yields naturally 125 GeV technidilaton with the couplings consistent with the current LHC data of the 125 GeV Higgs boson.

\subsection{Summary of previous lattice results}

In previous publications~\cite{Aoki:2013xza,Aoki:2013qxa,Aoki:2014oha} we have presented lattice results for $N_f=8$ QCD indicating salient features of walking dynamics, quite different from those of either our $N_f=12$ QCD data~\cite{Aoki:2012eq} (consistent with conformality) or our $N_f=4$ QCD data~\cite{Aoki:2013xza} (indicating a chirally-broken phase similarly to ordinary QCD).

We found~\cite{Aoki:2013xza} walking signals as {\it dual features} of spontaneous chiral symmetry breaking and simultaneously of approximate conformal behavior, depending on the mass region $m_f \leq 0.04$ and $m_f > 0.04$, respectively. In the latter case, the dynamically mass $m_D$ generated by the chiral symmetry breaking was estimated to be around $0.04$, roughly of order ${\cal O} (F)$, with $F=F_\pi(m_f=0)$ (the value of $F_\pi$ in the chiral limit) being estimated to be $F\simeq 0.03$ based on ChPT.

The former aspect was typically shown from the ratios $M_\rho/M_\pi, F_\pi/M_\pi$ growing towards the chiral limit  $m_f\rightarrow 0$, which is consistent with the chiral perturbation theory (ChPT) fit valid for $m_f\leq 0.04$, $M_\pi \rightarrow 0$, $F_\pi \rightarrow F\ne 0$, $M_\rho \rightarrow \ne 0$ and $\langle\overline{\psi} \psi \rangle \rightarrow \ne 0$ in a way to satisfy the Gell-Mann-Oakes-Renner (GMOR) relation.
Similar behavior was also observed in the $N_f=4$~\cite{Aoki:2013xza}, and is known to occur in ordinary QCD.
These features are consistent with lattice studies of the running coupling in $N_f=8$ QCD suggesting the absence of an IR fixed point~\cite{Appelquist:2007hu,Hasenfratz:2014rna}, though different conclusions are reached in Ref.~\cite{Ishikawa:2013wf}.

The latter feature, conformality, was demonstrated by the approximate hyperscaling relation valid for $m_f>0.04$,
similarly to $N_f=12$. However,
in contrast to our $N_f=12$ data~\cite{Aoki:2012eq} with the universal hyperscaling $M_H \sim m_f^{1/(1+\gamma)}$ (for $\gamma_m \simeq \gamma \simeq 0.4$), for all the quantities
(ratios between them are constant) in the whole range of $m_f$,
the hyperscaling relation in $N_f=8$ was {\it not universal},  with $\gamma \sim 1$
(a large anomalous dimension, as desired for walking technicolor) for most quantities,
with the notable exception of the pion mass $M_\pi$, with $\gamma \simeq 0.6$ (namely, more rapidly decreasing than other quantities, or the ratio $M_{H\ne\pi}/M_\pi$ rising, near the chiral limit as
mentioned above). These are in fact the walking signals mentioned before.
It was also contrasted to the $N_f=4$, where no approximate (even non-universal) hyperscaling relations hold at all.

It is remarkable that the LSD Collaboration~\cite{Appelquist:2014zsa}, using a different lattice action with domain wall fermions, has obtained $N_f=8$ results similar to ours---in particular, that the ratio $M_\rho/M_\pi$ grows when approaching the chiral limit.
Moreover, the data support non-universal hyperscaling with $\gamma \sim 1$ except for $M_\pi$ with $\gamma \sim 0.6$.
Furthermore, recent results by the LSD Collaboration~\cite{Appelquist:2016viq}, based on nHYP staggered fermions, are also very consistent with ours, with the ratio $M_\rho/M_\pi$ rising more prominently, up to $M_\rho/M_\pi \simeq 2$ (compared with our highest ratio $M_\rho/M_\pi\simeq 1.5$) when getting to smaller  $m_f$.

 We further found~\cite{Aoki:2014oha} a light flavor-singlet scalar meson with mass $M_\sigma$ comparable to the $M_\pi$, $M_\sigma\simeq M_\pi$.
 Such a light $\sigma$ appear similarly in $N_f=12$ but $M_\sigma \lesssim M_\pi$~\cite{Aoki:2013zsa,Fodor:2014pqa} and is very different from the ordinary QCD case $M_\pi < M_\sigma$~\cite{Kunihiro:2003yj}.
On the other hand, the lightness of $\pi$ and $\sigma$ in contrast to other states, e.g.\ $M_\sigma \simeq M_\pi < M_\rho$, in $N_f=8$ (together with the $m_f$ dependence of $M_\rho/M_\pi$ growing when approaching the chiral limit) is consistent with the pseudo NG boson nature of both states in the S$\chi$SB phase.
This is in contrast to $N_f=12$ QCD~\cite{Aoki:2013zsa} where their lightness is moderate, e.g. $M_\sigma \lesssim M_\pi \lesssim M_\rho$ (particularly for large $\beta =4.0$, see also Fig.3 of the latest update~\cite{Aoki:2015gea}), with the ratio $M_\rho/M_\pi, M_\pi/M_\sigma\, (\lesssim 1.2)$ being independent of $m_f$ all the way down to the lightest $m_f$ consistently with the universal hyperscaling in the conformal window.
It is also remarkable that this light flavor-singlet scalar meson, with a mass comparable to $M_\pi$, was confirmed recently by the LSD Collaboration~\cite{Appelquist:2016viq} at smaller fermion masses.

\subsection{Outline of this paper}

In this paper, we present updated results of Refs.~\cite{Aoki:2013xza,Aoki:2014oha}.
Several preliminary results were shown in Refs.~\cite{Aoki:2015jfa,Aoki:2016fxd}, together with the latest updated comparison to $N_f=12$~\cite{Aoki:2015gea} and $N_f=4$~\cite{Aoki:2015zny}.
We have generated more configurations at $\beta=6/g^2= 3.8$ with lattice volumes $(L,T) = (18,24), (24,32), (30,40), (36,48)$ and $(42,56)$, for various fermion masses.
Compared to our previous results in Refs.~\cite{Aoki:2013xza,Aoki:2014oha}, we have added new simulation points in the small mass region $m_f = 0.012$ and $0.015$ with $L = 42$ with 2200 and 4760 HMC trajectories.
We have now typically ten times more trajectories than the previous data for small masses.
The data analyses in this paper are based on the ``Large Volume Data Set'' to be shown in Table~\ref{tab:data_set}, which includes both new and old data.

We further confirm our previous discovery of a light flavor-singlet scalar, $\sigma$, $M_\sigma \simeq M_\pi$~\cite{Aoki:2014oha}, down to the smaller $m_f$ region.
Also the above-mentioned characteristic feature of lightness of  $M_\sigma, M_\pi$, i.e., $M_\sigma \simeq M_\pi < M_\rho$, in contrast to $M_\sigma \simeq M_\pi \simeq M_\rho$ in $N_f=12$,  now becomes more generic including other states:  $M_\sigma \simeq M_\pi < M_\rho, M_{a_0}, M_{a_1}, M_{b_1}, M_N$.

Given the amount of new results we present in this paper, we feel that the reader will benefit from a short summary of different sections.
This will allow interested readers to skip to the individual sections knowing what type of analysis we perform there.

In {\bf section~\ref{sec:sim}}, we describe our lattice setup.
This section also includes a study of the topological charge history, and the technical measurement details of two-point functions for flavor non-singlet mesons.

In {\bf section~\ref{sec:hadspec}} we present the results of the hadron spectrum.
We first focus on mesonic quantities such as $M_\pi$, $F_\pi$ and $M_\rho$.
A comparison with the spectrum obtained with the same lattice setup in the $N_f=12$ and $N_f=4$ theories is also reported.
The main aspects of this comparison include:
\begin{itemize}
\item Finite-volume effects are negligible for the largest volume data;
\item The taste symmetry breaking effects are negligible similarly to the $N_f=12$ and in contrast to $N_f=4$;
\item The updated ratios of $F_\pi/M_\pi$ and $M_\rho/M_\pi$ have a tendency to grow up towards the chiral limit consistently with $N_f=8$ being in the broken phase as in our previous publication~\cite{Aoki:2013xza}.
This is in sharp contrast to the $N_f=12$ data, which tend to flatten near the chiral limit~\cite{Aoki:2012eq,Aoki:2015gea}, but is similar to $N_f=4$ data, which follow chiral symmetry breaking predictions.
\end{itemize}
We also show the Edinburgh plot, $M_N/M_\rho$ versus $M_\pi/M_\rho$, and similar plots, $M_N/F_\pi$ versus $M_\pi/F_\pi$,
which compare favorably with the $N_f=4$ data and the ordinary QCD point.

{\bf Section~\ref{sec:chpt}} is devoted to the ChPT analysis of the hadron spectrum to attempt an extrapolation to the chiral limit.
The chiral extrapolation (without chiral logs) of $F_\pi$ gives a non-zero value $F=0.0212(12)$.
We show that ChPT with this value of $F$ is self-consistent, since the expansion parameter ${\cal X}=N_{f} [M_{\pi}/(4\pi F/\sqrt{2})]^{2}$ is of order ${\cal O}(1)$.
This is in stark contrast to the case of $N_f=12$ which has ${\cal X} \simeq 40$.
The chiral extrapolation of $M_\pi^2/m_f$ is non-zero, similarly to the $N_f=4$ data and consistent with our previous paper~\cite{Aoki:2013xza}.
We also check the chiral extrapolation of the chiral condensate $\langle \overline{\psi} \psi\rangle$ is non-zero, and coincides with that from the GMOR relation  $F^2 M_\pi^2/(4m_f)$ and also from  $F F_\pi M_\pi^2/(4m_f)$, another version of the GMOR relation.

In this section we also present our full numerical results, including  both the updated data of the NG boson pion, flavor non-singlet vector ($\rho$), and the new data on the flavor non-singlet scalar ($a_0$), flavor non-singlet axialvector ($a_1$ with $J^{PC}=1^{++}$ and $b_1$ with $J^{PC}=1^{+-}$), and the nucleon $N$.
Particularly, we give the chiral limit extrapolation of $M_\pi$ based on ChPT, and the linear extrapolation of the other quantities to the chiral limit, which is relevant to the discussions on the application to walking technicolor, which has $m_f=0$.
Notable in the chiral limit is that the flavor non-singlet chiral partners tend to be {somewhat more} degenerate, $M_{a_1}\sim \sqrt{2} M_\rho$, closer to the conformal window compared with ordinary QCD, while other chiral partners, $0=M_{\pi} \ll M_{a_0} (\simeq M_\rho) $, are clearly separated, consistently with the broken phase.

We estimate chiral logs, which are used to evaluate a systematic error on our chiral extrapolation.
The final results are $F = 0.0212(12)(^{+49}_{-71})$, $\left. \langle \overline{\psi} \psi \rangle\right|_{m_f\to 0} = 0.00022(4)(^{+22}_{-12})$, and $\frac{M_\rho}{F/\sqrt{2}} = 10.1(0.6)(^{+5.0}_{-1.9})$.
The estimated chiral limit values of $M_\rho,M_{a_0},M_{a_1},M_{b_1},$ and $M_N$ are given in Table X in units of $F$.

In {\bf section~\ref{sec:FSHS}} we report the hyperscaling analyses we use to test if the theory is in the conformal window.
We find that naive hyperscaling holds for quantities such as $F_\pi $ and $ M_\rho$ with $\gamma \simeq 1.0$ and $0.9$, respectively, while for $M_\pi$ it suggests $\gamma \simeq 0.6-0.7$.
This non-universality of the anomalous dimension is consistent with our findings reported in a previous paper~\cite{Aoki:2013xza}. However, the hyperscaling relation now holds down to smaller masses compared to Ref.~\cite{Aoki:2013xza}, where we find it breaking down for $m_f<0.05$.
This might be due to the drastically reduced chiral limit value of $F_\pi$ (and hence $m_D$ as well), so that the crossover point between the hyperscaling validity region and the  ChPT validity region discussed in Ref.~\cite{Aoki:2013xza}---if existed at all---may have been shifted to the
 smaller $m_f$ region in the new data. The non-universal hyperscaling is also seen for other quantities: most quantities show $\gamma \sim 1.0$ except for $M_\pi$ which has $\gamma \simeq 0.6-0.7$.

 We further estimate the ``effective mass anomalous dimension'' $\gamma_{\mathrm{eff}}(m_f)$, which is found to depend on the $m_f$ region, particularly near the chiral limit:
 the anomalous dimension for $M_\pi$ gradually increases from 0.6 to 0.7 as $m_f$ decreases, a tendency towards 1.0 which would coincide with the $M_\pi$ power behavior of the ChPT.

The Finite-Size Hyperscaling (FSHS) relation is analyzed, including systematics for various volume data.
The FSHS is reduced to the naive one for the infinite volume limit $L\rightarrow \infty$.
First we confirm the FSHS fit for individual quantities separately, similarly to the naive hyperscaling analysis.
The non-universality---or the dependence of $\gamma$ on the quantity considered---of the FSHS becomes more manifest than in the case of naive hyperscaling, with smaller statistical uncertainty due to the higher statistics by combining data from different volumes.
In particular, the FSHS fit to the $M_\pi$ data is rather bad with $\chi^2/{\rm dof} \sim18$ and with $\gamma \simeq 0.6$ sharply contrasted to other quantities with $\gamma \sim 1.0$.
We also check the validity of the simultaneous FSHS with a given universal $\gamma$ for different quantities.
Taking three typical quantities $M_\pi, F_\pi, M_\rho$,
the ``best fit'' is rather bad with $\gamma=0.687(2)$ and $\chi^2/{\rm dof}=105$.

To check the possibility that violations of the universality of the hyperscaling
are due to $m_f$ being far away from the chiral limit, we also check whether or not the simultaneous FSHS with a given universal $\gamma$ for different quantities can be obtained by including possible $m_f$ corrections.
The results are given in Table XIII.

Among others, we include the irrelevant operator with the coefficient $C^{M_H}_2$ in the correction factor $(1+C^{M_H}_2 (g) m_f^\omega)$ being the free parameter depending on the quantities $H$ and the gauge coupling $g$, which was motivated as a perturbation for $N_f=12$~\cite{Cheng:2013xha}  where $\omega$ was estimated using two-loop perturbation theory.
In the case at hand  $N_f=8$, the perturbative arguments are not reliable and we leave it a free parameter to fit to the data.
The results read $\omega=0.347(14),\, \gamma=1.108(48), \chi^2/{\rm dof}=1.05$, which appears reasonable. However, the corrections for $M_\pi$ are unnatural in the sense that the correction terms $m_f^\omega$ does not diminish all the way down to the smallest $m_f$ region due to the small power $\omega$,---i.e., they are no longer the mass corrections---and the correction to $M_\pi$ is particularly large, about 50\%.
This can be understood as the large corrections changing the divergent behavior of the ratios $F_\pi/M_\pi, M_\rho/M_\pi$ near the chiral limit into an artificial flat behavior in accord with the universality, particularly near the chiral limit.
Thus the universal hyperscaling in various versions does not hold  for $N_f=8$, in sharp contrast to $N_f=12$.

{\bf Section~\ref{sec:stringtension}} includes the results for the string tension obtained from the measurement of correlators of Wilson loops.
The string tension is measured by fitting the static potential and also via the Creutz ratio. We consider two fits, the quadratic fit $\sqrt{s}=A_2 m_f^2  + A_1 m_f + A_0$, with $A_0=0.058(4), \chi^2/{\rm dof}=0.99$, and the hyperscaling fit $\sqrt{s}=C  m_f^{1/(1+\gamma)}$, with $ \gamma=0.96(6), \chi^2/{\rm dof}=1.26$. Again we observe the dual features of the walking signals:
both the S$\chi$SB phase and conformal phase with $\gamma \simeq 1.0$ are consistent as far as the string tension alone is concerned. $\gamma\simeq 1.0$ is consistent with the spectrum except for $M_\pi$ again, non-universal hyperscaling.

{\bf Section~\ref{sec:flavorsinglet}} presents the highlight of this paper, the flavor-singlet scalar $\sigma$, with mass comparable to the pion, which is the updated version of Ref.~\cite{Aoki:2014oha}.
The advantage of the disconnected correlator $D(t)$ for extracting the $\sigma$ mass is emphasized. We estimate $M_\sigma$ from the disconnected correlator $2 D(t)$ with accuracy better than that from the full correlator, including new data at $m_f=0.012$ and partly new data at $m_f=0.015$ in addition to the old data in Ref.~\cite{Aoki:2014oha} (See Table XIV).
The resulting $M_\sigma$ is comparable to $M_\pi$ (Fig. 23), $M_\sigma \simeq M_\pi \ll M_\rho$, which is consistent with the previous one.

As to the chiral extrapolation, we use the leading order of the ``dilaton ChPT''~\cite{Matsuzaki:2013eva}, $M_\sigma^2= d_0 + d_1 M_\pi^2$, where $d_0=M_\sigma^2|_{m_f=0}$ and $d_1=(3-\gamma_m)(1+\gamma_m)/4\cdot (N_f F^2/F_\sigma^2) \simeq N_f F^2/F_\sigma^2$ (for $\gamma_m \simeq \gamma \simeq 1$), with $F_\sigma$ being the decay constant of the (pseudo) dilaton as $\sigma$, $\langle 0| \theta_{\mu\nu}|\sigma(q) \rangle= F_\sigma (q_\mu q_\nu- q^2 g_{\mu\nu})/3$.
We use the data  for $m_f\leq 0.03$ as in Ref.~\cite{Aoki:2014oha}, which is in accord with the ChPT fit range for other hadrons in Section IV.
The fit is $d_0=-0.0028(98)(^{+36}_{-354}), d_1=0.89(26)(^{+75}_{-11})$.
Although the error is so large that no definite conclusion can be drawn at this moment, the value of $d_0$ above is consistent with the identification of the 125 GeV Higgs as the technidilaton in the one-family walking technicolor, with $M_\sigma =125 {\rm GeV} \simeq v_{\rm EW}/2 = F/\sqrt{2}$, which would correspond to  $d_0 \sim  (F/\sqrt{2})^2 \simeq 0.0002$ (using $ F\simeq 0.02$ obtained in Section IV).

{\bf Section~\ref{sec:summary}} is devoted to a discussion and a summary.

\clearpage

\section{Lattice simulation setup}
\label{sec:sim}

\subsection{Lattice action and simulations}
\label{sec:action_and_param}

In a series of studies of SU(3) gauge theory with respect to its
properties at many flavors, we have been using the Highly Improved
Staggered Quarks (HISQ)~\cite{Follana:2006rc,Bazavov:2010ru}
for fermions in the fundamental representation.
Our studies intend to explore the theory space from
usual QCD to those in the conformal window, passing through the conformal phase
boundary. As the correct counting of the light degree of freedom
is important for such a study, good flavor symmetry is the
first priority in our choice of the action.
The HISQ action has been successful in usual QCD simulations~\cite{Bazavov:2010ru,Bazavov:2011nk}
at systematically
reducing the breaking of taste symmetry~\cite{Follana:2006rc},
which is a part of the flavor symmetry. Later in Sec.~\ref{sec:taste_breaking}
we will show the taste symmetry breaking effect in the pseudoscalar
meson masses.

A schematic expression of our action reads
\begin{equation}
 S = S_g(\beta)[U] + S_f^{\rm HISQ}(N_f,m_f)[U],
\end{equation}
with $S_g(\beta)[U]$ here being the tree-level Symanzik-improved
gauge action without tadpole improvement. It consists of the
$1\times 1$ plaquette and $2\times 1$ rectangular Wilson loops
made of the gauge link field $U_{x,\mu}\in {\rm SU}(3)$. The coupling is defined as
$\beta\equiv 6/g^2$,\footnote{This convention is different from the one
conventionally used for HISQ simulations in usual QCD, $\beta\equiv
10/g^2$.} with $g$ being the bare gauge
coupling. The fermion part reads
\begin{equation}
 S_f^{\rm HISQ}(N_f,m_f)[U] = \sum_{i=1}^{N_f/4}\overline{\chi}^i (D^{\rm HISQ}[U]+m_f)\chi^i,
\end{equation}
where the number of flavors in this study is $N_f=8$ for the main result,
with additional results for $N_f=4$ and $12$ for comparison. $\chi^i$ is the staggered
fermion field in the fundamental representation of the color SU(3) group,
of $i$-th species, with suppressed coordinate and color labels. $m_f$
is the bare staggered fermion mass common for all the species.
$D^{\rm HISQ}[U]$ is the massless staggered Dirac operator for HISQ~\cite{Follana:2006rc,Bazavov:2010ru},
which involves one and three link hopping terms  where
different levels of smeared link of $U_{x,\mu}$ enter,
to effectively reduce (but not completely remove) the taste-exchanging
one-loop $O(a^2)$ effects. Through this, the flavor symmetry is largely
improved. The mass correction to the Naik term is not included as our
interest is the system in the chiral limit.

The exact symmetry of this system at non-zero lattice spacing ($a\ne 0$) is the
${\rm U}(1)_V$ and spin-taste-diagonal axial symmetry ${\rm U}(1)_\epsilon$
for the $N_f=4$ (one species) case. In the continuum limit 
the full symmetry
{of $N_f=4$ QCD} should be recovered. For the $N_f=8$ and $12$
cases, the exact symmetries at $a\ne 0$ are extended to include
${\rm SU}(N_s)_{V-A}\times {\rm SU}(N_s)_{V+A}$\footnote{Note that due to the difference of staggered tastes in the
conserved vector and axialvector symmetry, $V-A$ and $V+A$ do not simply
correspond to left and right chiral symmetry.}
 with the number of species $N_s=2$ and $3$
for $N_f=8$ and $12$ respectively.\footnote{Such an extended symmetry
is useful, for example, to formulate a method to calculate the Peskin-Takeuchi $S$
parameter~\cite{Peskin:1990zt}
with staggered fermions~\cite{Aoki:2016bfp}.}
Restoring the taste symmetry in the continuum limit leads to the
restoration of full symmetry
{${\rm SU}(N_f)_L \times {\rm SU}(N_f)_R \times {\rm U}(1)_V$}.{\footnote{The $U(1)_A$ is broken by quantum anomaly.}}

Eventually we need to understand the dynamics
of theory at each $N_f$  in the limit of all fermions simultaneously vanishing
$m_f\to 0$, in the continuum $a\to 0$ and infinite volume limits.
For initial steps towards this ultimate goal, we fix the lattice spacing
by fixing the gauge coupling $\beta=3.8$~\cite{Aoki:2013xza}
for our main calculation in this study, which is $N_f=8$.
However, we examine the volume and mass systematically
to study the infinite volume and chiral limits.

For the study of finite size hyperscaling to test the conformal scenario,
it is advantageous to fix the space-time aspect ratio of the lattice,
so that the change of the system size is represented by one parameter, which is
either $L$ for the spatial or $T$ for the temporal size for
$L^3\times T$ lattice.
To this end we use volumes which satisfy $L/T=3/4$ for $N_f=8$ and
$12$, while the aspect ratio for $N_f=4$ is fixed to $L/T=2/3$.

The spatial size for the $N_f=8$ varies as $L=42$, 36, 30, 24, 18, and 12.
The $L=42$ lattice volume is new here, while the other volumes have
already been used either in our study of the flavor singlet scalar~\cite{Aoki:2014oha} or in the earlier publication on the ``walking
signals''~\cite{Aoki:2013xza}.
Among these, the majority of both the new data set and the ones already used in the scalar
study~\cite{Aoki:2014oha} have higher statics than those used in
Ref.~\cite{Aoki:2013xza}.
These ensembles, which are more important for this study than the other
old ensembles and are called the main ensembles, will be described in detail.
For the old ensembles we refer to Ref.~\cite{Aoki:2013xza}.

Table~\ref{tab:nf8stat} shows the statistics of our main ensembles of
$N_f=8$ in terms of maximum number of thermalized trajectories
used in this study.
$N_{str}$ shows the number of streams. In the multiple stream cases,
$N_{\rm Traj}^{\rm max}$ shows the total number over all streams.
For generating the gauge field ensembles, the hybrid Monte Carlo
(HMC) algorithm~\cite{Duane:1987de} with
Hasenbusch preconditioning~\cite{Hasenbusch:2001ne} is used.
HMC parameters, including those related to the preconditioning as well as the molecular
dynamics step size, are shown in Table~\ref{tab:nf8hmc}.
Through all parameter sets  the
Monte Carlo accept/reject step is placed at the end of each molecular
dynamics (MD) integration of unit time 1. Each of such a step
is conventionally called a trajectory.
\begin{table}
 \caption{Statistics of the main ensembles for $N_f=8$.
 $L$ and $T$ for the spatial and temporal size for $L^3\times T$
 lattice, staggered fermion mass $m_f$,
 number of HMC streams $N_{\rm str}$,
 and maximum number of thermalized trajectories
 $N_{\rm Traj}^{\rm max}$. Details of HMC parameters are shown in
 Table~\ref{tab:nf8hmc}. }
 \label{tab:nf8stat}
\begin{ruledtabular}

\begin{tabular}{cc l c r}
\multicolumn{1}{c}{$L$} &
\multicolumn{1}{c}{$T$} &
\multicolumn{1}{c}{$m_f$} &

\multicolumn{1}{c}{$N_{\rm str}$} &
\multicolumn{1}{c}{$N_{\rm Traj}^{\rm max}$} \\
\hline
42 & 56 & 0.012 & 2 &  4760 \\
   &    & 0.015 & 1 &  2200 \\
\hline
36 & 48 & 0.015 & 2 & 10800 \\
   &    & 0.02  & 1 &  9984 \\
   &    & 0.03  & 1 &  2000 \\
\hline
30 & 40 & 0.02  & 1 & 16000 \\
   &    & 0.03  & 1 & 33024 \\
   &    & 0.04  & 3 & 25600 \\
\hline
24 & 32 & 0.03  & 2 & 74752 \\
   &    & 0.04  & 2 &100352 \\
   &    & 0.06  & 1 & 39936 \\
   &    & 0.08  & 2 & 17408 \\
\hline
18 & 24 & 0.04  & 1 & 17920 \\
   &    & 0.06  & 1 & 17920 \\
   &    & 0.08  & 1 & 17920 \\
\end{tabular}
\end{ruledtabular}
\end{table}

The MILC code ver.~7 is used for the HMC evolution and
measurements on the obtained gauge fields.
Some modifications to the MILC code~\cite{MilcCodeV7} have been made to simulate without
the rational hybrid Monte Carlo, which is not needed for the values of $N_f$ we use,
as well as to speed up the fermion force computations and so on.

\begin{figure}[!tbp]
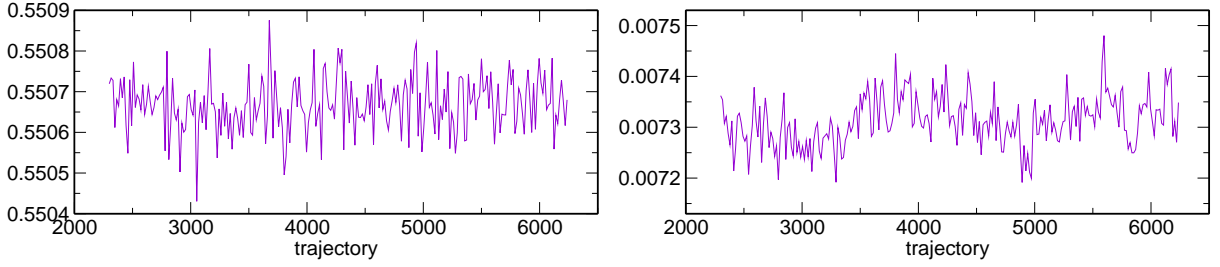
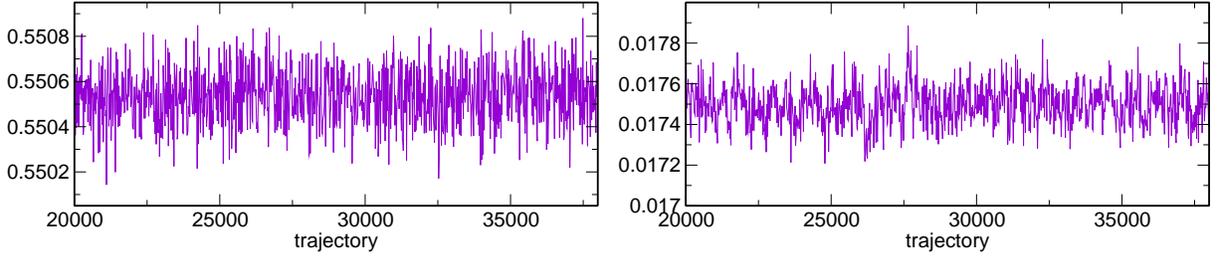
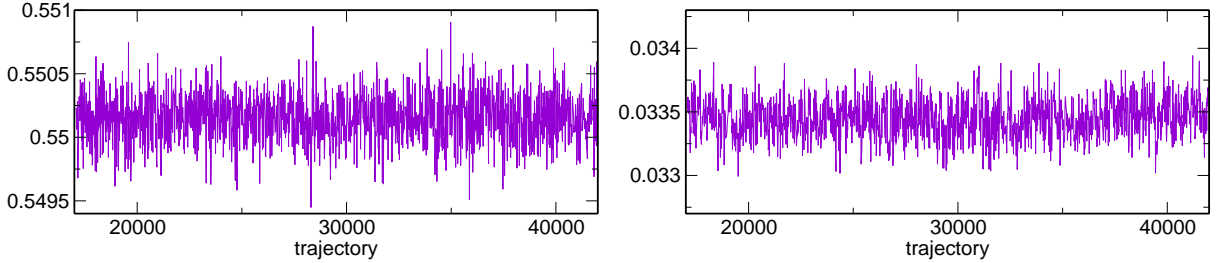

\begin{minipage}{\textwidth}
 \includegraphics[width=0.48\textwidth]{fig002a1.eps}\;
 \includegraphics[width=0.48\textwidth]{fig002a2.eps}
 \begin{center}
  (a) {$m_f=0.012$, $L=42$}
 \end{center}
\end{minipage}
 \vspace{12pt}

 \begin{minipage}{\textwidth}
  \includegraphics[width=0.48\textwidth]{fig002b1.eps}\;
  \includegraphics[width=0.48\textwidth]{fig002b2.eps}
  \begin{center}
   (b) {$m_f=0.03$, $L=30$}
  \end{center}
 \end{minipage}
 \vspace{12pt}

 \begin{minipage}{\textwidth}
  \includegraphics[width=0.48\textwidth]{fig002c1.eps}\;
  \includegraphics[width=0.48\textwidth]{fig002c2.eps}
  \begin{center}
   (c) {$m_f=0.06$, $L=24$}
  \end{center}
 \end{minipage}
 \vspace{8pt}

 \caption{Histories of the plaquette (left) and chiral condensate
 $\Sigma$ (right) for the three indicated ensembles.} \label{fig:bulk}
\end{figure}

For representative ensembles of the main ensemble set, we show how
typical bulk observables change with the Monte Carlo time.
The plaquette and chiral condensate are
shown for three ensembles: (a) $m_f=0.012$, $L=42$, (b) $m_f=0.03$,
$L=30$ and (c) $m_f=0.06$, $L=24$ in Fig.~\ref{fig:bulk}.

\begin{figure}[!tbp]
	\begin{minipage}{\textwidth}
		\includegraphics[width=0.5\textwidth]{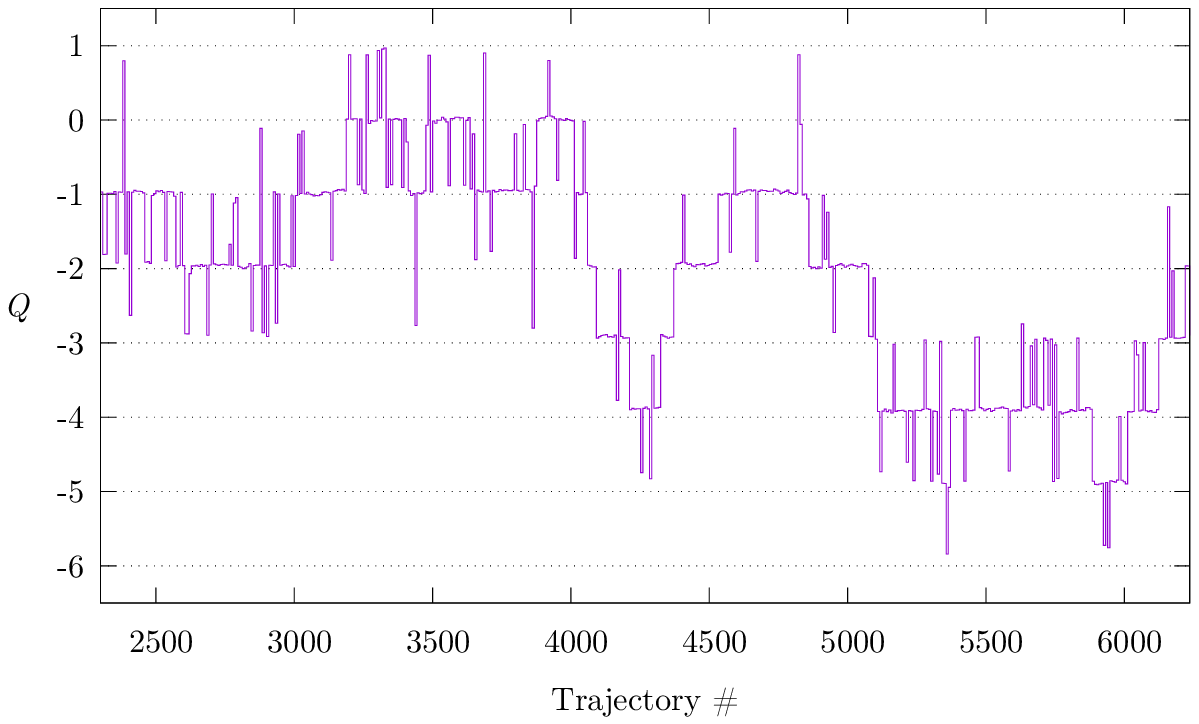}\includegraphics[width=0.5\textwidth]{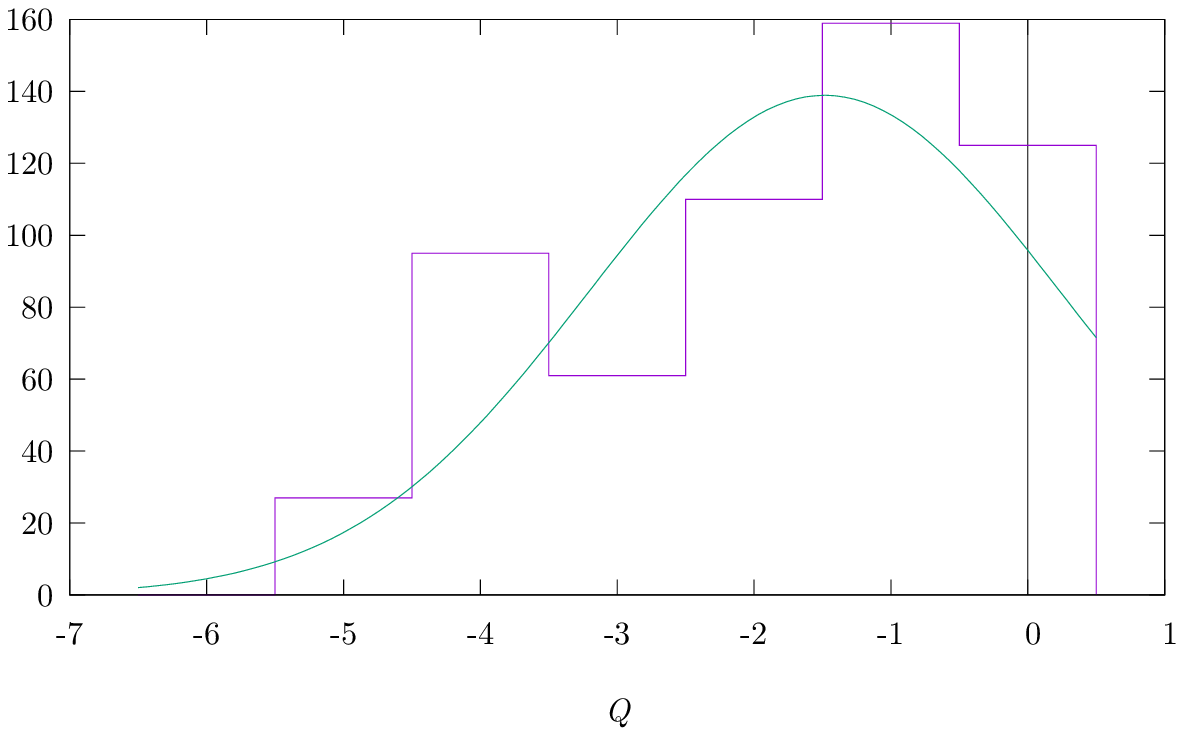}
                \begin{center}
		(a) {$m_f=0.012$, $L=42$}
                \end{center}
	\end{minipage}
	\vspace{12pt}

	\begin{minipage}{\textwidth}
		\includegraphics[width=0.5\textwidth]{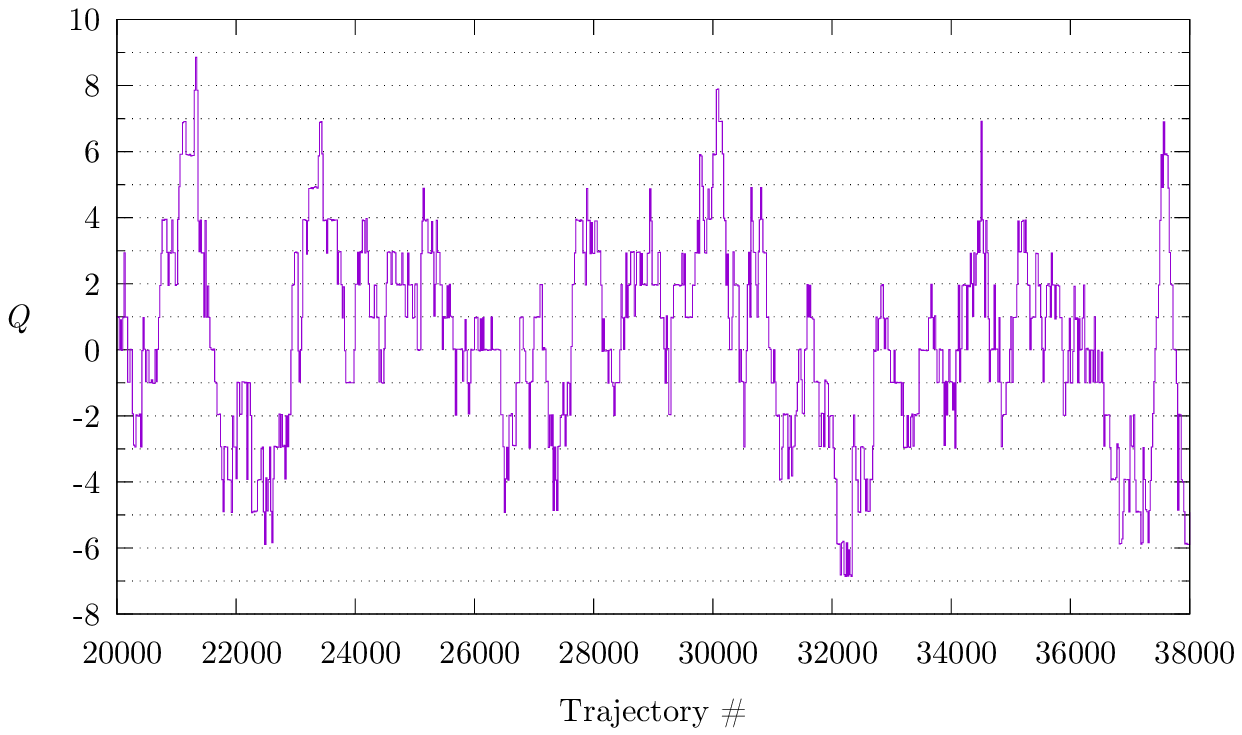}\includegraphics[width=0.5\textwidth]{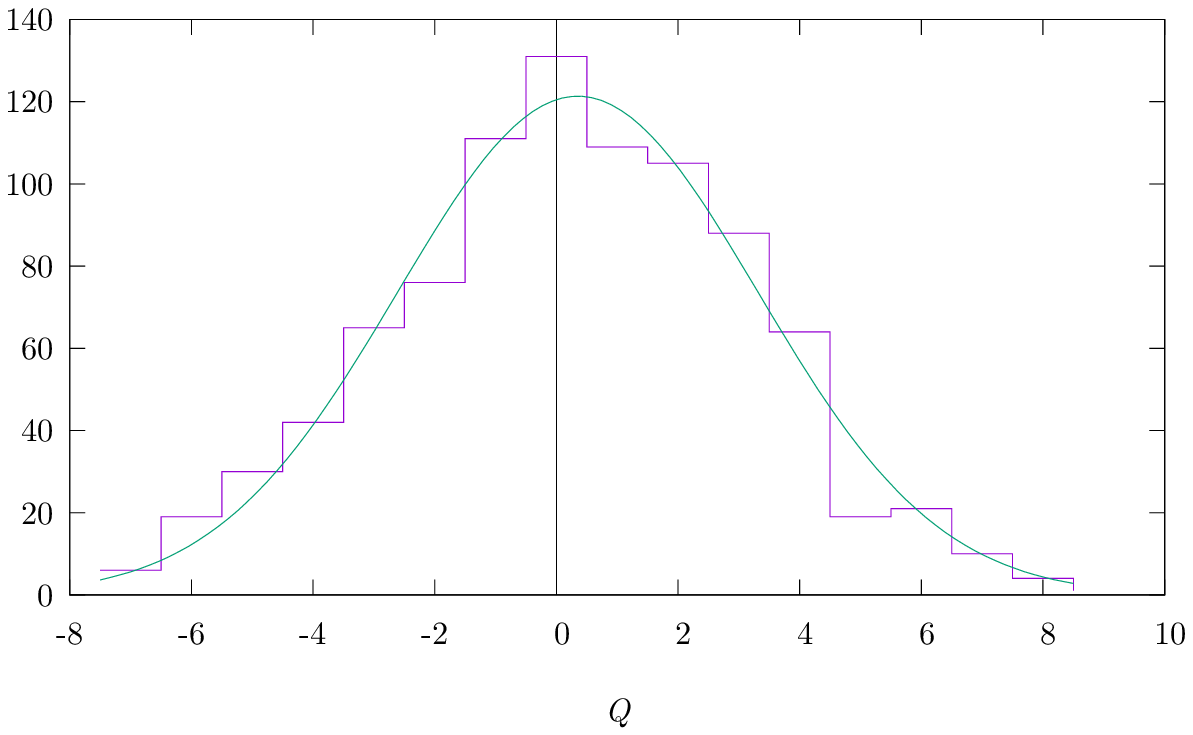}
                \begin{center}
		(b) {$m_f=0.03$, $L=30$}
                \end{center}
	\end{minipage}
	\vspace{12pt}

	\begin{minipage}{\textwidth}
		\includegraphics[width=0.5\textwidth]{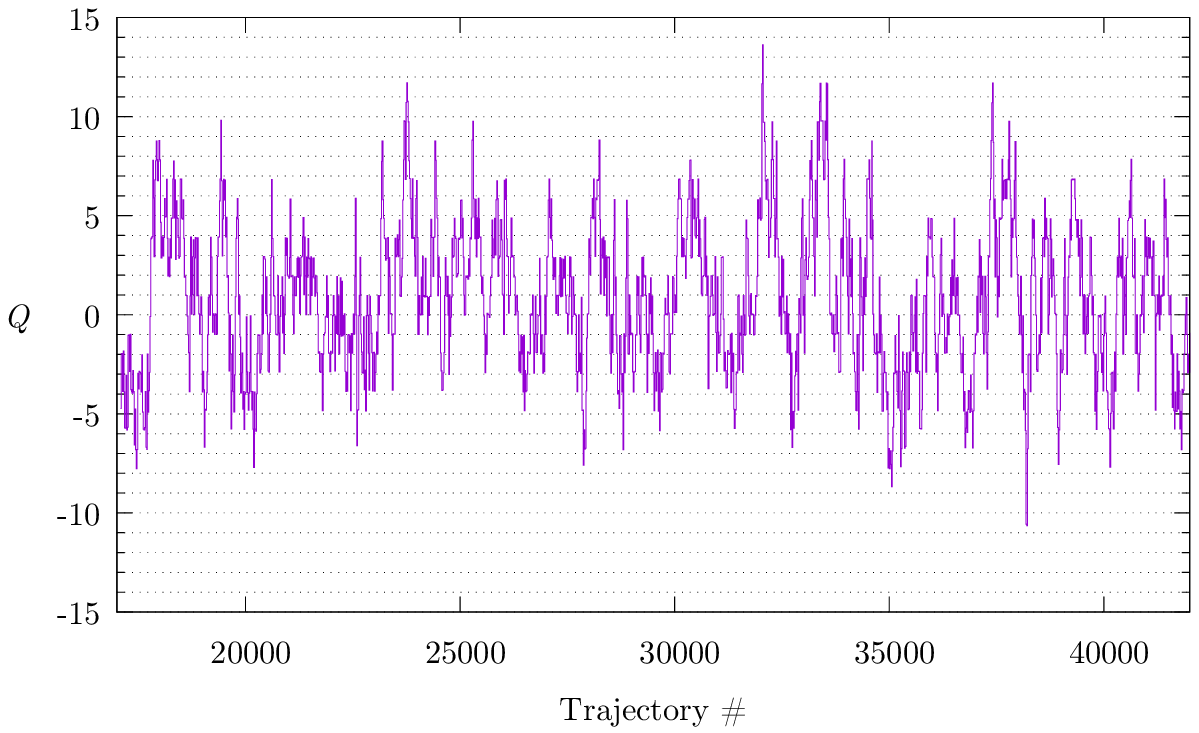}\includegraphics[width=0.5\textwidth]{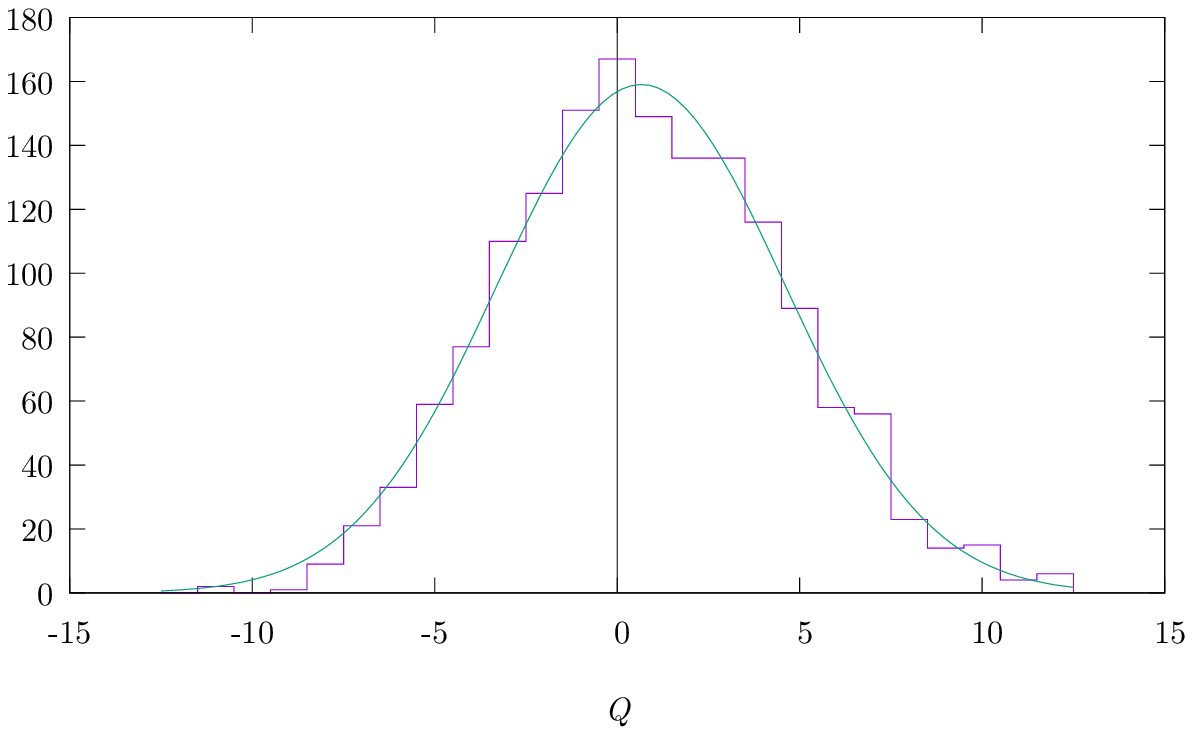}
                \begin{center}
		(c) {$m_f=0.06$, $L=24$}
                \end{center}
	\end{minipage}
	\vspace{8pt}

	\caption{Histories and histograms (with a Gaussian fit) of the topological charge $Q$ for the three indicated ensembles. The first shows frozen behavior, while the latter two show good coverage of topological sectors.}
	\label{fig:tophist}
\end{figure}

The topological charge typically develops the longest autocorrelation time among the
quantities of interest. On the same ensembles used in Fig.~\ref{fig:bulk}
the topological change ($Q$) history and its histogram are plotted in
Fig.~\ref{fig:tophist}. At the lightest masses ($m_f = 0.012$) freezing
behavior begins to manifest, although it still is moving through the
trajectories. Except this lightest mass the topology is moving well and
good sampling of topological sectors is observed.

\subsection{Extraction of mass and amplitude of composite states}

The analysis of the two-point functions of the gauge-invariant composite
operators used to calculate the spectrum in flavor-non-singlet channels
is described here, with a special emphasis on the staggered-fermion
specific definitions and treatments. The case for the
flavor-singlet scalar is discussed in Sec.~\ref{sec:flavorsinglet}.

The generic staggered bilinear operator, which is composed of the
staggered and anti-staggered fields in a unit hypercube, reads
\begin{equation}
 \overline{\chi}_i(y+A)(\Gamma\otimes\Xi)_{AB}\chi_j(y+B),
\end{equation}
where $y$ identifies the origin of the unit hypercube, and $A$ and $B$ are
displacement vectors from the origin to any point in the hypercube.
$\Gamma$ and $\Xi$ are the spin and flavor (taste) matrices.
The details of how these expressions work can be found, for example,
in Ref.~\cite{Gupta:1997nd}.
Let us here note that $i$ and $j$ are species indices, which can take
$i$, $j$ = $1$, $\cdots$, $N_s$, where $N_s=1$, $2$ or $3$ for $N_f=4$, $8$,
$12$ respectively. We note that there is a remarkable difference
between $N_f=4$ and $N_f=8$, $12$. The bilinear operator in $N_f=4$ can
 be made from only one staggered species.

For the flavor non-singlet meson channel we always use $i\ne j$
operators for $N_f=8$ and $12$, which prevent the contribution of
disconnected diagrams for two-point functions.
This ``trick'' cannot be used for the $N_f=4$ case. However,
as long as the taste non-singlet operators are concerned, we will not
include the disconnected contributions. Since the
disconnected pieces will not contribute in the continuum limit,
omitting them will introduce a lattice artifact which will vanish in
the continuum limit{,
thus is $O(a^2)$ at most, and is further reduced by the HISQ improvement.\footnote{In a later section we study the taste symmetry violation effect in the pion spectrum for $N_f=4$, which is an example of a similar $O(a^2)$ effect. Let us note that there is a systematic study~\cite{Bazavov:2011nk} of the taste violation using exactly the same action, but for real-world QCD ($N_f=2+1$), which is expected to have similar properties as $N_f=4$. There a similar lattice spacing as this study is shown to be well in the scaling region and the violation is far smaller than for other actions commonly used.}
}

For the pions we mainly use the exact Nambu-Goldstone (NG) channel,
\begin{equation}
 (\Gamma\otimes\Xi) = (\gamma_5\otimes\xi_5),
\end{equation}
which is associated with the exact ${\rm SU}(N_s)_{V-A}\times {\rm SU}(N_s)_{V+A}$ 
staggered chiral symmetry.
The operator reads
\begin{equation}
 \overline{\chi}_i(x)\chi_j(x)(-)^x,
  \label{eq:staggered_local_pi}
\end{equation}
and thus is local in $x$,
where $x$ runs through all the sites including all the corners in
hypercubes.
The pion mass is measured from the local-local two-point function,
with zero-momentum projection. The pion decay constant is measured using
the PCAC relation, which holds due to the exact symmetry and
correspondence of the continuum and lattice matrix
elements~\cite{Kilcup:1986dg},
\begin{equation}
 \langle 0 | \overline{d}\gamma_5 u(x) | \pi^{+}\rangle_{cont}
  \leftrightarrow
  \frac{1}{\sqrt{N_{fs}}}\langle 0 | \overline{\chi}_i \chi_j(x)(-)^x |
  \pi\rangle_{latt},
\end{equation}
where $N_{fs}=4$ which is the number of flavors per staggered species.
Our pion decay constant $F_\pi$ is calculated with the matrix element in the
right hand side, as
\begin{equation}
  \frac{1}{\sqrt{N_{fs}}}\langle 0 | \overline{\chi}_i \chi_j(n)(-)^n |
  \pi\rangle = \frac{M_\pi^2}{2m_f} F_\pi,
\end{equation}
with $M_\pi$ being the mass of the pion in NG channel.
This corresponds to the continuum definition,
\begin{equation}
\langle 0 | \overline{d}\gamma_5u(0) | \pi^+\rangle =
 \frac{M_\pi^2}{m_d+m_u} F_\pi,
\end{equation}
where $m_q$, with $q=u$ or $d$, is the quark mass associated with the
flavor $q$.
From this expression our pion decay constant can be understood as being normalized
with the $131$ MeV convention in usual QCD.

The staggered matrix element is calculated from the two point function amplitude at large
Euclidean time separation
\begin{equation}
 G_{PS}(t) = \sum_{\vec{x},\vec{x_0}} \langle
  \overline{\chi}_i(x)\chi_j(x)(-)^x \cdot
  \overline{\chi}_j(x_0)\chi_j(x_0)(-)^{x_0}\rangle,
\end{equation}
where $x=(\vec{x},t)$, $x_0=(\vec{x_0},0)$ written with the spatial and
temporal coordinate separately.
The contraction and zero-momentum projection at the source position $t=0$
use a stochastic estimator with single Gaussian random number.
In practice we average two-point functions with displaced source time
positions in addition to $t=0$ to effectively increase the statistics.\footnote{See for example the $N_{\rm meas}$ column in Table~\ref{tab:hadron_stat},
which shows the number of such displaced measurements.}

For large time separation, $G(t)$ will be dominated by the ground state.
With a finite temporal size of our lattice $T$, {
we often encounter the situation where the ground state dominance is
questionable. Therefore we use a method to extend the temporal lattice size
in the valence sector to be 2T. The method is combining the fermion
propagators with periodic and anti-periodic boundary conditions to make the
single fermion propagator for $0\le t<T$ with the sum and $T\le t<2T$ with
the difference of them. By this the resultant fermion propagator has a
periodicity of $2T$. As a result, the most distant source-sink separation
of the hadron two point function is made to $T$ from $T/2$ in the original,
which helps to access the $t$ range where the ground state dominates}
{(see, e.g., Ref.~\cite{Blum:2001xb})}. 
Let $G_{PS}(t)$ be the
pion two point function
after this manipulation. Its asymptotic form is then given as
\begin{equation}
 G_{PS}^{asym}(t) = C( e^{-M_\pi t}+e^{-M_\pi(2T-t)}) + B(-)^t,
\end{equation}
with $M_\pi$ being the mass of the pion in NG channel,
whose decay constant is calculated from the amplitude $C$.
The effect of the last term, which is constant but oscillating in $t$,
is substantial, especially at large number of flavors.
The existence of such a term can be understood as follows: 
the {fermion} and anti-{fermion} propagate in opposite 
direction from the source and meet together at the sink position after 
one moves through the boundary.{\footnote{
For non-staggered (such as Wilson) fermions{~\cite{Umeda:2007hy}}, such a wrap around contribution produces a constant term, because the length of the combined fermion lines are constant ($2T$) as a function of the sink position ($t$), {i.e., $e^{-m_f t} \times e^{-m_f (2T-t)} = e^{-2m_f T}$}. It is well-known that such a term exist and the effect is significant in high-temperature (real-world) QCD. Now, noting that the backward propagating fermion will have an opposite parity to the forward one, each one step move of $t$ direction of staggered fermion accompanies an oscillating sign. As a result, such a contribution will be proportional to $(-)^t$ rather than a plain constant. It is easy to see this effect in free field staggered fermions at the NG pion channel, where there exists no staggered parity partner thus no oscillating source exists otherwise. 
}}
As the number of flavors increases the {fermion} and anti-{fermion} are bound
more loosely due to color screening.

\begin{figure}[!tbp]
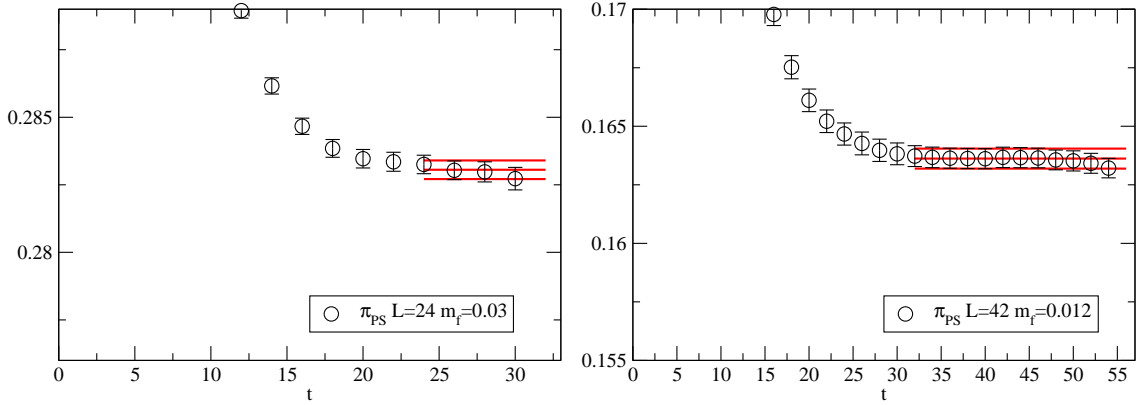

\includegraphics[width=0.45\textwidth]{fig004a.eps}\;
\includegraphics[width=0.45\textwidth]{fig004b.eps}
\caption{
 The effective mass $M^{\rm eff}(t)$ of NG pion calculated with
 $\tilde{G}_{\rm PS}(t)/\tilde{G}_{\rm PS}(t+2)$
 for $N_f=8$ $m_f=0.03$, $24^3\times 32$ (left) and $m_f=0.012$,
 $42^3\times 56$ (right), as typical examples.
 Red lines show the $t$-range of the global fit, the central value of the mass from
 the fit, and the jackknife error band.
 }
\label{fig:effm_pi}
\end{figure}

In practice we eliminate the effect of the $B$-term by taking the linear
combination of $G(t)$ with the nearest neighbor,
\begin{equation}
 \tilde{G}_{\rm H}^{(+)}(t)=\frac{1}{2} G_{\rm H}(t)
  + \frac{1}{4} G_{\rm H}(t-1) + \frac{1}{4} G_{\rm H}(t+1),
\label{eq:tildeHp}
\end{equation}
where $t$ is restricted to even number, and $H=PS$ in this case.
The asymptotic form of this correlation function at large $t$ is given by
\begin{equation}  \label{eq:PSasym}
\tilde{G}_{\rm PS}^{asym}(t)={\tilde C} \left (e^{-M_{\pi}t}  +
					  e^{-M_{\pi}(2T-t)} \right),
\end{equation}
where $2 {\tilde C} = C \left(1+ \cosh(M_\pi) \right)$.
In Fig.~\ref{fig:effm_pi} we show typical effective mass of the NG pion mass at $N_f=8$
extracted from the two neighboring points $t$ and $t+2$ using the
asymptotic form Eq.~(\ref{eq:PSasym}).
Fitting $\tilde{G}_{PS}(t)$ with Eq.~(\ref{eq:PSasym}) in the
$t$ range that shows a plateau of the effective mass
gives the mass $M_\pi$, and the decay constant from
\begin{equation}
 F_\pi^2 = \frac{4m_f^2}{M_\pi^3 [1+\cosh(M_\pi)]} \tilde{C}.
\end{equation}
The results are shown in the later sections.

Operators local in a staggered hypercube are always used for the other
flavor non-singlet hadrons.\footnote{Exceptions may apply when we study
the taste symmetry violation, for which we use all the taste
partners of the pion.}
We examine hadronic channels which couple to the following four
operators:
\[
\begin{array}{cccc}
 \mbox{name(H)} & \mbox{operator} & \mbox{state(1)} & \mbox{state(2)}\\
 PS   & (\gamma_5\otimes\xi_5) & \pi & -\\
 SC   & (\gamma_4\gamma_5\otimes\xi_4\xi_5) & \pi & a_0\\
 VT   & (\gamma_k\otimes\xi_k) & \rho & b_1\\
 PV   & (\gamma_k\gamma_4\otimes\xi_k\xi_4) & \rho & a_1\\
\end{array}
\]
Here conventional QCD state names are used to label the corresponding
states in the many-flavor system.
In this assignment state(1) always appears lighter than state(2)
{
which is the staggered parity partner of state(1).  With fixed time-slice operators the states (1) and (2) always mix~\cite{Golterman:1984cy, Golterman:1985dz} (see~\cite{Ishizuka:1993mt} for a good practical example). 
The asymptotic form of the zero spatial momentum two-point function reads
\begin{equation}
  G_H^{asym}(t) = C_1 (e^{-M_1 t} + e^{-M_1(2T-t)}) + C_2 (-)^t (e^{-M_2 t} + e^{-M_2(2T-t)}),
\end{equation}
where $M_i$ for $i=1$ and $2$ are the masses of the state($i$).}
In practice in the following sections, state(1) is extracted first
with a single exponential fit to $\tilde{G}_H$ (Eq.~(\ref{eq:tildeHp})),
which suppresses the effect of state(2),
{as well as other oscillating components, such as the $B$ term in Eq.(10)}.
The state(2) is then extracted from the negatively projected linear combination,
\begin{equation}
 \tilde{G}_{\rm H}^{(-)}(t)=\frac{1}{2} G_{\rm H}(t)
  - \frac{1}{4} G_{\rm H}(t-1) - \frac{1}{4} G_{\rm H}(t+1),
\label{eq:tildeHm}
\end{equation}
with the contribution of state(1) explicitly subtracted.
\begin{figure}[!tbp]
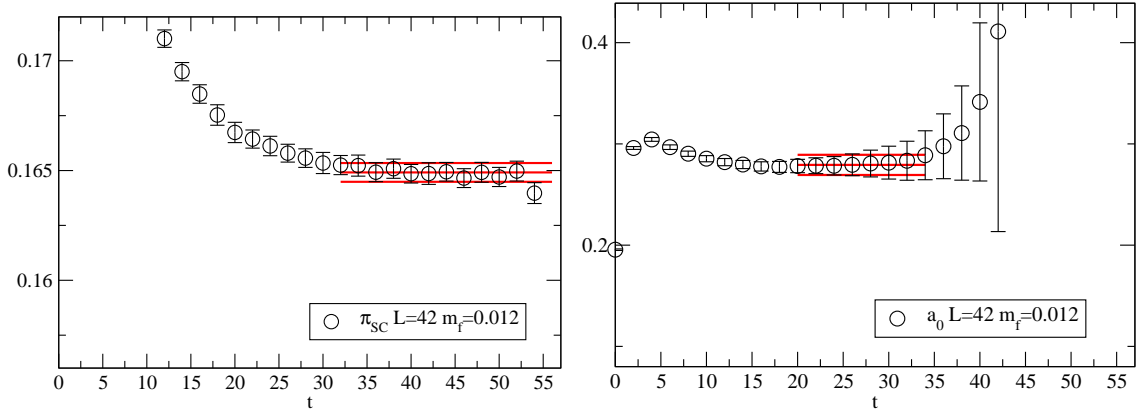

\includegraphics[width=0.45\textwidth]{fig005a.eps}\;\
\includegraphics[width=0.45\textwidth]{fig005b.eps}
\caption{
 The effective mass of $\pi_{SC}$ (left) and its staggered parity partner
 $a_0$ (right) for $N_f=8$ $m_f=0.012$, $42^3\times 56$.
}
\label{fig:effm_pisc_a0}
\end{figure}
For the non-NG and flavor non-singlet state we always use the
so-called corner source, where the fermion source vector takes
the unit value at the origin of every staggered hypercube and zero
otherwise. At the sink position, a zero momentum projection
is applied after taking the proper contraction for the staggered
bilinear operator.
We average two-point functions with displaced source time
positions in addition to $t=0$ to effectively increase the statistics
here as well.

\begin{figure}[!tbp]
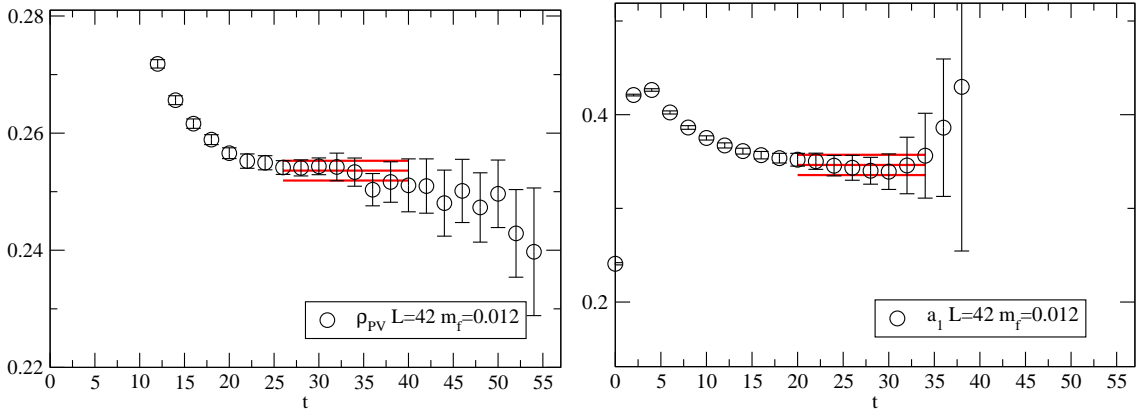

\includegraphics[width=0.45\textwidth]{fig006a.eps}\;\
\includegraphics[width=0.45\textwidth]{fig006b.eps}
\caption{
 The effective mass of $\rho_{PV}$ (left) and its staggered parity partner
 $a_1$ (right) for $N_f=8$ $m_f=0.012$, $42^3\times 56$.
}
\label{fig:effm_rhopv_a1}
\end{figure}

Fig.~\ref{fig:effm_pisc_a0} shows examples of the effective mass of
$\pi_{SC}$ and $a_0$ extracted this way. Similar examples for
$\rho_{PV}$ and $a_1$ are shown in Fig.~\ref{fig:effm_rhopv_a1}.

\begin{figure}[!tbp]
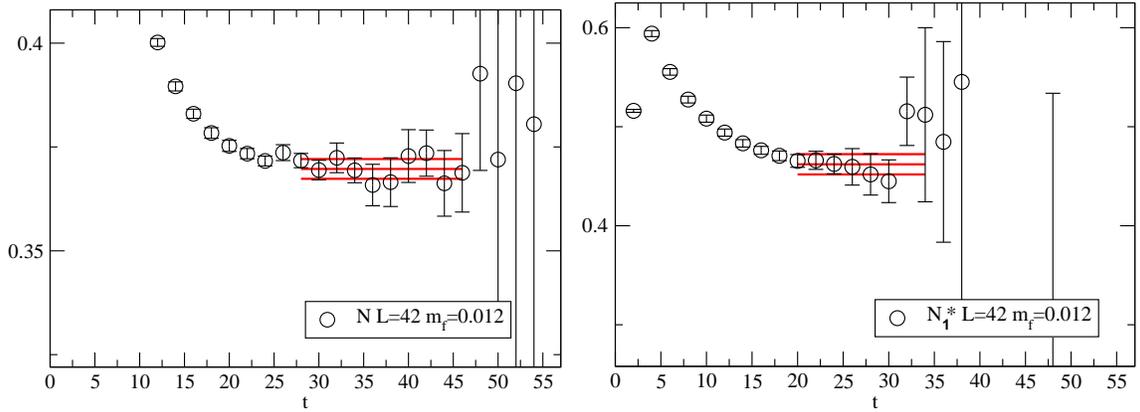

\includegraphics[width=0.45\textwidth]{fig007a.eps}\;\
\includegraphics[width=0.45\textwidth]{fig007b.eps}
\caption{
 The effective mass of $N$ (left) and its staggered parity partner
 $N_{\bf 1}^*$ (right) for $N_f=8$ $m_f=0.012$, $42^3\times 56$.
}
\label{fig:effm_N_N1s}
\end{figure}

For the nucleons we use the local operator
with three {fermion} degrees of freedom on the same point in the staggered
hypercube.
This operator interpolates the spin $1/2$ state in the {\bf 20}$_M$,
mixed symmetry irrep of SU(4) flavor symmetry, for positive parity and
that in the {\bf 4}$_A$, anti-symmetric irrep, for negative parity.
We refer to the former as $N$ and to the latter as $N_{\mathbf 1}^*$.\footnote{
The index ${\mathbf 1}$ indicates
the fact the lowest state among these is SU(3) flavor singlet in
usual QCD.
In Refs.~\cite{Ishizuka:1993mt,Golterman:1984dn} our $N_{\mathbf 1}^*$
is named as $\Lambda(1405)$. Here we adopted a different notation
to avoid possible confusion.
}
The nucleon mass is extracted in the same way as the case of
non-NG mesons, with a sign oscillation for the backward
propagating anti-nucleon signal through the anti-periodic temporal boundary.
The typical effective mass is shown in
Fig.~\ref{fig:effm_N_N1s}.

\clearpage

\section{Analysis of hadron mass spectrum}
\label{sec:hadspec}

Using the lattice gauge ensembles described in the previous section, we investigate the spectrum of typical hadrons in $N_f = 8$ QCD.
We first look at the pion decay constant ($F_{\pi}$), pion mass ($M_{\pi}$), rho meson mass ($M_{\rho}$) and the nucleon mass ($M_{N}$).
We then study finite-volume effects, taste symmetry breaking effects and mass ratios, comparing them with those in $N_f = 4$ and $N_f = 12$.

\subsection{Study of finite-volume effects}\label{subsec:mh_fv}

\begin{figure}[!tbp]
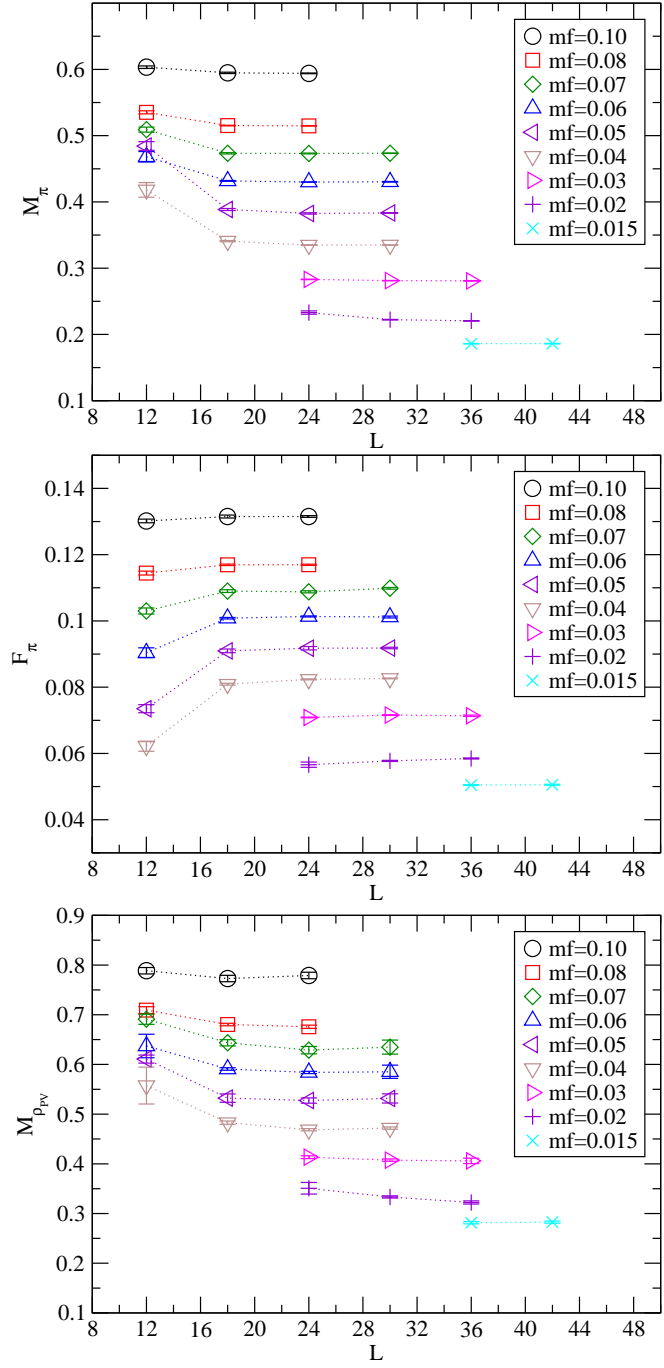

\makebox[.5\textwidth][r]{\includegraphics*[scale=.5]{fig008a.eps}}
\makebox[.5\textwidth][r]{\includegraphics*[scale=.5]{fig008b.eps}}
\makebox[.5\textwidth][r]{\includegraphics*[scale=.5]{fig008c.eps}}
\caption{
The lattice volume dependence of
$M_\pi$ (top), $F_\pi$ (middle) and $M_\rho$ (bottom) for various fermion masses $m_f$.
}
\label{fig:mpi-fpi-L}
\end{figure}

\begin{figure}[!tbp]
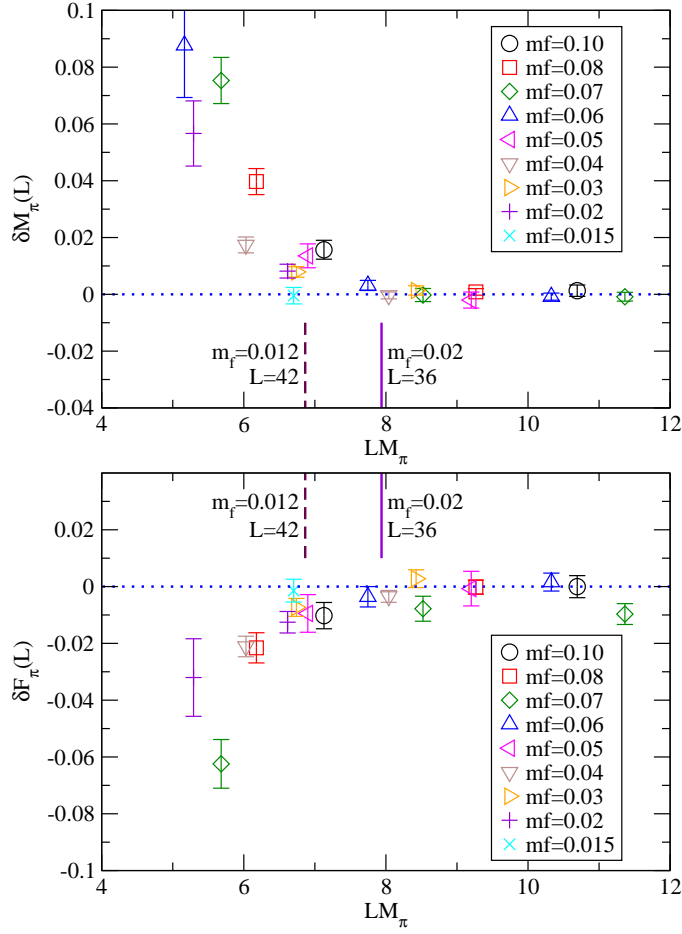

\makebox[.5\textwidth][r]{\includegraphics*[scale=.5]{fig009a.eps}}
\makebox[.5\textwidth][r]{\includegraphics*[scale=.5]{fig009b.eps}}
\caption{
The finite volume effects
$\delta M_\pi(L)$ (top) and $\delta F_\pi(L)$ (bottom)
defined in Eq.~(\ref{eq:fv_rel})
as a function of $LM_{\pi}$ for various fermion masses $m_f$.
}
\label{fig:mpi-fpi-deltaL}
\end{figure}

We evaluate finite-volume effects in our lattice gauge ensembles for $N_f = 8$.
To this end, we plot $M_\pi$, $F_\pi$, and $M_\rho$ as a function of the lattice volume $L$ for each fermion mass $m_f$ in Fig.~\ref{fig:mpi-fpi-L}.
Here $M_{\rho}$ represents the staggered PV vector mass and we will adopt this terminology in the following unless explicitly stated otherwise.
As shown in the figure, the spectrum on the largest two volumes is reasonably consistent for all $m_f$ except $m_f = 0.02$ for which some deviation between the two volumes is seen.
We quantify the finite volume effects by using
\begin{equation}
\delta M_\pi(L) = \frac{M_\pi(L)-M_\pi(L_{\rm max})}{M_\pi(L_{\rm max})}
\ \ {\rm and}\ \
\delta F_\pi(L) = \frac{F_\pi(L)-F_\pi(L_{\rm max})}{F_\pi(L_{\rm max})}
\ ,\label{eq:fv_rel}
\end{equation}
with $L_{\rm max}$ being the largest lattice volume at each $m_f$.
Figure~\ref{fig:mpi-fpi-deltaL} shows these quantities as a function of $L M_\pi(L)$.
For $m_f = 0.02$, we find $LM_{\pi}|_{L = L_{\rm max} = 36}\simeq 8$ (the solid vertical lines); for the somewhat larger masses $m_f = 0.03$ and $0.04$, both $\delta M_\pi(L)$ and $\delta F_\pi(L)$ become consistent with zero.
The finite volume effect for $m_f = 0.02$ around $LM_{\pi}\simeq 8$ would be further suppressed, since for a fixed $LM_{\pi}$, $\delta M_\pi(L)$ and $\delta F_\pi(L)$ tend to decrease with smaller $m_f$ as shown for other values of $LM_{\pi}$---for example, around $L M_\pi(L)\simeq 7$.
Such $m_f$ dependences of the finite volume effects may be a consequence of broken chiral symmetry for $N_f = 8$ with regards to the NLO-ChPT prediction~\cite{Gasser:1987zq}.

\begin{figure}[!tbp]
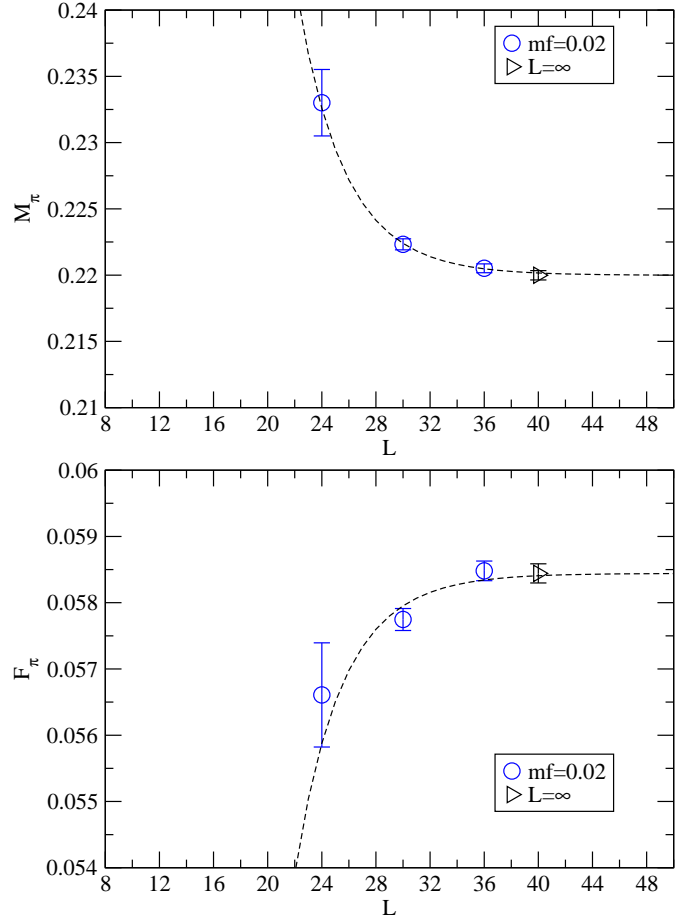

\makebox[.5\textwidth][r]{\includegraphics*[scale=.5]{fig010a.eps}}
\makebox[.5\textwidth][r]{\includegraphics*[scale=.5]{fig010b.eps}}

\caption{
The lattice volume dependence of
$M_\pi$ (top) and $F_\pi$ (bottom) at $m_f = 0.02$.
The black dashed line represents the ChPT motivated fit given by Eqs.~(\ref{eq:Mpi_L_chpt}) and (\ref{eq:Fpi_L_chpt}).
}
\label{fig:fit-mpi-fpi-L-mf0.02}
\end{figure}

Additionally, we fit the data for $M_\pi(L)$ and $F_\pi(L)$ at $m_f = 0.02$ using the following functions, which are inspired by ChPT~\cite{Gasser:1987zq, Luscher:1985dn}, as in Ref.~\cite{Aoki:2012eq},
\begin{eqnarray}
M_\pi(L) &=& M_\pi + c_{M_\pi} \frac{e^{-LM_\pi}}{(L M_\pi)^{3/2}}\ ,\label{eq:Mpi_L_chpt}\\
F_\pi(L) &=& F_\pi + c_{F_\pi} \frac{e^{-LM_\pi}}{(L M_\pi)^{3/2}}\ ,\label{eq:Fpi_L_chpt}
\end{eqnarray}
where $c_{M_\pi}$ and $M_\pi$ are the fit parameters of $M_\pi(L)$, and $c_{F_\pi}$ and $F_\pi$ are the fit parameters of $F_\pi(L)$.\footnote{$M_\pi$ in the $F_\pi(L)$ fit is fixed to the value estimated from the $M_\pi(L)$ fit.}
The fit results are plotted in Fig.~\ref{fig:fit-mpi-fpi-L-mf0.02} as a function of $L$.
The figure shows that the largest volume data agrees with the estimated result in the infinite-volume limit within the statistical error.
Therefore, we conclude that the finite-volume effects in the data at $L=36$ with $m_f = 0.02$ are negligible, as is the case for the largest volume data at all other values of $m_f$.

Data for the spectrum at the lightest fermion mass, $m_f = 0.012$, is available for only one volume, $L=42$, and we estimate its finite-volume effects by utilizing data at the second lightest mass, $m_f=0.015$, shown in Fig.~\ref{fig:mpi-fpi-deltaL}.
The value of $L M_\pi$ at $m_f = 0.012$ is highlighted with a dashed vertical line in the figure. Its value is similar to $L M_\pi$ of $m_f = 0.015$, where the relative differences $\delta M_\pi(L)$ and $\delta F_\pi(L)$ are consistent with zero.
Therefore, in the following sections, we assume that finite-volume effects at $m_f = 0.012$ are smaller than the statistical error.
The spectra $F_{\pi}$, $M_{\pi}$, $M_{\rho}$ (as well as $\langle\overline{\psi}\psi\rangle$, which will be investigated in the next section) are summarized in the tables in Appendix~\ref{sec:table_fpi_mpi}.

In the following sections,
we select the spectral data on the largest volumes at each $m_f$;
the finite volume effects are negligible for them as explained above.
Exceptions are made for $m_f = 0.015$, $0.03$, and $0.06$,
where we use the data on the second largest volume,
as significantly larger statistics can be utilized.
As shown in Fig.~\ref{fig:mpi-fpi-deltaL},
the finite volume effects~(\ref{eq:fv_rel}) for these data
are consistent with zero;
the corresponding data points on the figure
are the light blue cross ($m_f = 0.015$, $L = 36$),
orange right triangle at the rightmost ($m_f = 0.03$, $L = 30$),
and blue triangle at the rightmost ($m_f = 0.06$, $L = 24$).
From now on in this paper,
we shall refer to this data set as the ``Large Volume Data Set'',
which is summarized in Table~\ref{tab:data_set}.

\begin{table}[!tbp]
\begin{tabular}{ccccccccccc}\hline\hline
$L$ & 42 & 36 & 36 & 30 & 30 & 30 & 24 & 24 & 24 & 24 \\\hline
$m_f$ & 0.012 & 0.015 & 0.02 & 0.03 & 0.04 & 0.05 & 0.06 & 0.07 & 0.08 & 0.10 \\\hline\hline
\end{tabular}
\caption{
The ``Large Volume Data Set'' described in the text
and used in the following analyses.
}
\label{tab:data_set}
\end{table}

\clearpage

\subsection{Taste symmetry breaking effects}
\label{sec:taste_breaking}

We investigate the taste symmetry breaking effects
using our $N_f = 8$ QCD lattice ensemble with the HISQ action.
For the lattice coupling $\beta = 3.8$ used in this paper,
the taste symmetry breaking in $M_\pi$ (PS and SC channels) and $M_\rho$ (PV and VT channels)
was shown to be tiny in our previous work~\cite{Aoki:2013xza}.

\begin{figure}[!tbp]
\makebox[.5\textwidth][r]{\includegraphics*[scale=.5]{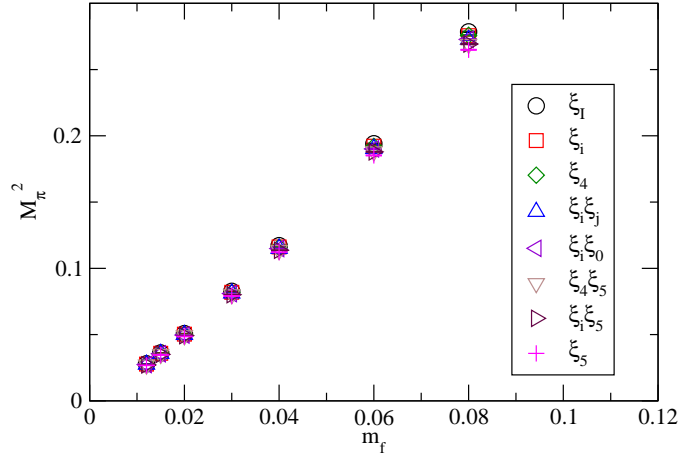}}
\caption{
The spectra for the NG pion and its taste partners, $M_{\pi_{\xi}}$ for $N_f = 8$.
}
\label{fig:mpi-taste-split}
\end{figure}

In the present paper, we extend the analysis to include all pion taste partners, $M_{\pi_\xi}$.
The results are tabulated in Table~\ref{tab:dif_taste_mpi} and shown in Fig.~\ref{fig:mpi-taste-split};
the taste partners are a taste-singlet ($\xi_I$), -vector ($\xi_i,\xi_4$), -tensor ($\xi_i\xi_j,\xi_i\xi_4$),
-axialvector ($\xi_i\xi_5,\xi_4\xi_5$), and a taste-pseudoscalar ($\xi_5$), where the last one
corresponds to the Nambu-Goldstone (NG) pion, $M_{\pi_{\xi}}|_{\xi = \xi_5} = M_{\pi}$.
At each fermion mass $m_f$, the spectra of $M_{\pi_{\xi}}$
are almost on top of each other,
and thus the taste symmetry breaking is confirmed to be small,
consistently with our previous findings~\cite{Aoki:2013xza}.

\begin{figure}[!tbp]
\makebox[.5\textwidth][r]{\includegraphics*[scale=.5]{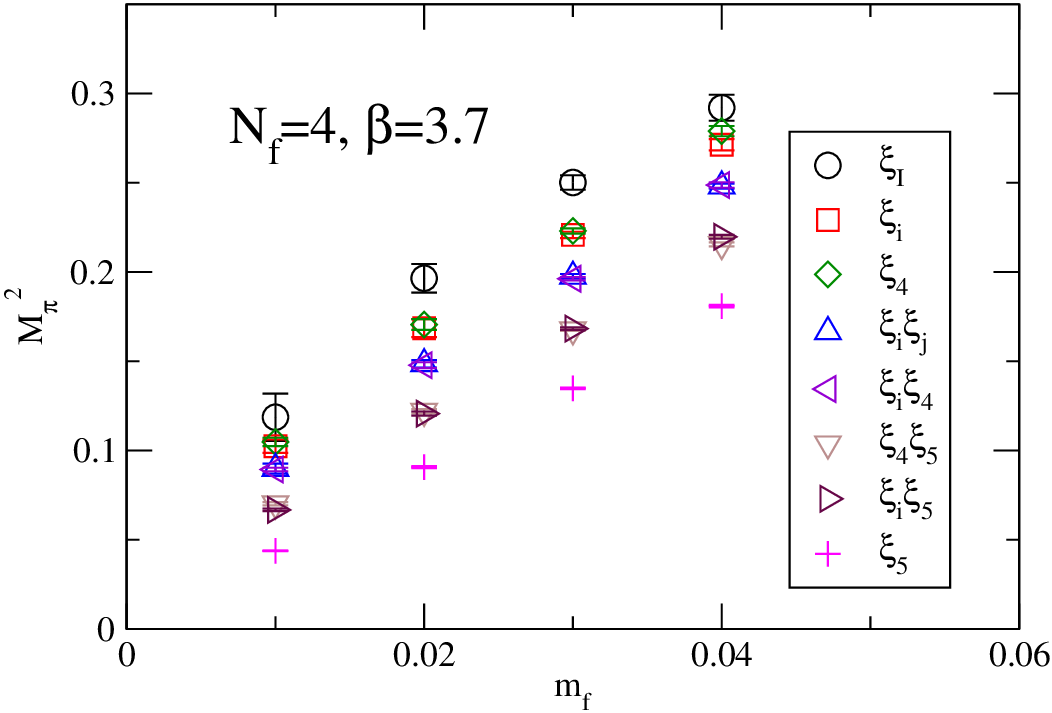}}
\caption{
The spectra for the NG pion and its taste partners, $M_{\pi_{\xi}}$ for $N_f = 4$.
}
\label{fig:nf4_taste_breaking}
\end{figure}

The taste symmetry violation in $N_f = 8$ QCD (Fig.~\ref{fig:mpi-taste-split})
looks quite different from that observed in usual QCD,
where much larger taste splitting is typically seen almost independently of $m_f$~\cite{Aubin:2004fs}.
In contrast to $N_f = 8$,
the taste symmetry breaking in $N_f = 4$ QCD is found to be closer to usual QCD,
as shown in Fig.~\ref{fig:nf4_taste_breaking}{.}
Thus, the tiny breaking of the taste symmetry found in $N_f = 8$ seems to be characteristic of the large number of flavors.
In fact, the taste symmetry breaking in $M_\pi$ (PS and SC channels) and $M_\rho$ (PV and VT channels)
is also tiny in $N_f = 12$~\cite{Aoki:2012eq}.

\begin{figure}[!tbp]
\makebox[.5\textwidth][r]{\includegraphics*[scale=.5]{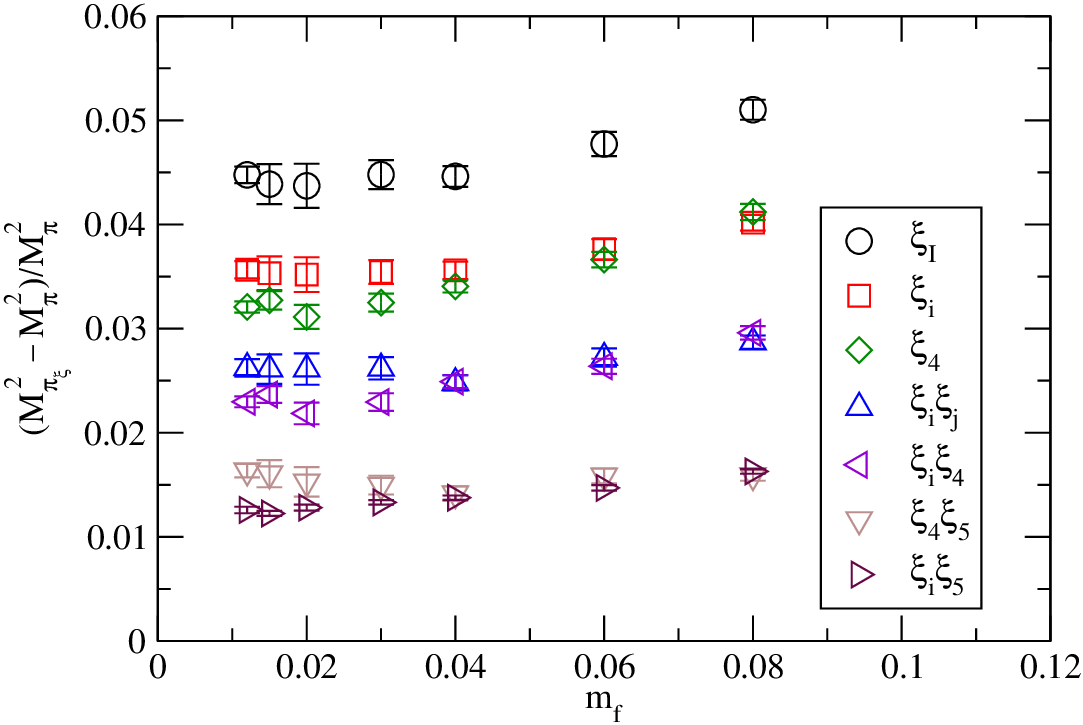}}
\caption{
The taste symmetry breaking $(M_{\pi_\xi}^2-M_{\pi}^2)/M_{\pi}^2$ for $N_f = 8$,
where $M_\pi$ and $M_{\pi_\xi}$ represent the mass of the NG pion
and its taste partner with an index $\xi$ , respectively.
}
\label{fig:mpi-taste-split_rel}
\end{figure}

The behavior of $N_f = 8$ taste symmetry breaking becomes more transparent
when differences from the NG pion, $M_{\pi_\xi}^2 - M_{\pi}^2$, are considered.
As shown in Fig.~\ref{fig:mpi-taste-split_rel},
the difference $M_{\pi_\xi}^2 - M_{\pi}^2$ is less than 6\% in units of $M_{\pi}^2$.
The ratio $(M_{\pi_\xi}^2 - M_{\pi}^2)/M_{\pi}^2$
slightly increases at larger $m_f$, while approaches to a constant at smaller $m_f$.
This implies that the taste symmetry breaking associated with $M_{\pi}$
tends to vanish toward the chiral limit.
A similar behavior was previously reported by Lattice Higgs Collaboration~\cite{Fodor:2009wk}.

\begin{figure}[!tbp]
\makebox[.5\textwidth][r]{\includegraphics*[scale=.5]{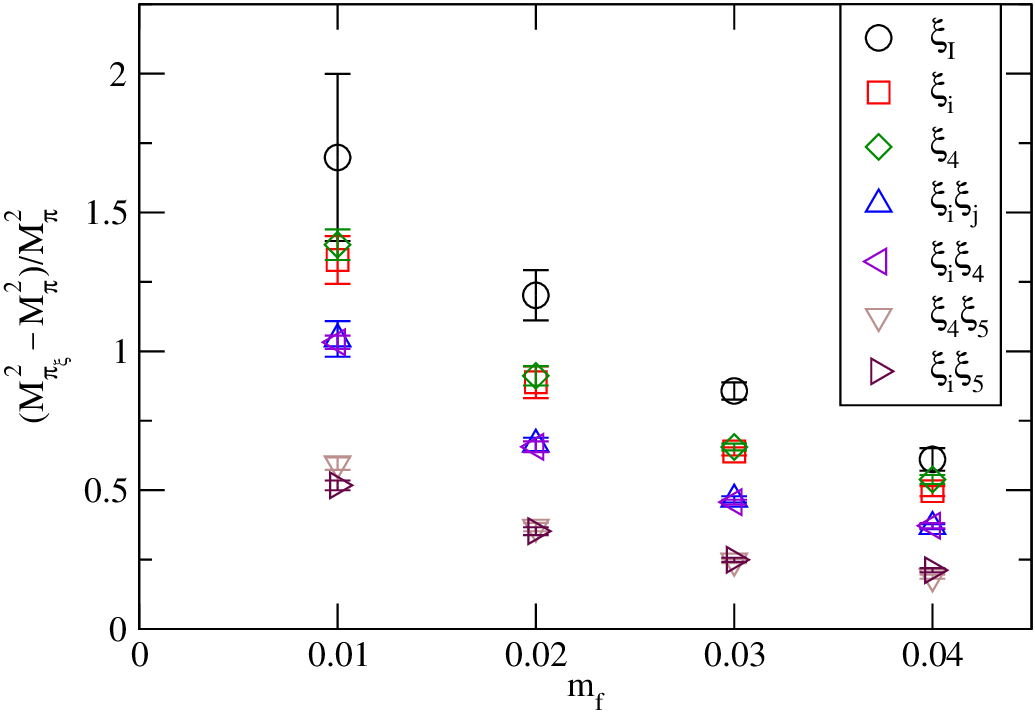}}
\caption{
The taste symmetry breaking $(M_{\pi_\xi}^2-M_{\pi}^2)/M_{\pi}^2$ for $N_f = 4$,
where $M_\pi$ and $M_{\pi_\xi}$ represent the mass of the NG pion
and its taste partner with an index $\xi$ , respectively.
}
\label{fig:nf4_taste-split_rel}
\end{figure}

The above features are different from standard knowledge of usual QCD;
the taste splitting increases with the lattice spacing,
$M_{\pi_\xi}^2 - M_{\pi}^2 = a^2\Delta_{\xi}$, where $\Delta_{\xi}$ is known to be almost independent of $m_f$
in usual QCD~\cite{Aubin:2004fs}. Therefore the ratio $(M_{\pi_\xi}^2 - M_{\pi}^2)/M_{\pi}^2$
with a ``fixed'' lattice spacing is expected to diverge as $m_f$ becomes smaller.
Such a divergent trend is clearly seen in the $N_f = 4$ case, as shown in Fig.~\ref{fig:nf4_taste-split_rel},
in contrast to $N_f = 8$.
In other words, the lack of divergence in $N_f = 8$ might be a consequence of
near-vanishing chiral dynamics. This subject will be further elaborated upon in Secs.~\ref{sec:chpt} and \ref{sec:FSHS}.

{
However, as we have only one lattice spacing for $N_f=8$, from these observations alone we cannot conclude if this apparent difference is due to a difference in the infrared dynamics. We will investigate various hadronic channels in more depth in the following sections. Although we will test only one or two taste partners in each channel, the results in the pion sector here lead to an expectation that the effects of taste symmetry breaking will be small for the mass range we simulate for the $N_f=8$ theory.
}

\begin{table}[!tbp]
\begin{tabular}{cccccccccc}\hline\hline
$m_f$ & $L$ & $\xi_5$ & $\xi_4\xi_5$ & $\xi_i\xi_5$ &
$\xi_i\xi_4$ & $\xi_i\xi_j$ & $\xi_4$ & $\xi_i$ & $\xi_I$
\\\hline
0.012 & 42 & 0.1636(4) & 0.1649(4) & 0.1646(4) & 0.1654(4) & 0.1657(4) &
0.1662(4) & 0.1665(4) & 0.1672(4)
\\
0.015 & 36 & 0.1862(3) & 0.1877(3) & 0.1873(3) & 0.1884(3) & 0.1886(4) &
0.1892(3) & 0.1895(4) & 0.1902(4)

\\
0.02  & 36 & 0.2205(4) & 0.2221(4) & 0.2219(4) & 0.2229(4) & 0.2233(3) &
0.2239(4) & 0.2243(4) & 0.2252(4)
\\
0.03  & 30 & 0.2812(2) & 0.2833(3) & 0.2831(2) & 0.2844(3) & 0.2849(3) &
0.2858(3) & 0.2862(3) & 0.2875(3)
\\
0.04  & 30 & 0.3349(2) & 0.3372(3) & 0.3372(2) & 0.3390(3) & 0.3390(3) &
0.3405(3) & 0.3408(3) & 0.3423(3)
\\
0.06  & 24 & 0.4303(3) & 0.4337(4) & 0.4335(3) & 0.4360(4) & 0.4362(4) &
0.4382(4) & 0.4384(4) & 0.4405(4)
\\
0.08  & 24 & 0.5147(3) & 0.5188(3) & 0.5189(3) & 0.5223(4) & 0.5221(4) &
0.5252(4) & 0.5250(4) & 0.5277(4)
\\\hline\hline
\end{tabular}
\caption{
The mass of the NG pion and the taste partners.
}
\label{tab:dif_taste_mpi}
\end{table}

\clearpage

\subsection{Hadron Mass Ratios}

{
The purpouse of this subsection is to give an overview of our 
hadron spectrum data using ratios of the hadron spectra before
carring out fit analyses, which will be discussed in the following sections.
}

\begin{figure}[!tbp]
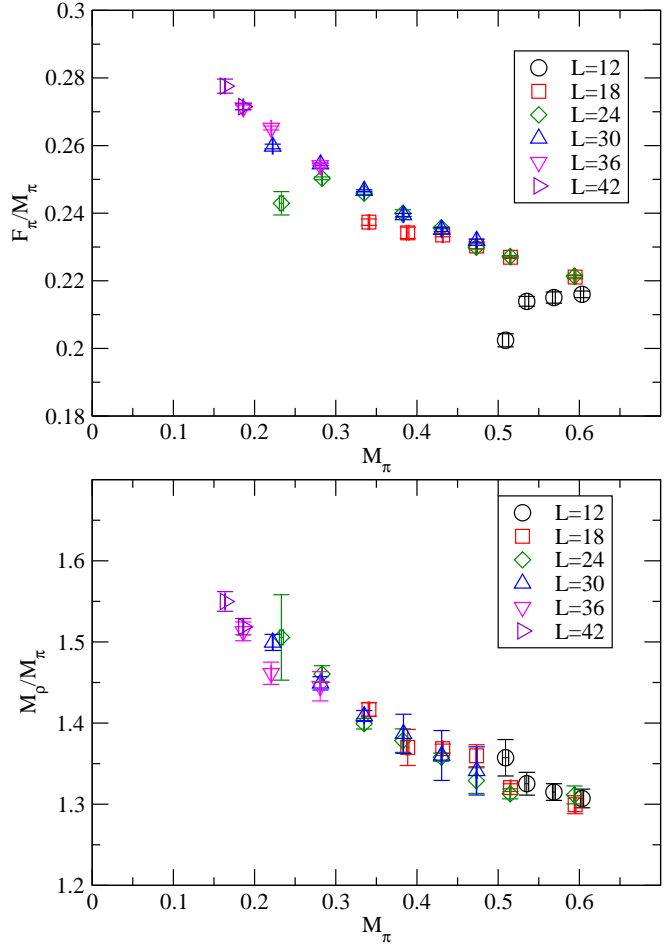

\makebox[.5\textwidth][r]{\includegraphics*[scale=.5]{fig015a.eps}}
\makebox[.5\textwidth][r]{\includegraphics*[scale=.5]{fig015b.eps}}
\caption{
$F_\pi/M_\pi$ (top) and $M_\rho/M_\pi$ (bottom).
}
\label{fig:rat_fpi_mpv}
\end{figure}

\begin{figure}[!tbp]
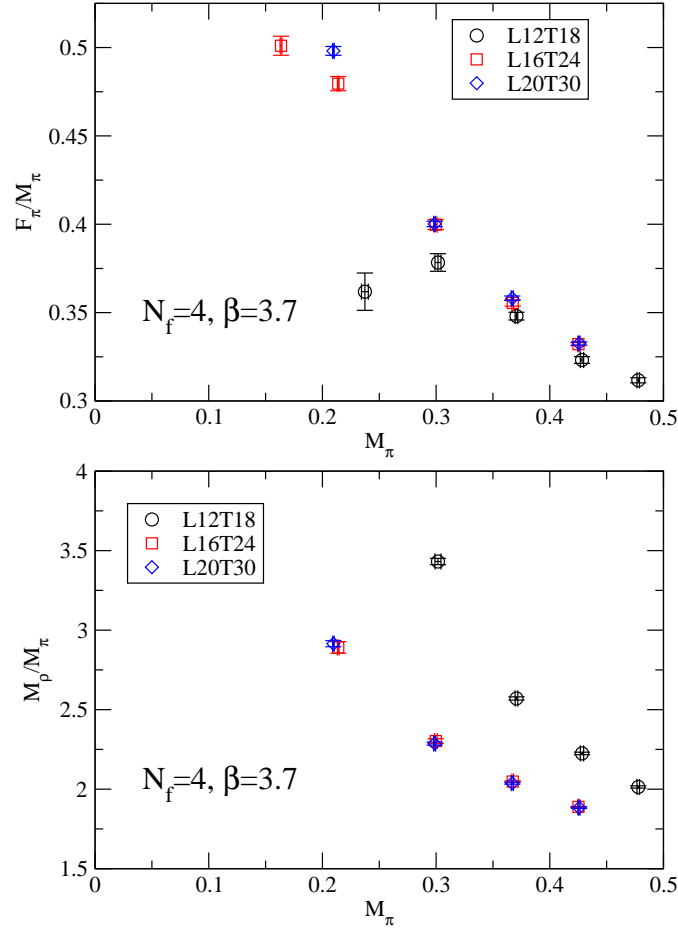

\makebox[.5\textwidth][r]{\includegraphics*[scale=.5]{fig016a.eps}}
\makebox[.5\textwidth][r]{\includegraphics*[scale=.5]{fig016b.eps}}
\caption{
$F_\pi/M_\pi$ (top) and $M_\rho/M_\pi$ (bottom) in $N_f = 4$ QCD.
}
\label{fig:rat_fpi_mpv_nf4}
\end{figure}

In Fig.~\ref{fig:rat_fpi_mpv},
we show the ratios $F_\pi/M_\pi$ and $M_\rho/M_\pi$ for $N_f = 8$ as a function of $M_{\pi}$
for various lattice volumes.
Up to some exceptions suffering from finite volume effects,
both ratios monotonically increase as $M_\pi$ decreases.
The present results are consistent with our previous work~\cite{Aoki:2013xza}
and add larger volume ($L = 42$) data in the small $M_{\pi}$ region,
where we confirm the increasing trend of the ratios still holds.
The aforementioned property of $N_f = 8$ is similar to $N_f = 4$ QCD shown in Fig.~\ref{fig:rat_fpi_mpv_nf4},
but different from $N_f = 12$ QCD~\cite{Aoki:2012eq,Aoki:2015gea}.
In the latter, the increasing trend ends up with the emergence of plateau at small $M_\pi$ region.

\begin{figure}[!tbp]
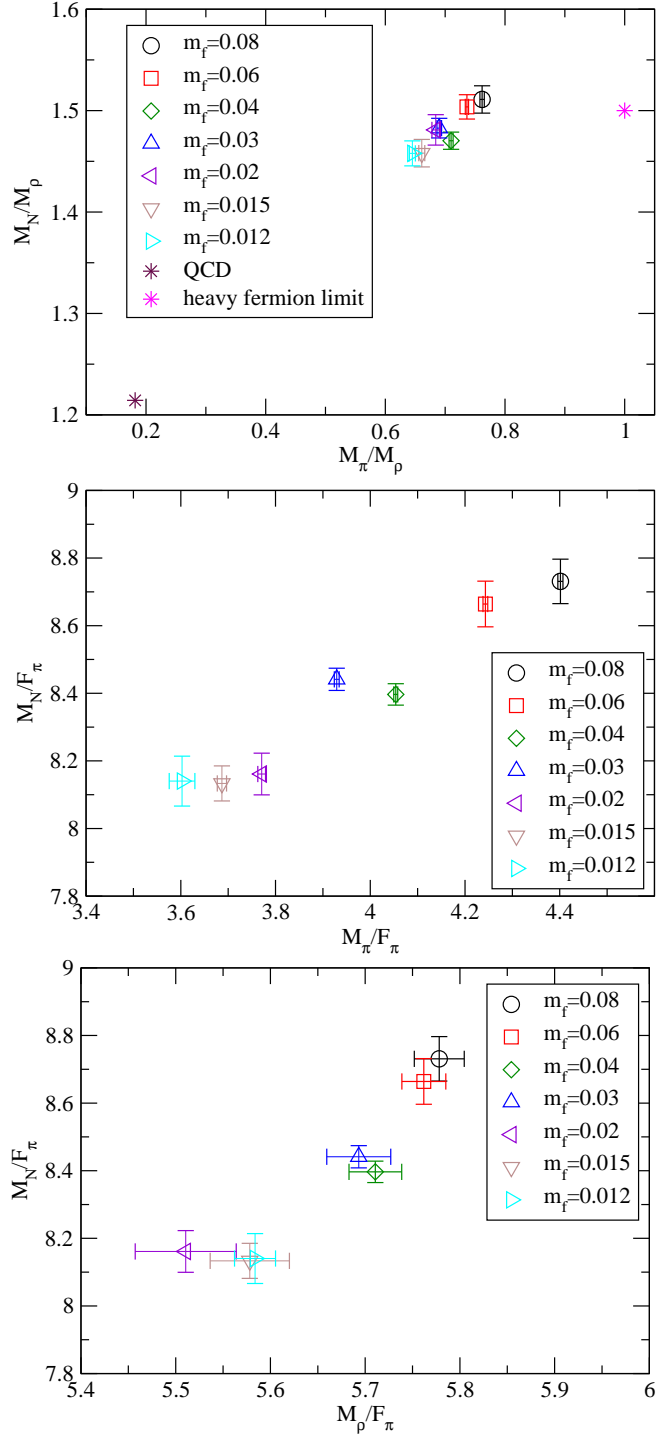

\makebox[.5\textwidth][r]{\includegraphics*[scale=.5]{fig017a.eps}}
\makebox[.5\textwidth][r]{\includegraphics*[scale=.5]{fig017b.eps}}
\makebox[.5\textwidth][r]{\includegraphics*[scale=.5]{fig017c.eps}}
\caption{
Edinburgh type plots: $M_N/M_\rho$ vs $M_\pi/M_\rho$ (top),
{$M_N/F_\pi$ vs $M_\pi/F_\pi$ (middle),} and
$M_N/F_\pi$ vs $M_\rho/F_\pi$ (bottom).
}\label{fig:rat_mn_mpv}
\end{figure}

We investigate the ratios of other spectra.
The top panel of Fig.~\ref{fig:rat_mn_mpv} shows an Edinburgh-type plot
with the Large Volume Data Set, together with
the infinite fermion mass limit and the usual QCD point.
We find that the $N_f = 8$ data differ from both QCD and heavy fermion limit.
The middle panel of Fig.~\ref{fig:rat_mn_mpv} is similar to the top panel,
but $F_\pi$ is used as the denominator of the ratios instead of $M_{\rho}$.
In the mass region of $0.02 \le m_f \le 0.08$, both ratios $M_N/F_\pi$ and $M_{\pi}/F_{\pi}$ show a decreasing trend as $m_f$ becomes smaller,
while only the former ratio becomes constant for the smallest three masses, $m_f = 0.012,\ 0.015$, and $0.02$;
the pion mass possesses the different $m_f$ dependence in the small $m_f$ region from $M_N$ and $F_{\pi}$.
When we replace the horizontal axis $M_{\pi}/F_{\pi}$ in the middle panel with $M_{\rho}/F_\pi$,
the pion mass $M_{\pi}$ is excluded from both the horizontal and vertical axes.
Then, the ratios in both axes ($M_{\rho}/F_\pi, M_N/F_\pi$) becomes the constant
at the smallest three masses (the bottom panel of Fig.~\ref{fig:rat_mn_mpv}).
This suggests that the $m_f$ dependence of $M_{\pi}$ {\em exceptionally} differs from the others.

\begin{figure}[!tbp]
\includegraphics[width=10cm]{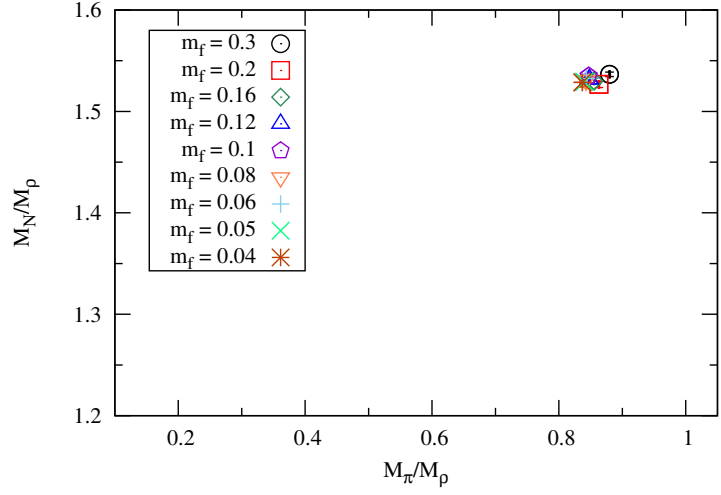}
\caption{
Edinburgh type plot, $M_N/M_\rho$ vs $M_\pi/M_\rho$, for $N_f = 12$ with $\beta = 4.0$.
The spectra obtained with the largest volume are selected at each $m_f$.
}\label{fig:edinburgh_nf12b0400}
\end{figure}

\begin{figure}[!tbp]
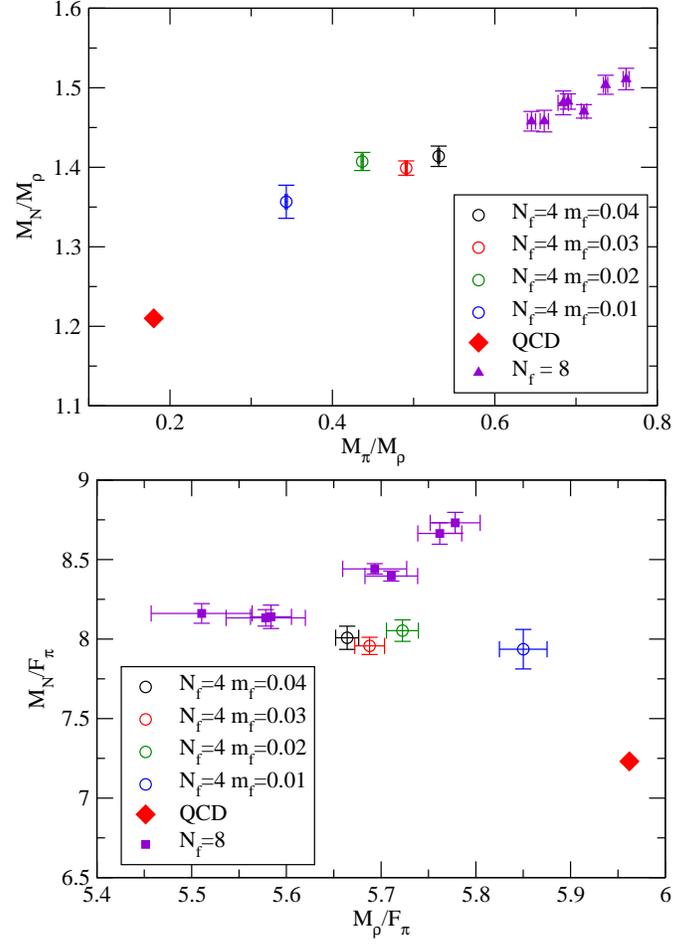

\makebox[.5\textwidth][r]{\includegraphics*[scale=.5]{fig019a.eps}}
\makebox[.5\textwidth][r]{\includegraphics*[scale=.5]{fig019b.eps}}
\caption{
Edinburgh type plots: $M_N/M_\rho$ vs $M_\pi/M_\rho$ (top) and
$M_N/F_\pi$ vs $M_\rho/F_\pi$ (bottom) in $N_f = 4$ QCD.
}\label{fig:rat_mn_mpv_nf4}
\end{figure}

For comparison, we show the $M_N/M_\rho$ versus $M_\pi/M_\rho$ for $N_f = 12$ with $\beta = 4.0$ in Fig.~\ref{fig:edinburgh_nf12b0400}.
The data almost stay at one point, indicating the conformal nature with no exceptional scaling in the spectra.
In Fig.~\ref{fig:rat_mn_mpv_nf4}, we compare the $N_f = 8$ and $4$ spectrum data in the Edinburgh type plots.
In the upper panel ($M_N/M_{\rho}$ vs. $M_{\pi}/M_{\rho}$), the $N_f = 4$ data approaches to the QCD point with decreasing $m_f$, while this is less clear in $N_f = 8$.
In the lower panel ($M_N/F_{\pi}$ vs. $M_{\rho}/F_{\pi}$), the $N_f = 4$ data points go closer to the QCD point, while the $N_f = 8$ data move in the opposite direction horizontally.
Thus, the scaling property of the $N_f = 8$ spectra differs from both those in $N_f = 12$ and $4$.

{
From the above analyses, we observe that $F_\pi/M_\pi$ and $M_\rho/M_\pi$
in $N_f=8$ QCD have a similar tendency to rise as the chiral limit is approached to $N_f = 4$ QCD, which is consistent behavior with
that observed in the chiral broken phase.
We also observe, however, that states other than $M_\pi$ exhibit scaling 
behavior in the small $m_f$ region.
The scaling is similar to the one expected in the conformal phase.
In the following sections, we further elaborate the $N_f = 8$ spectra 
by considering both chirally broken and conformal hypotheses.
}

\clearpage

\section{Chiral perturbation Theory analysis}
\label{sec:chpt}

In this section, we perform polynomial fits using the Large Volume
Data Set (as defined in Table~\ref{tab:data_set}),
under the assumption that $N_f = 8$ QCD is in the
chirally broken phase.
For this purpose, we focus on the smaller $m_f$ data,
$0.012 \le m_f \le 0.06$.
We check the validity of the assumption from
the values of physical quantities in the chiral limit, such as $F$,
and estimate their values, which would be helpful to predict hadron
masses in technicolor models.
In the last subsection, we estimate the chiral log correction in ChPT.

\subsection{$F_\pi$ and $M_\pi$}

\begin{figure}[!tbp]
\makebox[.5\textwidth][r]{\includegraphics*[scale=.5]{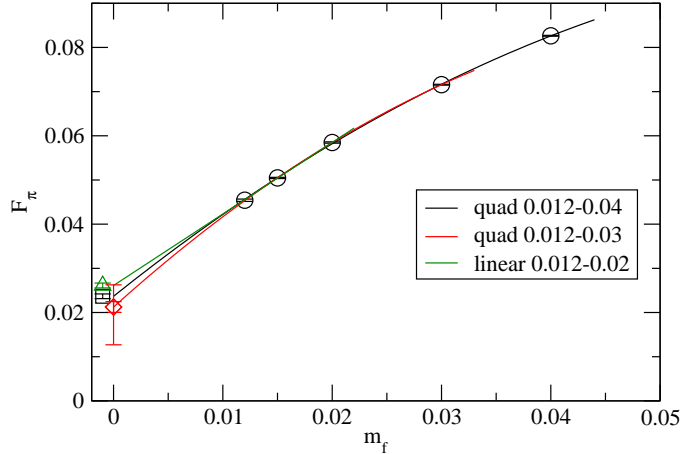}}
\caption{
$F_\pi$ as a function of $m_f$.
Curves are fit results with
polynomial function, Eq.~(\ref{eq:fpi_fit}).
``quad'' and ``linear'' denote quadratic and linear fit results, respectively.
Each fit result in the chiral limit is expressed by a colored symbol.
The {square and triangle are} shifted in the horizontal axis for clarity.
{The diamond symbol has two error bars: 
the outer represents the statistical and systematic uncertainties 
added in quadrature, 
while the inner error is only statistical.
The systematic error is discussed in Sec.~\ref{sec:chpt:log}.}
}
\label{fig:fit-fpi}
\end{figure}

\begin{figure}[!tbp]
\makebox[.5\textwidth][r]{\includegraphics*[scale=.5]{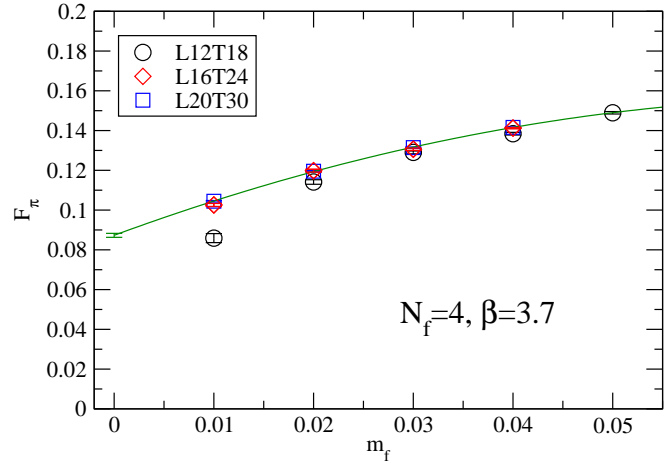}}
\caption{
$F_\pi$ as a function of $m_f$ on three volumes in $N_f = 4$ QCD
at $\beta = 3.7$.
The solid curve is a quadratic fit result with the largest volume data for
each $m_f$.
}
\label{fig:nf4_fit-fpi}
\end{figure}

Figure~\ref{fig:fit-fpi} presents the $m_f$ dependence of
$F_\pi$ in the small $m_f$ region.
A polynomial fit function is used, defined by
\begin{equation}
F_\pi = F + C_1 m_f + C_2 m_f^2.
\label{eq:fpi_fit}
\end{equation}
Linear ($C_2 = 0$) and quadratic ($C_2 \ne 0$) fits are carried out with
several fit ranges, as summarized in Table~\ref{tab:fit-fpi}.
The fit functions are regarded as NLO and NNLO ChPT predictions of $F_\pi$
without the chiral log terms.
The linear fit function of $m_f$ works well for the three lightest data,
while it does not work if the next-lightest $m_f$ data point is included in the fit.
The quadratic fit gives smaller $\chi^2/$dof, and
works up to $m_f \le 0.03$.
All the results of $F$ in the reasonable fits are nonzero,
as shown in Fig.~\ref{fig:fit-fpi}.
This is a similar property to that observed in our $N_f = 4$ data as presented
in Fig.~\ref{fig:nf4_fit-fpi}.

The expansion parameter of ChPT in $N_f$ flavor QCD~\cite{Soldate:1989fh,Chivukula:1992gi,Harada:2003jx}
is defined as
\begin{equation}
\mathcal{X} =N_f \left( \frac{M_\pi}{4 \pi F/\sqrt{2}} \right)^2 ,
\label{eq:X}
\end{equation}
and this quantity is required to not be too large, $\mathcal{X} < O(1)$.
The values of $\mathcal{X}$ for the maximum and minimum $m_f$
in the fit are evaluated in each fit result, which are
shown in Table~\ref{tab:fit-fpi}.

\begin{table}[!tbp]
\caption{Results of a chiral fit of $F_\pi$ with
$F_\pi=F+C_1 m_f + C_2m_f^2$ for various fit ranges.
Asterisks~($^*$) denote linear fits.
$m_f^{\min}$ and $m_f^{\max}$ denote the minimum and maximum
$m_f$ in each fit range, respectively.
\label{tab:fit-fpi}
}
\begin{ruledtabular}

\begin{tabular}{llllll}

\multicolumn{1}{c}{fit range ($m_f$)} &
\multicolumn{1}{c}{$F$} &
\multicolumn{1}{c}{$\mathcal{X}(m_f^{\min})$} &
\multicolumn{1}{c}{$\mathcal{X}(m_f^{\max})$} &
\multicolumn{1}{c}{$\chi^2/{\rm dof}$} &
\multicolumn{1}{c}{dof} \\
\hline
0.012-0.02$^*$ & 0.02612(55) & 3.978(17) &  7.22(31) &  0.43 & 1 \\
0.012-0.03$^*$ & 0.02953(24) & 3.111(53) &  9.19(15) & 23.8  & 2 \\\hline
0.012-0.03     & 0.0212(12)  & 6.01(70)  & 17.8(2.1) &  0.31 & 1 \\
0.012-0.04     & 0.02368(54) & 4.84(22)  & 20.29(92) &  2.58 & 2 \\
0.012-0.05     & 0.02435(41) & 4.57(16)  & 25.10(85) &  3.00 & 3 \\
0.012-0.06     & 0.02633(30) & 3.911(90) & 27.02(61) & 14.4  & 4 \\

\end{tabular}
\end{ruledtabular}
\end{table}

\begin{figure}[!tbp]
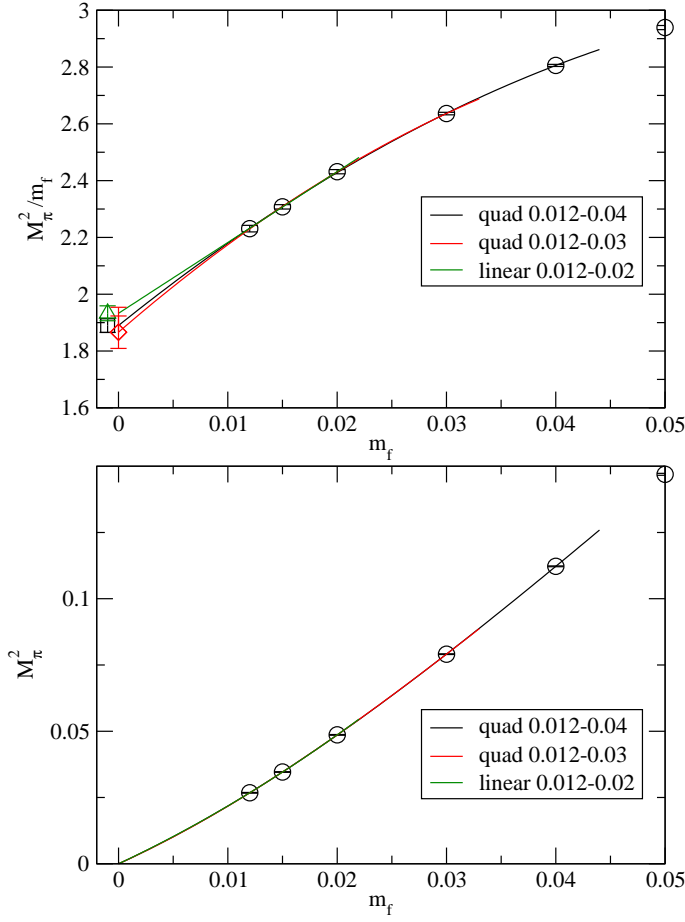

\makebox[.5\textwidth][r]{\includegraphics*[scale=.5]{fig022a.eps}}
\makebox[.5\textwidth][r]{\includegraphics*[scale=.5]{fig022b.eps}}
\caption{
$M_\pi^2/m_f$ (top) and $M_\pi^2$ (bottom) as a function of $m_f$.
Solid lines are fit results using polynomial functions.
``Quad'' and ``linear'' denote quadratic and linear fit results, respectively.
In the top panel, each fit result in the chiral limit
is expressed by a colored symbol.
The {square and triangle are} shifted to 
the negative direction on the horizontal axis for clarity.
{The diamond symbol has two error bars: the outer
represents the statistical and systematic uncertainties 
added in quadrature, 
while the inner error is only statistical.
The systematic error is discussed in Sec.~\ref{sec:chpt:log}.}
}
\label{fig:fit-mpi2_ov_mf}
\end{figure}

\begin{figure}[!tbp]
\makebox[.5\textwidth][r]{\includegraphics*[scale=.5]{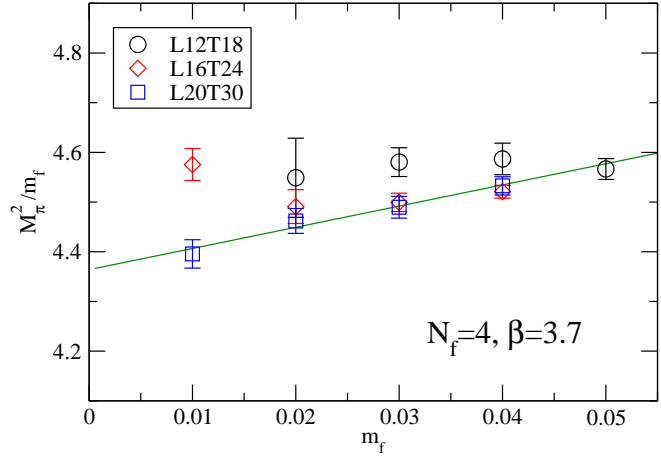}}
\caption{
$M_\pi^2/m_f$ as a function of $m_f$ on three volumes in $N_f = 4$ QCD
at $\beta = 3.7$.
The solid curve is the result of a linear fit using the largest volume data in
each $m_f$.
}
\label{fig:nf4_mpi2_ov_mf}
\end{figure}

The $m_f$ dependence of $M_\pi^2/m_f$ is plotted in the top panel of
Fig.~\ref{fig:fit-mpi2_ov_mf} with the fit function
\begin{equation}\label{eq:MpiOverMf_fit}
\frac{M_\pi^2}{m_f} = C_0 + C_1 m_f + C_2 m_f^2.
\end{equation}
Since the ratio approaches a constant towards the chiral limit,
$M_\pi^2$ would vanish in the chiral limit.
Due to the visible curvature of the ratio,
higher order terms than a linear $m_f$ term
are necessary to explain our data
in contrast to the $N_f = 4$ case, where $M_\pi^2/m_f$
in the largest volumes at each $m_f$ is reasonably expressed
by a linear function of $m_f$ as shown in Fig.~\ref{fig:nf4_mpi2_ov_mf}.
The polynomial fits are carried out with several fit ranges,
as tabulated in Table~\ref{tab:fit-mpi2}.
The linear fits work in the smaller $m_f$ range, $0.012 \le m_f \le 0.03$.
The quadratic fits give reasonable values of $\chi^2$/dof in a
wider $m_f$ range, $0.012 \le m_f \le 0.06$, than in the linear fit.
The $m_f$ dependence of $M_\pi^2$ and the fit results are
plotted in the bottom panel of Fig.~\ref{fig:fit-mpi2_ov_mf}.

The above analyses for $F_\pi$ and $M_\pi^2/m_f$ show that
our data can be explained by polynomial functions of $m_f$,
which would be regarded as the ChPT formula without log terms,
in the smaller $m_f$ region.

While in our previous work~\cite{Aoki:2013xza} we took
the fit results with $0.015 \le m_f \le 0.04$ data for our central values,
after accumulating more statistics and including data at even smaller $m_f$,
we choose the quadratic fit results with
$0.012 \le m_f \le 0.03$ data,
whose values of $\chi^2$/dof are reasonable,
as the central values in this work.
{
Our central values for $F$ and $M_\pi^2/m_f$ in the chiral limit are
\begin{equation}
F = 0.0212(12),\ \ \ \ 
\left.\frac{M_\pi^2}{m_f}\right|_{m_f \to 0} = 1.866(57),
\end{equation}
where the errors are only statistical.
We will discuss a systematic error of $F$ coming from 
the logarithmic correction in Sec.~\ref{sec:chpt:log}.}
In analyses for other physical quantities as shown in the following
subsections, we evaluate their central values with the same $m_f$ range.

\begin{table}[!tbp]
\caption{Results of a chiral fit of $M_\pi^2/m_f$ with
$M_\pi^2/m_f= C_0 + C_1 m_f + C_2m_f^2$ for various fit ranges.
Asterisks~($^*$) denote linear fits.
\label{tab:fit-mpi2}
}
\begin{ruledtabular}
\begin{tabular}{llll}

\multicolumn{1}{c}{fit range ($m_f$)} &
\multicolumn{1}{c}{$C_0$} &
\multicolumn{1}{c}{$\chi^2/{\rm dof}$} &
\multicolumn{1}{c}{dof} \\
\hline
0.012-0.02$^*$ & 1.933(26)  &  0.23 & 1 \\
0.012-0.03$^*$ & 1.981(12)  &  2.13 & 2 \\
0.012-0.04$^*$ & 2.0282(83) & 12.2  & 3 \\\hline
0.012-0.03     & 1.866(57)  &  0.04 & 1 \\
0.012-0.04     & 1.890(24)  &  0.12 & 2 \\
0.012-0.05     & 1.896(18)  &  0.12 & 3 \\
0.012-0.06     & 1.934(13)  &  2.57 & 4 \\

\end{tabular}
\end{ruledtabular}
\end{table}

\clearpage

\subsection{Chiral condensate and GMOR relation}

\begin{figure}[!tbp]
\makebox[.5\textwidth][r]{\includegraphics*[scale=.5]{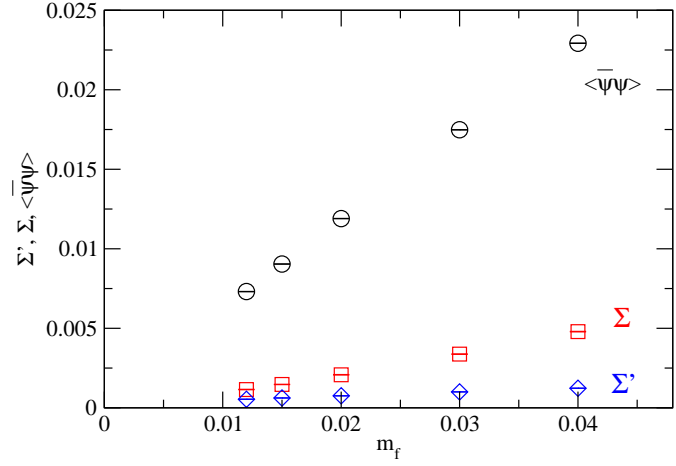}}
\caption{
$\langle \overline{\psi}\psi\rangle$, $\Sigma = F_\pi^2 {M}_\pi^2/4 m_f$,
and $\Sigma^\prime = F F_\pi M_\pi^2/4 m_f$ as a function of $m_f$,
as defined by Eqs.~(\ref{eq:pbp_direct}), (\ref{eq:gmor_sig_p}),
and (\ref{eq:gmor_sig}), respectively.
}
\label{fig:pbp-gmor}
\end{figure}

The chiral condensate, $\langle \overline{\psi} \psi \rangle$,
in each flavor is measured by the trace of the inverse
Dirac operator, divided by a factor of four corresponding to
the number of tastes, as
\begin{equation}
\langle \overline{\psi} \psi \rangle =
\frac{{\rm Tr}\left[D^{-1}_{\rm HISQ}(x,x)\right]}{4}
\ ,\label{eq:pbp_direct}
\end{equation}
where $D_{\rm HISQ}$ is the Dirac operator of the HISQ action.
The Ward-Takahashi identity for the chiral symmetry tells us the quantities
\begin{align}
&\Sigma^{\prime}(m_f) \equiv \frac{FF_\pi M_\pi^2}{4m_f}
\ ,\label{eq:gmor_sig_p}\\
&\Sigma(m_f) \equiv \frac{F_\pi^2 M_\pi^2}{4m_f}
\ ,\label{eq:gmor_sig}
\end{align}
with $F = F_{\pi}|_{m_f\to 0}$ being identical to the chiral condensate
in the chiral limit $\langle\overline{\psi}\psi\rangle|_{m_f\to 0}$, through
the Gell-Mann-Oakes-Renner (GMOR) relation,
\begin{equation}
\left.\langle \overline{\psi} \psi \rangle\right|_{m_f\to 0}
= \frac{B F^2}{2},
\label{eq:GMOR}
\end{equation}
where $B = M_\pi^2/2m_f$ in the chiral limit,
corresponding to $C_0/2$ in Table~\ref{tab:fit-mpi2}.
In this subsection, we estimate the chiral condensate
in the chiral limit from the above quantities 
{using polynomial fits.}
The $m_f$ dependence for $\langle\overline{\psi}\psi\rangle$,
$\Sigma^\prime$, and $\Sigma$ in the small $m_f$ region is
shown in Fig.~\ref{fig:pbp-gmor}.

\begin{figure}[!tbp]
\makebox[.5\textwidth][r]{\includegraphics*[scale=.5]{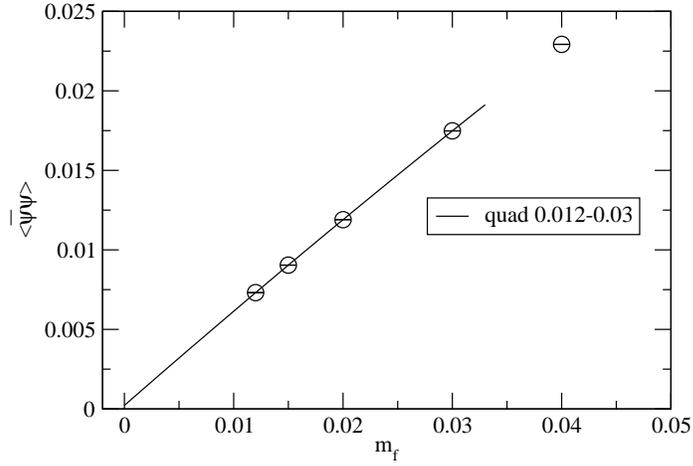}}
\caption{
$\langle \overline{\psi}\psi\rangle$ as a function of $m_f$.
The solid curve is a quadratic fit result.
}\label{fig:fit-pbp}
\end{figure}

We first discuss $\langle\overline{\psi}\psi\rangle$.
The data of $\langle \overline{\psi} \psi \rangle$
depend almost linearly on $m_f$.
The linear term in $\langle \overline{\psi} \psi \rangle$
contains a UV power divergence, $m_f/a^2$, which vanishes in the chiral limit.
To estimate $\langle \overline{\psi} \psi \rangle$ in the chiral limit,
the data are fitted by linear and quadratic functions of $m_f$,
whose results are summarized in Table~\ref{tab:fit-pbp}.
The quadratic fit result in $0.012 \le m_f \le 0.03$
plotted in Fig.~\ref{fig:fit-pbp} shows that
the value in the chiral limit is much smaller than the measured values.
This is due to the large linear term.
In the smaller $m_f$ region,
both the linear and quadratic fits work well,
and give nonzero chiral condensate in the chiral limit.

Using the fit results for $F_\pi$ and $M_\pi$ in Tables~\ref{tab:fit-fpi} and
\ref{tab:fit-mpi2}, respectively,
we calculate the right hand side of Eq.~(\ref{eq:GMOR})
in each fit range for the two fit forms.
The values are compared to the fit results of
$\langle \overline{\psi} \psi \rangle$ in Table~\ref{tab:fit-pbp}.
While in the linear fit result with the smallest fit range
the two values are inconsistent,
the three quadratic fit results,
whose values of $\chi^2/$dof are reasonable,
agree well with those from the GMOR relation.

\begin{figure}[!tbp]
\makebox[.5\textwidth][r]{\includegraphics*[scale=.5]{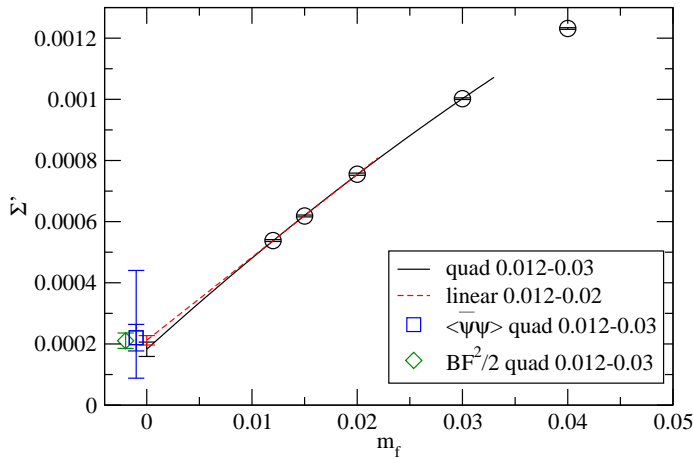}}
\caption{
$\Sigma^{\prime} = FF_\pi {M}_\pi^2/4 m_f$
as a function of $m_f$.
The solid curves are polynomial fit results.
``Quad'' and ``linear'' denote quadratic and linear fit results, respectively.
Each result in the chiral limit is expressed by a colored symbol.
Square and diamond represent results for
$\langle \overline{\psi} \psi \rangle$ and GMOR relation
in the chiral limit, respectively.
They are shifted to the negative direction on the horizontal
axis for clarity.
{The square symbol has two error bars: the outer
represents the statistical and systematic uncertainties 
added in quadrature, 
while the inner error is only statistical.
The systematic error is discussed in Sec.~\ref{sec:chpt:log}.}
}\label{fig:fit-gmor^p}
\end{figure}

The chiral limit value of $\Sigma^{\prime}$ is estimated from
a quadratic fit in $0.12 \le m_f \le 0.03$,
whose value is consistent with the ones obtained by
$\langle \overline{\psi} \psi \rangle$ and $BF^2/2$
as shown in Fig.~\ref{fig:fit-gmor^p}.
A linear fit with the range of $0.012$--$0.02$ also works, and
gives a consistent result with the quadratic fit,
as presented in the second column of Table~\ref{tab:fit-gmor}.
The chiral limits of $\Sigma^{\prime}$ determined by the wider fit ranges
give somewhat smaller values than those obtained by Eq.~(\ref{eq:GMOR}).
The difference becomes larger as a larger $m_f$ is included into the fit,
and would be attributed to higher-order $m_f$ effects.

\begin{figure}[!tbp]
\makebox[.5\textwidth][r]{\includegraphics*[scale=.5]{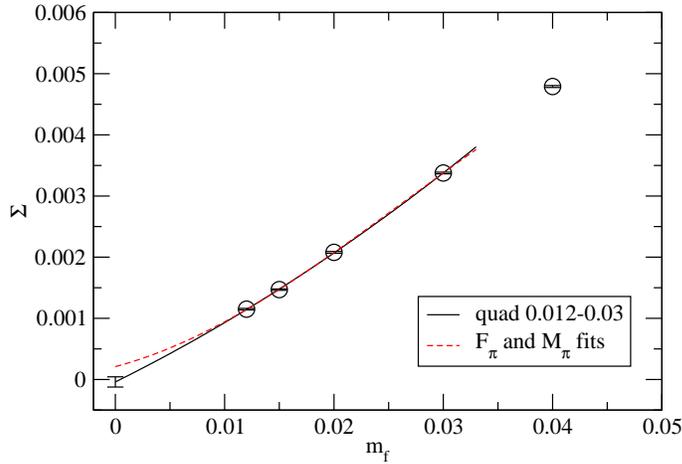}}
\caption{
$\Sigma = F_\pi^2 {M}_\pi^2/4 m_f$
as a function of $m_f$.
The solid and dashed curves are a quadratic fit result
and expected result respectively from each fit of $F_\pi$ and $M_\pi$.
}
\label{fig:fit-gmor}
\end{figure}

As shown in Fig.~\ref{fig:fit-gmor} and
the fourth column of Table~\ref{tab:fit-gmor},
the chiral limit value of $\Sigma$ is inconsistently smaller than those for
$\langle\overline{\psi}\psi\rangle$, $BF^2/2$, and $\Sigma^{\prime}$.
This would not be surprising:
as we have seen in the chiral extrapolation of $F_\pi$ (Table~\ref{tab:fit-fpi}),
the $m_f^2$ term is required to perform a reasonable fit of the $F_{\pi}$ data
in $0.012 \le m_f \le 0.03$,
and thus, $m_f^3$ and $m_f^4$ terms would be necessary
to capture the $m_f$ dependence of $\Sigma\propto F_{\pi}^2$
in the same fit range.
The quadratic fit lacks such higher-order terms.
In other words, even our smallest fit range $0.012$--$0.03$
would not be small enough
for the quadratic chiral extrapolation of $\Sigma$.
The result in Fig.~\ref{fig:fit-gmor} supports this expectation,
as the quadratic fit curve in the smaller $m_f$ region
deviates from the $m_f$ dependence of $\Sigma$ expected from
the fit results for $F_\pi$ and $M_\pi$.

Although the extrapolation of $\Sigma$ has the above difficulties,
{we observe
the consistency among the chiral limits of
$\langle\overline{\psi}\psi\rangle$, $BF^2/2$, and $\Sigma^{\prime}$.
Our central value of the chiral condensate is
determined from the chiral extrapolation of 
$\langle \overline{\psi} \psi \rangle$ presented in 
Table~\ref{tab:fit-pbp}, whose value is
\begin{equation}
\left.\langle \overline{\psi} \psi \rangle\right|_{m_f\to 0}
= 0.000221(43),
\end{equation}
where the error is only statistical.
A systematic error of the chiral condensate coming from
the logarithmic correction will be discussed in Sec.~\ref{sec:chpt:log}.
The positive value of the chiral condensate is consistent with 
the property expected in the chirally broken phase.}
For future work, it is important to confirm that
the chiral limit of $\Sigma$ becomes consistent with the other results
by adding more data points in the small $m_f$ region.

\begin{table}[!tbp]
\caption{Chiral fit result of
$\langle \overline{\psi} \psi \rangle$ with
$\langle \overline{\psi} \psi \rangle=C_0+C_1 m_f + C_2 m_f^2$
in various fit ranges.
$BF^2/2$ is evaluated using the results in
Tables~\ref{tab:fit-fpi} and \ref{tab:fit-mpi2}.
Asterisks~($^*$) denote linear fits.
\label{tab:fit-pbp}
}
\begin{ruledtabular}
\begin{tabular}{lldcl}

\multicolumn{1}{c}{fit range ($m_f$)} &
\multicolumn{1}{c}{$C_0$} &
\multicolumn{1}{c}{$\chi^2/{\rm dof}$} &
\multicolumn{1}{c}{dof} &
\multicolumn{1}{c}{$BF^2/2$} \\
\hline
0.012-0.02$^*$ & 0.000436(19)  &  0.92 & 1 & 0.000330(15)  \\
0.012-0.03$^*$ & 0.0005867(84) & 37.4  & 2 & 0.0004319(74) \\\hline
0.012-0.03     & 0.000221(43)  &  0.54 & 1 & 0.000211(25)  \\
0.012-0.04     & 0.000255(18)  &  0.65 & 2 & 0.000265(12)  \\
0.012-0.05     & 0.000263(15)  &  0.63 & 3 & 0.000281(10)  \\
0.012-0.06     & 0.000313(10)  &  5.97 & 4 & 0.0003352(79) \\

\end{tabular}
\end{ruledtabular}
\end{table}

\begin{table}[!tbp]
\caption{The polynomial fit results of
$\Sigma(m_f) = F_\pi^2 {M}_\pi^2/4 m_f$
and
$\Sigma^{\prime}(m_f) = FF_\pi {M}_\pi^2/4 m_f$.
The $C_0$ corresponds to the chiral limit values of them:
$\Sigma$ or $\Sigma^{\prime} = C_0 + C_1 m_f + C_2 m_f^2$.
The results with(without) an asterisk~($^*$) denote linear(quadratic) fits.
\label{tab:fit-gmor}
}
\begin{ruledtabular}
\begin{tabular}{l cc cc c}

&
\multicolumn{2}{c}{$\Sigma^\prime$} &
\multicolumn{2}{c}{$\Sigma$} &
\\
\cline{2-3} \cline{4-5}
\multicolumn{1}{c}{fit range ($m_f$)} &
\multicolumn{1}{c}{$C_0$} &
\multicolumn{1}{c}{$\chi^2/{\rm dof}$} &
\multicolumn{1}{c}{$C_0$} &
\multicolumn{1}{c}{$\chi^2/{\rm dof}$} &
\multicolumn{1}{c}{dof} \\
\hline
0.012-0.02$^*$ & 0.000212(15) & 0.06 & $-$0.000257(37) & 4.12 & 1 \\
0.012-0.03$^*$ & 0.000233(14) & 3.28 & $-$0.000378(18) & 9.29 & 2 \\
\hline
0.012-0.03     & 0.000183(24) & 0.54 & $-$0.000039(84) & 1.38 & 1 \\
0.012-0.04     & 0.000189(15) & 0.34 & $-$0.000108(45) & 1.17 & 2 \\
0.012-0.05     & 0.000186(13) & 0.31 & $-$0.000175(38) & 3.04 & 3 \\
0.012-0.06     & 0.000206(13) & 3.29 & $-$0.000159(27) & 2.37 & 4 \\

\end{tabular}
\end{ruledtabular}
\end{table}

\clearpage

\subsection{Other hadron masses}

\begin{figure}[!tbp]
\makebox[.5\textwidth][r]{\includegraphics*[scale=.5]{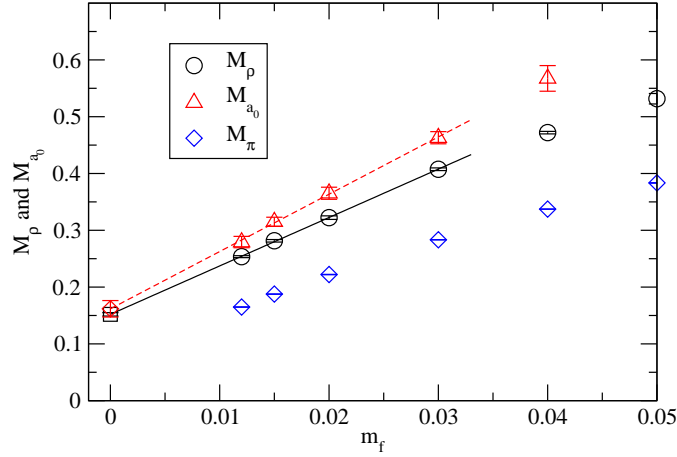}}
\caption{
$M_\rho$ and $M_{a_0}$ as a function of $m_f$, together with $M_\pi$.
Solid and dashed lines express the linear fit results for $M_\rho$ and $M_{a_0}$,
respectively.
}
\label{fig:fit-pv}
\end{figure}

\begin{figure}[!tbp]
\makebox[.5\textwidth][r]{\includegraphics*[scale=.5]{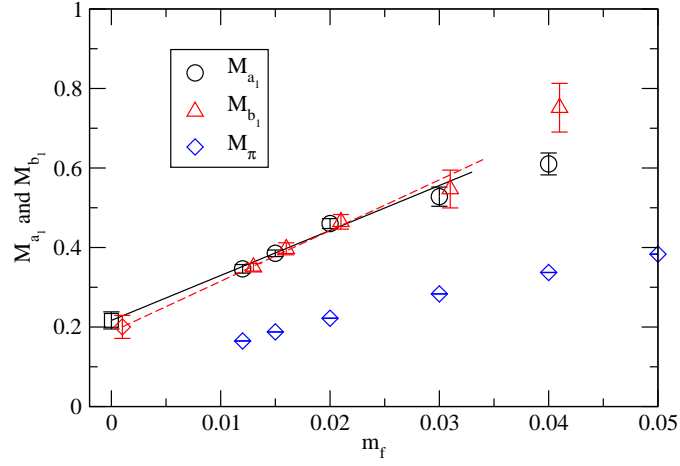}}
\caption{
As Fig.~\ref{fig:fit-pv}, but for $M_{a_1}$ and $M_{b_1}$.
Triangle and diamond symbols are shifted to the positive direction on the horizontal
axis for clarity.
}
\label{fig:fit-a0-a1-b1}
\end{figure}

\begin{figure}[!tbp]
\makebox[.5\textwidth][r]{\includegraphics*[scale=.5]{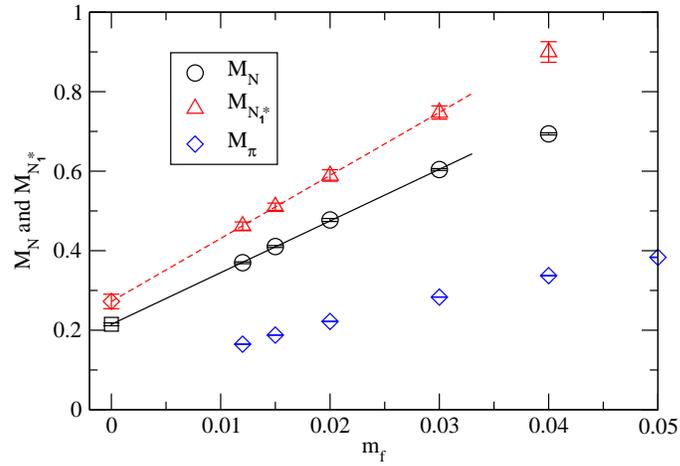}}
\caption{
Same as Fig.~\ref{fig:fit-pv}, $M_N$ and $M_{N_{\bf 1}^*}$.
}
\label{fig:fit-N-Nstr}
\end{figure}

We extrapolate the masses of other hadrons, such as $\rho$ and $N$,
to the chiral limit.
Since the data for the hadrons have larger error than
the ones for $F_\pi$ and $M_\pi$,
linear fits work in the small $m_f$ region, $0.012 \le m_f \le 0.03$,
where the quadratic fits for $F_\pi$ and $M_\pi^2/m_f$ give
reasonable $\chi^2/$dof.
The fit results are summarized in Table~\ref{tab:fit-other_hadron},
and plotted in Figs.~\ref{fig:fit-pv},
\ref{fig:fit-a0-a1-b1}, and~\ref{fig:fit-N-Nstr}.

While $M_\rho$ and $M_{a_0}$ at each $m_f$ are different,
the linear fit results coincide within the error
as shown in Fig.~\ref{fig:fit-pv}.
The near degeneracy of $\rho$ and $a_0$ in the chiral limit
was also observed in Ref.~\cite{Appelquist:2014zsa}.
$a_1$ and $b_1$ are almost degenerate at each $m_f$,
and also in the chiral limit, as shown in Fig.~\ref{fig:fit-a0-a1-b1}.
This property in the chiral limit is roughly consistent with usual QCD, where
$a_1$ and $b_1$ are almost degenerate at the physical $m_f$.
Note that $N$ is also almost degenerate to $a_1$ and $b_1$ at each $m_f$
as well as in the chiral limit, as shown
in Tables in Appendix~\ref{sec:table_had_mass} and
Table~\ref{tab:fit-other_hadron}.
This degeneracy in the chiral limit is
a different property from usual QCD at the physical $m_f$.
A similar trend for $M_{a_1}$ and $M_N$
is observed in Ref.~\cite{Appelquist:2016viq}.

We also carry out quadratic fits for all the hadron masses
with several wider fit ranges,
and find that those results with reasonable $\chi^2$/dof
are nonzero in the chiral limit.
Therefore, all the hadron masses in the chiral limit
are nonzero in this analysis,
which is consistent with what would be expected for the chirally broken phase.

\begin{figure}[!tbp]
\makebox[.5\textwidth][r]{\includegraphics*[scale=.5]{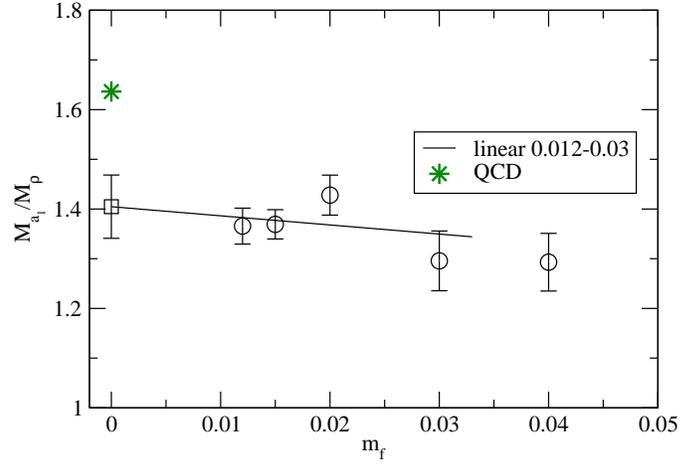}}
\caption{
Mass ratio $M_{a_1}/M_\rho$ as a function of $m_f$.
The solid line expresses a linear fit result.
The star symbol represents the ratio in usual QCD.
}
\label{fig:fit-mass-ratios}
\end{figure}

We investigate the ratio of masses of the parity partners,
$\rho$ and $a_1$.
Figure~\ref{fig:fit-mass-ratios} shows that
the mass ratio has milder $m_f$ dependence than
each hadron mass in the numerator and denominator.
The linear fit result of the data in $0.012 \le m_f \le 0.03$
in the figure, as tabulated in Table~\ref{tab:fit-other_hadron},
shows that the ratio in the chiral limit is different from unity, and
is smaller than the one in usual QCD, which is 1.636.
The parity partners $\rho$ and $a_1$ are expected to be degenerate in
the chiral unbroken phase.
This property will be discussed in the discussion subsection, Sec.~\ref{subsec:chptdiscussion}.

\begin{table}[!tbp]
\caption{Chiral fit result of hadron mass $M_H$
with
$M_H = C_0+C_1 m_f$ using fit range $0.012 \le m_f \le 0.03$
for $H = \rho, a_0, a_1, b_1, N,$ and $N_{\bf 1}^*$.
\label{tab:fit-other_hadron}
}
\begin{ruledtabular}
\begin{tabular}{cddc}

\multicolumn{1}{c}{$H$} &
\multicolumn{1}{c}{$C_0$} &
\multicolumn{1}{c}{$\chi^2/{\rm dof}$} &
\multicolumn{1}{c}{dof} \\
\hline
$\rho$ & 0.1520(30) & 0.36 & 2 \\
$a_0$  & 0.162(14)  & 0.12 & 2 \\
$a_1$  & 0.217(22)  & 1.81 & 2 \\
$b_1$  & 0.200(29)  & 0.52 & 2 \\
$N$    & 0.2148(35) & 0.40 & 2 \\
$N_{\bf 1}^*$  & 0.272(18)  & 0.03 & 2 \\

\end{tabular}
\end{ruledtabular}
\end{table}

\begin{table}[!tbp]
\caption{Chiral fit result of mass ratio of
parity partners $M_{a_1}/M_\rho$
with
$M_{a_1}/M_\rho = C_0+C_1 m_f$
using fit range $0.012 \le m_f \le 0.03$.
\label{tab:fit-mass-ratios}
}
\begin{minipage}{.5\textwidth}
\begin{ruledtabular}
\begin{tabular}{ccc}

$C_0$ & $\chi^2/{\rm dof}$ & dof \\
\hline
1.405(64) & 1.66 & 2 \\

\end{tabular}
\end{ruledtabular}
\end{minipage}
\end{table}

\clearpage

\subsection{Estimate of chiral log correction}
\label{sec:chpt:log}

The logarithmic correction to the chiral fits for $F_\pi$ and $M_\pi$ are
estimated in the same way as in our previous work~\cite{Aoki:2013xza}.
In next-to-leading order (NLO) ChPT,
the logarithmic $m_f$ dependence
in $M_\pi^2/m_f$ and $F_\pi$ is predicted~\cite{Gasser:1983yg} as,
\begin{eqnarray}
\frac{M_\pi^2}{m_f} &=& 2B\left( 1 + \frac{x}{N_f}\log(x) + c_3 x\right)
\label{eq:chpt_mpi}
\\
F_\pi &=& F\left(1 - \frac{N_f\, x}{2}\log(x) + c_4 x\right),
\label{eq:chpt_fpi}
\end{eqnarray}
where $x = 4 B m_f / ( 4\pi F)^2$,
and $B, F, c_3$ and $c_4$ are the low energy constants.
Even in the lighter $m_f$ region,
the data for both $M_\pi^2/m_f$ and $F_\pi$ show no
such logarithmic dependences, as shown in the previous subsections.

The size of the logarithmic correction in $F$ and $B$ is estimated by
matching the quadratic fit results to the NLO ChPT formulae at values of
$m_f$ such that ${\cal X}=1$, with ${\cal X}$ defined in Eq.~(\ref{eq:X}),
where $F$ should read the re-estimated one in this analysis.
The results of the analysis are presented in Appendix~\ref{app:log}.
The correction reduces $F$ by about 30\% from the quadratic fit,
{while the effect of the correction is small in $B$.}

The results for $F$, {$B$,}
and the chiral condensate
at the chiral limit in this work are
\begin{eqnarray}
F &=& 0.0212(12)(^{+49}_{-71}),\\
B &=& 0.933(29)(^{+33}_{-0}),\\
\left. \langle \overline{\psi} \psi \rangle\right|_{m_f\to 0} &=&
0.00022(4)(^{+22}_{-12}),
\end{eqnarray}
where the first and second errors are
statistical and systematic ones, respectively.
{
These results including the systematic errors are plotted in 
Figs.~\ref{fig:fit-fpi}, \ref{fig:fit-mpi2_ov_mf}, and \ref{fig:fit-gmor^p}.}
For all the quantities,
the central values come from the quadratic fit
with a fit range $0.012 \le m_f \le 0.03$,
and the upper systematic error is estimated from
the difference of the central values between
the quadratic fit and the linear fit with $0.012 \le m_f \le 0.02$.
{
The fit results for $F$, $2B(=C_0$ in the table), and the chiral condensate 
are tabulated in Tables~\ref{tab:fit-fpi}, \ref{tab:fit-mpi2},
and \ref{tab:fit-pbp}, respectively.
The lower systematic error of $F$ 
comes from the logarithmic correction in NLO ChPT.
For the chiral condensate, the lower systematic error is estimated from
the difference between the central value and $BF^2/2$ with the
logarithmic correction.
}

It would be useful to estimate physical quantities
in units of $F$,
because in technicolor models
$F$ is related to the weak scale,
\begin{equation}
\sqrt{N_{\rm D}} F/\sqrt{2} = 246\ {\rm GeV},
\label{eq:weak_scale}
\end{equation}
where $N_{\rm D}$ is the number of the fermion weak doublets
as $1 \le N_{\rm D} \le N_f/2$.
The ratios for all the hadron masses, tabulated in
Table~\ref{tab:fit-other_hadron}, to $F$ in the chiral limit
are summarized in Table~\ref{tab:R_mH_F},
where the systematic error comes from the one in $F$.
From our result,
the ratio $M_\rho/F$ in the chiral limit
is given as
\begin{equation}
\frac{M_\rho}{F/\sqrt{2}} = 10.1(0.6)(^{+5.0}_{-1.9}).
\label{eq:rho_f_ratio}
\end{equation}
If one chooses the one-family model with 4 weak-doublets, i.e., $N_{\rm D}=4$
in Eq.~(\ref{eq:weak_scale}),
$M_\rho$ corresponds to 1.0--1.9 TeV.

\begin{table}[!tbp]
\caption{
Ratios of $\sqrt{2}M_H/F$ with $H = \rho, a_0, a_1, b_1, N,$ and $N_{\bf 1}^*$.
The first and second errors are statistical and systematic errors.
\label{tab:R_mH_F}
}
\begin{minipage}{.5\textwidth}
\begin{ruledtabular}
\begin{tabular}{@{\hspace{1em}}cd@{\hspace{3em}}}
$\rho$        & 10.1~(0.6)~({}^{+5.0}_{-1.9}) \\
$a_0$         & 10.8~(1.1)~({}^{+5.4}_{-2.0}) \\
$a_1$         & 14.4~(1.7)~({}^{+7.2}_{-2.7}) \\
$b_1$         & 13.3~(2.1)~({}^{+6.6}_{-2.5}) \\
$N$           & 14.3~(0.9)~({}^{+7.1}_{-2.7}) \\
$N_{\bf 1}^*$ & 18.1~(1.6)~({}^{+9.0}_{-3.4}) \\
\end{tabular}
\end{ruledtabular}
\end{minipage}
\end{table}

\clearpage

\subsection{Discussion}
\label{subsec:chptdiscussion}

{
The chiral limit extrapolation of the spectrum of our data in Table VIII  indicates non-zero masses $M_{H\ne \pi} \ne 0$ with characteristic ratios: 
\begin{equation}
M^2_{a_1}/M^2_\rho \simeq (1.43)^2\simeq 2\, \,, \quad   M^2_\rho/M^2_{a_0}  \simeq \, (0.94)^2 \simeq 1. 
\label{chirallimitdata}
\end{equation}

The first relation is a clear signal of the spontaneously broken NG phase, since it is nothing but the famous 
Weinberg mass relation~\cite{Weinberg:1967kj}
 in ordinary QCD. It follows critically from the inequality of the vector and axial vector current correlators, typically the
 Weinberg's spectral function sum rules (SRs)
 \begin{align}
 	F_\rho^2 &= (F_\pi/\sqrt{2})^2 + F_{a_1}^2\;, \tag{SR1} \\
	F_\rho^2 M_\rho^2 &= F_{a_1}^2 M_{a_1}^2\;, \tag{SR2}
\end{align}
combined  with the Kawarabayashi-Suzuki-Riazuddin-Fayyazuddin (KSRF) relation $F_\rho^2=2 (F_\pi/\sqrt{2})^2$. If the chiral symmetry were not spontaneously broken, there would be no $\pi$ pole contribution to the axial vector current and hence $(F_\pi/\sqrt{2})^2$ term would be missing in SR1; this would imply $F_\rho^2=F_{a_1}^2$. SR2 would then conclude $M_\rho^2=M_{a_1}^2$---i.e.,  the Wigner phase, as would be expected in a linear sigma model, with degenerate massive parity doubling; this sharply contrasts with our result $M_{a_1}^2 \simeq 2 M_\rho^2$.\footnote{In a walking theory 
there actually is no reason for SR2  to be valid, since  $\gamma_m \simeq 1$  yields a slower damping UV behavior $\sim ( \langle \bar q q\rangle^{(R)})^2\cdot q^{2\gamma_m}/q^6\sim 1/q^4$ of the difference between the vector and axial vector current correlators, instead of $1/q^6$ in the QCD (where $\gamma_m\simeq 0$). The KSRF relation may also change in walking theories, as shown in  the Hidden Local Symmetry framework~\cite{Harada:2003jx}. Nevertheless Eq.~(\ref{chirallimitdata})---the same as the Weinberg mass relation---can also follow in walking theories in the NG phase, 
without recourse to Weinberg's SR2 and the KSRF relation, as will be described below. 
}
  
The second relation in Eq.~(\ref{chirallimitdata}) is a novel result also consistent with the broken phase.  The unbroken chiral symmetry would be consistent with a linear sigma model for $N_f>2$ case, particularly when $N_f \gg N_c$ as is the case in our study. The chiral partner should be  
the parity-doubling flavor-non-singlet {$N_f^2-1$ pairs} $(\pi, a_0)$---instead of  $(\pi,\sigma)$ in the $N_f=2$ case---which are only half
 of the 
 {singlet/non-singlet $2 N_f^2$} $q {\bar q}$ bound states, excluding the other half $(a_0, \eta)$. Then the unbroken chiral symmetry created by the parity doubling
would imply the degeneracy $M_\pi^2=M_{a_0}^2$, in sharp contrast to our result $M_\pi^2\ll M_{a_0}^2 \,(\simeq M_\rho^2)$.
 
 To further understand the second relation $M_{a_0}^2 \simeq M_\rho^2$ together with the first one in Eq.~(\ref{chirallimitdata}), we recall the once-fashionable ``representation mixing''~\cite{Gilman:1967qs,Weinberg:1969hw}, in which resonance saturation  of the Adler-Weisberger sum rule (which is obtained for the spontaneously broken chiral algebra in the infinite momentum frame) occurs. A modern formulation of this method is called ``Mended Symmetry''~\cite{Weinberg:1990xn}, which targets ordinary QCD (and its simple scaled-up version of technicolor, but not walking technicolor).
In contrast to our study of large $N_f (\gg N_c)$ QCD as a walking theory, the analysis in~\cite{Weinberg:1990xn}  is crucially based on the large $N_c (\gg N_f=2)$ limit, with singlet-non-singlet degeneracy (the ``nonet scheme''), and the $N_f=2$ peculiarity of pseudo-real fermion representations,
 ${\underbar 2}^*\simeq {\underbar 2}$, in such a way that the unbroken chiral partner  in the linear sigma model can be identified as $(\pi,\sigma)$. 
In our case  $N_f=8 \ne 2$, on the other hand,  the chiral partner of $\pi$ is $a_0$ but not the flavor-singlet scalar  $\sigma$ as mentioned above, although  our data for $N_f=8$ imply that $M_\sigma \simeq M_\pi$ (See the results in the later section). 
 
Consider two one-particle states $\alpha$ and $\beta$, with collinear momenta ${\vec p}=(p^+=\frac{p^0+p^3}{\sqrt{2}},p^1,p^2)$ and ${\vec q}=(q^+,q^1,q^2)$, and helicity $\lambda, \lambda^\prime$, respectively. The axial charge in the infinite momentum frame (or equivalently the light-like axial charge), has a matrix element between $\alpha$ and $\beta$ which coincides with the Weinberg's $X$-matrix of axial charge (an analogue of the $g_A$ for the nucleon matrix)~\cite{Weinberg:1969hw}: 
\begin{eqnarray}
\langle {\vec p},\lambda, \alpha|{\hat Q}_{5 a}|{\vec q},\lambda^\prime, \beta \rangle&=&\langle {\vec p},\lambda, \alpha|\int dx^- dx^1dx^2 J^+_{5 a}(x)|{\vec q},\lambda^\prime, \beta \rangle\nonumber \\
&=& (2\pi)^3 \delta^{(3)}({\vec p}- {\vec q})\cdot 2 p^+\delta_{\lambda \, \lambda^\prime} [X_a (\lambda)]_{\alpha\,\beta}
\,,
\end{eqnarray}
where $J^+_{5 a}=(J^0_{5a}+J^3_{5a})/\sqrt{2}$. Like the nucleon $g_A$ term, the $X_a(\lambda)$ has no $\pi$  pole term (which would have the form $\sim (p^+ - q^+)/[(p-q)^2 -M^2_\pi]=0$)
for the collinear momentum, even in the chiral limit $M^2_\pi \rightarrow 0$. (The absence of the $\pi$ pole in the corresponding light-like charge was rigorously shown in the original paper of the discrete light-cone quantization~\cite{Maskawa:1975ky}, and hence 
gives well-defined classification algebra, even in the spontaneously broken phase~\cite{Yamawaki:1998cy}.)
On the other hand, the absence of the $\pi$ pole term means that it does not commute with ${\rm Mass}^2=M^2= P_\mu^2=2 P_+P_- - P_1^2- P_2^2$.\footnote{
The current conservation balances the non-pole term, $X$ matrix ($g_A$), with the pole term $\pi$ emission vertex ($G_{\pi NN}$), which yields the Goldberger-Treiman relation.
}  Namely, the physical states (mass eigenstates) are not in the irreducible representation in general,  but in a mixed representation.

For the helicity $\lambda=0$ states, physical flavor-non-singlet meson states  fall into four possible chiral representations~\cite{Weinberg:1969hw,Ida:1973ec}: For even normality ($P=(-1)^J$) we have 
 $|\rho\rangle = |(N_f^2 -1, 1) + (1, N_f^2-1) \rangle$,  $|a_0\rangle=| (N_f, N_f^*) + (N_f^*,N_f)\rangle$,  while for odd normality ($P=-(-1)^J$)
we have  $\pi$ and $a_1$ as admixtures of  $|(N_f^2-1, 1) - (1,N_f^2-1)\rangle$ and $|(N_f, N_f^*)- (N_f^*,N_f)\rangle$. Since normality $P\cdot (-1)^J$ commutes with $M^2$, we have 
\begin{eqnarray}
M^2_\rho &=& M^2_{(N_f^2 -1, 1) + (1, N_f^2-1)}=M^2_{(N_f^2 -1, 1) - (1, N_f^2-1)}=M^2_\pi \cos^2\theta + M^2_{a_1} \sin^2\theta  \nonumber\\
M^2_{a_0} &=& M^2_{(N_f, N_f^*) + (N_f^*,N_f)}=M^2_{(N_f, N_f^*) - (N_f^*,N_f)}=M^2_\pi \sin^2\theta + M^2_{a_1} \cos^2\theta  \,,
\label{repmix}
\end{eqnarray}
which also yields a $\theta$-independent relation
\begin{equation}
	M^2_\rho+M^2_{a_0}=M^2_\pi+ M^2_{a_1}.
\label{eq:angleind}
\end{equation}
Thus for the broken phase $M^2_\pi \ll M^2_{H\ne\pi}$, we have $M^2_{a_1}\simeq M^2_\rho+M^2_{a_0}$. More specifically, if ``ideal mixing'' $\tan^2 \theta=1$ is imposed in Eq.\ (\ref{repmix}),
then setting $M_\pi^2\simeq 0$ just reproduces 
our data $ M^2_\rho \simeq  M^2_{a_0}\simeq M^2_{a_1}/2$ (i.e. Eq.\ (\ref{chirallimitdata})).

It is tempting to compare this with our data in $N_f=12$ \cite{Aoki:2013zsa}
  (see also the updated results: Figs.\ \ref{fig:nf12_ratio_mrho}, \ref{fig:nf12_ratio_a0-pi}, \ref{fig:nf12_ratio_a1-a0}, and \ref{fig:nf12_ratio_a1-rho} in Appendix \ref{app:nf12}):
\begin{eqnarray}
 M^2_\rho/M^2_\pi &\simeq& (1.2)^2\,  \ll M^2_{a_0}/M^2_\pi\simeq (1.4)^2 \,,\nonumber\\
M^2_{a_1}/M^2_{a_0} &\simeq& (1.05)^2\ll M^2_{a_1}/M^2_\rho \simeq (1.25)^2.  
\label{Nf12data}
\end{eqnarray}   
Here, all the masses---including $\pi$---obey the {\it universal} hyperscaling relation with $\gamma_m \simeq 0.4$, consistent with the conformal window (in contrast to $N_f=8$, as will be discussed in the next section).
The parity-doubling degeneracy $M^2_{a_0}/M^2_\pi=M^2_{a_1}/M^2_\rho=1$ is again badly broken, similarly to $N_f=8$, but for a different reason. 
In the conformal window, there are no bound states in the exact chiral limit $m_f\equiv 0$; for this reason, states are often dubbed ``unparticles'', and bound states are possible only when the explicit mass $m_f\ne0$ exists, so that the chiral symmetry is essentially broken explicitly. The phase is a weakly interacting Coulomb phase where the non-relativistic bound state mass is roughly twice the current quark mass, in conformity with hyperscaling: $M_H \sim 2\, m_f^{(R)} \sim m_f^{1/(1+\gamma_m)}$ (see discussions in the Introduction).\footnote{The above mass ratios are in rough agreement  with the S-wave degeneracy $\rho/\pi$ (up to
spin-spin interaction splitting) and P-wave degeneracy $a_1/a_0$, in contrast to $N_f=8$. 
} 
Thus even without spontaneous breaking and the NG $\pi$ pole,
the representation mixing should take the same form as Eq.~(\ref{repmix}).
Then
the angle-independent relation (Eq.~(\ref{eq:angleind}) is again expected, and is indeed in rough agreement with $N_f=12$ data Eq.~(\ref{Nf12data}). In this case an alternative Wigner phase known as ``Vector Manifestation''~\cite{Harada:2003jx}, with $M^2_\rho/M^2_\pi= M^2_{a_1}/M^2_{a_0}=1$ {($\theta=0$)}, seems better than the conventional linear sigma type Wigner phase with massive parity-doubling degeneracy, $M^2_{a_0}/M^2_\pi=M^2_{a_1}/M^2_\rho=1$($\theta=\pi/2$).

}

\clearpage

\section{Hyperscaling Analyses}
\label{sec:FSHS}

In the previous section we have performed ChPT based analyses for the $N_f = 8$ hadron spectra and suggested that the system is consistent with being in the S$\chi$SB phase.
In this section we adopt a conformal hyperscaling ansatz as an alternative hypothesis.
We highlight characteristic properties of the $N_f = 8$ spectra distinct from those in $N_f = 12$ QCD as well as $N_f = 4$ QCD.

In Sec.~\ref{subsec:hs}, we analyze the spectrum data with the naive hyperscaling ansatz and evaluate a (would-be) mass anomalous dimension ($\gamma$).
Next in Sec.~\ref{subsec:geff}, we shed light on the fate of $\gamma$ near the chiral limit via the {\em effective} mass anomalous dimension ($\gamma_{\mathrm{eff}}$) defined as a function of bare fermion mass $m_f$.
Finally in Sec.~\ref{subsec:fshs}, we further elaborate the scaling properties of the spectra based on the Finite-Size Hyper-Scaling (FSHS) analyses.

\subsection{Hyperscaling Fit}\label{subsec:hs}

If $N_f = 8$ QCD is in the conformal window,
hadron mass spectra $M_H$ should scale as
\begin{align}
M_{H} = C^{M_H} m_f^{1/(1 + \gamma)}\ ,\label{eq:hs_naive}
\end{align}
for sufficiently small $m_f$ in the continuum and thermodynamic limits.
The critical exponent $\gamma$ is known as the mass anomalous dimension associated with the infrared fixed point (IRFP).
While the coefficient $C^{M_H}$ may be operator dependent, $\gamma$ should be universal.
If the $N_f = 8$ system is in the hadronic phase but near to the conformal window,
we expect that the scaling law~(\ref{eq:hs_naive}) approximately holds
with a ``would-be'' mass anomalous dimension, which may lose the robust universality.
This naive expectation is based on past Schwinger-Dyson studies~\cite{Aoki:2012ve}
and is supported by our previous work~\cite{Aoki:2013xza}.

\begin{figure}[!tbp]
\begin{center}
\includegraphics[width=7.5cm]{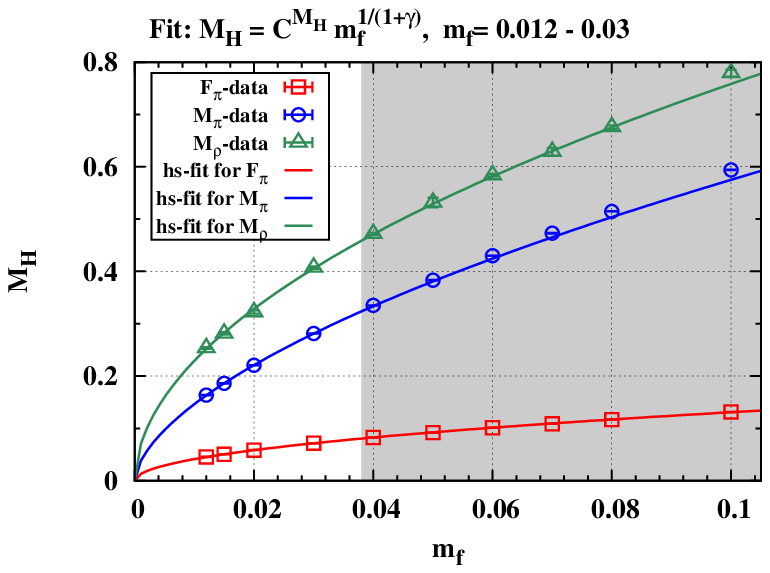}
\includegraphics[width=7.5cm]{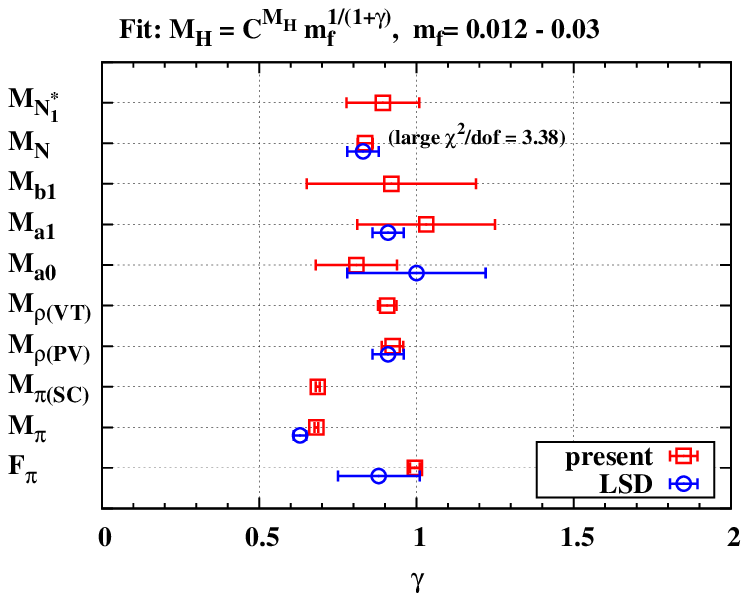}
\caption{
Left:
The hyperscaling fit~(\protect\ref{eq:hs_naive}) for $F_{\pi},\ M_{\pi}$ or, $M_{\rho}$ data.
The fit range is set to $m_f = 0.012 - 0.03$; the shaded region has been excluded.
Right:
The mass anomalous dimension $\gamma$ obtained by
the hyperscaling fit~(\protect\ref{eq:hs_naive}) for various observables (red squares).
The fit range is set to $m_f = 0.012 - 0.03$.
See also Table~\protect\ref{tab:gam_hs_ca00a00_nf08b0380_obsall_dat10}.
For comparison, we have quoted the results reported by the LSD Collaboration (blue circles):
fit results for the mass range of $0.015-0.03$ with d.o.f. = 2.
in Table X in Ref.~\cite{Appelquist:2014zsa}.
}\label{fig:hs_nf08b0380}
\end{center}
\end{figure}

We adopt the conformal hyperscaling ansatz~(\ref{eq:hs_naive}),
and investigate the same spectrum data set as the previous section
(the Large Volume Data Set, Table~\ref{tab:data_set}).
We first select three observables $F_{\pi}$, $M_{\pi}$, and $M_{\rho}$
which were investigated in the previous work~\cite{Aoki:2013xza},
and fit their $m_f$ dependences with Eq.~(\ref{eq:hs_naive})
for the mass range $m_f = 0.012 - 0.03$.
Figure~\ref{fig:hs_nf08b0380} shows the three observables
as a function of $m_f$ with hyperscaling fit lines.
The fit works with relatively small $\chi^2/\mathrm{dof}$,
but $\gamma$ is found to be operator dependent:
$(\gamma,\chi^2/\mathrm{dof}) = (0.995(15), 0.65),\ (0.682(6), 1.74)$, and $(0.924(34), 2.98)$
for $F_{\pi}$, $M_{\pi}$, and $M_{\rho}$, respectively.
The non-universal property of $\gamma$ holds for other fit ranges as shown
in Table~\ref{tab:gam_hs_ca00a00_nf08b0380_obs3a_datm}.
We find that the fit quality for $M_{\pi}$ becomes worse significantly for wider fit ranges.
Thus, $N_f = 8$ QCD spectra partly show conformal-like scaling
but something different from a universal one.

The results explained above are, in principle, consistent with our previous work~\cite{Aoki:2013xza},
while there are some modifications to be noted:
the hyperscaling ansatz~(\ref{eq:hs_naive}) in our previous work
failed to explain $F_{\pi}$ data in the small mass region,
and this trend has disappeared in the present work.
The modifications result from the updates in the small mass region;
we have added new data point at $m_f = 0.012$,
and two to ten times larger statistics are accumulated for $m_f = 0.015 - 0.03$
for which the central values of the spectra have been modified around one percent or less,
slightly beyond the statistical errors in several cases.
However, the main conclusion in the previous work~\cite{Aoki:2013xza}
(the non-universal $\gamma$ and the large $\chi^2/{\mathrm{dof}}$ for $M_{\pi}$)
remains true in the present study independently of the above modifications.

A relevant question is how such a non-universal hyperscaling law has emerged
in the spectrum data of $N_f = 8$ QCD.
One possibility is that $N_f = 8$ QCD is in the chirally broken phase
but the system is very close to the conformal window,
and the system still possesses a remnant of the conformal dynamics.
Another possibility is that
$N_f = 8$ QCD is in the conformal window
and the conformal dynamics is contaminated by explicit breaking effects,
such as lattice spacing, finite $m_f$, and lattice volume $L$ effects.

\begin{table}[!tbp]
\caption{
The mass anomalous dimension $\gamma$ obtained by
the hyperscaling fit~(\protect\ref{eq:hs_naive}) for $F_{\pi}$, $M_{\pi}$, or $M_{\rho}$
in various fit ranges.
\label{tab:gam_hs_ca00a00_nf08b0380_obs3a_datm}
}
\begin{ruledtabular}

\begin{tabular}{l c ld ld ld}
\multicolumn{1}{c}{fit range ($m_f$)} &
\multicolumn{1}{c}{dof} &
\multicolumn{1}{c}{$\gamma(F_{\pi})$} &
\multicolumn{1}{c}{$\chi^2/\mathrm{dof}(F_{\pi})$} &
\multicolumn{1}{c}{$\gamma(M_{\pi})$} &
\multicolumn{1}{c}{$\chi^2/\mathrm{dof}(M_{\pi})$} &
\multicolumn{1}{c}{$\gamma(M_{\rho})$} &
\multicolumn{1}{c}{$\chi^2/\mathrm{dof}(M_{\rho})$} \\
\hline
$0.012 - 0.03$ & $2$ & $0.995(15)$ & $0.65$ & $0.682(06)$ & $1.74$ & $0.924(34)$ & $2.98$ \\
$0.012 - 0.04$ & $3$ & $0.997(10)$ & $0.45$ & $0.668(04)$ & $4.20$ & $0.918(22)$ & $2.00$ \\
$0.012 - 0.05$ & $4$ & $1.006(09)$ & $1.59$ & $0.666(03)$ & $3.58$ & $0.917(22)$ & $1.51$ \\
$0.012 - 0.06$ & $5$ & $0.999(07)$ & $1.63$ & $0.652(02)$ & $10.72$ & $0.913(15)$ & $1.22$ \\
$0.012 - 0.07$ & $6$ & $1.003(06)$ & $1.73$ & $0.649(02)$ & $10.83$ & $0.915(14)$ & $1.06$ \\
$0.012 - 0.08$ & $7$ & $0.999(05)$ & $1.58$ & $0.638(02)$ & $19.26$ & $0.919(12)$ & $0.94$ \\
$0.012 - 0.10$ & $8$ & $0.992(05)$ & $2.41$ & $0.630(02)$ & $28.07$ & $0.908(12)$ & $1.67$ \\
\hline
$0.06 - 0.10$ & $2$ & $0.962(18)$ & $3.71$ & $0.584(05)$ & $3.25$ & $0.844(53)$ & $2.87$ \\
$0.05 - 0.10$ & $3$ & $0.950(15)$ & $2.96$ & $0.586(04)$ & $2.30$ & $0.850(51)$ & $1.97$ \\
$0.04 - 0.10$ & $4$ & $0.981(08)$ & $3.84$ & $0.605(03)$ & $8.07$ & $0.886(30)$ & $1.65$ \\
$0.03 - 0.10$ & $5$ & $0.988(06)$ & $3.42$ & $0.613(02)$ & $11.40$ & $0.904(22)$ & $1.48$ \\
$0.02 - 0.10$ & $6$ & $0.992(05)$ & $3.04$ & $0.619(02)$ & $16.33$ & $0.886(18)$ & $1.63$ \\
$0.015 - 0.10$ & $7$ & $0.991(05)$ & $2.61$ & $0.627(02)$ & $24.12$ & $0.891(15)$ & $1.43$ \\

\end{tabular}
\end{ruledtabular}
\end{table}

The $m_f$ and $L$ effects will be investigated in the later subsections,
and we shall here focus on the lattice spacing {artifacts}.
The important update from our previous work
is the collection of $\gamma$ from states other than $(F_{\pi},M_{\pi},M_{\rho})$.
The results are tabulated in Table~\ref{tab:gam_hs_ca00a00_nf08b0380_obsall_dat10}
and compared with those reported by LSD Collaboration~\cite{Appelquist:2014zsa}
in the right panel of Fig.~\ref{fig:hs_nf08b0380}.
In the latter, the domain wall fermion was adopted, in contrast to our choice (HISQ action).
Figures show that two different actions result in a consistent $\gamma$
with similar observable dependences.
This suggests that a non-universal $\gamma$
appears independently of lattice spacing artifacts.

\begin{table}[!tbp]
\caption{
The mass anomalous dimension $\gamma$ obtained by
the hyperscaling fit~(\protect\ref{eq:hs_naive})
for various observables.
The fit range is set to $m_f = 0.012 - 0.03$.
\label{tab:gam_hs_ca00a00_nf08b0380_obsall_dat10}
}
\begin{minipage}{.5\textwidth}
\begin{ruledtabular}
\begin{tabular}{lcc}
$M_H$ & $\gamma$ ($m_f = 0.012 - 0.03$) & $\chi^2/\mathrm{dof}$ \\
\hline
$F_{\pi}$                & $0.995(015)$ & $0.65$ \\
$M_{\pi}$                & $0.682(006)$ & $1.74$ \\
$M_{\pi \mathrm{(SC)}}$  & $0.686(006)$ & $1.68$ \\
$M_{\rho \mathrm{(PV)}}$ & $0.924(034)$ & $2.98$ \\
$M_{\rho \mathrm{(VT)}}$ & $0.907(029)$ & $2.25$ \\
$M_{a_0}$                & $0.809(129)$ & $0.08$ \\
$M_{a_1}$                & $1.031(219)$ & $0.89$ \\
$M_{b_1}$                & $0.920(269)$ & $0.15$ \\
$M_{N}$                  & $0.837(024)$ & $3.38$ \\
$M_{N_{\bf 1}^*}$                & $0.893(116)$ & $0.47$ \\
\end{tabular}
\end{ruledtabular}
\end{minipage}
\end{table}

\subsection{Effective mass anomalous dimension}\label{subsec:geff}

The mass anomalous dimension $\gamma$ toward the chiral limit $m_f\to 0$
is particularly interesting when considering applications to walking technicolor models.
To shed light on the chiral limit from the available data,
we investigate the {\em Effective Mass Anomalous Dimension} ($\gamma_{\mathrm{eff}}(m_f)$)
which is evaluated as follows;
first, we divide the fermion mass range of
the large volume data set (Table~\ref{tab:data_set}) into sub-blocks
with sequential three fermion masses,
and then, we fit the spectra in each sub-block
with the naive hyperscaling ansatz~(\ref{eq:fshs_naive}).
The exponent $\gamma$ is determined in each sub-block
as a function of $m_f$, giving $\gamma_{\mathrm{eff}}(m_f)$.

\begin{figure}[!tbp]
\begin{center}
\includegraphics[width=7.5cm]{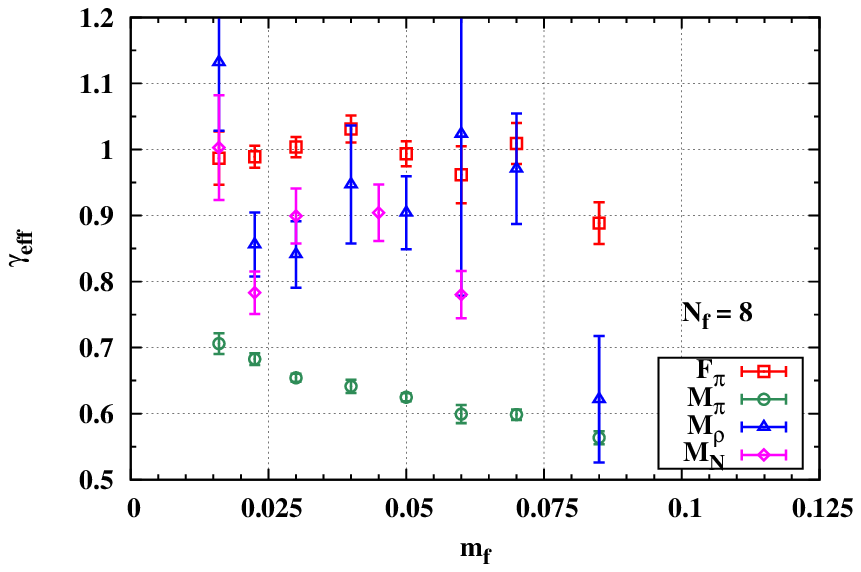}
\includegraphics[width=7.5cm]{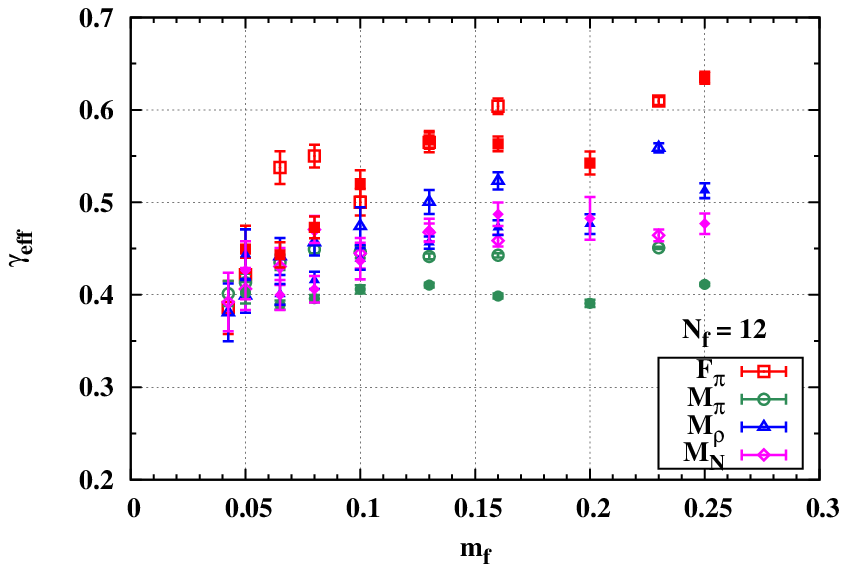}
\caption{
The effective mass anomalous dimensions $\gamma_{\mathrm{eff}}$ obtained by
the hyperscaling fit~(\protect\ref{eq:hs_naive})
for $N_f = 8$ (left, see Table~\protect\ref{tab:nf8b3.8gamma} for details)
and $12$ (right, see Tables~\protect\ref{tab:nf12b4gamma} and \protect\ref{tab:nf12b3.7gamma} for details)
spectrum data.
In the right panel, the open/filled symbols represent the results with $\beta = 3.7/4.0$.
The fit range is slid by keeping the fit degrees of freedom 1.
The value of $m_f$ in the horizontal axis is
the average of the maximum and minimum among three fermion masses.
}\label{fig:geff}
\end{center}
\end{figure}

In the left panel of Fig.~\ref{fig:geff},
we show $\gamma_{\mathrm{eff}}$ evaluated from
the data set of $F_{\pi}$, $M_{\pi}$, $M_{\rho}$ as a function of $m_f$.
Here we also include the nucleon mass $M_N$.
The horizontal axis is the average of the maximum and minimum among three fermion masses in each sub-block.
$\gamma_{\mathrm{eff}}$ for $M_{\pi}$ (green circles) clearly increases with decreasing $m_f$
and it appears to approach $\sim 1$, implying the dynamics of the broken chiral symmetry;
if a system is in the chirally broken phase,
the ChPT predicts $M_{\pi}\propto m_f^{1/2}$, which is identical to $\gamma_{\mathrm{eff}} = 1$.
$\gamma_{\mathrm{eff}}$ should be contrasted to the $N_f = 12$ results shown in the right panel;
$\gamma_{\mathrm{eff}}$ for $M_{\pi}$ never approaches 1
and all $\gamma_{\mathrm{eff}}$ meet at smaller $m_f$,
indicating conformal dynamics with a universal $\gamma\sim 0.4$.

The available data for $N_f = 8$ would not be enough to exclude a conformal scenario;
there is a possibility that all $\gamma_{\mathrm{eff}}$ meet somewhere near $1$
toward the chiral limit. In addition, all $\gamma_{\mathrm{eff}}$ except for $M_{\pi}$ should blow up
toward the chiral limit, which would be a smoking gun of
chiral symmetry breaking and has not been observed yet.
As such, to get more conclusive statement, we need additional data in the smaller mass region,
which is considered as a target for future work.

\subsection{Finite-Size Hyperscaling Analyses}\label{subsec:fshs}

\subsubsection{Preliminaries}

\begin{figure}[!tbp]
\begin{center}
\includegraphics[width=7.5cm]{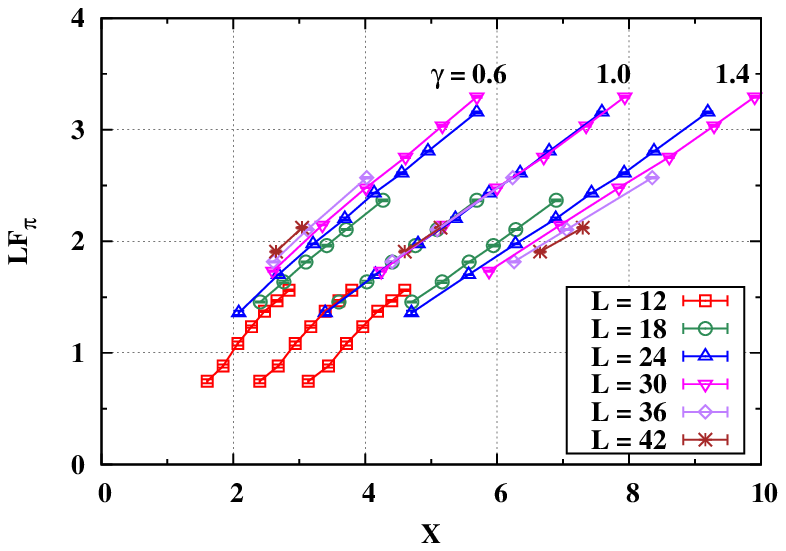}
\includegraphics[width=7.5cm]{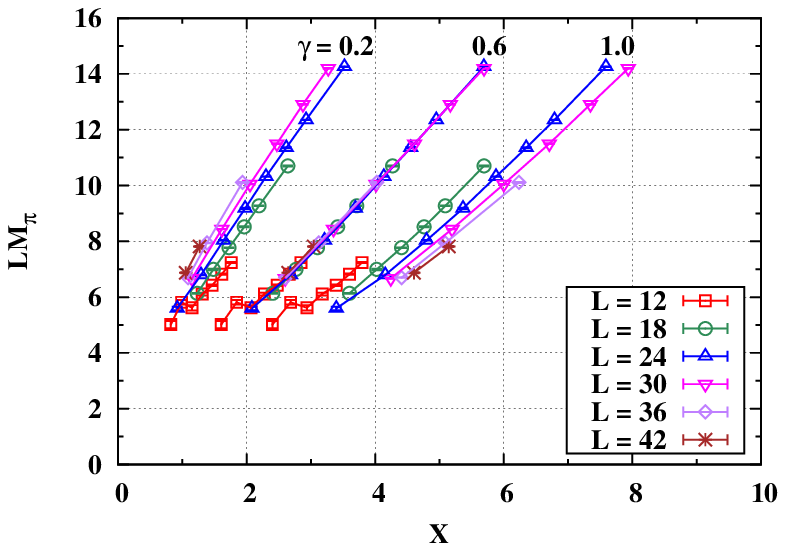}
\includegraphics[width=7.5cm]{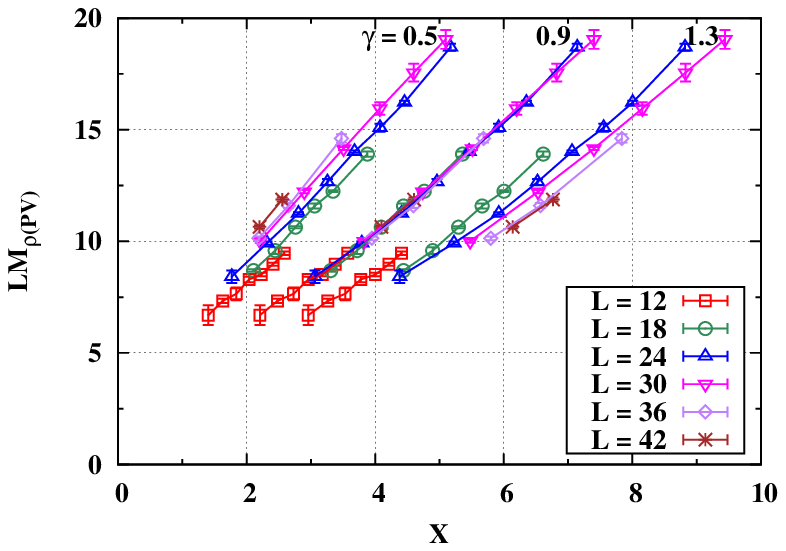}
\includegraphics[width=7.5cm]{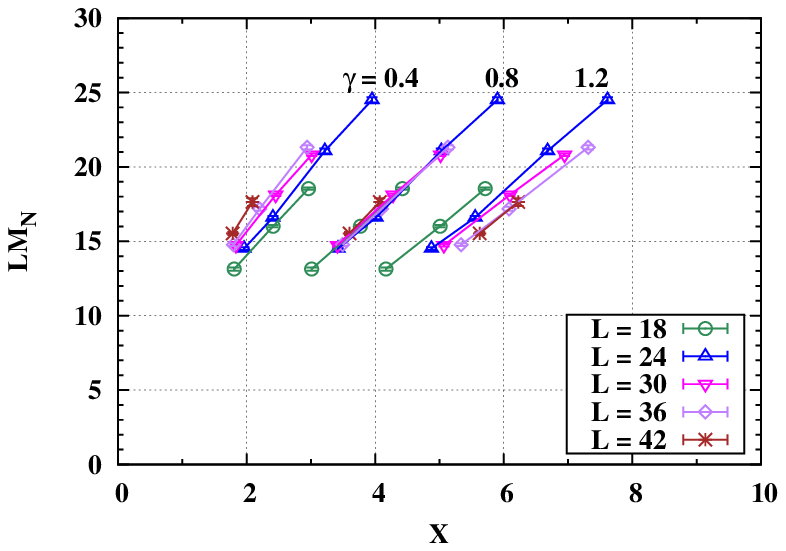}
\caption{
The FSHS test for $N_f = 8$:
the normalized mass spectra
$LF_{\pi}$ (upper left),
$LM_{\pi}$ (upper right),
$LM_{\rho}$ (lower left), and
$LM_{N}$ (lower right)
are plotted as a function of the scaling variable $X=Lm_f^{1/(1+\gamma)}$
for selected $\gamma$.
}\label{fig:fshs_gamxx_nf08b0380_obs3}
\end{center}
\end{figure}

We shall now upgrade the hyperscaling ansatz~(\ref{eq:hs_naive})
to take account of effects of the finite lattice volume $L$.
In many-flavor QCD theories having an IRFP, the fermion mass $m_f$ and the gauge coupling
act as relevant and irrelevant operators respectively, in the renormalization group (RG) flow.
For a sufficiently small $m_f$ and large $L$,
the RG~\cite{DelDebbio:2010ze} dictates the finite-size hyperscaling (FSHS) law,
\begin{align}
& LM_H = F(X,A_{\mathrm{crr}})\ , \label{eq:fshs_x}\\
& X\equiv Lm_f^{1/(1+\gamma)}\ .\label{eq:x}
\end{align}
{
At the IRFP ($m_f \to 0$) in the conformal window,
the function $F$ depends on only the scaling variable $X$.
The $A_{\mathrm{crr}}$ denotes corrections to this scaling, {\em i.e.},
effects of an irrelevant operator and/or a chiral symmetry breaking.
In general, $F$ is an arbitrary function of the scaling variable $X$,
and in practice, one needs to specify its functional expression.
The most probable argument is that the $F(X, A_{\mathrm{crr}})$ should reproduce
the infinite volume hyperscaling formula (\ref{eq:hs_naive}) in the thermodynamic limit ($L\to\infty$),
which indicates the asymptotic formula,}
\begin{align}
LM_H = F(X,A_{\mathrm{crr}})|_{L=\text{large}} \to C_0^{M_H} + C_1^{M_H} X\ .\label{eq:fshs_naive}
\end{align}
{
A $F(X,A_{\mathrm{crr}})$ beyond the asymptotic expression is not well known.
If one wants to include a non-linear terms of $X$, one needs some assumption for its expression,
which leads to a source of theoretical ambiguities. 
In this subsection to avoid such ambiguities, 
we consider only the linear $X$ ansatz in Eq.~(\ref{eq:fshs_naive}).
This strategy is based on the following observation; as shown in Fig.~\ref{fig:fshs_gamxx_nf08b0380_obs3},
the $F_{\pi}$ (upper left), $M_{\pi}$ (upper right), $M_{\rho}$ (lower left), and $M_N$ (lower right),
approximately align for $\gamma = 1.0$, $0.6$, $0.9$, and $0.8$, respectively,
and thus linearly depends on $X$ up to anomalous behavior at small $X$.
This motivates the use of linear $X$ ansatz without the small $X$ data.
In practice, the small $X$ non-linearity are excluded by selecting the spectral data with
the parameter sets $(L,m_f)$ satisfying $LM_{\pi} > 8$
and with $\{(L,m_f)=(42,0.012),(42,0.015),(36,0.015),(36,0.02)\}$.
We refer to the data set as the {\em ``FSHS-'' Large Volume Data Set}, which will be used in the following analyses.
Details of the data selection scheme are summarized in Appendix~\ref{app_subsec:gams}.
}

\begin{figure}[!tbp]
\begin{center}
\includegraphics[width=7.5cm]{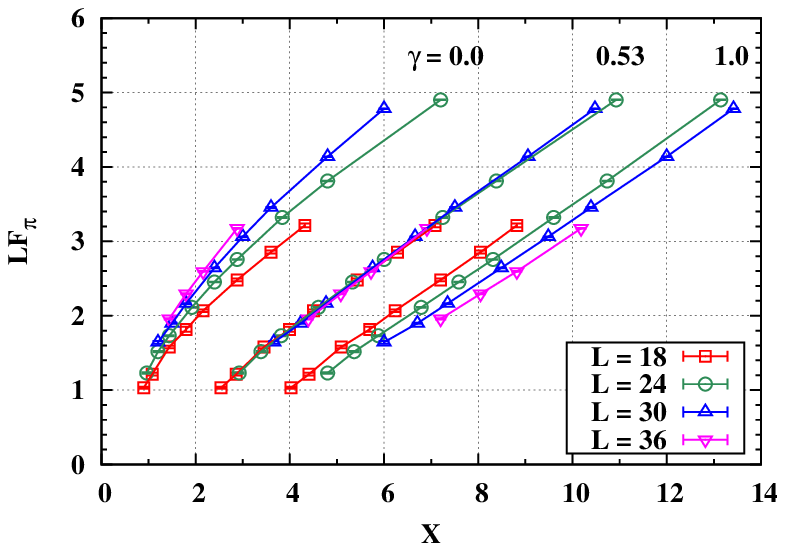}
\includegraphics[width=7.5cm]{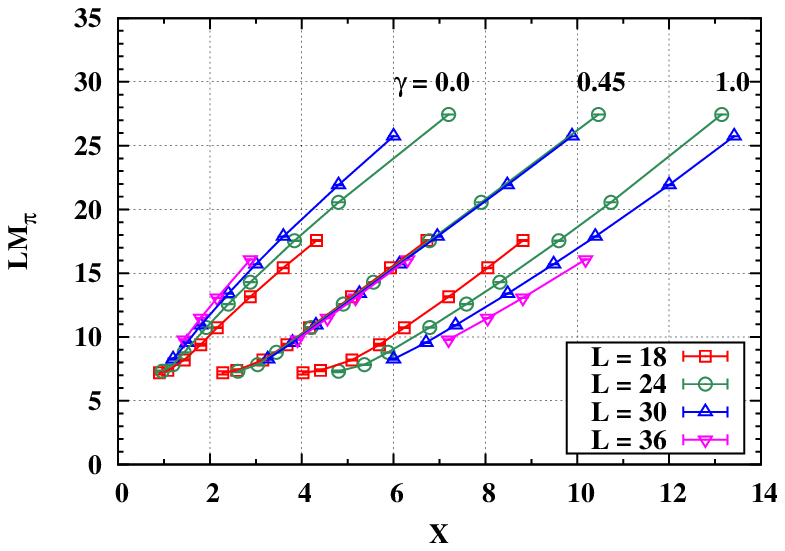}
\includegraphics[width=7.5cm]{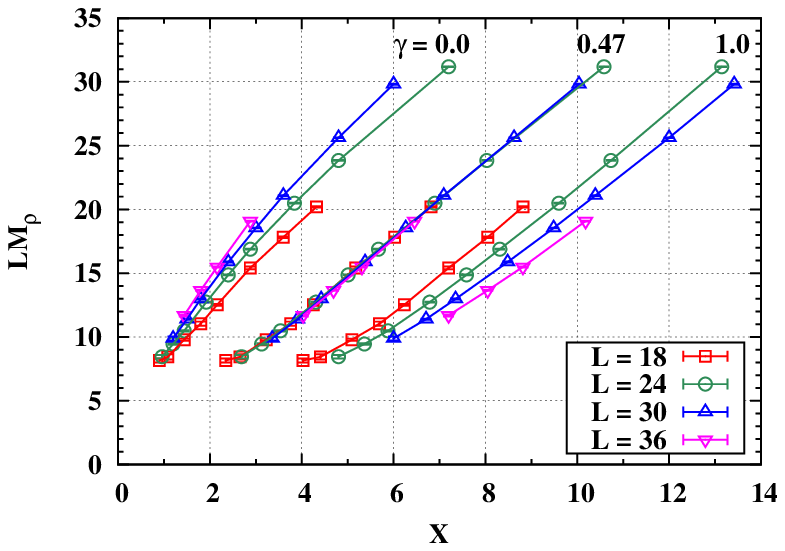}
\includegraphics[width=7.5cm]{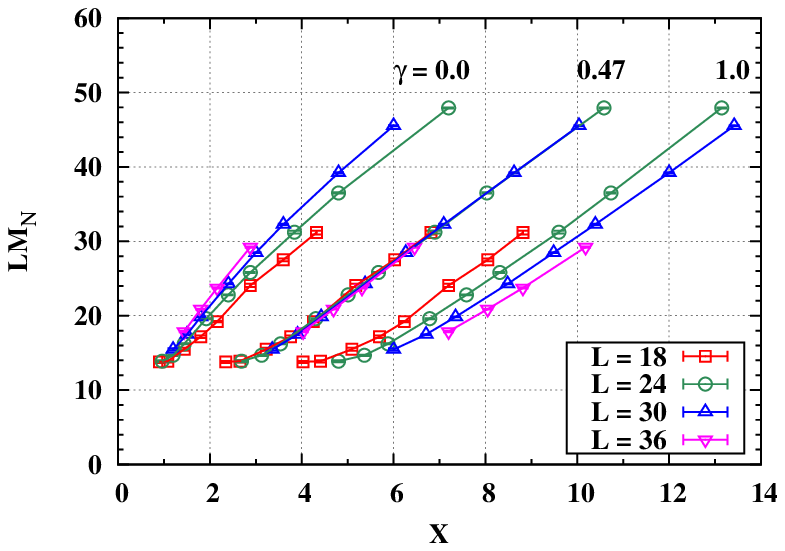}
\caption{
The FSHS test for $N_f = 12$ spectra with $\beta = 4.0$;
the normalized mass spectra
$LF_{\pi}$ (upper left),
$LM_{\pi}$ (upper right),
$LM_{\rho}$ (lower left), and
$LM_{N}$ (lower right)
are plotted as a function of the scaling variable $X=Lm_f^{1/(1+\gamma)}$ for selected $\gamma$.
The spectral data are found in Appendix~\protect\ref{app:nf12}.
}\label{fig:fshs_gamxx_nf12}
\end{center}
\end{figure}

\begin{figure}[!tbp]
\begin{center}
\includegraphics[width=5cm]{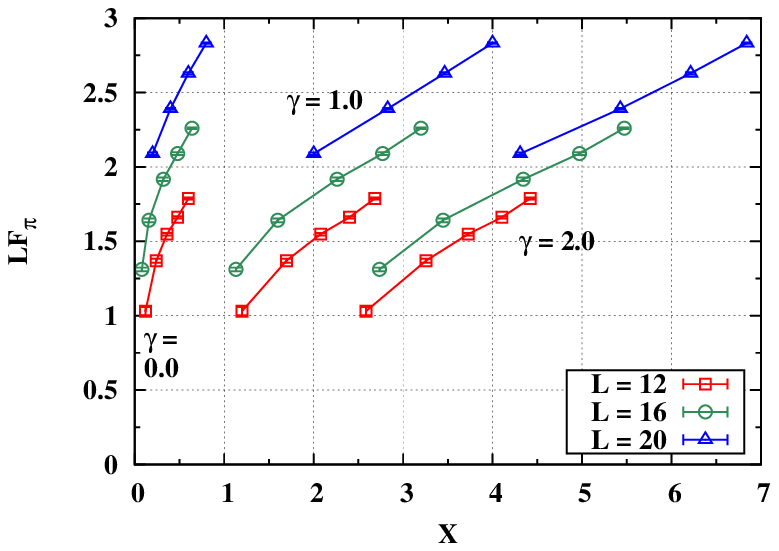}
\includegraphics[width=5cm]{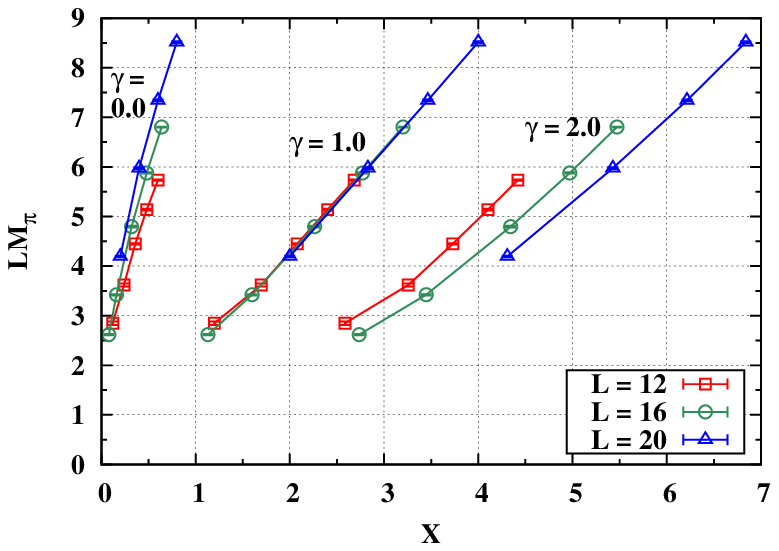}
\includegraphics[width=5cm]{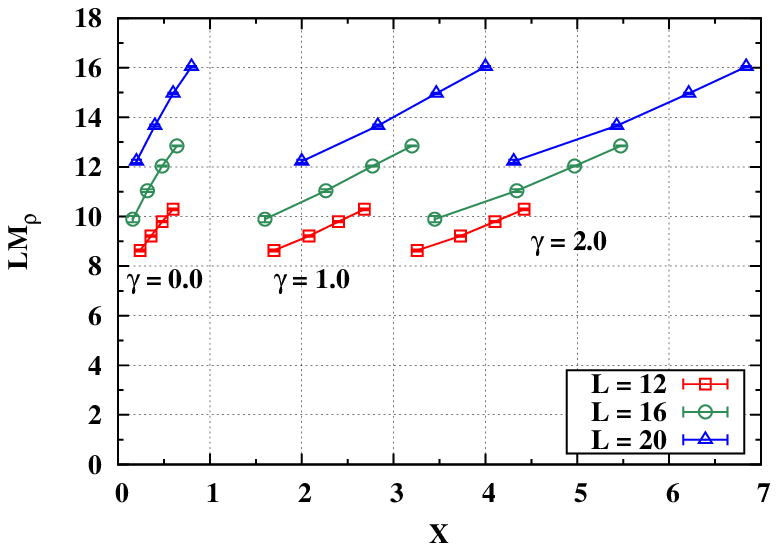}
\caption{
The FSHS test for $N_f = 4$;
the normalized mass spectra
$LF_{\pi}$ (upper left),
$LM_{\pi}$ (upper right), and
$LM_{\rho}$ (lower left)
are plotted as a function of the scaling variable $X=Lm_f^{1/(1+\gamma)}$ for
selected $\gamma$.
}\label{fig:fshs_gamxx_nf04b0370}
\end{center}
\end{figure}

{
By construction, the FSHS-Large Volume Data Set should approximately scale as Eq.~(\ref{eq:fshs_naive}).
This does not necessarily mean the complete conformal nature of $N_f = 8$ QCD since the optimal ${\gamma}$
obtained from the alignment depends on the observable. This will be the focus of the next subsection
in terms of the correction terms $A_{\mathrm{crr}}$.
For comparison, we show the same figure for $N_f = 12$ in Fig.~\ref{fig:fshs_gamxx_nf12},
where the optimal $\gamma$ with the alignment shows a much milder dependence on the observables.
The comparison to the $N_f = 4$ case (Fig.~\ref{fig:fshs_gamxx_nf04b0370}) is also interesting;
the $F_{\pi}$ and $M_{\rho}$ for $N_f = 4$ {\em never} show any alignment-like behavior
within the unitary band $\gamma \in [0,2]$ and thus the conformal invariance is completely spoiled
by the chiral symmetry breaking. For $M_{\pi}$, the alignment takes place at $\gamma = 1.0$
which is consistent to the leading-order ChPT prediction: $M_{\pi}\propto m_f^{1/2}$.
}

{
The FSHS-Large Volume Data Set does not necessarily exclude the spectra with small $m_f$
as long as the linear X dependence holds.}
This is in contrast to the previous work~\cite{Aoki:2013xza}
where the spectrum data with $m_f < 0.05$ was excluded in the FSHS analysis.
As will be shown later,
the update of the data selection scheme leads to only a minor modification to the results.

\begin{figure}[!tbp]
\begin{center}
\includegraphics[width=8cm]{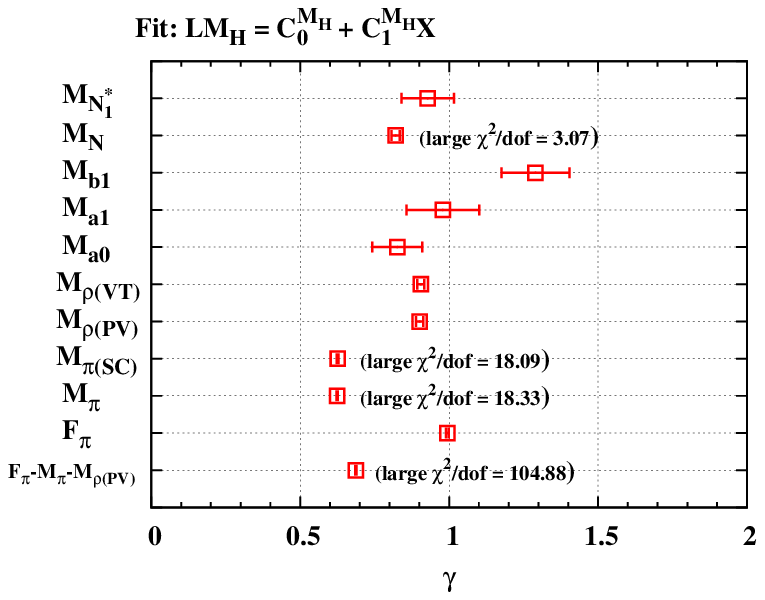}
\caption{
The mass anomalous dimension $\gamma$ obtained by
FSHS fits for various observables with the ansatz~(\protect\ref{eq:fshs_naive}).
See also Table~\protect\ref{tab:gam_fshs_xx_nf08b0380_obs3a_datMLp678x}.
}\label{fig:gam_fshs_nf08b0380_obsall_datLMpx}
\end{center}
\end{figure}

In Fig.~\ref{fig:gam_fshs_nf08b0380_obsall_datLMpx},
we show the value for $\gamma$ obtained by the FSHS fit for various quantities.
(See Table~\ref{tab:gam_fshs_crr_nf08b0380_obs3a_datMLp678x} for numerical details.)
The results are similar to those obtained in the naive hyperscaling fits.
The observable dependences of $\gamma$ remain even in the FSHS with finite volume effects
being considered, and rather become manifest with smaller statistical uncertainty
owing to the larger number of degrees of freedom.
It is also important that
the pion mass $M_{\pi}$ does not respect the FSHS, as indicated by
the considerably large $\chi^{2}/\mathrm{dof}\sim 18$.
Thus, the incompleteness of the conformal dynamics
seems to be a generic feature of $N_f = 8$ QCD independently of
the finite size effects.

\begin{table}[!tbp]
\caption{
The mass anomalous dimension $\gamma$ obtained by
the FSHS fit~(\protect\ref{eq:fshs_naive}) for various observables
with FSHS-Large Volume data set.
}\label{tab:gam_fshs_crr_nf08b0380_obs3a_datMLp678x}
\begin{minipage}{.5\textwidth}
\begin{ruledtabular}
\begin{tabular}{lcc}
$M_H$ & $\gamma$ & $\chi^2/\mathrm{dof}$ \\
\hline
$F_{\pi}$                & $0.994(05)$ & $1.83$  \\
$M_{\pi}$                & $0.624(02)$ & $18.33$ \\
$M_{\pi \mathrm{(SC)}}$  & $0.626(03)$ & $18.09$ \\
$M_{\rho \mathrm{(PV)}}$ & $0.901(12)$ & $1.32$  \\
$M_{\rho \mathrm{(VT)}}$ & $0.905(12)$ & $1.16$  \\
$M_{a_0}$                & $0.826(84)$ & $0.69$  \\
$M_{a_1}$                & $0.979(22)$ & $1.68$  \\
$M_{b_1}$                & $1.290(14)$ & $0.91$  \\
$M_{N}$                  & $0.820(14)$ & $3.07$  \\
$M_{N_{\bf 1}^*}$                & $0.928(88)$ & $1.19$  \\
\hline
$F_{\pi}$-$M_{\pi}$-$M_{\rho \mathrm{(PV)}}$ & $0.687(2)$ & $104.88$ \\
\end{tabular}
\end{ruledtabular}
\end{minipage}
\end{table}

\subsubsection{Simultaneous fit with {Naive} FSHS ansatz}\label{ssub:fshs_naive}

So far, we have investigated the FSHS scaling for each observable individually,
and obtained the observable-dependent $\gamma$.
We shall now perform a simultaneous FSHS fit,
where we construct the {\em combined} spectral data including various observables
and {\em impose} a universal $\gamma$ among them.
We select the three observables $\{F_{\pi},M_{\pi},M_{\rho}\}$ as the combined spectra;
the data $\{F_{\pi},M_{\pi},M_{\rho}\}$ are available for the widest $m_f$ range
from the FSHS-Large Volume data set, and the data quality is better than the other observables.
Another practical reason for the choice is that
those three observables were used in the previous study~\cite{Aoki:2013xza}
and allows us to compare the present result with the previous one.

{
Our focus is whether the FSHS fit $LM_H = F(X, A_{\mathrm{crr}})$ works or fails
for a common $\gamma$ with or without correction terms $A_{\mathrm{crr}}$.
In this subsection and the following subsections, 
we consider three fit models, $F(X, A_{\mathrm{crr}})$, which are summarized in 
Table~\ref{tab:fshs_fit_model}.
Our FSHS fit is carried out in the following procedure;
\begin{enumerate}
\item
Construct normalized-combined spectral data $\{LF_{\pi}, LM_{\pi}, LM_{\rho}\}(L, m_f)$.
\item
We perform fits for the combined data by adopting each ansatz summarized in Table XIV.
The would-be mass anomalous dimension $\gamma$ ($\in X = L m_f^{1/(1+{\gamma})}$) is a common fit parameter.
In the models with correction terms shown in Table~\ref{tab:fshs_fit_model},
the second exponent $\alpha$ or $\omega$ is also a common fit parameter
(For $\alpha$, we consider also fixed cases).
In contrast, the coefficients $C_i^{M_H}$ are the observable dependent fit parameters.
\item
From each fit, we obtain the $\gamma$ and the other fit parameters with ${\chi}^2/\mathrm{dof}$.
A reasonable fit quality with small ${\chi}^2/\mathrm{dof}$ indicates the existence of a universal $\gamma$.
In this case, if the correction term $A_{\mathrm{crr}}$ is much smaller than the leading term $C_1^{M_H} X$,
the universal $\gamma$ can be identified with the mass anomalous dimension and
the system is interpreted to be in the conformal window.
As will be shown later (Sec.~\ref{subsec:discuss}), this is not the case for $N_f = 8$.
\item
Solve the FSHS formula $LM_H = F(X,A_{\mathrm{crr}})$ in terms of the $X$ formally:
\begin{align}
X = F^{-1}(LM_H, A_{\mathrm{crr}}) (\equiv Y)\ ,\label{eq:xy}
\end{align}
The explicit form of $Y$ in each model is summarized in the third column of Table~\ref{tab:fshs_fit_model}.
The left-hand side $X$ is determined by the lattice parameters $(L, m_f)$ and fit results of $\gamma$.
The right-hand side $Y$ is obtained from $(L, m_f, M_H)$ and results of the fit
parameters other than $\gamma$. The fit quality becomes visible in the $X-Y$ plane;
if a fit model works/fails, spectral data points align/misalign on the $Y = X$ line.
\end{enumerate}
}

\begin{table}[!tbp]
\caption{
The summary table of FSHS fit models:
$LM_H = F(X,A_{\mathrm{crr}})$ with $X = Lm_f^{1/(1 + \gamma)}$.}\label{tab:fshs_fit_model}
{
\begin{ruledtabular}
\begin{tabular}{lcc}
Model Name &
$F(X, A_{\mathrm{crr}})$ &
$Y = F^{-1}(LM_H, A_{\mathrm{crr}})$ \\
\hline
Naive FSHS &
$C_0^{M_H} + C_1^{M_H} X$ &
$(LM_H - C_0^{M_H})/C_1^{M_H}$ \\
FSHS with Power Crr. &
$C_0^{M_H} + C_1^{M_H} X + C_2^{M_H} L m_f^{\alpha}$ &
$(LM_H - C_0^{M_H} - C_2^{M_H} L m_f^{\alpha})/C_1^{M_H}$ \\
FSHS with RG. Crr. &
$(1 + C_2^{M_H}m_f^{\omega})(C_0^{M_H} + C_1^{M_H} X)$ &
$(LM_H / (1 + C_2^{M_H}m_f^{\omega}) - C_0^{M_H}) / C_1^{M_H}$
\end{tabular}
\end{ruledtabular}
}
\end{table}

In this subsection,
we carry out the simultaneous fit with the leading-order FSHS ansatz~(\ref{eq:fshs_naive}),
for which we have,
\begin{align}
Y = \frac{LM_H - C_0^{M_H}}{C_1^{M_H}}\ .
\label{eq:fshs_naive_xy}
\end{align}
In the left panel of Fig.~\ref{fig:fshs_nf08b0380_obx3a_datLMpx},
we plot the combined data $\{F_{\pi}, M_{\pi}, M_{\rho}\}$ in the $X - Y$ plane.
The data are scattered around the $X = Y$ fit line beyond the statistical errors,
and thus the FSHS fit~(\ref{eq:fshs_naive}) fails:
$(\gamma,\chi^2/\mathrm{dof}) = (0.687(2),104.88)$.
The result corroborates the lack of a universal $\gamma$.

\subsubsection{FSHS simultaneous fit with power-law correction}\label{ssub:fshs_pow}

In Sec.~\ref{subsec:geff},
we have observed the sizable $m_f$ dependence of
the effective mass anomalous dimension $\gamma_{\mathrm{eff}}(m_f)$
evaluated from the pion mass spectrum.
This is indicative of the necessity of
a mass modified FSHS to describe the mass spectra.
In this regard, we consider the FSHS ansatz~\cite{Aoki:2013xza,Aoki:2012eq},
\begin{align}
LM_H = F(X,A_{\mathrm{crr}}) = C_0^{M_H} + C_1^{M_H} X + C_2^{M_H}Lm_f^{\alpha}\ .\label{eq:fshs_pow}
\end{align}
Here, the linear term of $Lm_f^{\alpha}$ can be regarded as
the effect of the correction $A_{\mathrm{crr}}$.
Among various choices of the second exponent $\alpha$,
we consider two possibilities: $\alpha = 1$ and $2$.
In the former case, the correction term
compensates the conformal symmetry breaking due to finite $m_f$.
In the latter case, the correction term accounts for $\mathrm{O}(a^2)$ discretization effects.

We show the fit result for $\alpha = 1$
in the middle panel of Fig.~\ref{fig:fshs_nf08b0380_obx3a_datLMpx},
where the vertical axis is
\begin{align}
Y = \frac{LM_H - C_0^{M_H} - C^{M_H}_2 Lm_f^{\alpha}}{C_1^{M_H}}
\ ,\label{eq:fshs_pow_xy}
\end{align}
followed by the definition~(\ref{eq:xy}) applied to the present ansatz (\ref{eq:fshs_pow}).
The figure shows that the data distribute on the $Y=X$ fit line
and thus the fit quality is greatly improved owing to the power correction term:
\begin{align}
(\gamma,\chi^2/{\mathrm{dof}}) = (0.929(14), 2.03)\ (\alpha = 1)\ .\label{eq:res_fshs_pow_a10}
\end{align}
We will discuss the interpretation for this result in the next subsection.
For $\alpha = 2$, the fit quality becomes better than that in the leading-order FSHS,
but still far from the acceptable level: $(\gamma,\chi^2/\mathrm{dof}) = (0.770(04), 21.32)$.
Thus the lattice spacing correction does not play an important role.
This is consistent with what we have speculated in the previous section (Fig.~\ref{fig:hs_nf08b0380})
by the coincidence of $\gamma$ in different lattice actions.

Here are three remarks in order.
First, if we treat the second exponent $\alpha$ as a fit parameter rather than a constant,
the FSHS fit~(\ref{eq:fshs_pow}) leads to a successful result:
\begin{align}
(\gamma,\alpha) = (1.109(49), 0.821(22))\ ,\quad
\chi^2/{\mathrm{dof}} = 1.05\ .\label{eq:res_fshs_pow_best}
\end{align}
The interpretation will be discussed in the following subsection, Sec.~\ref{subsec:discuss}.
Second, the results with the power-law correction explained above are
qualitatively consistent to our previous results~\cite{Aoki:2013xza}.
And finally, we have also fitted using the Schwinger-Dyson (SD) motivated ansatz~\cite{Aoki:2012ve},
where the second exponent is fixed as $\alpha = (3 - 2\gamma)/(1 + \gamma)$.
We find that the $\chi^2$ over the parameter space $\{\gamma,C_0^{M_H},C_1^{M_H},C_2^{M_H}\}$
contains approximately flat directions due to the parameter redundancy;
for example, for $\gamma = 1$, the $C_1^{M_H}$- and $C_2^{M_H}$-terms are identical,
which gives rise to the flat direction parametrized by $C_1^{M_H} + C_2^{M_H} = const$.
For $\gamma$ not far from $1$, the flat direction remains, at least approximately,
and prevents us from precisely determining the parameters.
Here, we report the rough estimate of the fit result, $\gamma \sim 0.8$,
for which the fit quality is not of an acceptable level, $\chi^2/{\mathrm{dof}} \sim 5$.
When we consider only the spectrum data with $m_f \geq 0.05$ in the FSHS-Large Volume Data Set
as done in our previous work~\cite{Aoki:2013xza}, the result becomes closer to the previous one.

\subsubsection{Simultaneous fit by FSHS with RG-motivated correction}\label{ssub:fshs_ah}

If some of irrelevant operators in the RG are nearly marginal,
the correction term $A_{\mathrm{crr}}$
in the FSHS formula~(\ref{eq:fshs_x}) may be dominated by the irrelevant operator $g$.
From this viewpoint, we adopt the FSHS ansatz
proposed by the recent lattice work Ref.~\cite{Cheng:2013xha},
\begin{align}
LM_H = F(X,A_{\mathrm{crr}}) = (1+C_2^{M_H}(g)m_f^{\omega})(C_0^{M_H} + C_1^{M_H} X)\ .\label{eq:fshs_ah}
\end{align}
Here, the $A_{\mathrm{crr}}$ has been set to be the $C_2^{M_H}(g)$ terms
which are responsible for the irrelevant operator $g$.
In lattice gauge theories, the operator $g$ is the gauge coupling,
or equivalently, the lattice spacing effect.

Although the FSHS analyses in the present study
do not combine the data computed at different lattice spacings,
it is still interesting to consider the ansatz~(\ref{eq:fshs_ah}) in an
alternate formulation. If
we assume the $X$-term is dominant and the others are subdominant in Eq.~(\ref{eq:fshs_ah}),
\begin{align}
C_1^{M_H}X\gg C_0^{M_H}, \ \ 
1 \gg C_2^{M_H}m_f^{\omega}\ ,\label{eq:fshs_ah_ordcnt}
\end{align}
and rewrite the ansatz~(\ref{eq:fshs_ah}) within the next-to-leading order,
\begin{align}
&F(X,A_{\mathrm{crr}})
\simeq C_0^{M_H} + C_1^{M_H}X + \bar{C}_2^{M_H}(g)Lm_f^{\bar{\alpha}}\ ,\label{eq:fshs_ah2}\\
&\bar{C}_2^{M_H}\equiv C_1^{M_H}C_2^{M_H}\ ,\quad
\bar{\alpha}\equiv\frac{1}{1 + \gamma} + \omega\ ,\label{eq:c12}
\end{align}
then the expression is analogous with FSHS including power correction term~(\ref{eq:fshs_pow}).
Thus, the ansatz~(\ref{eq:fshs_ah}) may be regarded as a modified version of (\ref{eq:fshs_pow})
with respect to the renormalization group argument.

In contrast to the original work~\cite{Cheng:2013xha} for $N_f = 12$ QCD,
the value of the second exponent $\omega$ cannot be determined by the two-loop beta function
in the present study, since an IRFP does not appear in $N_f = 8$ QCD---at least in the two-loop approximation.
We treat $\omega$ as a fit parameter, and thus two exponents $(\gamma,\omega)$
are determined by fitting the spectra.
As shown in Appendix~\ref{app_subsec:fshs_rg},
we have also performed a global parameter search for $\omega$,
and confirmed that the $\omega$ obtained by the fit realizes the global minimum of $\chi^2/\mathrm{dof}$.

\begin{figure}[!tbp]
\begin{center}
\includegraphics[width=5cm]{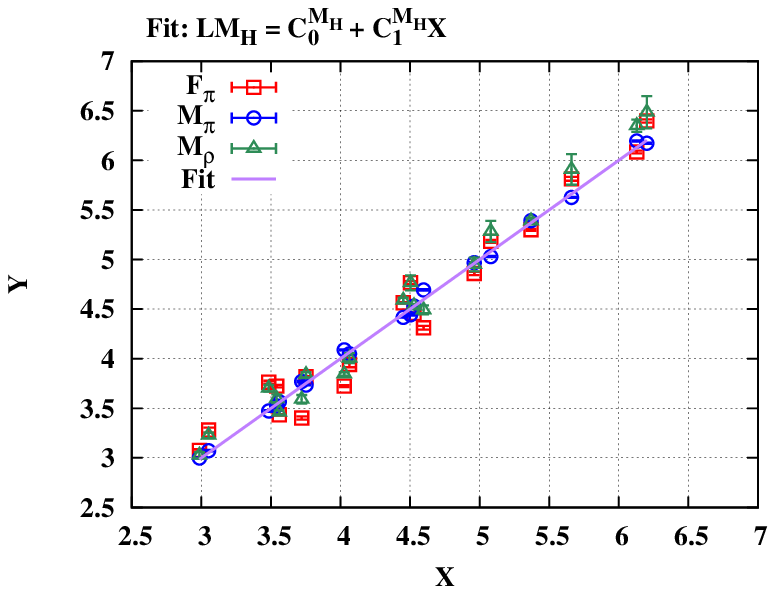}
\includegraphics[width=5cm]{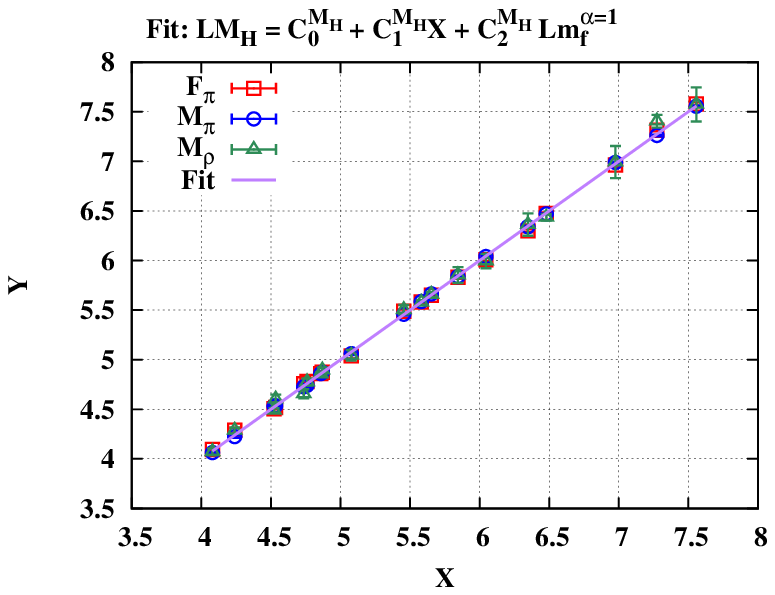}
\includegraphics[width=5cm]{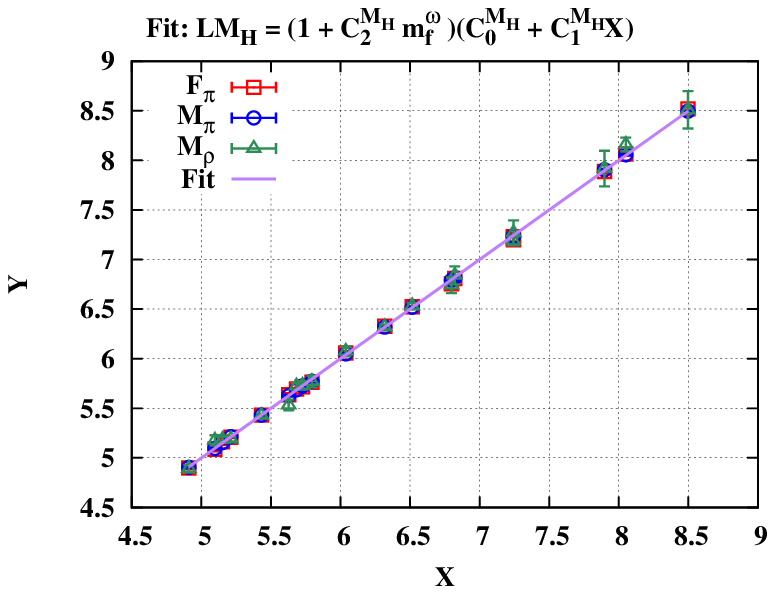}
\end{center}
\caption{
The simultaneous FSHS fit
of combined data for $\{F_{\pi},\ M_{\pi},\ M_{\rho}\}$.
For $\gamma$ obtained from the fits
(see, Table~\protect\ref{tab:gam_fshs_nf08b0380_obx3a_datLMpx}),
the data
$\{F_{\pi} \text{(red boxes)}, M_{\pi} \text{(green circles)}, M_{\rho} \text{(blue triangles)}\}$
are plotted in the $X-Y$ plane where $X = Lm_f^{1/(1 + \gamma)}$
and $Y$ is defined by Eq.~(\protect\ref{eq:fshs_naive_xy}),
(\protect\ref{eq:fshs_pow_xy}), or (\protect\ref{eq:fshs_ah_xy})
depending on the fit ansatz. For details, see the text.
The simultaneous fit line (solid purple) is given by $Y=X$.
Left: Naive FSHS~(\protect\ref{eq:fshs_naive}).
Middle: FSHS with power-law correction~(\protect\ref{eq:fshs_pow})
for the fixed exponent of $\alpha = 1$.
Right: FSHS with RG-motivated correction~(\protect\ref{eq:fshs_ah}).
}\label{fig:fshs_nf08b0380_obx3a_datLMpx}
\end{figure}

For the combined data $\{F_{\pi},M_{\pi},M_{\rho}\}$
constructed from the FSHS-Large Volume Data Set,
the RG-motivated FSHS fit~(\ref{eq:fshs_ah}) results
in the right panel of Fig.~\ref{fig:fshs_nf08b0380_obx3a_datLMpx},
where the vertical axis $Y$ reads
\begin{align}
Y = \frac{1}{C_1^{M_H}}\Bigl(\frac{LM_H}{1+C_2^{M_H}(g)m_f^{\omega}}-C_0^{M_H}\Bigr)
\ ,\label{eq:fshs_ah_xy}
\end{align}
followed by the definition~(\ref{eq:xy}) applied to the present ansatz (\ref{eq:fshs_ah}).
The data points distribute on the $Y=X$ fit line with the nice fit quality,
\begin{align}
(\gamma,\omega) = (1.108(48), 0.347(14))\ ,\quad \chi^2/{\mathrm{dof}} = 1.05\ .\label{eq:fshs_ah_res}
\end{align}
We will discuss the interpretation for this result in the next subsection.
The $\bar{\alpha}$ defined in Eq.~(\ref{eq:c12}) is found to be
\begin{align}
\bar{\alpha} = 0.821(18)\ .\label{eq:fshs_ah_abar}
\end{align}
As expected from the similarity between Eqs.~(\ref{eq:fshs_pow}) and (\ref{eq:fshs_ah2}),
the RG-motivated FSHS results, Eqs.~(\ref{eq:fshs_ah_res}) and (\ref{eq:fshs_ah_abar}),
are almost identical to those in the FSHS with the power-law correction~(\ref{eq:res_fshs_pow_best}).

\subsection{Discussions}\label{subsec:discuss}

\begin{figure}[!tbp]
\begin{center}
\includegraphics[width=5cm]{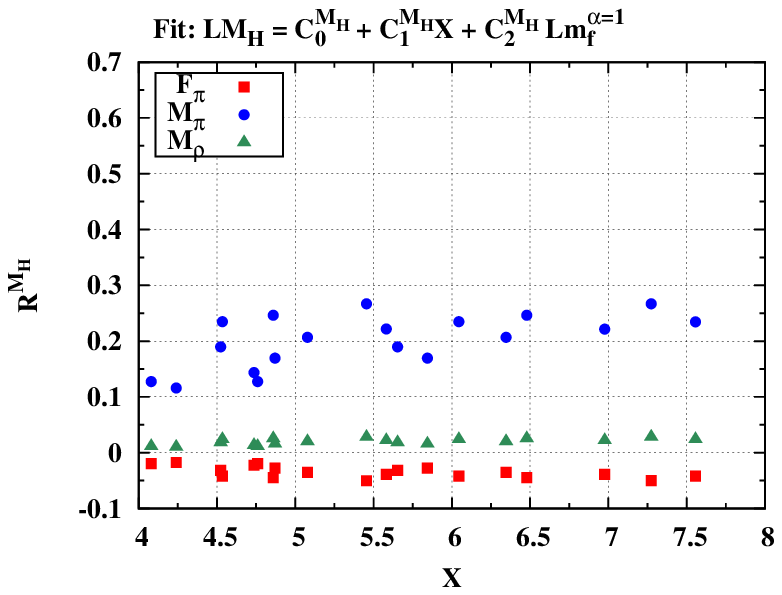}
\includegraphics[width=5cm]{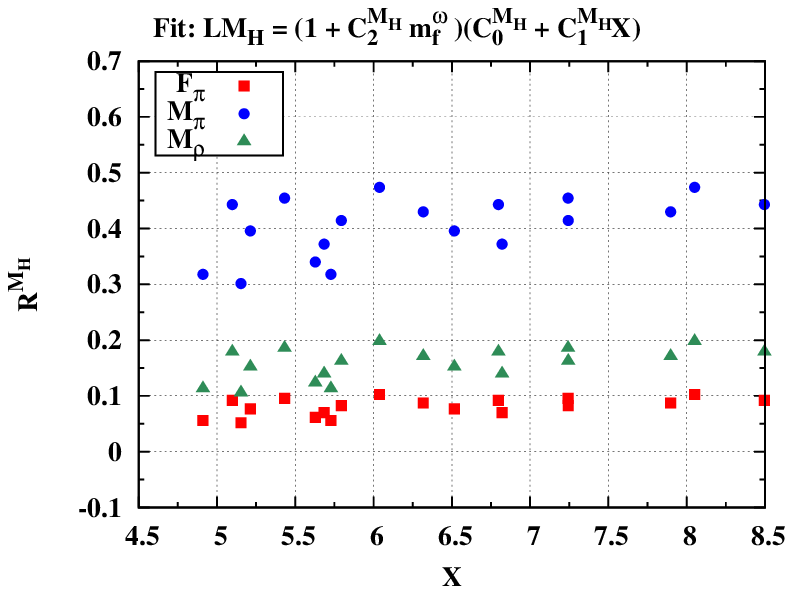}
\includegraphics[width=5cm]{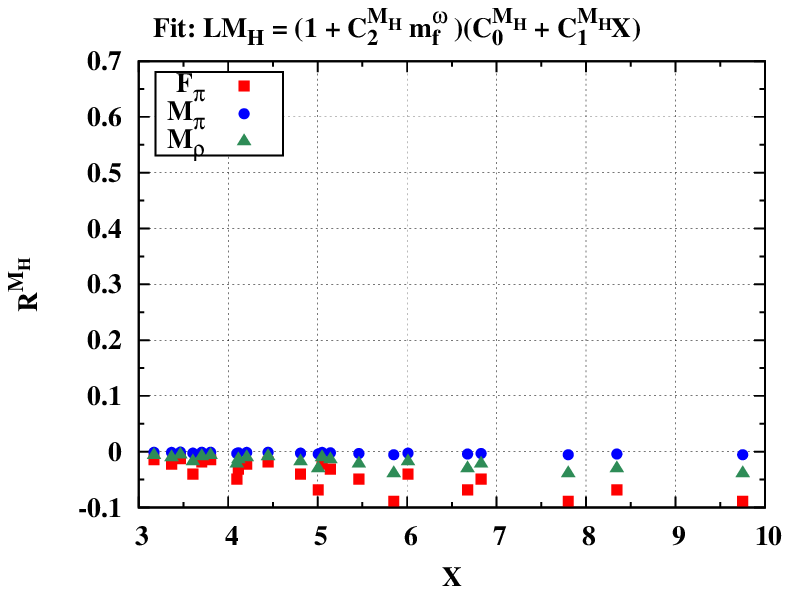}
\caption{
The strength of the correction term $R^{M_H}$ defined in Eq.~(\protect\ref{eq:r_fshs_crr}).
Left: In the case of FSHS with power-law correction ($\alpha = 1$) in $N_f = 8$ QCD.
Middle: In the case of FSHS with RG-motivated correction in $N_f = 8$ QCD.
Right: In the case of FSHS with RG-motivated correction in $N_f = 12$ QCD with $\beta = 3.7$.
}\label{fig:fshs_nf08b0380_obx3a_datLMpx_crr}
\end{center}
\end{figure}

We summarize the mass anomalous dimension
obtained so far from the FSHS simultaneous fits in Table~\ref{tab:gam_fshs_nf08b0380_obx3a_datLMpx}.
The fits have achieved an acceptable $\chi^2/{\mathrm{dof}}$
for the FSHS ansatz with power-law and RG-motivated corrections.
We discuss an interpretation for these results by using the ratio
\begin{align}
R^{M_H}(X)\equiv
\begin{cases}
C_2^{M_H}Lm^{\alpha}/LM_H
& \text{for Eq.~(\protect\ref{eq:fshs_pow})}\ ,\\
C_2^{M_H}m^{\omega}(C_0^{M_H}+C_1^{M_H}X)/LM_H
& \text{for Eq.~(\protect\ref{eq:fshs_ah})}\ .
\end{cases}
\label{eq:r_fshs_crr}
\end{align}
The ratio $R^{M_H}$ quantifies the strength of the correction terms.
In Fig.~\ref{fig:fshs_nf08b0380_obx3a_datLMpx_crr},
we show the $R^{M_H}$ as a function of $X=Lm_f^{1/(1+\gamma)}$.
The left and middle panels correspond to
the cases of the FSHS with power-law ($\alpha = 1$) and RG-motivated corrections, respectively.
In both cases, the correction associated with pions $R^{M_H=M_{\pi}}$
is considerably larger than those of the other observables.
Obviously, only the pion mass $M_{\pi}$ possesses
a different $m_f$ dependence from the others, which we have already seen as
the blowing up of $M_H/M_{\pi}$ at small $m_f$ (Fig.~\ref{fig:rat_fpi_mpv})
in the context of the chiral symmetry breaking in the previous section.

For the RG-motivated corrections (middle panel),
the correction associated with pions $R^{M_H=M_{\pi}}$
becomes almost fifty percent of the total fit function.
The exponent $\gamma$ obtained through such a fit
would not be regarded as a mass anomalous dimension any more.
The same problem potentially exists in the case of the power-law correction term (left panel):
the hierarchy of the leading and correction terms are less problematic for $\alpha = 1$,
but it increases for the best-fit value of $\alpha=0.821(22)$
to the same level as the RG-motivated FSHS case.

The above results suggest that
the fit has just {\em parametrically} absorbed the existing chiral dynamics
(the blowing up of $M_H/M_{\pi}$) into the large correction term associated with pion
to reconcile the $N_f = 8$ spectra with the universal $\gamma$.
In contrast in $N_f = 12$ QCD
(the right panel of Fig.~\ref{fig:fshs_nf08b0380_obx3a_datLMpx_crr}),
the correction terms stay subdominant,
and the correction associated with pions $R^{M_H = M_{\pi}}$
is the same order as the others.
These are regarded as the distinct properties of
the existence of the conformal theory
in the chiral limit of $N_f = 12$ QCD.

To summarize,
the universal $\gamma$ which has been achieved by the correction terms in $N_f = 8$ QCD
does not necessarily mean the theory lies within the conformal window,
and the analyses of $R^{M_H}$ rather implies that the chiral symmetry is broken---though the smoking gun for the symmetry breaking is still missing.
It is probable that $N_f = 8$ system is on the border of the chirally broken and conformal phases,
and thus $N_f = 8$ QCD is a fascinating theory as a candidate for a walking
technicolor model.

\begin{table}[!tbp]
\caption{
The mass anomalous dimension $\gamma$ obtained by
simultaneous FSHS fits~(\protect\ref{eq:fshs_naive}) -- (\protect\ref{eq:fshs_ah})
for combined data $\{F_{\pi},\ M_{\pi},\ M_{\rho}\}$.
The left-most column represents the fit ansatz.
\label{tab:gam_fshs_nf08b0380_obx3a_datLMpx}
}
\begin{ruledtabular}
\begin{tabular}{ldd}
\multicolumn{1}{c}{Fit Ansatz} &
\multicolumn{1}{c}{$\gamma$} &
\multicolumn{1}{c}{$\chi^2/\mathrm{dof}$} \\
\hline
Naive FSHS                        & $0.687(02)$ & $104.88$ \\
Power Crr. ($\alpha = 1$)         & $0.929(14)$ & $2.03$ \\
Power Crr. ($\alpha = 2$)         & $0.770(04)$ & $21.32$ \\
Power Crr. ($\alpha = 0.821(22)$) & $1.109(49)$ & $1.05$ \\
RG. Crr.   ($\omega = 0.347(14)$) & $1.108(48)$ & $1.05$ \\
\end{tabular}
\end{ruledtabular}
\end{table}

\clearpage

\section{Analysis of the string tension}
\label{sec:stringtension}

We investigate the string tension between two static quarks in $N_f = 8$ QCD from Wilson loops,
using the gauge ensembles stored for the study of the hadron mass spectra.
The fermion masses $m_f$ and lattice volumes $L$ are selected
to include the ``Large Volume Data Set'' (as defined in Table~\ref{tab:data_set})
used in the analyses of the hadron mass spectra.
There is one exception:
to obtain a better signal,
the lattice volume $L = 18$ is used
for the largest mass $m_f = 0.1$ instead of $L = 24$,
due to the significantly increased statistics available.
Finite volume effects are expected to be negligible
since the value of $LM_{\pi} = 10.7$ obtained for $(m_f, L) = (0.1,18)$
is in the safe region in Fig.~\ref{fig:mpi-fpi-deltaL}.

In order to obtain a better signal at large distance,
temporal link variables are HYP2 smeared in the spatial direction,
and spatial link variables are APE smeared in the spatial direction.
We have adopted one of the standard parameter sets
(see Refs.~\cite{Albanese:1987ds} and \cite{Hasenfratz:2001hp,DellaMorte:2003mn,DellaMorte:2005nwx} for details):
\begin{align}
&\text{HYP2 smearing: }
(\alpha_1,\alpha_2,\alpha_3) = (1.0,1.0,0.5)\ ,\label{eq:hyp2}\\
&\text{APE smearing: }
(N_{\mathrm{APE}},\alpha_{\mathrm{APE}}) = (20,0.5)\ .\label{eq:ape}
\end{align}

\begin{figure}[!tbp]
\begin{center}
\includegraphics[width=5cm]{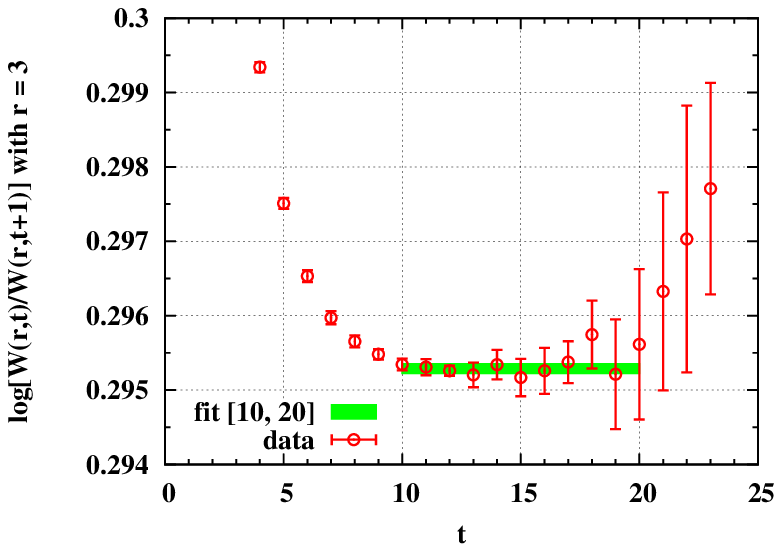}
\includegraphics[width=5cm]{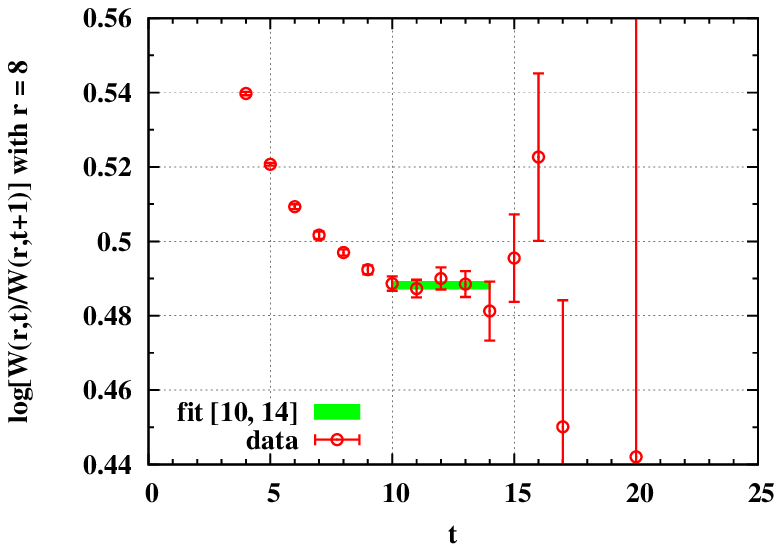}
\includegraphics[width=5cm]{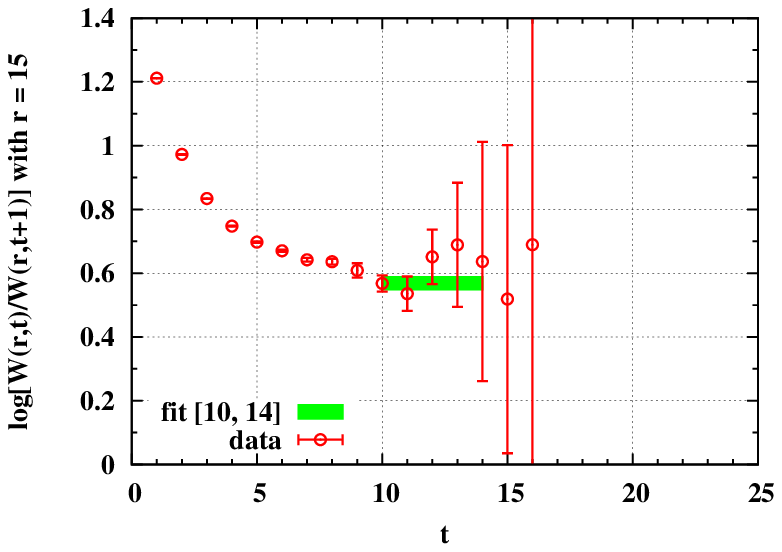}
\caption{
The Wilson loop potential $\mathcal{V}(r,t)$ given in Eq.~(\protect\ref{eq:meff})
for $m_f=0.012$ with $L=42$
as a function of time $t$
for selected spatial extensions $r=3$ (left), $8$ (middle), and $15$ (right).
The green band represents the results of the fit over the plateau.
}\label{Fig:meff_nf08b0380m0012L42T56HB4}
\end{center}
\end{figure}

We briefly explain the procedure to evaluate
the string tension from the measured Wilson loops.
Let us consider the potential associated with Wilson loops,
\begin{align}
\mathcal{V}(r,t)
&=
\bigg\langle\log
\frac{W(r,t)}{W(r,t+1)}
\bigg\rangle\ ,\label{eq:meff}
\end{align}
where the $W(r,t)$ represents the Wilson loop with extension $r\times t$,
and the bracket $\langle\cdots\rangle$ indicates taking a jackknife average.
For sufficiently large $t$, $\mathcal{V}(r,t)$ becomes constant,
which is interpreted as the potential between two static fermions,
\begin{align}
V(r)=\mathcal{V}(r,t)|_{t\in\text{plateau region}}\ .
\end{align}
In Fig.~\ref{Fig:meff_nf08b0380m0012L42T56HB4},
we show the $\mathcal{V}(r,t)$ obtained for $m_f = 0.012$ as a function of temporal extension $t$
for the selected spatial extension $r = 3$, $8$, and $15$.
In the left panel ($r = 3$),
the plateau appears at $10 \leq t \lesssim 20$,
and we determine the static fermion potential $V(r=3)$
from a constant fit over the plateau.
We repeat the same procedure for the other $r$
by keeping the lower edge of the fit range $t = 10$.
For a larger $r$, the plateau tends to diminish as seen
in the $r = 8$ (middle panel) and $15$ (right panel) cases,
where the statistical uncertainty rapidly increases.

\begin{figure}[!tbp]
\begin{center}
\includegraphics[width=7.5cm]{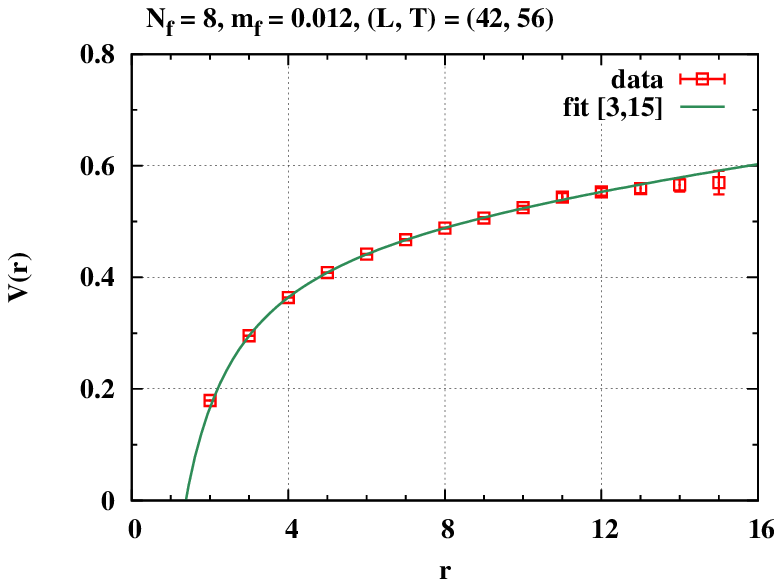}
\includegraphics[width=7.5cm]{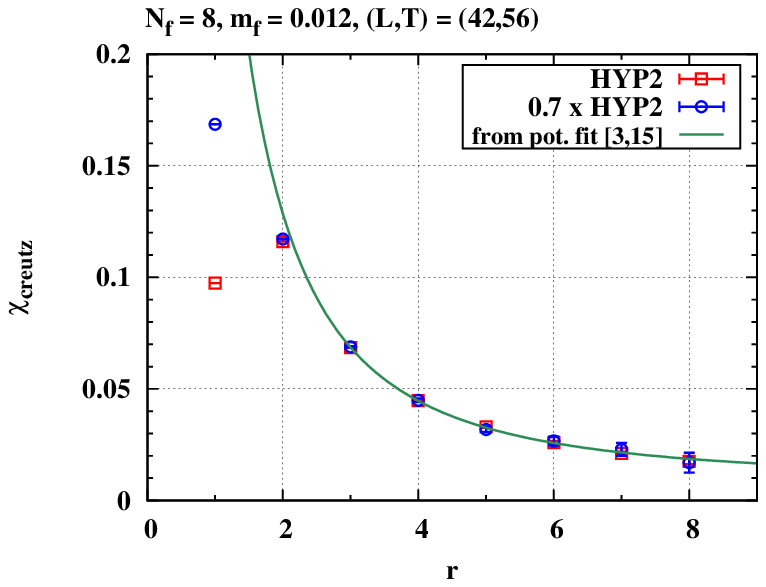}
\caption{
The static fermion potential (left) and the Creutz ratio (right)
as a function of spatial distance $r$ for $m_f=0.012$ with $L=42$.
In the left panel, the solid green line represents the fit line using data at $r = 3 - 15$.
Substituting the obtained $(\alpha,s)$ into Eq.~(\ref{eq:rc_ansatz}),
we have obtained the solid green line in the right panel.
}\label{Fig:pot1_nf08b0380m0012L42T56HB4_smr01}
\end{center}
\end{figure}

We extract the string tension $s$ (in lattice units)
by fitting the static potential with the ansatz
\begin{align}
V(r) = v_0 - \frac{\alpha}{r} + s\cdot r\ ,\label{eq:pot}
\end{align}
with $(v_0, \alpha, s)$ being fit parameters.
The static potential for each jackknife bin is fitted separately,
to obtain the per-bin string tension,
from which we evaluate the central value of the square root of the string tension $(\sqrt{s})$
and its statistical uncertainty $(\delta\sqrt{s})$.
In the left panel of Fig.~\ref{Fig:pot1_nf08b0380m0012L42T56HB4_smr01},
we show the static potential $V(r)$ in the case of $(m_f,L) = (0.012,42)$.
The solid green line represents the fit line for $r \in [3,15]$, giving $\sqrt{s} = 0.0931(25)$.
We repeat the above procedure for various fit ranges over $r$,
and select the $(\sqrt{s},\delta \sqrt{s})$
obtained from the widest fit range for which a reasonable $\chi^2/\mathrm{dof}$ holds.
The fits with other ranges give similar values,
from which we pick out the largest and smallest string tensions.
The differences between the largest/smallest one and the above central value
are used to quantify the systematic uncertainty.
Data for $\sqrt{s}$ including both the statistical and systematic uncertainties are summarized in Table~\ref{Tab:s_pot_nf08b0380}.

The lower bound of the fit range is set to avoid smearing artifacts.
The string tension $s$ is responsible for the large-distance behavior of the static potential,
while the smearing affects short distance scales.
Therefore, one expects that the $s$ is independent of the smearing.
In practice, however, the fit quality is affected by the smearing artifacts,
which prevent a precise determination of $s$.
Thus, we need to specify the smearing-free region.
To this end, we utilize the Creutz ratio,
\begin{align}
\chi_{\mathrm{creutz}}(r)
&=
\bigg\langle
\log
\frac{W(r,t)W(r+1,t+1)}{W(r,t+1)W(r+1,t)}
\bigg\rangle\bigg|_{t\in\text{plateau region}}\ .\label{eq:rc_eff}
\end{align}
The behavior of the Creutz ratio is expressed by using
$(\alpha,s)$ appearing in the potential fit function (Eq.~(\ref{eq:pot})) as
\begin{align}
\chi_{\mathrm{creutz}}(r) = s + \alpha\Bigl(\frac{1}{r} - \frac{1}{r + 1}\Bigr)
\ .\label{eq:rc_ansatz}
\end{align}
This expression reduces to the string tension $s$ itself at $r\to\infty$.
We note that the constant term $v_0$ in the static potential ansatz (Eq.~(\ref{eq:pot}))
is sensitive to the smearing, and it has been canceled out in Eq.~(\ref{eq:rc_ansatz}).
Thus, smearing artifacts would appear only in the Coulombic term coefficient $\alpha$ in Eq.~(\ref{eq:rc_ansatz})
and becomes negligible with increasing $r$.

In the right panel of Fig.~\ref{Fig:pot1_nf08b0380m0012L42T56HB4_smr01},
we compare the $\chi_{\mathrm{creutz}}(r)$ obtained by using different smearing levels
in the case of $(m_f,L) = (0.012,42)$:
one (the red squares) is the same as explained in Eqs.~(\ref{eq:hyp2}) and (\ref{eq:ape}),
while the other (the blue circles) is 70\% as strong;
i.e. the smearing parameters $\alpha_{1,2,3,\mathrm{APE}}$ in Eqs.~(\ref{eq:hyp2}) and (\ref{eq:ape})
are multiplied by $0.7$.
At $r = 1$, the results of the two smearing levels yield totally different $\chi_{\mathrm{creutz}}(r)$
and thus the artifacts dominate.
Therefore, we exclude the data at $r=1$ in the following analyses.
At $r = 2$, the difference between the two smearings becomes invisible
within the resolution of the figure, but still exists beyond the statistical errors.
Accordingly, the static potential fit (Eq.~(\ref{eq:pot})) including $r=2$ data
gives a large $\chi^2/\mathrm{dof}$ in most cases.
For $r \geq 3$, the two smearings give a consistent $\chi_{\mathrm{creutz}}$ within the statistical errors.
This holds independently of $(m_f,L)$, and
guarantees that the $s$ obtained by fitting the data at $r \geq 3$
with the potential fit ansatz (Eq.~(\ref{eq:pot})) should be free from smearing artifacts.
The solid green line in the right panel corresponds to Eq.~(\ref{eq:rc_ansatz})
with $(\alpha,s)$ being specified to those obtained in the potential fit (Eq.~(\ref{eq:pot})).
The line agrees with all data at $r\geq 3$,
confirming that the potential fit results and the Creutz ratio data are consistent.

\begin{figure}[!tbp]
\begin{center}
\includegraphics[width=8.0cm]{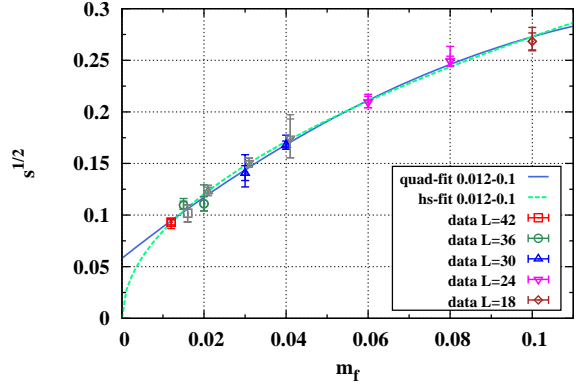}
\caption{
The square root of the string tension in lattice units
as a function of fermion mass.
The inner error-bars represent the statistical uncertainty.
In the outer error-bars, the statistical and systematic uncertainty
are added in quadrature.
The solid light blue and dashed light green lines show respectively
the fit results with the quadratic and conformal ansatz described in the text.
In the fits, only the statistical errors are taken into account.
The gray (slightly shifted) symbols were not used in the fit
but are shown to confirm that the finite volume effects are negligible.
}\label{Fig:s_pot_nf08b0380}
\end{center}
\end{figure}

We perform the above procedures at each fermion mass in the range $m_f = 0.012 - 0.1$.
The results for $\sqrt{s}$ are summarized in Table~\ref{Tab:s_pot_nf08b0380}.
In Fig.~\ref{Fig:s_pot_nf08b0380}, we plot $\sqrt{s}$ as a function of fermion mass,
with fit lines for the quadratic ansatz $\sqrt{s} = A_2\cdot m_f^2 + A_1\cdot m_f + A_0$
and the hyperscaling ansatz $\sqrt{s} = C\cdot m_f^{1/(1+\gamma)}$.
In the fits, only statistical errors are taken into account.
The gray (slightly shifted) symbols in the figure
are obtained from the data with smaller lattice volumes or statistics and are not used in the quadratic/hyperscaling fits,
but are shown to confirm that the finite volume effects are negligible.
The fitted parameters are found to be
\begin{align}
&A_0 = 0.058(4)\ ,\ \chi^2/\text{dof} = 0.99\ (\text{quadratic fit})\ ,\\
&\gamma = 0.96(6)\ ,\ \chi^2/\text{dof} = 1.26\ (\text{hyperscaling fit})\ .
\end{align}
On the one hand,
the quadratic fit works well with reasonable $\chi^2/\mathrm{dof}$
and gives a finite intercept of $A_0$, which implies that chiral symmetry is broken~\cite{Hasenfratz:2010fi}.
On the other hand, the hyperscaling fit also works, and results in the large mass anomalous dimension:
$\gamma\sim \mathcal{O}(1)$.
These properties suggest that $N_f = 8$ system is at the border of the chirally broken and conformal phases.
It is remarkable that the $\gamma$ obtained here is similar to
those obtained by the various hadron spectra (Table~\ref{tab:gam_hs_ca00a00_nf08b0380_obsall_dat10}),
with the exception of the pion mass case.
Although it seems to be difficult to discriminate between the chirally broken and conformal scenarios
by the string tension alone, the results are complementary to and consistent with
the hadron spectra, and thus support the walking scenario suggested in the previous sections.

\begin{table}[!tbpt]
\caption{The square root of the string tension $\sqrt{s}$
for various fermion masses $m_f$. In the third column,
the first and second brackets represent the statistical and systematic
errors, respectively.
As indicated in the right-most column,
some data have not been used in the quadratic/hyperscaling fits
since they have been determined in the smaller lattice volumes or the statistics are not satisfactory.
They correspond to the gray symbols in Fig.~\protect\ref{Fig:s_pot_nf08b0380}.
}\label{Tab:s_pot_nf08b0380}
\begin{center}
\begin{tabular}{lllll}
\hline\hline
$m_f$ & $L$ & $\sqrt{s}$ & $\chi^2/\mathrm{dof}$ & comment\\
\hline
$0.012$ & $42$ & $0.0931(25)({}^{\ 7}_{56})$    & $0.557$ & \\
$0.015$ & $42$ & $0.1018(82)({}^{10}_{30})$     & $0.352$ & not used in fit\\
$0.015$ & $36$ & $0.1096(39)({}^{52}_{\ 3})$    & $0.126$ & \\
$0.020$ & $36$ & $0.1110(69)({}^{170}_{\ \ 7})$ & $0.39$  & \\
$0.020$ & $30$ & $0.1230(40)({}^{45}_{\ 2})$    & $0.455$ & not used in fit\\
$0.030$ & $30$ & $0.1406(72)({}^{164}_{113})$   & $0.489$ & \\
$0.030$ & $24$ & $0.1508(43)({}^{17}_{\ 1})$    & $0.045$ & not used in fit\\
$0.040$ & $30$ & $0.1678(38)({}^{88}_{12})$     & $0.452$ & \\
$0.040$ & $24$ & $0.1743(188)({}^{131}_{\ 0})$  & $0.108$ & not used in fit\\
$0.060$ & $24$ & $0.2096(55)({}^{49}_{\ 8})$    & $0.197$ & \\
$0.080$ & $24$ & $0.2492(47)({}^{133}_{\ 13})$  & $0.768$ & \\
$0.100$ & $18$ & $0.2683(82)({}^{107}_{\ 39})$  & $0.293$ & \\
\hline\hline
\end{tabular}
\end{center}
\end{table}

\clearpage

\section{Light Flavor-Singlet Scalar Meson}
\label{sec:flavorsinglet}

In this section we study the flavor-singlet scalar meson $\sigma$.
In the previous sections we have shown that $N_f=8$ QCD shows signals of walking behavior,
being consistent with spontaneous chiral symmetry breaking
and also having (non-universal) hyperscaling with $\gamma \sim 1$.
When chiral symmetry is spontaneously broken
by the condensate $\langle \overline{\psi} \psi\rangle \ne 0$,
the scale symmetry is also spontaneously broken by the same condensate in the vacuum.
This produces a composite dilaton, the NG boson of the spontaneously broken scale symmetry.
In fact, the scale symmetry is also explicitly broken by the dynamically
generated mass scale $m_D$ associated with the same condensate,
and the dilaton would be a pseudo dilaton having a small mass of order
${\cal O} (m_D) (\ll \Lambda_{\rm QCD})$ in the chiral limit $m_f=0$.
In such a case the flavor-singlet scalar meson, as a pseudo dilaton, is expected to be light.
In the case of walking technicolor, such a pseudo dilaton, dubbed a technidilaton,
may be identified with the 125 GeV Higgs boson discovered at LHC.
Thus it is very important to investigate such a possibility of a light flavor-singlet scalar meson
in a fully nonperturbative manner on the lattice, particularly for $N_f=8$.
For this purpose we study the flavor-singlet scalar meson $\sigma$ in $N_f=8$, and $N_f=12$ and 4 as well for comparison.
In $N_f=12$ QCD we have found the flavor-singlet scalar mass
to be lighter than the NG-pion~\cite{Aoki:2013zsa}.
The existence of a light composite scalar has been observed by
several lattice groups with different lattice actions~\cite{Fodor:2014pqa, Brower:2015owo}.
Although the $N_f=12$ theory is likely in the conformal phase and not a candidate
technicolor model, this result suggests that the conformal dynamics
may play a role for obtaining a light composite scalar
whose properties are quite different from usual QCD.
In fact, we have previously measured the mass of the flavor-singlet scalar
in $N_f=8$ QCD on the lattice,
and found that the scalar is as light as the pion at the simulated fermion masses~\cite{Aoki:2014oha, Appelquist:2016viq},
which could be the first evidence of a candidate for the composite Higgs as a technidilaton,
since $N_f=8$ QCD has been considered as a good candidate for the walking technicolor model.
We explore the $\sigma$ mass in a region of lighter fermion masses 
with respect to our previous paper~\cite{Aoki:2014oha},
so that we can study the chiral behavior of the scalar mass in detail.
In the following, we explain the simulation setup and the methods for the flavor-singlet scalar mass
measurement, and show the results for the correlation functions, and $M_\sigma$ as a
function of $m_f$.
We discuss the chiral behavior of the $\sigma$ mass,
from which we will obtain the mass in the chiral limit.

\subsection{Measurement setup}
We carry out simulations of the SU(3) gauge theory with eight fundamental fermions,  and
calculate the mass of the flavor-singlet scalar $M_\sigma$ at six fermion masses
($m_f$=0.012, 0.015, 0.02, 0.03, 0.04, 0.06),
on five different lattice volumes $(L=18, 24, 30, 36, 42)$.
We accumulate $4,000-100,000$ trajectories
after more than 1,500 trajectories for thermalization.
The total number of the configurations as well as the simulation parameters
are shown in Table~\ref{tab:simulations}.

For the calculation of the two-point correlation functions of the flavor-singlet scalar operator,
we use the local fermionic bilinear operator
with the taste-spin structure $({\bf 1} \otimes {\bf 1})$,
\begin{equation}
\mathcal{O}_{S}(t)=\sum_i \sum_{x} \overline{\chi}_i(x,t)\chi_i(x,t),
 \label{eq:O}
\end{equation}
where $i$ runs through different staggered fermion species, $i=1,2$
and summation over a time slice $t$ is taken for zero momentum projection.
Using this operator,
we write the correlator  as $\langle O_{S}(t) O_{S}(0)\rangle \propto 2D(t) - C(t)$,
where $C(t)$ and $D(t)$ are the connected and the vacuum subtracted
disconnected correlators, respectively.
The factor $2$ in the disconnected correlator
arises from the number of species.
To calculate the disconnected piece of the two-point functions,
we need the inverse of the Dirac operator for all space-time points.
We employ a stochastic noise method with a variance reduction technique based
on the axial Ward-Takahashi identity~\cite{Venkataraman:1997xi},
which has been applied in the literature~\cite{Venkataraman:1997xi, Gregory:2007ev, McNeile:2012xh, Jin:2012dw}.
We use 64 random sources spread in spacetime
and color spaces for this noise-reduction method.
For staggered fermions, the interpolating operator in Eq.~(\ref{eq:O})
can also couple to the state with $(\gamma_4 \gamma_5 \otimes \xi_4\xi_5)$,
which is the staggered parity partner of $\sigma$,
a flavor non-singlet pseudoscalar.
The asymptotic behavior of the correlators of the flavor-singlet scalar is given by
\begin{eqnarray}
2D(t)-C(t) &=& A_\sigma(t)+(-1)^t A_{\pi_{\overline{SC}}} (t),
 \label{eq:st}
\end{eqnarray}
where $A_H(t)=A_H (e^{-M_H t}+ e^{-M_H(T-t)})$,
and the state $\pi_{\overline{SC}}$ is a species-singlet pseudoscalar with taste-spin structure
$(\gamma_4 \gamma_5 \otimes \xi_4\xi_5)$.
The connected piece $C(t)$ behaves as
	\begin{eqnarray}
-C(t) &=& A_{a_0}(t)+(-1)^t A_{\pi_{SC}} (t),
 \label{eq:ct}
\end{eqnarray}
where the states $a_0$ and $\pi_{SC}$ are the flavor non-singlet scalar
and the flavor non-singlet species non-singlet pseudoscalar, respectively.
Then the disconnected piece $2D(t)$ can be written as
\begin{equation}
2D(t)= A_{\sigma}(t) - A_{a_0}(t)
+(-1)^t (A_{\pi_{SC}} (t)-A_{\pi_{\overline{SC}}} (t)).
 \label{eq:dt}
\end{equation}
If the flavor symmetry is exact, the masses of both the flavor non-singlet pseudoscalar $\pi_{\overline{SC}}$
and $\pi_{SC}$ are degenerate, and their amplitudes coincide;
this means $A_{\pi_{SC}} (t) = A_{\pi_{\overline{SC}}}(t)$,
since the disconnected piece of the flavor non-singlet channel disappears in the taste symmetric limit.
The contribution of the opposite parity state in the staggered fermion can be suppressed by
applying the positive parity projection,
$C_+(t) \equiv 2C(t) + C(t+1) + C(t-1)$ at even $t$.
Another projection
$C_-(t) \equiv 2C(t) - C(t+1) - C(t-1)$ at even $t$
is also defined to maximize the opposite parity contribution $\pi_{SC}$.

\begin{figure}[!tbp]
  \includegraphics*[height=5.cm]{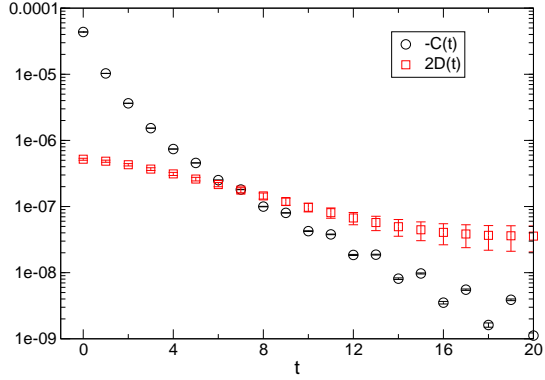}
  \caption{\label{fig:corr}
  Connected $-C(t)$ and disconnected correlators $2D(t)$ for $L=30$, $m_f=0.02$.
 }
\end{figure}

\begin{figure}[!tbp]
  \includegraphics*[height=5cm]{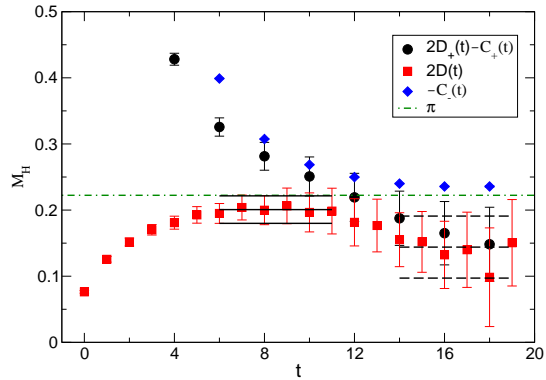} \quad
  \caption{\label{fig:meff}
Effective mass for $L =30$, $m_f=0.02$, from correlators
using the projection explained in the text. }
\end{figure}

A typical result for $D(t)$ and $C(t)$ is shown in Fig.~\ref{fig:corr}.
In the figure, $D(t)$ behaves as a smooth function of $t$
in contrast to $C(t)$, which has an oscillating behavior.
This result means the taste symmetry breaking effect on
$A_{\pi_{SC}} (t)$ and $A_{\pi_{\overline{SC}}}(t)$
is negligible in the parameter region we simulate.
The effective masses
of $2D_+(t)-C_+(t)$, $D(t)$, and $C_-(t)$ are shown in Fig.~\ref{fig:meff}.
Since the combination $2D_+(t) - C_+(t)$ at large $t$ is dominated by $2D(t)$,
the effective mass of the $2D_+(t)-C_{+}(t)$ at large $t$ becomes consistent with
the one obtained from $D(t)$.
An advantage of using $D(t)$ in extracting $M_\sigma$ is that
that the plateau of $D(t)$ appears at earlier $t$ than
that of the $2D_+(t) - C_+(t)$,
which enables us to determine $M_\sigma$ with better accuracy.
This earlier plateau happens to appear in the mass parameter we simulate,
which might be caused by a reasonable cancellation between
the contributions of $A_{a_0}(t)$ and excited states of $\sigma$.
It is also shown that the effective masses of $D(t)$ as well as $2D_+(t)-C_+(t)$ are
smaller than that of $M_\pi$, as plotted in the figure.
Due to the smallness of $M_\sigma$ compared to other hadron masses,
the exponential damping of $D(t)$ is milder than that in usual QCD.
It helps to prevent the rapid degradation of the signal-to-noise ratio.

\begin{figure}[!tbp]
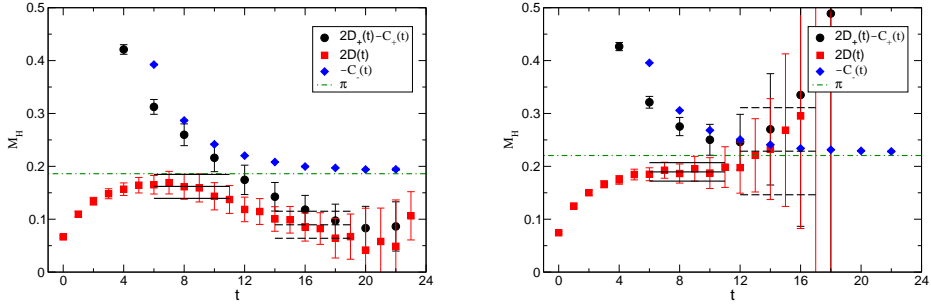

  \includegraphics*[height=4.cm]{fig045a.eps} \quad \quad
  \includegraphics*[height=4.cm]{fig045b.eps}
  \caption{\label{fig:meff_36}
  Effective mass for $L =36$,
  $m_f=0.015$ (left) and $m_f=0.020$ (right),
  from correlators using the projection explained in the text. }
\end{figure}

We fit $D(t)$ with the assumption of a single scalar propagation.
The fit range is $[t_{min},t_{max}]=[6,11]$ for all the simulation parameters
for which we find an effective mass plateau.
In order to estimate a systematic uncertainty
coming from the fixed fitting range effect, we also fit with a later $t$ region,
with the same number of data points,
as shown in Fig.~\ref{fig:meff}.
We quote the fit result with fixed $t$ range as a central value,
and estimate a systematic error as the difference of the values obtained by differing fit ranges.
All the results are tabulated in Table~\ref{tab:simulations}.
It should be noted that, in somewhat smaller mass region,
an additional effective mass plateau seems to appear at later $t$ region,
whose mass is below the one obtained in the region at small $t$.
In the later time region, however, the effective masses
are not stable, with larger error in $D(t)$,
so that more data are required for a better identification of the ground state mass.
We find that the results with two different fit ranges
are consistent with each other except for $L=36, m_f=0.015$,
whose result is shown in Fig.~\ref{fig:meff_36}.

{
While in the current analysis we determine $M_\sigma$ from the earlier
plateau of $D(t)$, and estimate the systematic error from the later one,
a more reliable result can be obtained from a plateau of the full correlator
$2D_+(t)-C_+(t)$, if the statistics is sufficient to obtain a clear signal
in the large $t$ region.
Therefore, it is an important future work to compare our results with
the ones from more reliable calculations with much larger statistics
to examine whether our determination of $M_\sigma$ is reasonable or not.
}

\begin{table}[!tbp]
   \caption{\label{tab:simulations} Simulation parameters for $N_f=8$ QCD at $\beta=3.8$.
$N_{\rm cf}$ ($N_{\rm st}$) is the total number of gauge configurations (Markov chain streams).
The second error of $M_\sigma$ is a systematic error arising from the choice of fit range.
The data with ($^\dagger$) and ($^*$) indicate a new result, and an update
from the previous result~\cite{Aoki:2014oha},
respectively.
   }
\begin{ruledtabular}
\begin{tabular}{l c r l l}
\multicolumn{1}{c}{$m_f$} &
\multicolumn{1}{c}{$L^3 \times T$} &
\multicolumn{1}{c}{$N_{\rm cf}$~[$N_{\rm st}$]} &
\multicolumn{1}{c}{$M_\sigma$} &
\multicolumn{1}{c}{$L M_\sigma$} \\
\hline
     0.012$^\dagger$ &  $42^3\! \times\! 56$ & 2300~[2] & 0.151(15)($^{\ 0}_{25}$)& $6.3(6)(^{\ \ 0}_{1.1})$\\
     0.015$^*$ & $36^3\! \times\! 48$ &  5400~[2] & 0.162(23)($^{\ 0}_{73}$)  & $5.8(8)(^{\ \ 0}_{2.6})$\\
     0.02 & $36^3\! \times\! 48$ &  5000~[1] & 0.190(17)($^{39}_{\ 0}$) & $6.8(6)(^{1.4}_{\ \ 0})$\\
     0.02 & $30^3\! \times\! 40$ &  8000~[1] & 0.201(21)($^{\ 0}_{60}$) & $6.0(6)(^{\ \ 0}_{1.8})$\\
     0.03 & $30^3\! \times\! 40$ & 16500~[1] & 0.282(27)($^{24}_{\ 0}$) & $8.5(8)(^{7}_{0})$\\
     0.03 & $24^3\! \times\! 32$ & 36000~[2] & 0.276(15)($^{6}_{0}$) & $6.6(4)(^{1}_{0})$\\
     0.04 & $30^3\! \times\! 40$ & 12900~[3] & 0.365(43)($^{17}_{\ 0}$) & $11.0(1.3)(^{0.5}_{\phantom{0.}0})$\\
     0.04 & $24^3\! \times\! 32$ & 50000~[2] & 0.322(19)($^{8}_{0}$) & $7.7(5)(^{2}_{0})$\\
     0.04 & $18^3\! \times\! 24$ &  9000~[1] & 0.228(30)($^{\ 0}_{16}$) & $4.1(5)(^{0}_{3})$\\
     0.06 & $24^3\! \times\! 32$ & 18000~[1] & 0.46(7)($^{12}_{\ 0}$) & $11.0(1.7)(^{2.8}_{\phantom{2.}0})$\\
     0.06 & $18^3\! \times\! 24$ &  9000~[1] & 0.386(77)($^{12}_{\ 0}$) & $7.0(1.4)(^{2}_{0})$\\
   \end{tabular}
\end{ruledtabular}
\end{table}

\begin{figure}[!tbp]
  \includegraphics*[height=6cm]{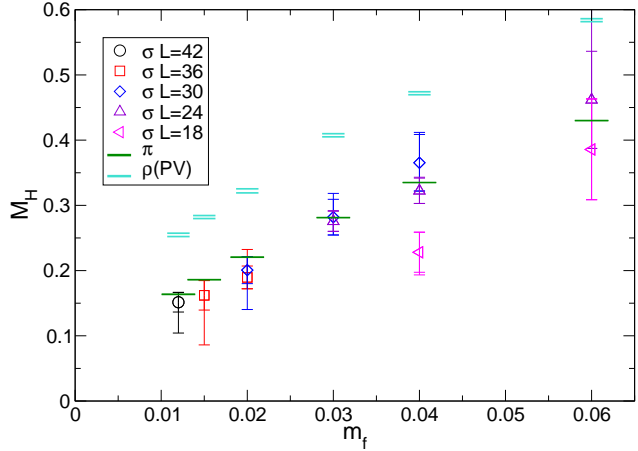} \ \
  \caption{\label{fig:m_mf}
  The fermion mass dependence of the mass of the flavor-singlet scalar $M_\sigma$.
  Masses of the NG-pion $\pi$ and vector meson $\rho$(PV) mass are also shown.
  The outer error represents the statistical and systematic uncertainties added in quadrature,
  while the inner error is only statistical.
  }
\end{figure}

\subsection{Chiral extrapolation}

The scalar is as light as $\pi$ in our fermion mass range.
Fig.~\ref{fig:m_mf} shows the results for the scalar mass
compared with $M_\pi$ and $M_\rho$.
To estimate the impact of finite size effects,
we show the values of $LM_\sigma$ in Table~\ref{tab:simulations}.
As we see, the data at the largest two volumes are consistent with each other
for $m_f \geq 0.02$.
An estimate of $LM_\sigma$ suggests that
the finite size effect on $M_\sigma$ is negligible in the data with $LM_\sigma \geq 6$,
so that finite size effects for $m_f=0.015$ and $m_f=0.012$,
where only a single volume is available, are inferred to be negligible in our statistics.

\begin{figure}[!tbp]
  \includegraphics*[height=6.cm]{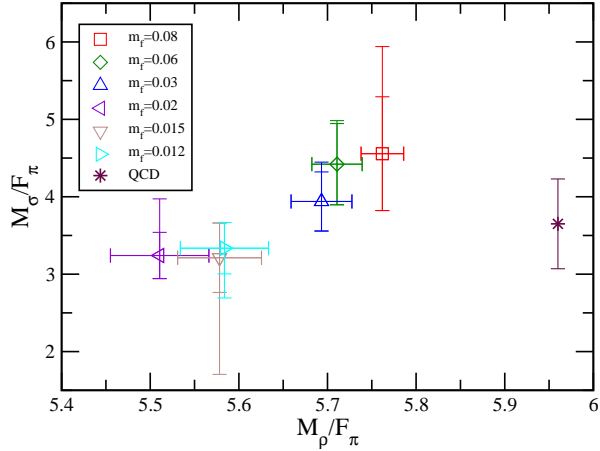}
  \caption{\label{fig:msigma-fpi}
Edinburgh type plots: $M_\sigma/F_\pi$ as a function of $M_\rho/F_\pi$.
The outer error represents the statistical and systematic uncertainties added in quadrature,
while the inner error is only statistical.
 }
\end{figure}

First we would like to study the chiral behavior of $M_\sigma$
from dimensionless ratios of spectral quantities.
Fig.~\ref{fig:msigma-fpi} shows $M_\sigma/F_\pi$ as a function of $M_\rho/F_\pi$
on the largest volume at each mass.
For reference, we also plot these quantities in QCD with
$f_0(500)$ as a candidate for the lowest scalar bound state.
The ratio $M_\sigma/F_\pi$ has a mild $m_f$ dependence
at smaller fermion masses,
and its value is close to the one in usual QCD,
while there are large
statistics and systematics uncertainties.

Assuming this theory is in the chirally broken phase,
we discuss
the chiral limit extrapolation of $M_\sigma$.
While we have a light scalar with mass comparable to $M_\pi$,
the validity of chiral perturbation theory is intact by introducing
a dilaton field which is a pseudo Nambu-Goldstone boson of the scale symmetry
(``Dilaton ChPT'' (DChPT)~\cite{Matsuzaki:2013eva}).

{Treating the light scalar as a dilaton field in the effective theory is, in general, just a particular assumption for its potential and possible interactions. However, for a vector-like gauge theory, such as the one we study here on the lattice ($N_f=8$ QCD), there cannot be a light flavor-singlet scalar other than a dilaton.
The global symmetry breaking patterns in the chiral limit of vector-like gauge theories, dictated by the Vafa-Witten theorem, are of the symmetric coset G/H~\cite{Kosower:1984aw,Peskin:1980gc}, none of which has a NG boson scalar that can be identified with the SM Higgs.
In fact, NG bosons in the symmetric coset G/H have no odd-numbered vertices, such as ``3-pion'' vertex (see e.g. Ref.~\cite{Bando:1987br}), and cannot decay into a pair of other NG bosons in the same coset G/H (which we usually think to be absorbed into W/Z). Thus such NG bosons in the symmetric coset cannot be identified with the 125 GeV Higgs, which has been established by the LHC experiments to decay into a pair of longitudinal W/Z bosons~\cite{Aad:2015mxa,Khachatryan:2014kca}. 

To find a workaround for the Higgs to W/Z coupling and still identify the light scalar Higgs as an NG boson of the same G/H as those absorbed into W and Z bosons is not an easy task. A UV completion for such a G/H is only realized in chiral gauge theories in accord with the Vafa-Witten theorem which is crucially based on the positivity of measure in the lattice regularization. Patterns of symmetry breaking in chiral gauge theories, however, have only been analyzed by the most-attractive-channel (MAC), a perturbative one-gauge-boson exchange picture. A full nonperturbative treatment on the lattice is not available yet. (Moreover, a chiral gauge theory is a necessary condition, but not a sufficient one, to avoid the symmetric coset, which is remains still a possibility, e.g., ${\rm SU}(5)/{\rm SO}(5)$, in chiral gauge theories, although symmetric coset in chiral gauge theories could be viable under more involved assumptions such as the vacuum misalignment to develop the Higgs VEV, etc.)

We therefore proceed our chiral extrapolation analysis of the light flavor-singlet scalar mass using DChPT.}
At leading order in DChPT, the $\sigma$ mass is given by
\begin{equation}
M_\sigma^2 = d_0 +d_1 M_\pi^2,
\label{eq:dn}
\end{equation}
where $d_0=M_\sigma^2|_{m_f=0}$,
and $d_1=\frac{(3-\gamma_m)(1+\gamma_m)}{4}\frac{N_f F^2}{F_\sigma^2}$.
The $\gamma_m$ is an effective mass anomalous dimension in the walking regime,
and $F$ and $F_\sigma$ are the decay constants of $\pi$ and $\sigma$ in the chiral limit, respectively.

\begin{figure}[!tbp]
  \includegraphics*[height=6cm]{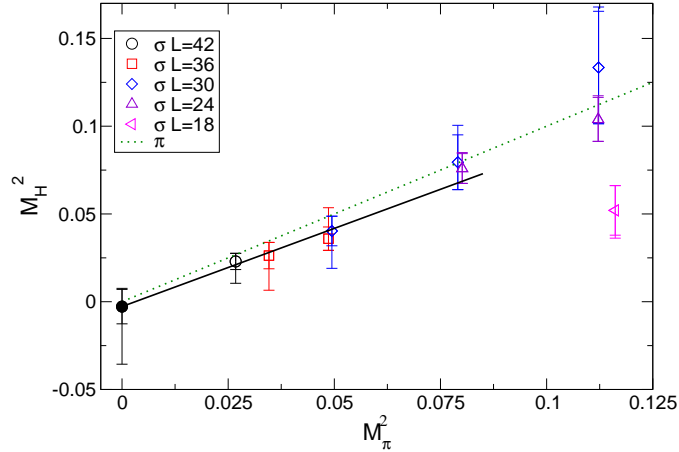} \ \
  \caption{\label{fig:m2_mpi2_fit}
  NG-Pion mass dependence of the mass of the flavor-singlet scalar.
  The outer error represents the statistical and systematic uncertainties added in quadrature,
  while the inner error is only statistical.
  Results of the chiral extrapolation by the DChPT are plotted by the solid line and full circle.
  }
\end{figure}

In the following fit analyses,
we shall use the lightest four data points ($m_f < 0.03$) with the leading order dilaton mass fit function.
We carry out the fit with a data set in which all the four data are chosen at the largest volume with
the fixed $t$ range, and obtain a result as a central value.
The effect of the systematic errors in $M_\sigma$
is taken into account in the chiral extrapolation fit
by varying the data set used.
We consider all the combinations of the data set where each of the data
is chosen as the result from the first fit range or the second one.
Thus we have in total sixteen data sets for the lightest four fermion masses.
We carry out fits for all the data sets, quoting the maximum difference of $M_\sigma$
as a measure of the systematic error due to the choice of the plateau
in the effective mass.
Each chiral extrapolation is performed with the statistical errors only.
The fit results are shown in Fig.~\ref{fig:m2_mpi2_fit}.
The fit of the chiral extrapolation gives a reasonable $\chi^2/$dof=0.40, and
the value of $M_\sigma^2$ in the chiral limit $d_0=-0.0028(98)(^{\phantom{3}36}_{354})$,
where the first and second errors are statistical and systematic, respectively.
Thanks to a higher precision of this result compared to Ref.~\cite{Aoki:2013zsa}, we now obtain a value of $d_0$ closer to zero.
From the linear slope $d_1$, we can read off the value of $F_\sigma$.
The DChPT fit gives $d_1=0.89(26)(^{75}_{11})$.
If the effective mass anomalous dimension is $\gamma_m \sim 1$,
we can obtain $F_\sigma \sim \sqrt{N_f} F$ with $d_1 \sim 1$,
which is consistent with another calculation of the dilaton decay constant
via the scalar decay constant
and the Ward-Takahashi identity of the scale transformation~\cite{Aoki:2015jfa}.
It is also noted that the result for $d_1$ is quite different from usual QCD,
where a larger slope is observed for $M_\pi > 670$ MeV~\cite{Kunihiro:2003yj}.

\begin{figure}[!tbp]
  \includegraphics*[height=6cm]{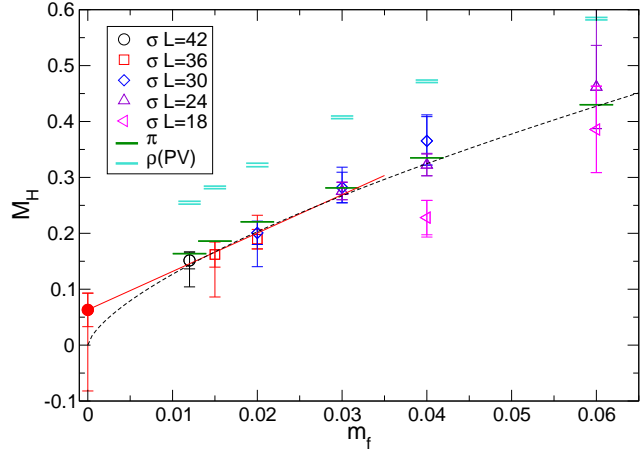} \ \
  \caption{\label{fig:m_mf_fit}
  Fermion mass dependence of the mass of the flavor-singlet scalar.
  Masses of the NG-pion $\pi$
  and vector meson ($\rho$(PV)) are also shown.
  The solid and dashed lines represent results of linear and
  hyperscaling fits respectively (fit range: $m_f=0.012 - 0.03$).
  The outer error represents the statistical and systematic uncertainties added in quadrature,
  while the inner error is only statistical.
  }
\end{figure}

We can also fit $M_\sigma$ with an empirical form, $M_\sigma=c_0+c_1 m_f$,
using the same data set.
The fit result is shown in Fig.~\ref{fig:m_mf_fit}.
The chiral fit of $m_f$ also gives a reasonable $\chi^2/$dof $\sim 0.40$,
$c_0=0.063(30)(^{\phantom{14}4}_{142})$.
This result is consistent with that of DChPT.
Although our result for the scalar mass has a sizeable error, and
the chiral limit is probably far from our current simulation regions,
both the fit results of $d_0$ and $c_0$ suggest the possibility to reproduce the Higgs boson
with mass $125$ GeV
via a scale setting of $F/\sqrt{2} \sim 123$ GeV
in the one-family technicolor model with four weak-doublets.
We have estimated a small value of $F_\pi = 0.0212(12)(^{49}_{71})$ in the chiral limit,
so that a small value of $c_0 \sim 0.015$ would be required to
be consistent with the composite Higgs in the one-family model.
Using the fit result for $c_0$,
we obtain $M_\sigma/(F/\sqrt{2}) = 4.2(2.0)(^{1.4}_{9.5})$ in
 the chiral limit. Thus the value of $M_\sigma$ is comparable
 to $F_\pi$ even in the chiral limit, while the error is large.
This result is encouraging for obtaining
a composite Higgs boson with mass $125$ GeV.

We note that the hyperscaling fit in the conformal hypothesis,
$M_\sigma=c_0 m_f^{1/(1+\gamma)}$,
works in the smaller mass region.
The fit result is shown in Fig.~\ref{fig:m_mf_fit}.
The conformal fit gives a $\chi^2$/dof $=0.60$, and $\gamma=0.47(33)(^{\phantom{8}9}_{80})$.
This should be compared with the result in the right panel of Fig.~\ref{fig:hs_nf08b0380}.
This behavior---that both the fits of the (D)ChPT and hyperscaling with a large
mass anomalous dimension work in an appropriate mass region---matches the one seen in the spectra of other hadrons.
It is quite different from usual QCD and
could be a signal of the walking gauge theory.
An important  future direction is to obtain a precise value of $M_\sigma$
in the chiral limit, which will be useful to study
if this theory really exhibits the desired walking behavior,
and reproduce the Higgs boson with $125$ GeV mass.
For this purpose,
we need more data at lighter fermion masses with larger volumes.

\subsection{Dilaton decay constant calculation}

From a phenomenological point of view,
the technidilaton decay constant (denoted here as $F_\sigma$)
is an important parameter,
since $F_\sigma$ controls all the technidilaton's couplings to SM particles.

The dilaton decay constant is defined as
$\langle 0 | \mathcal{D}^\mu(x)  |\sigma(p) \rangle =- i F_\sigma p^\mu e^{-ipx}$,
from which we also obtain
$\langle 0 | \partial_\mu \mathcal{D}^\mu(0)  |\sigma(0) \rangle =- F_\sigma M_\sigma^2$.
Therefore the dilaton decay constant can be directly calculated
from the matrix element of the dilatation current.
However, the dilatation current is rather difficult to construct on the lattice,
since it contains a power divergence that needs to be subtracted.
Instead, we consider an alternative way to estimate it from
a relation between
the scalar decay constant $F_S$ and $F_\sigma$ obtained in the continuum theory.
Here the scalar decay constant $F_S$ is defined as
the scalar operator matrix element,
\begin{eqnarray}
\langle 0 | m_f O_{S}(0,0) |\sigma(0) \rangle =F_S M_\sigma^2,
\label{eq:Os}
\end{eqnarray}
where $\mathcal{O}_{S}$ is the flavor-singlet scalar bilinear operator
$\mathcal{O}_{S}(x, t) = \sum_i \overline{\chi}_i(x,t)\chi_i(x,t)$.
Thus the above matrix element can be calculated from
the two-point correlation function ($C_\sigma(t)$) on the lattice.
We note that this quantity is renormalization group invariant and a physical quantity.
Following the argument based on the continuum theory~\cite{Bando:1986bg},
we obtain a relation
\begin{eqnarray}
F_S F_\sigma M_\sigma^2=- \Delta_{\overline{\psi}\psi} m_f
\sum_i^{N_f}
\left\langle  \overline{\psi}_i \psi_i \right\rangle,
\label{eq:Fs}
\end{eqnarray}
where $\Delta_{\overline{\psi}\psi}$ is the scaling dimension of ${\overline{\psi}\psi}$
and $N_f$ total number of fermions.\footnote{
This relation can be derived by using the Ward-Takahashi relation for the dilatation current
in the continuum theory. For the detail of the derivation, see~\cite{Aoki:2015jfa}.}
(Dividing both sides by $m_f$ leads to the relation obtained at $m_f=0$~\cite{Matsuzaki:2012xx}.)

We note that this relation holds in the continuum theory
with infrared conformality
by saturating the Ward-Takahashi identity for the dilatation current
by the single pole dominance of $\sigma$ as a dilaton
in the spontaneously broken phase of the scale symmetry, as well as the chiral symmetry.
We here assume that this relation remains valid also on the lattice up to discretization effects,
in the same spirit as our analysis of $M_\sigma$ and
$F_\sigma$ using Eq.~(\ref{eq:dn}). The method here is a semi-direct estimate of
$F_\sigma$ alternative to that based on Eq.~(\ref{eq:dn}).

\begin{figure}[!tbp]
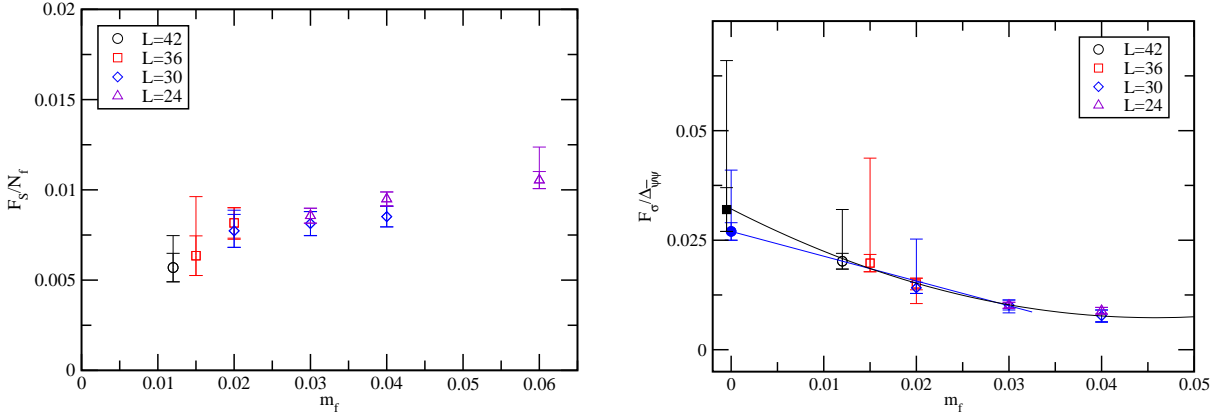

\begin{center}
\includegraphics[width=3in,clip]{fig050a.eps}
\hspace{5mm}
\includegraphics[width=3in,clip]{fig050b.eps}
\end{center}
\caption{
Left: Scalar decay constant of the flavor-singlet scalar.
Right: Dilaton decay constant from a semi-direct method.
The blue and black lines show the chiral fits
using linear and quadratic polynomials in $m_f$,
with the lightest 4 and 5 data points used.
The outer error represents the statistical and systematic uncertainties added in quadrature,
while the inner error is only statistical.
}
\label{fig:f_sigma}
\end{figure}

The result for $F_S$ is summarized in the left panel of Fig.~\ref{fig:f_sigma}, where we obtain a signal for $F_S$ as the same statistical accuracy as the $\sigma$ mass.
We estimate a systematic uncertainty coming from the choice of fitting range, which is also shown in the figure.
On the right panel of Fig.~\ref{fig:f_sigma}, the result of $F_\sigma$ from the semi-direct estimate Eq.~(\ref{eq:Fs}) is shown.
For the chiral condensate, we use its chiral limit value to avoid large lattice artifacts.
We also carry out chiral extrapolation fits, whose results are also shown in the figure.
In the chiral limit, we obtain $\frac{F_\sigma}{\Delta_{\overline\psi \psi}} \sim 0.03$.
Given that $\Delta_{\overline{\psi} \psi} =3-\gamma \simeq 2, (\gamma \simeq 1)$, we have shown how two different methods, DChPT (Eq.~(\ref{eq:dn})) and this method, give a consistent result for $F_\sigma$.

\subsection{Phenomenological implications for dark matter physics}

One interesting phenomenological implication of the measurement of $F_\sigma$ is an application to dark matter (DM) direct detection.
In technicolor models, there is a good candidate for composite DM: a neutral baryonic bound state made of constituent (possibly charged) technifermions.
As shown below, the coupling between SM particles and the DM as well as the mass of the DM are constrained by direct detection experiments.
The scattering rate of DM with heavy nuclei in detectors is an important parameter for experiments, a dominant contribution to which is the Higgs (scalar)-mediated spin-independent process.
Using $F_\sigma$ we discuss the detectability of DM from that process.

We consider a DM effective theory including a dilaton, based on DChPT~\cite{Matsuzaki:2013eva}.
Since the DM is the lightest technibaryon, the extension to the baryon sector of the DChPT is straightforward.
In the leading order the dilaton field can only couple through the nucleon mass term as
\begin{eqnarray}
\mathcal{L} = \overline{N}(x)(i\gamma_\mu \partial^\mu - \chi(x) M_N )N(x),
\end{eqnarray}
where $\chi(x) = e^{\sigma(x)/F_\sigma}$, $N(x)$ is the baryonic DM field,  and $M_N$ is its mass in the chiral limit.
The parameter $M_N$ explicitly breaks the scale symmetry, and the (pseudo) dilaton acts on this term to make the action scale invariant.
Then the dilaton-DM effective coupling ($y_{\overline{N}N\sigma}$) is uniquely determined as $y_{\overline{N}N\sigma}=M_N/F_\sigma$.
Regarding to the SM sector, we also use the dilaton effective theory~\cite{Matsuzaki:2012vc} to determine the coupling of $\sigma$ and a target nucleus.
Combining both SM and technicolor sectors, the cross section with a SM nucleus $B$ for the spin-independent part is given as $\sigma_{SI} = \frac{M_R(B, N)^2}{\pi}(Z f_p+(A-Z)f_n)^2$, with $M_R(B,N)=(M_B M_N)/(M_B+M_N)$, where $M_{B}$ is the mass of the target nucleus, and $Z$ and $A-Z$ are the total number of the protons ($p$) and neutrons ($n$) in the nucleus ($A$ is the mass number).
The parameter $f_{(n,p)}$ is defined as $f_{(n, p)}=\frac{M_B}{\sqrt{2} M_\sigma^2} \frac{y_{\overline{N}N\sigma}}{F_\sigma} (3-\gamma)\left( \sum_{q=u,d,s} f_{T_q}^{(n,p)}+\frac{2}{9}f_{T_G}^{(n,p)}  \right)$, where $f_{T_q}^{(n,p)}$ is the nucleon $\sigma$-term of the light quarks ($q=u,d,s$), and $f_{T_G}^{(n,p)}$ is that of the heavy quarks.\footnote{ It is possible that technifermions can be charged under SM color, so that there may exist additional contributions to the nucleon $\sigma$-term from the technifermions.
In this analysis, we omit these contributions for simplicity. }

\begin{figure}[!tbp]
\begin{center}
\includegraphics[width=3in,clip]{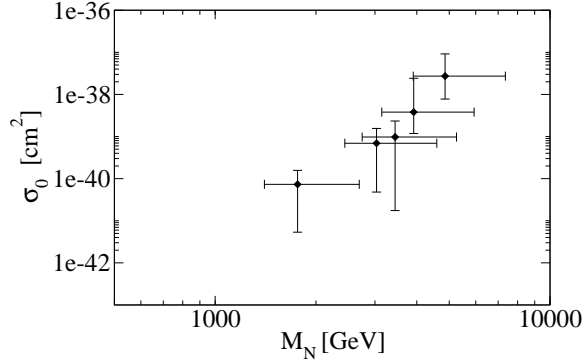}
\end{center}
\caption{
$\sigma_0$ [cm$^2$] as a function of $M_N$ [GeV].
The results for $m_f=0.030, 0.020, 0.015, 0.012$, and the chiral limit are shown from upper-right to lower-left.
Both the statistical and systematic errors are included.
The experimentally allowed region is below the plotted window.
The current experimental bound for $\sigma_0$ approximately is $10^{-45}$[cm$^2$] in this mass range under the assumption that the dark matter interacting in the detector is a thermal relic.
}
\label{fig:sigma0}
\end{figure}

Here we show our numerical results of the DM cross section.\footnote{ A similar analysis on the lattice has been performed for a different composite DM model based on strong dynamics~\cite{Appelquist:2014jch}.}
We use the lattice results of the dilaton decay constant $(F_\sigma)$ obtained from the previous section and nucleon mass, while the scalar mass $M_\sigma$ is fixed to its experimental value ($125$ GeV) in this analysis.
To set the scale, we use the relation $\sqrt{N_f/2}F_\pi/\sqrt{2}=246$ GeV.
To compare with experiment, we use the cross section per nucleon ($\sigma_0$) instead of $\sigma_{SI}$, which is defined as $\sigma_0 = \sigma_{SI}\frac{M_R(N,n)^2}{A^2 M_R(N,B)^2}$.
The result is shown in Fig.~\ref{fig:sigma0}.
According to DM direct detection experiments (see e.g.~\cite{Akerib:2013tjd, Akerib:2016vxi,Tan:2016zwf} for recent experimental results) our values for $\sigma_0$ are excluded under the assumption that the $N_f=8$ technibaryon is the major component of the dark matter relic density.\footnote{See Ref.~\cite{Fodor:2016wal} for a similar example in the context of composite baryonic dark matter.}
We note that there exist other contributions to the DM cross section, e.g. gauge boson mediated interaction, and higher order terms, which might affect the DM cross section.
It would be interesting to investigate these contributions.

\clearpage

\section{Summary and Discussion}
\label{sec:summary}

In QCD with $N_f=8$, we have confirmed the general structure of the walking signals in the spectrum observed in our previous results~\cite{Aoki:2013xza}, with dual features such as satisfying both ChPT and (non-universal) hyperscaling relations.
We have also confirmed the outstanding discovery of a light flavor-singlet scalar $\sigma$, with mass comparable to that of the pion $\pi$~\cite{Aoki:2014oha}, extending the studied parameter space down to $m_f=0.012$ and to the larger volume $(L,T)=(42,56)$.
We have studied more spectral quantities, including $a_0$, $a_1$, $b_1$ and $N$, in addition to those presented in Refs.~\cite{Aoki:2013xza,Aoki:2014oha} (namely $M_\pi$, $F_\pi$, $M_\rho$ and $M_\sigma$), with higher statistics.
Typically ten times as many trajectories as the previous data were used, for all but the $\sigma$.

We paid particular attention to the systematic comparison of the $N_f=8$ data with our $N_f=12$ data (shown to be consistent with the conformal window) and $N_f=4$ data (consistent with S$\chi$SB phase without remnants of conformality) obtained with the same lattice setup.

We performed a ChPT analysis of $M_\pi$ and $F_\pi$ for $m_f\leq 0.03$ with the estimate of the chiral log used to evaluate the systematic error:
 \begin{eqnarray}
 F &=& 0.0212(12)(^{+49}_{-71}),\\
\left. \langle \overline{\psi} \psi \rangle\right|_{m_f\to 0} &=&
0.00022(4)(^{+22}_{-12})\,,
\end{eqnarray}
which is consistent with the GMOR relation in the S$\chi$SB phase.
The chiral limit value of $\rho$ mass in units of $F/\sqrt{2}$ is $\frac{M_\rho}{F/\sqrt{2}} = 10.1(0.6)(^{+5.0}_{-1.9})$.

On the other hand, the hyperscaling relation holds in various intervals of $m_f$, including the lightest $m_f=0.012$, with $\gamma \sim 1$ for all the quantities (including the string tension) as anticipated for the walking technicolor.
A notable exception to this is $M_\pi$, for which hyperscaling is valid only in a restricted range of masses (see Table \ref{tab:gam_hs_ca00a00_nf08b0380_obs3a_datm}), and Finite-Size Hyperscaling (FSHS) (cfg. Fig.~\ref{fig:gam_fshs_nf08b0380_obsall_datLMpx}) gives $\gamma \simeq 0.6$ with a large $\chi^2/\textrm{dof}\sim18$.
This is consistent with the NG-boson nature of $\pi$, whose mass obeys the ChPT relation (see Eq.~(\ref{eq:MpiOverMf_fit}))
\begin{equation}
  M_\pi^2 = C_\pi m_f +C_\pi^\prime m_f^2+\cdots\, ,
\end{equation}
which would imitate the hyperscaling relation $M_\pi^2 \sim m_f^{2/(1+\gamma)}$ only locally---in a very narrow $m_f$ range---such that $\gamma\sim 0$ for larger $m_f$ with $C_\pi m_f \ll C_\pi^\prime m_f^2$, and $\gamma\sim 1$ for smaller $m_f$ with $C_\pi m_f \gg C_\pi^\prime m_f^2$, in perfect consistency with the $m_f$-dependence of $\gamma_{\rm eff}(m_f)$ for $M_\pi$ (see Fig.~\ref{fig:geff}).
The average result $\gamma \sim 0.6$ is just in between these two extremes, showing a sharp distinct behavior of $M_\pi$ compared with all other quantities.
This is also compared with $N_f=12$ in Fig.~\ref{fig:geff}.
Overall, a characteristic hyperscaling fit of our data gives:
\begin{eqnarray}
\gamma\sim 1\,,
\quad \gamma \simeq 0.6 \,\,(M_\pi)\,.
\end{eqnarray}
Thus we find that the hyperscaling relation in $N_f=8$ is non-universal, for both the naive hyperscaling and the FSHS (see Fig.~\ref{fig:geff} and Fig.~\ref{fig:fshs_gamxx_nf12}). This is in sharp contrast to $N_f=12$ where we find near universal hyperscaling with $\gamma\sim 0.4$, including $M_\pi$.
It addition, the ratio of quantities with respect to $M_\pi$, such as $M_\rho/M_\pi$, is increasing even at the smallest $m_f$ ($M_\rho/M_\pi \leq 1.54$), without indications of plateaux. This trend is not observed in $N_f=12$, while it is a characteristic of $N_f=4$ ratios.

The $N_f=8$ result  is also contrasted to the $N_f=4$ data where all the quantities but $M_\pi$ do not obey the hyperscaling relation at all.
They have no remnants of conformality, while only $M_\pi$ imitates the hyperscaling with $\gamma=1$, which is actually nothing but the ChPT formula with  $C_0 m_f \gg C_1 m_f^2$ (see Fig.~\ref{fig:fshs_gamxx_nf04b0370}).

We further confirmed our previous discovery of a light flavor-singlet scalar, $\sigma$, $M_\sigma \simeq M_\pi$~\cite{Aoki:2014oha}, even at smaller $m_f$.
Also the hierarchy of masses $M_\sigma \simeq M_\pi < M_\rho$, in contrast to $M_\sigma \simeq M_\pi \simeq M_\rho$ in $N_f=12$,  now became more generic including other states:
 \begin{equation}
 M_\sigma \simeq M_\pi < M_\rho, M_{a_0}, M_{a_1}, M_{b_1}, M_N\,.
 \end{equation}
The chiral limit extrapolation value using dilaton-ChPT~\cite{Matsuzaki:2013eva}  is  more consistent with the identification of $\sigma$ with the 125 GeV Higgs than the previous one, with improved error bars and a central value closer to zero.

Although we observed data consistent with ChPT for $m_f<0.03$, with $F\simeq 0.02$, the $m_f$ region we studied seems to be still too far from the chiral limit to establish 
whether the theory is in the S$\chi$SB
or the conformal phase, if the $N_f=8$ data are analyzed in 
isolation from $N_f=12$ and $N_f=4$ data.
We found no decisive evidence for the S$\chi$SB phase, such as the chiral log effects in ChPT and the obvious breakdown of the hyperscaling (or divergent $\gamma$ for spectrum other than $M_\pi$) as we expect in $m_f\ll m_D$.{\footnote{The Pagels-Stokar formula, together with the SD equation, implies $m_D \sim 2 F \sim 0.04$ 
for $N_c=3$, while $m_f^{(R)} =Z_m^{-1} m_f$ may be estimated as 
$\sim (M_\rho-M_\rho|_{m_f=0})/2 \sim 0.05 (m_f=0.012)$--0.16 $(m_f=0.04)$ 
(See Tables~\ref{tab:fit-other_hadron} and \ref{tab:hadron_spectI:L42}--\ref{tab:hadron_spectI:L12}). It is clear that our data are not in the near chiral limit where $m_f^{(R)} \ll m_D$.
}}
Nor did we observe any clear evidence of the conformal phase: i.e., the hyperscaling, if existed at all, is not universal---in particular for $M_\pi$---and $M_{H\ne \pi,\sigma}/M_\pi$ is increasing down to $m_f=0.012$, in contradiction to what is expected in the conformal phase.
In addition, we observed no near degeneracy of the chiral partners, $M_{a_0}/M_\pi\simeq M_\rho/M_\pi > 1$, $M_{a_1}/M_\rho \simeq M_{a_1}/M_{a_0} > 1$, in contrast to our expectation for the conformal phase without S$\chi$SB.

Although the decisive conclusion has yet to be drawn about the discrimination between the S$\chi$SB and the conformal phases, what we observed is fairly consistent with expected signals of a walking theory: having light $\pi$ and $\sigma$ as pseudo NG bosons, and being in the S$\chi$SB phase (as indicated by the ChPT fit) together with remnants of the conformal window (non-universal hyperscaling and non-degeneracy of the chiral partners), sharply distinct from our data for the $N_f=12$ theory (definitely consistent with the conformal phase and in disagreement with the S$\chi$SB phase), and those on the $N_f=4$ which obviously signal the S$\chi$SB phase.

As we noted in Ref.~\cite{Aoki:2014oha}, if $\sigma$ is the pseudo dilaton, then $M_\sigma$ is expected to become bigger than $M_\pi$ when we get to near the chiral limit, $m_f\ll m_D$.
This is because $\sigma$ as a pseudo dilaton would have chiral limit mass $M_\sigma \ne 0$ due to the trace anomaly generated by $m_D$, as was mentioned before, and is estimated to be~\cite{Matsuzaki:2015sya} of order $M_\sigma^2 ={\cal O}(m_D^2)\gg M_\pi^2 ={\cal O} (m_D m_f)$ for $m_f\ll m_D$.
It is also phenomenologically crucial to have this chiral limit behavior in order for $\sigma$ to be a viable candidate for the 125 GeV Higgs in walking technicolor with $m_f=0$.

In this sense the recent LSD results~\cite{Appelquist:2016viq}, in a region of smaller $m_f$ than ours, show a similar tendency: $M_\sigma \simeq M_\pi$, but not $M_\sigma >M_\pi$, suggesting we may still be some distance away from the chiral limit.
Data at even smaller $m_f$ are needed to further establish the chiral behavior of $M_\sigma$.

All the couplings of the technidilaton as a composite Higgs are described by the effective field theory respecting all symmetries of the
underlying theory.
In the case at hand---the walking theory---these are the chiral and scale symmetries.
Possible explicit breakings such as $m_f$ effects and the trace anomaly are also controlled in terms of the spurion fields. The relevant effective field theory is dilaton ChPT~\cite{Matsuzaki:2013eva} as a straightforward scale-symmetric extension of conventional ChPT.
(See also~\cite{Matsuzaki:2015sya} and references therein.)

The ``Higgs potential'' for the dilaton field $\sigma(x)$ is uniquely determined by the trace anomaly in terms of just two parameters $M_\sigma$ and $F_\sigma$:
\begin{equation}
V(\sigma) =-\frac{F_\sigma^2}{4} M_\sigma^2 \chi^4 \left( \log \frac{\chi}{S} - \frac{1}{4}\right)\,
= - \frac{M_\sigma^2 F_\sigma^2}{16} +\frac{1}{2}M_\sigma^2\,\sigma^2 +\frac{4}{3} \frac{M_\sigma^2}{F_\sigma} \,\sigma^3
+ 2 \frac{M_\sigma^2}{F_\sigma^2}\, \sigma^4
+ \cdots
\,
\label{dilatonpotential}
\end{equation}
where $\chi(x)=\exp(\sigma(x)/F_\sigma)$ transforms as $\delta_D \chi= (1+ x^\mu\partial_\mu) \chi(x)$ $(\delta_D \sigma= F_\sigma + x^\mu\partial_\mu {\sigma}$) under scale transformations, and so does the spurion field $S(x)$
($\langle S\rangle=1$), hence
$\langle \delta_D V(\sigma)\rangle  =M_\sigma^2 F_\sigma^2 \langle \chi\rangle/4 =M_\sigma^2 F_\sigma^2/4 =\langle \theta^\mu_\mu\rangle $ in accordance with the PCDC in the underlying theory.
Thus measuring $F_\sigma$ as well as $M_\sigma$ on the lattice determines completely the Higgs potential for the 125 GeV Higgs as the technidilaton.
Also all the $\sigma$ couplings to the SM particles are determined by $F_\sigma$ (up to some nonperturbative contributions to the couplings of the technidilaton with the SM gauge boson pairs).

From the lattice data on the $\sigma$ mass through Eq.~(\ref{eq:dn}) based on dilaton-ChPT, we can read off not only  the chiral limit value of $M_\sigma$ given by $d_0$ but also the decay constant $F_\sigma$ from the slope $d_1=[(3-\gamma_m)(1+\gamma_m)/4] N_fF^2/F_\sigma^2 \simeq N_f F^2/F_\sigma^2$ (for $\gamma_m \sim1$).

Our data suggest that $F_\sigma \sim \sqrt{N_f} F$ from $d_1\sim1$.
A similar result has also been obtained by a different method~\cite{Aoki:2015jfa}.
In the case of the one-family model, this would imply $F_\sigma \sim {2} v_{\rm EW}$, which would be {somewhat smaller} compared with the favorable value {$F_\sigma \sim 3.7 v_{\rm EW}$ 
(for $N_c = 3, N_f = 8$)}
for accounting for the LHC 125 GeV Higgs data~\cite{Matsuzaki:2012xx}.
Considering that our data have significant uncertainties and are still far from the chiral limit, however, it may be a bit premature to
draw a definite conclusion for phenomenology.
Particularly, we should look at the relevant slope $d_1$ as well as $d_0$ in the region where $M_\sigma^2>M_\pi^2$, which is not available in our present lattice setting.

The $S$ parameter~\cite{Peskin:1990zt,Holdom:1990tc,Golden:1990ig} is usually a challenge for the walking technicolor based on the large $N_f$ QCD, since the large $N_f$ factor (more precisely, a large number of electroweak doublets $N_{\rm D}=N_f/2$ if all flavors carry the electroweak charges) enhances the $S$ parameter from the pure technicolor sector at least in perturbative calculations and/or a simple scale up of QCD.\footnote{
However, the large $S$ parameter from the technicolor sector as it stands is not necessarily in conflict with the experimental
value of the $S$ from the electroweak precision measurements, since it
can easily be canceled by the strong mixing with the SM fermion contribution
through the ETC  interactions, as was demonstrated in  the Higgsless model~\cite{Cacciapaglia:2004rb,Foadi:2004ps,Chivukula:2005xm}. It is also trivial to avoid the large $N_f$ factor by restricting
to only one doublet  carrying the electroweak charges, with the rest of the $N_f$ flavors being electroweak singlets, or vector-like.
}
Thus fully nonperturbative calculations on the lattice provide important constraints on the model building for walking technicolor.
The $S$ parameter in $N_f=8$ QCD on the lattice has been measured by the LSD Collaboration~\cite{Appelquist:2014zsa} based on domain wall fermions and shows some reduction of $S$ near the chiral limit, up to large errors and possible finite volume effects.

We measured the $S$ parameter (the preliminary results are given in Ref.~\cite{Aoki:2016bfp}),  based on our earlier observation~\cite{Aoki:lat2013} that $N_f=8$ staggered fermions have exact chiral symmetry and hence can give a well-defined $S$ parameter.
We observed some reduction of $S$ for smaller $m_f$, similarly to the LSD results. This however cannot be discriminated from the finite volume effects at this moment.
More careful analysis of the finite volume effects is required before drawing conclusions on the $S$ parameter.

The flavor-singlet pseudoscalar meson is an analogue of the $\eta^\prime$ in ordinary QCD and we call it $\eta^\prime$ here.
In $N_f=8$ there exist 64 NG bosons for the S$\chi$SB of the ${\rm U}(8)_L\times {\rm U}(8)_R$ symmetry, which explicitly breaks down to  ${\rm SU}(8)_L\times {\rm SU}(8)_R \times {\rm U}(1)_V$ due to the ${\rm U}(1)_A$ anomaly so that only $\eta^\prime$ becomes massive within the technicolor dynamics alone.
In large $N_f$ QCD we may consider the (``anti-Veneziano'') limit $N_f/N_c=$fixed $\gg 1$ in the large $N_c$ limit (with $N_c\alpha=$ fixed).
Consider the anomalous chiral Ward-Takahashi identity:~\cite{Matsuzaki:2015sya}
\begin{eqnarray}
 N_F F_\pi^2 M_{\eta^\prime}^2 &=& {\cal F.T.} \left\langle T\left(\partial^\mu A^0_\mu(x) \cdot \partial^\mu A^0_\mu(0)\right)\right\rangle \nonumber\\
 &=&
{\cal F.T.} \left\langle T\left(N_F \frac{\alpha}{4\pi} G^{\mu\nu} {\tilde G}_{\mu\nu} (x)\cdot  N_F \frac{\alpha}{4\pi} G^{\mu\nu} {\tilde G}_{\mu\nu}(0)
\right)\right\rangle
\nonumber\\
 &\sim& N_F^2\alpha^2 \times \left[
 N_C^2 \,\, ({\rm gluon\,\, loop})  +
 N_C^3 N_F \, \alpha^2 \,\ ({\rm inside \,\,fermion\,\,loop})
 \right]\,.
 \end{eqnarray}
In the usual Veneziano limit $N_f/N_c \ll 1$ we have the vanishing $\eta^\prime$ mass as $M_{\eta^\prime}^2/M_\rho^2 \sim (N_f/N_c)  \ll 1$ (pseudo NG boson), while for $N_f/N_c \gg1$ we would have $M_{\eta^\prime}^2/M_\rho^2 \sim (N_f/N_c)^2 \gg 1$.
Thus a non-perturbative understanding of the flavor-singlet pseudoscalar meson spectra in many-flavor QCD would be interesting.
In fact we have been studying the $\eta'$ mass in this model using a topological charge density operator $(q(x))$ constructed
from gauge link variables.
There is an advantage to using the gluonic operator over a fermionic one, since the gluonic operator does not directly couple to the lighter flavored pseudoscalar ($\pi$), and thus a better signal without pion contamination would be expected.
The gradient flow method~\cite{Luscher:2010iy} is also employed to improve the statistical accuracy.
We calculate the point-point correlation function $\langle q(x) q(y)\rangle$ for various flow time $t$.
Our (preliminary) lattice data suggests a large $\eta'$ mass of $M_{\eta^\prime}/M_\rho\simeq 3$~\cite{Aoki:2015jfa}.
In the future it would be desirable to apply the method above to study the gluonic correlation function in the flavor singlet scalar channel.

We have sampled the topological charge history for our data and found that the evolution slows down at smaller $m_f$ and larger $L$.
At no point does the evolution become sufficiently non-ergodic that we are concerned about the reliability of our results.
The topological charge and susceptibility may be studied in greater detail to compare and contrast theories with different $N_f$; preliminary results of this have been presented in Ref.~\cite{Aoki:2016yrm}, and full results will be deferred to a forthcoming publication.

\acknowledgments

We thank Marc Wagner for providing us the code for the measurements of the Wilson loops.
Numerical calculations have been carried out
on the high-performance computing systems at KMI (${\Large\varphi}$), 
at the Information Technology Center in Nagoya University (CX400), 
and at the Research Institute for Information Technology in Kyushu University (CX400 and HA8000).
This study also uses computational resources of the HPCI system provided by
the Research Institute for Information Technology in Kyushu University
through the HPCI System Research Projects 
(Project ID: hp140152, hp150157, hp160153).
This work utilizes the JLDG constructed over the SINET4 and SINET5 of NII.
This work is supported by the JSPS Grants-in-Aid for Scientific Research (S) No. 22224003, 
(C) No. 16K05320 (Y.A.), {(C) No. 15K05049 (K.N.),} 
for Young Scientists (A) No.16H06002 (T.Y.), (B) No.25800138 (T.Y.), 
(B) No.25800139 (H.O.), (B) No.15K17644 (K.M.),
and also by the MEXT Grants-in-Aid for Scientific Research on
Innovative Areas No.25105011 (M.K.).
K.M. is supported by the OCEVU Labex (ANR-11-LABX-0060) and the A*MIDEX project (ANR-11-IDEX-0001-02),
funded by the ``Investissements d'Avenir'' French government program and managed by the ANR.
The work of H.O. and E.R is supported by the RIKEN Special Postdoctoral Researcher program.
This work was performed in part under the auspices of the
U.S. Department of Energy by Lawrence Livermore National Laboratory
under contract~{DE-AC52-07NA27344}.

\clearpage

\appendix
\section{Parameters of hybrid Monte Carlo for the main ensembles}
\label{sec:appendix-hmc}

In Sec.~\ref{sec:action_and_param} a description of our main ensemble is
provided.
Here more detailed parameters for each ensemble are given.
Table~\ref{tab:nf8hmc} shows
for each stream in each ensemble, the molecular
dynamics time step size, values of the masses for the Hasenbusch
preconditioning if applicable, and the maximum number of thermalized
trajectories.
One trajectory amounts to a molecular dynamics evolution for one
unit time and successive accept-reject step.

\begin{table}
 \caption{Parameters of the main ensembles for $N_f=8$.
 $L$ and $T$ for the spatial and temporal size for $L^3\times T$
 lattice, staggered fermion mass $m_f$, molecular dynamics time step
 $\Delta\tau$, number of masses for the Hasenbusch preconditioning
 $N_{m_{\rm H}}$,  values of Hasenbusch masses $m_{\rm H}^i$,
 and maximum number of thermalized trajectories $N_{\rm Traj}^{\rm max}$
 are shown for each ``stream''.
}
 \label{tab:nf8hmc}
\begin{ruledtabular}
\begin{tabular}{cc l l l c r r r r c}
\multicolumn{1}{c}{$L$} &
\multicolumn{1}{c}{$T$} &
\multicolumn{1}{c}{$m_f$} &
\multicolumn{1}{c}{$\Delta\tau$} &
\multicolumn{1}{c}{$N_{m_{\rm H}}$} &
\multicolumn{1}{c}{$m_{\rm H}^1$} &
\multicolumn{1}{c}{$m_{\rm H}^2$} &
\multicolumn{1}{c}{$m_{\rm H}^3$} &
\multicolumn{1}{c}{$m_{\rm H}^4$} &
\multicolumn{1}{c}{$N_{\rm Traj}^{\rm max}$} &
\multicolumn{1}{c}{stream}\\

\hline
42 & 56 & 0.012 & 0.004    & 4  & 0.2 & 0.4 & 0.6 & 0.8 &  4440 & 1 \\
   &    &       & 0.004    & 4  & 0.2 & 0.4 & 0.6 & 0.8 &   320 & 2 \\
\cline{3-11}
   &    & 0.015 & 0.005    & 4  & 0.2 & 0.4 & 0.6 & 0.8 &  2200 & 1 \\
\hline
36 & 48 & 0.015 & 0.006667 & 2  & 0.3 & 0.8 & -   & -   & 10048 & 1 \\
   &    &       & 0.006667 & 2  & 0.3 & 0.8 & -   & -   &   752 & 2 \\
\cline{3-11}
   &    & 0.02  & 0.01     & 2  & 0.3 & 0.8 & -   & -   &  9984 & 1 \\
\cline{3-11}
   &    & 0.03  & 0.01     & 2  & 0.3 & 0.8 & -   & -   &  2000 & 1 \\
\hline
30 & 40 & 0.02  & 0.01     & 1  & 0.8 & -   & -   & -   & 16000 & 1 \\
\cline{3-11}
   &    & 0.03  & 0.0125   & 2  & 0.6 & 0.8 & -   & -   & 33024 & 1 \\
\cline{3-11}
   &    & 0.04  & 0.01     & 1  & 1.0 & -   & -   & -   & 14528 & 1 \\
   &    &       & 0.01     & 1  & 1.0 & -   & -   & -   &  4544 & 2 \\
   &    &       & 0.01     & 1  & 0.8 & -   & -   & -   &  6528 & 3 \\
\hline
24 & 32 & 0.03  & 0.015625 & 1  & 0.8 & -   & -   & -   & 27648 & 1 \\
   &    &       & 0.015625 & 1  & 0.8 & -   & -   & -   & 47104 & 2 \\
\cline{3-11}
   &    & 0.04  & 0.01     & 0  & -   & -   & -   & -   & 29696 & 1 \\
   &    &       & 0.015625 & 1  & 0.8 & -   & -   & -   & 70656 & 2 \\
\cline{3-11}
   &    & 0.06  & 0.016667 & 1  & 0.8 & -   & -   & -   & 39936 & 1 \\
\cline{3-11}
   &    & 0.08  & 0.0125   & 0  & -   & -   & -   & -   &  1216 & 1 \\
   &    &       & 0.016667 & 1  & 0.8 & -   & -   & -   & 16192 & 2 \\
\hline
18 & 24 & 0.04  & 0.0125   & 0  & -   & -   & -   & -   & 17920 & 1 \\
\cline{3-11}
   &    & 0.06  & 0.0125   & 0  & -   & -   & -   & -   & 17920 & 1 \\
\cline{3-11}
   &    & 0.08  & 0.0125   & 0  & -   & -   & -   & -   & 17920 & 1 \\
\end{tabular}
\end{ruledtabular}
\end{table}

\clearpage

\section{Topological charge measurement and gradient flow}
\label{sec:appendix-flow}

\newcommand \mf {m_{\textnormal{f}}}
\newcommand \su[1]{\mathrm{SU}(#1)}
\newcommand \Nf {N_{\textnormal{f}}}
\newcommand \dee {\mathrm{d}}
\newcommand \Dee {\mathrm{D}}
The topological charge is expected to be quantized to integer values, and to have a symmetric distribution about zero. As the continuum and chiral limits are approached, the Monte Carlo experiences a critical slowing down phenomenon where the autocorrelation time of the topological charge diverges; in extreme cases, the charge becomes frozen at a single value for many thousand units of Monte Carlo time. Since sampling a single or small range of topological charge gives a bias to the measured quantities, it is therefore important to investigate the history of the topological charge, to verify that a good sampling of topological sectors is being made.

The topological charge is defined in the continuum as
\begin{equation}
	Q = \int \dee^4x q(x)\;,\quad q(x)=\frac{1}{32\pi^2}\epsilon_{\mu\nu\rho\sigma}F_{\mu\nu}F_{\rho\sigma}\;.
\end{equation}
To find the lattice equivalent, we replace the integral by a sum, and obtain the equivalent of the field strength $F_{\mu\nu}$ by taking the path-ordered product of link variables around a clover-shaped path. We encounter a problem however when we apply this to gauge configurations as produced by a typical Monte Carlo process: namely that ultraviolet fluctuations dominate over the topological contribution to the charge. In principle these cancel out across the lattice volume, but in practice the topological contribution is smaller than the precision error of the UV fluctuations, and so the signal is lost.

We therefore need to suppress these contributions. Smoothing methods (for example, cooling~\cite{Teper:1985rb} and link smearing~\cite{Bonnet:2001rc}) have historically been used successfully. More recently, the gradient flow, as suggested by L\"uscher~\cite{Luscher:2010iy}, has gained in popularity to its greater physical motivation and connection with the continuum physics. In this work we use the latter method.

\subsubsection{Gradient flow}
The gradient flow defines a flowed field $B_\mu(t,x)$ at flow time $t$ as
\begin{align}
	\frac{\dee}{\dee t}B_\mu &= \Dee_\nu G_{\nu\mu}\, \\
	\left. B_\mu \right|_{t=0} &= A_\mu\;,\\
	G_{\mu\nu} &= \partial_\mu B_\nu - \partial_\nu B_\mu + [B_\mu, B_\nu]\;,\\
	\Dee_\mu &= \partial_\mu + \left[B_\mu, \frac{\dee}{\dee t}\right]\;,
\end{align}
where the flow starts at $B_\mu(0,x) = A_\mu(x)$, the physical gauge field.

Integrating numerically from $A_\mu$ allows calculation of $B_\mu(t,x)$ at arbitrary $t$. We do this using the Runge-Kutta-like scheme also outlined by L\"uscher~\cite{Luscher:2010iy}:
\begin{align}
	B_{t+\epsilon/3} &= \exp\left[\tfrac{1}{4} \epsilon Z(B_t)\right]B_t \;,\\
	B_{t+2\epsilon/3} &= \exp \left[ \tfrac{8}{9} \epsilon Z(B_{t+\epsilon/3}) - \tfrac{17}{36}\epsilon Z(B_t)\right]B_{t+\epsilon/3}\;,\\
	B_{t+\epsilon/3} &= \exp \left[ \tfrac{3}{4} \epsilon Z(B_{t+2\epsilon/3}) - \tfrac{8}{9} \epsilon Z(B_{t+\epsilon/3})\right.+ \\
	&\qquad \quad + \left. \tfrac{17}{36}\epsilon Z(B_t)\right] B_{t+2\epsilon/3}\;.
\end{align}

The characteristic smoothing radius of the flow is $\sqrt{8t}$. The gauge configurations tested for this work are, at a minimum, flowed to $\sqrt{8t} = L/2$, where $L$ is the spatial extent of the lattice; flowed configurations are saved at $\sqrt{8t}=L/6,L/4,L/2$ for later analysis. Beyond this point, the code algorithmically determines when the topological charge may be considered stable as the flow time changes, and stops the flow at that point.

\subsubsection{Scale setting}
It is also possible to use the gradient flow to define a scale. Two such scales have been proposed: $t_0$ by L\"uscher~\cite{Luscher:2010iy} and $w_0$ by the BMW collaboration~\cite{Borsanyi:2012zs}. Both of these are based on the behavior of $E = \frac{1}{4}G_{\mu\nu}G_{\mu\nu}$, as discretized by either of two methods---via the average plaquette, or via constructing a symmetric four-plaquette clover operator for $G_{\mu\nu}$; these definitions become equivalent in the continuum limit.

$t_0$ is defined as the flow time at which $t^2 \langle E(t)\rangle = c$, where $c$ is some appropriately-chosen constant given the physics of interest; for QCD, $c=0.3$ is generally taken. Meanwhile, for $w_0$, the function $W(t) = t\frac{\dee}{\dee t}\left[t^2 \langle E(t)\rangle\right]$ is used in almost the same way: again, we look at where $W(t)=c$, but now take this time as $t=w_0^2$---i.e. $t_0$ has mass dimension $-2$, while $w_0$ has mass dimension $-1$.

Values for $t_0$ and $w_0$ for $N_f = 4$ and $8$ are tabulated in Table~\ref{tab:t0w0}.

\begin{table}[!tbp]
\begin{ruledtabular}
\begin{tabular}{c ccc cccc}

$N_f$ & $L$ & $T$ & $m_f$ & $t_0^{\textnormal{plaq.}}$ & $t_0^{\textnormal{sym.}}$ & $w_0^{\textnormal{plaq.}}$ & $w_0^{\textnormal{sym.}}$ \\
\hline
4 & 20 & 30 & 0.01 & 0.8490(7) & 1.1381(8) & 1.1607(9) & 1.1612(9) \\
4 & 20 & 30 & 0.02 & 0.8117(6) & 1.0911(8) & 1.1045(8) & 1.1084(8) \\
4 & 20 & 30 & 0.03 & 0.7800(7) & 1.0524(8) & 1.0623(8) & 1.0685(8) \\
4 & 20 & 30 & 0.04 & 0.7491(6) & 1.0146(7) & 1.0240(6) & 1.0323(6) \\
\hline
8 & 42 & 56 & 0.012 & 4.1292(48) & 4.7543(55) & --- & --- \\
8 & 36 & 48 & 0.015 & 3.8798(58) & 4.4556(66) & 3.5982(65) & 3.5848(68) \\
8 & 36 & 48 & 0.02 & 3.5186(44) & 4.0304(50) & 3.1247(48) & 3.1122(47) \\
8 & 30 & 40 & 0.03 & 2.9567(29) & 3.3842(33) & 2.5600(28) & 2.5494(27) \\
8 & 30 & 40 & 0.04 & 2.5787(35) & 2.9563(39) & 2.2365(29) & 2.2281(29) \\
8 & 24 & 32 & 0.05 & 2.3138(62) & 2.6599(71) & --- & --- \\
8 & 24 & 32 & 0.06 & 2.0750(6) & 2.3932(7) & 1.8436(5) & 1.8397(5) \\
8 & 24 & 32 & 0.07 & 1.8906(31) & 2.1885(34) & 1.7050(24) & 1.7035(23) \\
8 & 18 & 24 & 0.08 & 1.7421(48) & 2.0241(54) & 1.5971(36) & 1.5976(36) \\
8 & 24 & 32 & 0.08 & 1.7528(22) & 2.0362(24) & 1.6064(17) & 1.6067(16) \\
8 & 18 & 24 & 0.10 & 1.5259(24) & 1.7876(26) & 1.4469(18) & 1.4510(18) \\
8 & 24 & 32 & 0.10 & 1.5188(18) & 1.7789(19) & 1.4397(14) & 1.4439(14) \\
8 & 12 & 16 & 0.12 & 1.363(90) & 1.610(98) & --- & --- \\
8 & 12 & 16 & 0.16 & 1.115(48) & 1.341(55) & --- & --- \\
\end{tabular}
\end{ruledtabular}
\caption{
Numbers for the gradient flow scales $t_0$ and $w_0$, as defined in the text. The ``plaq.'' and ``sym.'' refer respectively to the single-plaquette and symmetric four-plaquette definitions for $G_{\mu\nu}$.}
\label{tab:t0w0}
\end{table}

\clearpage

\section{Summary Tables for Hadron Mass Spectra I:
$F_{\pi},M_{\pi},M_{\rho}$, and $\langle\overline{\psi}\psi\rangle$.}
\label{sec:table_fpi_mpi}

In this appendix results for the basic hadron spectra are summarized.
The parameters for the measurement in the updated simulations
from our previous paper~\cite{Aoki:2013xza} are summarized in
Table~\ref{tab:hadron_stat}.
Tables~\ref{tab:hadron_spectI:L42}--\ref{tab:hadron_spectI:L12}
present the results for $F_\pi, M_\pi, M_{\pi{\rm (SC)}}, M_{\rho{\rm (PV)}},
M_{\rho{\rm (VT)}}$, and $\langle \overline{\psi}\psi\rangle$
on each volume.

\begin{table}[!tbp]
\caption{
Numbers for trajectories ($N_{\rm Traj}$),
stream ($N_{\rm str}$),
configuration ($N_{\rm conf}$),
for spectrum measurement in each parameter.
The bin size of jackknife analysis ($N_{\rm bin}$) and
number of measurements per configuration ($N_{\rm meas}$)
are also summarized.
}
\label{tab:hadron_stat}
\begin{ruledtabular}
\begin{tabular}{cc l r c c r r r}
\multicolumn{1}{c}{$L$} &
\multicolumn{1}{c}{$T$} &
\multicolumn{1}{c}{$m_f$} &
\multicolumn{1}{c}{$N_{\rm Traj}$} &
\multicolumn{1}{c}{$N_{\rm str}$} &

\multicolumn{1}{c}{$N_{\rm conf}$} &
\multicolumn{1}{c}{$N_{\rm bin}$} &
\multicolumn{1}{c}{$N_{\rm meas}$} \\
\hline
42 & 56 & 0.012 &  4760 & 2 & 1190 & 476 &14 \\
42 & 56 & 0.015 &  2200 & 1 &  550 & 200 & 7 \\
36 & 48 & 0.015 & 10800 & 2 & 1350 & 200 & 6 \\
36 & 48 & 0.02  &  9984 & 1 &  312 & 256 & 6 \\
36 & 48 & 0.03  &  2000 & 1 &  500 & 200 & 6 \\
30 & 40 & 0.02  & 16000 & 1 &  500 & 320 & 6 \\
30 & 40 & 0.03  & 33024 & 1 &  516 & 256 & 6 \\
30 & 40 & 0.04  & 25600 & 3 &  400 & 256 & 6 \\
24 & 32 & 0.03  & 74752 & 2 &  584 & 512 & 8 \\
24 & 32 & 0.04  &100352 & 2 &  392 &1024 & 8 \\
24 & 32 & 0.06  & 39936 & 1 &  312 & 512 & 8 \\
24 & 32 & 0.08  & 17408 & 2 &  272 & 256 & 8 \\
18 & 24 & 0.04  & 17920 & 1 &  280 & 256 & 6 \\
18 & 24 & 0.06  & 17920 & 1 &  280 & 256 & 6 \\
18 & 24 & 0.08  & 17920 & 1 &  280 & 256 & 6 \\

\end{tabular}
\end{ruledtabular}
\end{table}

\begin{table}[!tbp]
\caption{
$L^3\times T = 42^3 \times 56$.
}
\label{tab:hadron_spectI:L42}
\begin{ruledtabular}
\begin{tabular}{lllllll}
\multicolumn{1}{c}{$m_f$} &
\multicolumn{1}{c}{$F_\pi$} &
\multicolumn{1}{c}{$M_\pi$} &
\multicolumn{1}{c}{$M_{\pi{\rm (SC)}}$} &
\multicolumn{1}{c}{$M_{\rho{\rm (PV)}}$} &
\multicolumn{1}{c}{$M_{\rho{\rm (VT)}}$} &
\multicolumn{1}{c}{$\langle \overline{\psi}\psi\rangle$} \\
\hline
0.012 & 0.04542(27) & 0.16362(43) & 0.16491(43) & 0.2536(17) & 0.2522(15) &
0.0073110(76) \\
0.015 & 0.05054(15) & 0.18614(44) & 0.18747(45) & 0.2827(21) & 0.2815(19) &
0.0090454(42)
\\
\end{tabular}
\end{ruledtabular}
\end{table}

\begin{table}[!tbp]
\caption{
$L^3\times T = 36^3 \times 48$.
}
\label{tab:hadron_spectI:L36}
\begin{ruledtabular}
\begin{tabular}{lllllll}
\multicolumn{1}{c}{$m_f$} &
\multicolumn{1}{c}{$F_\pi$} &
\multicolumn{1}{c}{$M_\pi$} &
\multicolumn{1}{c}{$M_{\pi{\rm (SC)}}$} &
\multicolumn{1}{c}{$M_{\rho{\rm (PV)}}$} &
\multicolumn{1}{c}{$M_{\rho{\rm (VT)}}$} &
\multicolumn{1}{c}{$\langle \overline{\psi}\psi\rangle$} \\
\hline
0.015 & 0.05047(14) & 0.18606(31) & 0.18769(35) & 0.2815(23) & 0.2813(20) &
0.0090392(53)
\\
0.02  & 0.05848(15) & 0.22052(33) & 0.22217(35) & 0.3223(31) & 0.3234(26) &
0.0119000(65)
\\
0.03  & 0.07137(20) & 0.28084(39) & 0.28271(44) & 0.4059(54) & 0.4021(46) &
0.0174826(91)
\\
\end{tabular}
\end{ruledtabular}
\end{table}

\begin{table}[!tbp]
\caption{
$L^3\times T = 30^3 \times 40$.
Dagger ($^\dagger$) denotes data that have not been updated since
our previous paper~\cite{Aoki:2013xza}.
}
\label{tab:hadron_spectI:L30}
\begin{ruledtabular}
\begin{tabular}{lllllll}
\multicolumn{1}{c}{$m_f$} &
\multicolumn{1}{c}{$F_\pi$} &
\multicolumn{1}{c}{$M_\pi$} &
\multicolumn{1}{c}{$M_{\pi{\rm (SC)}}$} &
\multicolumn{1}{c}{$M_{\rho{\rm (PV)}}$} &
\multicolumn{1}{c}{$M_{\rho{\rm (VT)}}$} &
\multicolumn{1}{c}{$\langle \overline{\psi}\psi\rangle$} \\
\hline
0.02  & 0.05775(17) & 0.22232(42) & 0.22411(49) & 0.3334(22) & 0.3326(21) &
0.0118837(54)
\\
0.03  & 0.07157(10) & 0.28122(24) & 0.28334(27) & 0.4075(24) & 0.4071(20) &
0.0174824(37)
\\
0.04  & 0.08264(10) & 0.33501(21) & 0.33729(27) & 0.4719(23) & 0.4709(22) &
0.0229218(42)
\\
0.05$^\dagger$ & 0.09182(23) & 0.38336(48) & 0.3859(5) & 0.5317(92) &
0.5302(80) & 0.028219(11) \\
0.06$^\dagger$ & 0.10118(28) & 0.43035(44) & 0.4332(4) & 0.585(13)  &
0.589(12)  & 0.033437(13) \\
0.07$^\dagger$ & 0.10985(25) & 0.47347(42) & 0.4769(4) & 0.635(14)  &
0.633(12)  & 0.038555(13)
\\
\end{tabular}
\end{ruledtabular}
\end{table}

\begin{table}[!tbp]
\caption{
$L^3\times T = 24^3 \times 32$.
Dagger ($^\dagger$) denotes data, which have not been updated from
our previous paper~\cite{Aoki:2013xza}.
}
\label{tab:hadron_spectI:L24}
\begin{ruledtabular}
\begin{tabular}{lllllll}
\multicolumn{1}{c}{$m_f$} &
\multicolumn{1}{c}{$F_\pi$} &
\multicolumn{1}{c}{$M_\pi$} &
\multicolumn{1}{c}{$M_{\pi{\rm (SC)}}$} &
\multicolumn{1}{c}{$M_{\rho{\rm (PV)}}$} &
\multicolumn{1}{c}{$M_{\rho{\rm (VT)}}$} &
\multicolumn{1}{c}{$\langle \overline{\psi}\psi\rangle$} \\
\hline
0.02$^\dagger$ & 0.05661(79) & 0.2330(25)  & 0.2367(37) & 0.351(12)  &
0.346(11)  & 0.011881(28) \\
0.03  & 0.07085(11) & 0.28306(34) & 0.28525(51) & 0.4134(29) & 0.4119(26) &
0.0174647(35)
\\
0.04  & 0.08235(14) & 0.33487(35) & 0.33751(43) & 0.4686(22) & 0.4677(21) &
0.0229131(34)
\\
0.05$^\dagger$ & 0.09176(51) & 0.3826(10)  & 0.3851(11) & 0.5274(54) &
0.5228(53) & 0.028248(23) \\
0.06  & 0.10134(16) & 0.43001(29) & 0.43369(35) & 0.5839(23) & 0.5840(21) &
0.0334455(46)
\\
0.07$^\dagger$ & 0.10879(32) & 0.47307(61) & 0.4767(7) & 0.6288(74) &
0.6345(75) & 0.038532(14) \\
0.08  & 0.11696(18) & 0.51479(31) & 0.51882(34) & 0.6758(33) & 0.6755(30) &
0.0435570(69)
\\
0.10$^\dagger$ & 0.13152(26) & 0.59401(55) & 0.5987(6) & 0.7790(65) &
0.7760(68) & 0.053338(12)
\\
\end{tabular}
\end{ruledtabular}
\end{table}

\begin{table}[!tbp]
\caption{
$L^3\times T = 18^3 \times 24$.
Dagger ($^\dagger$) denotes data that have not been updated from
our previous paper~\cite{Aoki:2013xza}.
}
\label{tab:hadron_spectI:L18}
\begin{ruledtabular}
\begin{tabular}{lllllll}
\multicolumn{1}{c}{$m_f$} &
\multicolumn{1}{c}{$F_\pi$} &
\multicolumn{1}{c}{$M_\pi$} &
\multicolumn{1}{c}{$M_{\pi{\rm (SC)}}$} &
\multicolumn{1}{c}{$M_{\rho{\rm (PV)}}$} &
\multicolumn{1}{c}{$M_{\rho{\rm (VT)}}$} &
\multicolumn{1}{c}{$\langle \overline{\psi}\psi\rangle$} \\
\hline
0.04  & 0.08090(28) & 0.34083(91) & 0.3436(12)  & 0.4829(33) & 0.4827(31) &
0.022844(18)
\\
0.05$^\dagger$ & 0.09096(57) & 0.38856(15) & 0.3908(19) & 0.5323(84) &
0.5248(79) & 0.028177(28) \\
0.06  & 0.10082(23) & 0.43170(63) & 0.43533(79) & 0.5908(25) & 0.5912(24) &
0.0334381(99)
\\
0.07$^\dagger$ & 0.10899(42) & 0.4734(10)  & 0.4777(13) & 0.6436(63) &
0.6398(64) & 0.038493(27) \\
0.08  & 0.11694(23) & 0.51524(50) & 0.51939(59) & 0.6804(25) & 0.6792(23) &
0.0435545(80)
\\
0.10$^\dagger$ & 0.13151(44) & 0.5948(11)  & 0.5993(12) & 0.7729(65) &
0.7698(56) & 0.053338(35)
\\
\end{tabular}
\end{ruledtabular}
\end{table}

\begin{table}[!tbp]
\caption{
$L^3\times T = 12^3 \times 16$.
Dagger ($^\dagger$) denotes data that have not been updated from
our previous paper~\cite{Aoki:2013xza}.
}
\label{tab:hadron_spectI:L12}
\begin{ruledtabular}
\begin{tabular}{lllllll}
\multicolumn{1}{c}{$m_f$} &
\multicolumn{1}{c}{$F_\pi$} &
\multicolumn{1}{c}{$M_\pi$} &
\multicolumn{1}{c}{$M_{\pi{\rm (SC)}}$} &
\multicolumn{1}{c}{$M_{\rho{\rm (PV)}}$} &
\multicolumn{1}{c}{$M_{\rho{\rm (VT)}}$} &
\multicolumn{1}{c}{$\langle \overline{\psi}\psi\rangle$} \\
\hline
0.04$^\dagger$ & 0.0622(15) & 0.4181(110)& 0.4397(113) & 0.5574(370)   & 0.5389(360) & 0.02167(4) \\
0.05$^\dagger$ & 0.0735(12) & 0.4844(65)  & 0.4987(72)    & 0.6108(81)     & 0.6157(71)   & 0.02704(5) \\
0.06$^\dagger$ & 0.0904(15) & 0.4681(79)  & 0.4664(120)  & 0.6372(237)   & 0.6343(218)& 0.03269(6) \\
0.07$^\dagger$ & 0.1030(9)   & 0.5091(38)  & 0.5168(64)     & 0.6910(102)   & 0.6882(87)  & 0.03802(7) \\
0.08$^\dagger$ & 0.1144(6)   & 0.5352(23)  & 0.5439(26)     & 0.7093(69)     & 0.7031(53)  & 0.04328(6) \\
0.09$^\dagger$ & 0.1222(7)   & 0.5686(17)  & 0.5774(24)     & 0.7478(54)     & 0.7449(54)  & 0.04826(7) \\
0.10$^\dagger$ & 0.1302(6)   & 0.6033(19)  & 0.6116(23)     & 0.7886(65)     & 0.7866(61)  & 0.05319(6) \\
0.12$^\dagger$ & 0.1442(4)   & 0.6694(12)  & 0.6760(13)     & 0.8556(43)     & 0.8517(43)  & 0.06265(4) \\
0.14$^\dagger$ & 0.1565(3)   & 0.7384(11)  & 0.7460(13)     & 0.9321(38)     & 0.9317(39)  & 0.07181(4) \\
0.16$^\dagger$ & 0.1676(2)   &0.8056(8)     & 0.8142(9)        & 1.0032(29)     & 1.0029(26)  & 0.08059(3)
\\
\end{tabular}
\end{ruledtabular}
\end{table}

\clearpage

\section{Summary Tables for Hadron Mass Spectra II:
$M_{a_0},M_{a_1},M_{b_1},M_{N}$, and $M_{N_{\bf 1}^*}$}
\label{sec:table_had_mass}

This appendix summarizes the masses for $a_0$, $a_1$, $b_1$, $N$, and
$N_{\bf 1}^*$.
Tables~\ref{tab:hadron_spectII:L42}--\ref{tab:hadron_spectII:L18}
present those results on each volume.

\begin{table}[!tbp]
\caption{
$L^3\times T = 42^3 \times 56$.
}
\label{tab:hadron_spectII:L42}
\begin{ruledtabular}
\begin{tabular}{llllll}
\multicolumn{1}{c}{$m_f$} &
\multicolumn{1}{c}{$M_{a_0}$} &
\multicolumn{1}{c}{$M_{a_1}$} &
\multicolumn{1}{c}{$M_{b_1}$} &
\multicolumn{1}{c}{$M_N$} &
\multicolumn{1}{c}{$M_{N_{\bf 1}^*}$} \\
\hline
0.012 & 0.279(10) & 0.346(11) & 0.3503(95) & 0.3697(24) & 0.462(10)
\\
0.015 & 0.310(10) & 0.387(10) & 0.429(21)  & 0.4200(62) & 0.542(20)
\\
\end{tabular}
\end{ruledtabular}
\end{table}

\begin{table}[!tbp]
\caption{
$L^3\times T = 36^3 \times 48$.
}
\label{tab:hadron_spectII:L36}
\begin{ruledtabular}
\begin{tabular}{llllll}
\multicolumn{1}{c}{$m_f$} &
\multicolumn{1}{c}{$M_{a_0}$} &
\multicolumn{1}{c}{$M_{a_1}$} &
\multicolumn{1}{c}{$M_{b_1}$} &
\multicolumn{1}{c}{$M_N$} &
\multicolumn{1}{c}{$M_{N_{\bf 1}^*}$} \\
\hline
0.015 & 0.3151(79) & 0.3854(82) & 0.397(15) & 0.4105(25) & 0.5106(92)
\\
0.02  & 0.365(11)  & 0.460(12)  & 0.465(18) & 0.4773(37) & 0.589(14)
\\
0.03  & 0.480(39)  & 0.572(23)  & 0.535(33) & 0.5921(41) & 0.712(17)
\\
\end{tabular}
\end{ruledtabular}
\end{table}

\begin{table}[!tbp]
\caption{
$L^3\times T = 30^3 \times 40$.
}
\label{tab:hadron_spectII:L30}
\begin{ruledtabular}
\begin{tabular}{llllll}
\multicolumn{1}{c}{$m_f$} &
\multicolumn{1}{c}{$M_{a_0}$} &
\multicolumn{1}{c}{$M_{a_1}$} &
\multicolumn{1}{c}{$M_{b_1}$} &
\multicolumn{1}{c}{$M_N$} &
\multicolumn{1}{c}{$M_{N_{\bf 1}^*}$} \\
\hline
0.02  & 0.3670(78) & 0.443(14) & 0.471(17) & 0.4913(28) & 0.6163(71)
\\
0.03  & 0.463(11)  & 0.528(24) & 0.547(48) & 0.6042(23) & 0.748(17)
\\
0.04  & 0.567(23)  & 0.610(27) & 0.752(61) & 0.6939(26) & 0.900(26)
\\
\end{tabular}
\end{ruledtabular}
\end{table}

\begin{table}[!tbp]
\caption{
$L^3\times T = 24^3 \times 32$.
}
\label{tab:hadron_spectII:L24}
\begin{ruledtabular}
\begin{tabular}{llllll}
\multicolumn{1}{c}{$m_f$} &
\multicolumn{1}{c}{$M_{a_0}$} &
\multicolumn{1}{c}{$M_{a_1}$} &
\multicolumn{1}{c}{$M_{b_1}$} &
\multicolumn{1}{c}{$M_N$} &
\multicolumn{1}{c}{$M_{N_{\bf 1}^*}$} \\
\hline
0.03  & 0.4597(73) & 0.504(14) & 0.546(24) & 0.6048(37) & 0.709(20)
\\
0.04  & 0.540(12)  & 0.676(39) & 0.579(36) & 0.6917(46) & 0.854(54)
\\
0.06  & 0.634(22)  & 0.726(27) & 0.764(59) & 0.8781(69) & 1.062(58)
\\
0.08  & 0.789(45)  & 0.948(67) & 0.885(73) & 1.0212(76) & 1.37(13)
\\
\end{tabular}
\end{ruledtabular}
\end{table}

\begin{table}[!tbp]
\caption{
$L^3\times T = 18^3 \times 24$.
}
\label{tab:hadron_spectII:L18}
\begin{ruledtabular}
\begin{tabular}{llllll}
\multicolumn{1}{c}{$m_f$} &
\multicolumn{1}{c}{$M_{a_0}$} &
\multicolumn{1}{c}{$M_{a_1}$} &
\multicolumn{1}{c}{$M_{b_1}$} &
\multicolumn{1}{c}{$M_N$} &
\multicolumn{1}{c}{$M_{N_{\bf 1}^*}$} \\
\hline
0.04  & 0.515(14) & 0.656(13) & 0.657(19) & 0.7298(61) & 0.882(15)
\\
0.06  & 0.651(19) & 0.686(42) & 0.722(56) & 0.8897(44) & 1.040(25)
\\
0.08  & 0.795(29) & 1.064(77) & 0.804(65) & 1.0301(43) & 1.201(48)
\\
\end{tabular}
\end{ruledtabular}
\end{table}

\clearpage

\section{Chiral log correction}
\label{app:log}

Since the strategy to estimate the chiral log corrections
to the low energy constants $F$ and $B$
is the same as in Appendix C of Ref.~\cite{Aoki:2013xza},
we show only the results using the data in this paper.

\begin{figure}[!tbp]
\makebox[.5\textwidth][r]{\includegraphics[scale=.5]{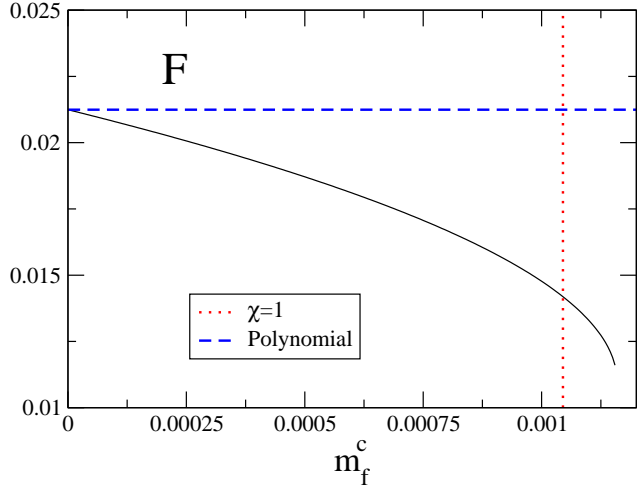}}
\caption{
The solid curve is the result for $F$ as a function of the matching point $m_f^c$.
The dashed and dotted lines represent $m_f^c$, where ${\mathcal X}=1$,
and the polynomial fit result tabulated in Table~\ref{tab:fit-fpi}, respectively.
}
\label{fig:log-F}
\end{figure}

\begin{figure}[!tbp]
\makebox[.5\textwidth][r]{\includegraphics[scale=.5]{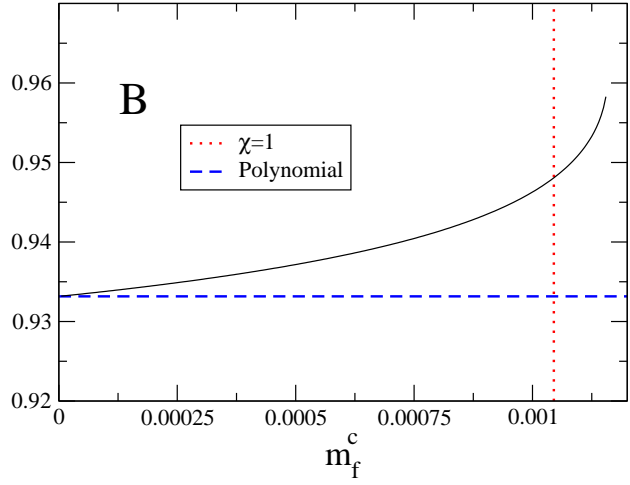}}
\caption{
As Fig.~\ref{fig:log-F}, but for $B$.
The polynomial fit result corresponds to $C_0/2$ in Table~\ref{tab:fit-fpi}.
}
\label{fig:log-B}
\end{figure}

\begin{figure}[!tbp]
\makebox[.5\textwidth][r]{\includegraphics[scale=.5]{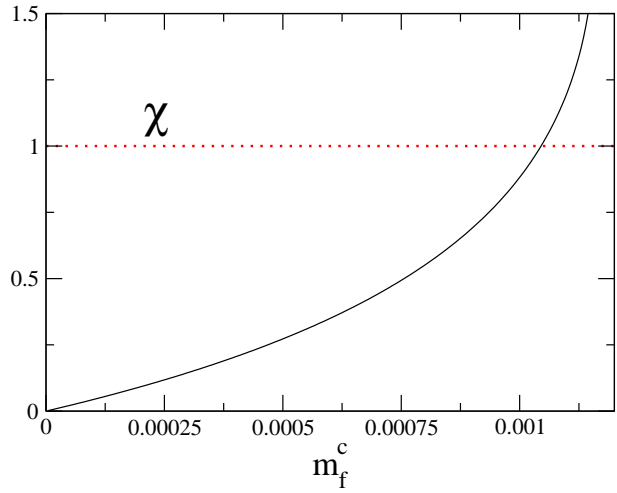}}
\caption{
The expansion parameter of ChPT ${\mathcal X}$ as a function of $m_f^c$.
The dotted line represents ${\mathcal X}=1$.
}
\label{fig:log-X}
\end{figure}

Figures~\ref{fig:log-F} and \ref{fig:log-B} show the
results for $F$ and $B$ as a function of the matching point $m_f^c$,
respectively.
At each $m_f^c$ the expansion parameter of ChPT $\mathcal{X}$
is calculated by $F$ and $B$ as plotted in Fig.~\ref{fig:log-X}.
We obtain $\mathcal{X}=1$ at $m_f^c = 0.001045$, where $F=0.142$
in Fig.~\ref{fig:log-F}.
This value is used to estimate the size of the chiral log correction of $F$.

\begin{figure}[!tbp]
\makebox[.5\textwidth][r]{\includegraphics[scale=.5]{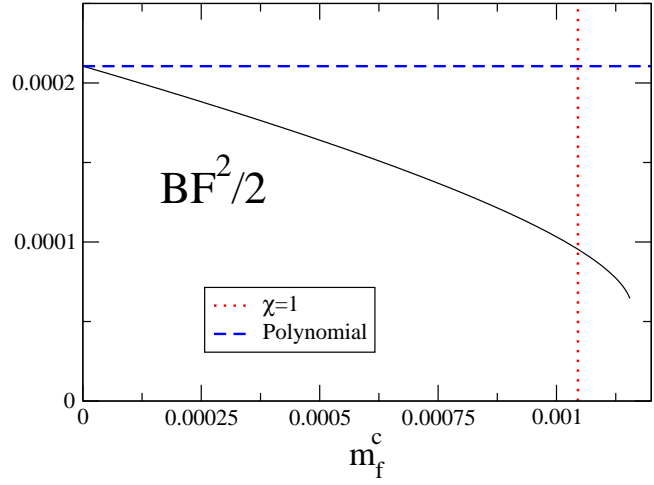}}
\caption{
As Fig.~\ref{fig:log-F}, but for $BF^2/2$.
The polynomial fit result is tabulated in Table~\ref{tab:fit-pbp}.
}
\label{fig:log-GMOR}
\end{figure}

We also estimate the size of the correction in the chiral condensate
through the GMOR relation Eq.~(\ref{eq:GMOR}).
The $m_f^c$ dependence of the quantity is presented in
Fig.~\ref{fig:log-GMOR}.
At $m_f^c = 0.001045$ we obtain roughly half of the quadratic fit result.

\clearpage

\section{Hyperscaling analyses}

\subsection{Effective mass anomalous dimension}\label{app_subsec:geff}
In Sec.~\ref{subsec:geff}, we investigated the effective mass anomalous dimension $\gamma_{\mathrm{eff}}$.
In Table~\ref{tab:nf8b3.8gamma},
we show the numerical results of $\gamma_{\mathrm{eff}}$ for $N_f = 8$.
For $\{F_{\pi},M_{\pi},M_{\rho}\}$,
the Large Volume Data Set explained in Sec.~\ref{subsec:mh_fv} is used to obtain the $\gamma_{\mathrm{eff}}$.
For $M_{N}$,
the available data are limited to the updated ensemble, for which the $\gamma_{\mathrm{eff}}$ is computed.
In Tables~\ref{tab:nf12b4gamma} and \ref{tab:nf12b3.7gamma},
we show the $\gamma_{\mathrm{eff}}$ for $N_f = 12$ with $\beta = 4.0$ and $3.7$, respectively.
The $N_f = 12$ spectrum data are summarized in Appendix~\ref{app:nf12},
from which the results obtained at the largest volumes are used to obtain the $\gamma_{\mathrm{eff}}$.
The $\gamma_{\mathrm{eff}}$ shown in this appendix are shown in Fig.~\ref{fig:geff}.

\begin{table}[!tbp]
\caption{
The effective mass anomalous dimensions $\gamma_{\mathrm{eff}}$ obtained by
a hyperscaling fit~(\protect\ref{eq:hs_naive}) of the $N_f = 8$ spectrum data
and shown in the left panel of Fig.~\protect\ref{fig:geff}.
For details, see text in Sec.~\protect\ref{subsec:geff}.
}\label{tab:nf8b3.8gamma}
\begin{ruledtabular}
\begin{tabular}{c cc cc cc cc}
\multicolumn{1}{c}{fit range} &
\multicolumn{2}{c}{$M_\pi$}   &
\multicolumn{2}{c}{$M_\rho$}  &
\multicolumn{2}{c}{$F_\pi$}   &
\multicolumn{2}{c}{$M_N$}    \\
\multicolumn{1}{c}{$m_f$} &
\multicolumn{1}{c}{$\gamma$} &
\multicolumn{1}{c}{$\chi^2/\text{dof}$} &
\multicolumn{1}{c}{$\gamma$} &
\multicolumn{1}{c}{$\chi^2/\text{dof}$} &
\multicolumn{1}{c}{$\gamma$} &
\multicolumn{1}{c}{$\chi^2/\text{dof}$} &
\multicolumn{1}{c}{$\gamma$} &
\multicolumn{1}{c}{$\chi^2/\text{dof}$} \\
\hline
0.012-0.02 & 0.706(16) & 0.65  & 1.132(104) & 0.0007 & 0.987(40) & 1.25  & 1.003(79) & 0.78 \\
0.015-0.03 & 0.676(7)  & 0.69  & 0.856(49)  & 2.74   & 0.989(16) & 0.60  & 0.783(32) & 1.23 \\
0.02-0.04  & 0.654(6)  & 1.66  & 0.841(50)  & 2.25   & 1.004(15) & 0.019 & 0.899(42) & 9.63 \\
0.03-0.05  & 0.647(8)  & 0.25  & 0.947(89)  & 0.069  & 1.031(21) & 3.15  & -         & -    \\
0.03-0.06  & -         & -     & -          & -      & -         & -     & 0.904(43) & 9.39 \\
0.04-0.06  & 0.625(6)  & 3.64  & 0.904(55)  & 0.012  & 0.994(19) & 6.62  & -         & -    \\
0.04-0.08  & -         & -     & -          & -      & -         & -     & 0.780(36) & 1.05 \\
0.05-0.07  & 0.599(14) & 0.62  & 1.024(247) & 0.070  & 0.962(43) & 7.39  & -         & -    \\
0.06-0.08  & 0.598(8)  & 0.54  & 0.971(84)  & 0.13   & 1.009(31) & 3.55  & -         & -    \\
0.07-0.10  & 0.563(10) & 0.45  & 0.622(96)  & 0.73   & 0.888(32) & 0.31  & -         & -    \\

\end{tabular}
\end{ruledtabular}
\end{table}

\begin{table}[!tbp]
\caption{
The effective mass anomalous dimension $\gamma_{\mathrm{eff}}$ obtained by
a hyperscaling fit~(\protect\ref{eq:hs_naive})
of the $N_f = 12$ spectrum data at $\beta = 4$.
The corresponding figure is found in the right panel of Fig.~\protect\ref{fig:geff}.
For details, see text in Sec.~\protect\ref{subsec:geff}.
}\label{tab:nf12b4gamma}
\begin{ruledtabular}
\begin{tabular}{c cc cc cc cc}
\multicolumn{1}{c}{fit range} &
\multicolumn{2}{c}{$M_\pi$}   &
\multicolumn{2}{c}{$M_\rho$}  &
\multicolumn{2}{c}{$F_\pi$}   &
\multicolumn{2}{c}{$M_N$}    \\
\multicolumn{1}{c}{$m_f$} &
\multicolumn{1}{c}{$\gamma$} &
\multicolumn{1}{c}{$\chi^2/\text{dof}$} &
\multicolumn{1}{c}{$\gamma$} &
\multicolumn{1}{c}{$\chi^2/\text{dof}$} &
\multicolumn{1}{c}{$\gamma$} &
\multicolumn{1}{c}{$\chi^2/\text{dof}$} &
\multicolumn{1}{c}{$\gamma$} &
\multicolumn{1}{c}{$\chi^2/\text{dof}$} \\
\hline
0.04-0.06 & 0.402(12) & 0.007 & 0.443(27) & 0.03 & 0.449(25) & 1.0     & 0.427(31)    & 0.009 \\
0.05-0.08 & 0.389(5) & 0.60    & 0.400(11) & 2.0    & 0.443(14) & 0.9  & 0.400(0.016) & 0.3   \\
0.06-0.1  & 0.396(4) & 3.0     & 0.416(9) & 4.9      & 0.473(12) & 6.5  & 0.406(14)    & 0.6   \\
0.08-0.12 & 0.406(4) & 0.004  & 0.439(12) & 0.7    & 0.520(14) & 0.2   & 0.437(20)    & 0.1   \\
0.1-0.16  & 0.410(3) & 0.2     & 0.456(6) & 2.4      & 0.568(10) & 3.9  & 0.470(13)    & 0.4   \\
0.12-0.2  & 0.399(3) & 8.8     & 0.473(8) & 0.05    & 0.563(25) & 4.7   & 0.487(13)    & 0.05  \\
0.16-0.24 & 0.391(4) & 1.7     & 0.476(10) & 0.6   & 0.543(13) & 0.4   & 0.483(23)    & 1.2   \\
0.2-0.3   & 0.411(2) & 0.03    & 0.513(8) & 0.1      & 0.635(7) & 0.06 & 0.477(11)    & 1.1   \\

\end{tabular}
\end{ruledtabular}
\end{table}

\begin{table}[!tbp]
\caption{
The effective mass anomalous dimension $\gamma_{\mathrm{eff}}$ obtained by
a hyperscaling fit~(\protect\ref{eq:hs_naive})
of the $N_f = 12$ spectrum data at $\beta = 3.7$.
The corresponding figure is found in the right panel of Fig.~\protect\ref{fig:geff}.
For details, see text in Sec.~\protect\ref{subsec:geff}.
}\label{tab:nf12b3.7gamma}
\begin{ruledtabular}
\begin{tabular}{c cc cc cc cc}
\multicolumn{1}{c}{fit range} &
\multicolumn{2}{c}{$M_\pi$}   &
\multicolumn{2}{c}{$M_\rho$}  &
\multicolumn{2}{c}{$F_\pi$}   &
\multicolumn{2}{c}{$M_N$}    \\
\multicolumn{1}{c}{$m_f$} &
\multicolumn{1}{c}{$\gamma$} &
\multicolumn{1}{c}{$\chi^2/\text{dof}$} &
\multicolumn{1}{c}{$\gamma$} &
\multicolumn{1}{c}{$\chi^2/\text{dof}$} &
\multicolumn{1}{c}{$\gamma$} &
\multicolumn{1}{c}{$\chi^2/\text{dof}$} &
\multicolumn{1}{c}{$\gamma$} &
\multicolumn{1}{c}{$\chi^2/\text{dof}$} \\
\hline

0.035-0.05 & 0.401(14) & 0.07 & 0.381(31) & 0.2    & 0.386(29) & 1.0    & 0.392(32)  & 0.3    \\
0.04-0.06  & 0.413(10) & 0.8  & 0.399(18) & 0.08   & 0.422(18) & 0.01   & 0.406(23)  & 0.0004 \\
0.05-0.08  & 0.435(8)  & 0.6  & 0.445(20) & 1.1    & 0.538(18) & 6.3    & 0.431(20)  & 0.3    \\
0.06-0.1   & 0.449(4)  & 0.4  & 0.456(14) & 0.09   & 0.550(12) & 1.2    & 0.471(15)  & 0.8    \\
0.08-0.12  & 0.446(7)  & 0.9  & 0.474(21) & 0.6    & 0.500(14) & 0.5    & 0.445(17)  & 3.2    \\
0.1-0.16   & 0.442(3)  & 0.03 & 0.500(13) & 0.0008 & 0.565(11) & 7.6    & 0.4675(97) & 5.7    \\
0.12-0.2   & 0.443(2)  & 0.01 & 0.523(9)  & 2.2    & 0.604(8)  & 0.2    & 0.4585(64) & 7.2    \\
0.16-0.3   & 0.451(1)  & 2.8  & 0.559(5)  & 0.4    & 0.610(4)  & 0.0007 & 0.4643(63) & 24.4   \\

\end{tabular}
\end{ruledtabular}
\end{table}

\subsection{FSHS analyses for $N_f = 8$ with various cuts}\label{app_subsec:gams}

In Table~\ref{tab:gam_fshs_xx_nf08b0380_obs3a_datMLp678x},
we tabulate the mass anomalous dimension $\gamma$ obtained by adopting
the linear FSHS ansatz~(\ref{eq:fshs_naive}) for various data selection schemes:
$LM_{\pi} > 6,7,8$ and FSHS-Large Volume Data Set.
{We investigate which data set achieves linear dependence on $X$.
We focus on the $\gamma$ for $F_{\pi}$ and $M_{\rho}$ for which the fit works in most cases.
The scheme $LM_{\pi} > 7$ and $LM_{\pi} > 8$ give a statistically equivalent $\gamma$,
and hence, the selection $LM_{\pi} > 8$ excludes enough the non-linearity at small X.
Then, the results by the FSHS-Large Volume data set is consistent to those in the data set with $LM_{\pi} > 8$,
and therefore, it would also be free from the non-linearity.}

{
For all selection schemes, the linear FSHS fit for $M_{\pi}$ fails.
Therefore, the non-linearlity typically appearing at small $X$ would not be the reason
for the deviation from the linear ansatz (\ref{eq:fshs_naive}), in the case of $M_{\pi}$.
A possible origin of the deviation is chiral dynamics as discussed in Sec.~\ref{subsec:discuss}.}

\begin{table}[!tbp]
\caption{
The mass anomalous dimension $\gamma$ obtained by
FSHS fits~(\protect\ref{eq:fshs_naive}) for $F_{\pi}$, $M_{\pi}$, or $M_{\rho}$ data.
The fits are performed independently in each operator.
The left-most column represents the data selection scheme in terms of $LM_{\pi}$ values.
\label{tab:gam_fshs_xx_nf08b0380_obs3a_datMLp678x}
}
\begin{ruledtabular}

\begin{tabular}{l c @{\hspace{1em}} ll ll ll}
\multicolumn{1}{c}{Data Set ($m_f$)} &
\multicolumn{1}{c}{dof} &
\multicolumn{1}{c}{$\gamma(F_{\pi})$} &
\multicolumn{1}{c}{$\chi^2/\mathrm{dof}(F_{\pi})$} &
\multicolumn{1}{c}{$\gamma(M_{\pi})$} &
\multicolumn{1}{c}{$\chi^2/\mathrm{dof}(M_{\pi})$} &
\multicolumn{1}{c}{$\gamma(M_{\rho})$} &
\multicolumn{1}{c}{$\chi^2/\mathrm{dof}(M_{\rho})$} \\
\hline
$LM_{\pi} > 6$ &       $25$ & $1.002(04)$ & $3.37$ & $0.613(02)$ & $19.39$ & $0.873(09)$ & $1.94$ \\
$LM_{\pi} > 7$ &       $16$ & $0.995(04)$ & $1.79$ & $0.615(02)$ & $12.91$ & $0.870(12)$ & $1.53$ \\
$LM_{\pi} > 8$ &       $12$ & $0.988(06)$ & $2.10$ & $0.611(02)$ & $6.43 $ & $0.884(17)$ & $1.11$ \\
FSHS-Large Vol. Data & $16$ & $0.994(05)$ & $1.83$ & $0.624(02)$ & $18.33$ & $0.901(12)$ & $1.32$ \\

\end{tabular}
\end{ruledtabular}
\end{table}

\subsection{Global parameter search in Renormalization group (RG)-motivated FSHS}\label{app_subsec:fshs_rg}

In Sec.~\ref{ssub:fshs_ah}, we have investigated the spectrum data by using
the RG-motivated FSHS ansatz, where we have two exponents $(\gamma,\omega)$
in the fit function~(\ref{eq:fshs_ah}).
This fit provided the results shown in the last line in Table~\ref{tab:gam_fshs_nf08b0380_obx3a_datLMpx}.
It is important to confirm that the exponents have been obtained from the global minimum in the parameter space.

\begin{figure}[!tbp]
\begin{center}
\includegraphics[width=7.5cm]{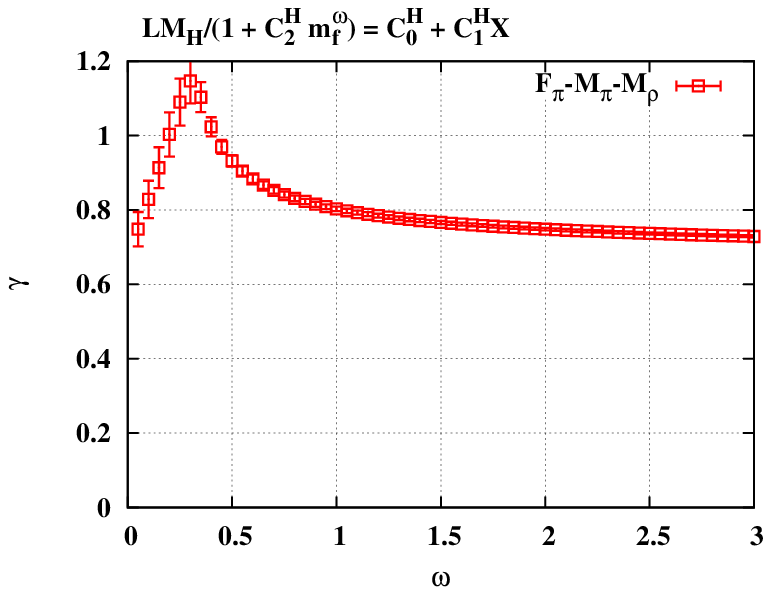}
\includegraphics[width=7.5cm]{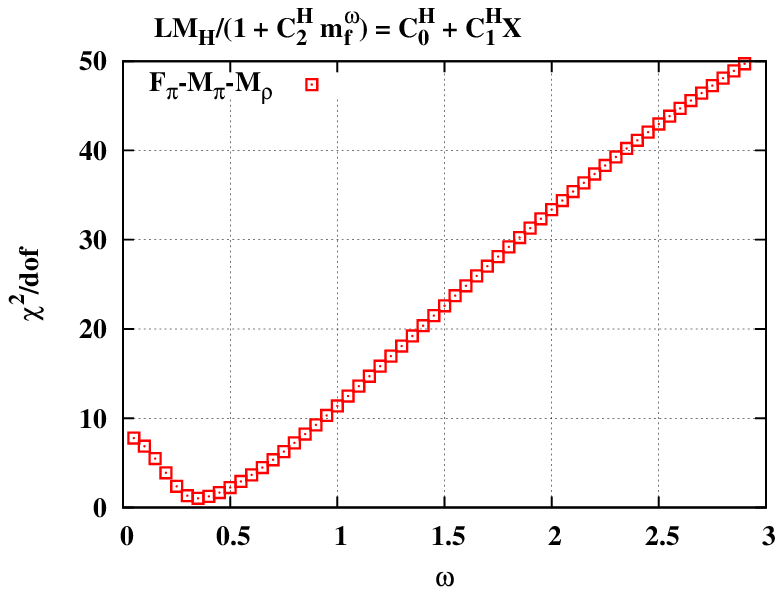}
\caption{
Left:
The mass anomalous dimension $\gamma$
obtained by the simultaneous fit for combined data $\{F_{\pi},\ M_{\pi},\ M_{\rho}\}$
as a function of the second exponent $\omega$.
Right: The corresponding $\chi^2/\mathrm{dof}$.
}\label{fig:fshs_ah0_axxxx_nf08b0380_obx3a_datLMpx}
\end{center}
\end{figure}

In Fig.~\ref{fig:fshs_ah0_axxxx_nf08b0380_obx3a_datLMpx},
we show $\gamma$ as a function of $\omega$ (left panel) and the corresponding $\chi^2/{\mathrm{dof}}$ (right panel).
$\gamma$ is well determined
without any instabilities,
and the $\chi^2/\mathrm{dof}$ has a clear minimum around $\omega\sim 0.35$,
which is consistent to what we have found by treating $\omega$ as a fit parameter
in Sec.~\ref{ssub:fshs_ah}: $\omega\simeq 0.347(14)$.
For larger $\omega$, the fit ansatz~(\ref{eq:fshs_ah}) reduces into the naive one~(\ref{eq:fshs_naive}),
which is shown to give a large $\chi^2/\mathrm{dof}\simeq 104.88$ (Sec.~\ref{ssub:fshs_naive}).
Consistently,
we observe a larger $\chi^2/\mathrm{dof}$ with increasing $\omega$
in the right panel of Fig.~\ref{fig:fshs_ah0_axxxx_nf08b0380_obx3a_datLMpx}.

\clearpage

\section{$N_f=12$}
\label{app:nf12}

We have simulated $N_f=12$ QCD at $\beta=4$ and $3.7$.
In this Appendix we show a brief summary
of results for the spectrum of $N_f=12$ QCD.

We have accumulated additional statistics for $L=24$ and $30$
since the previous result shown in Ref.~\cite{Aoki:2012eq},
and also have new data at a larger volume of $L=36$.
The simulation parameters and results for the hadron spectrum
at $\beta=4$ and $3.7$ are summarized in Tables~\ref{tab:b4}, \ref{tab:b37}
\ref{tab:b4-2}, and \ref{tab:b37-2}
(For $L=18$, see Ref.~\cite{Aoki:2012eq}).

\begin{figure}[!tbp]
\includegraphics[width=7.5cm,clip]{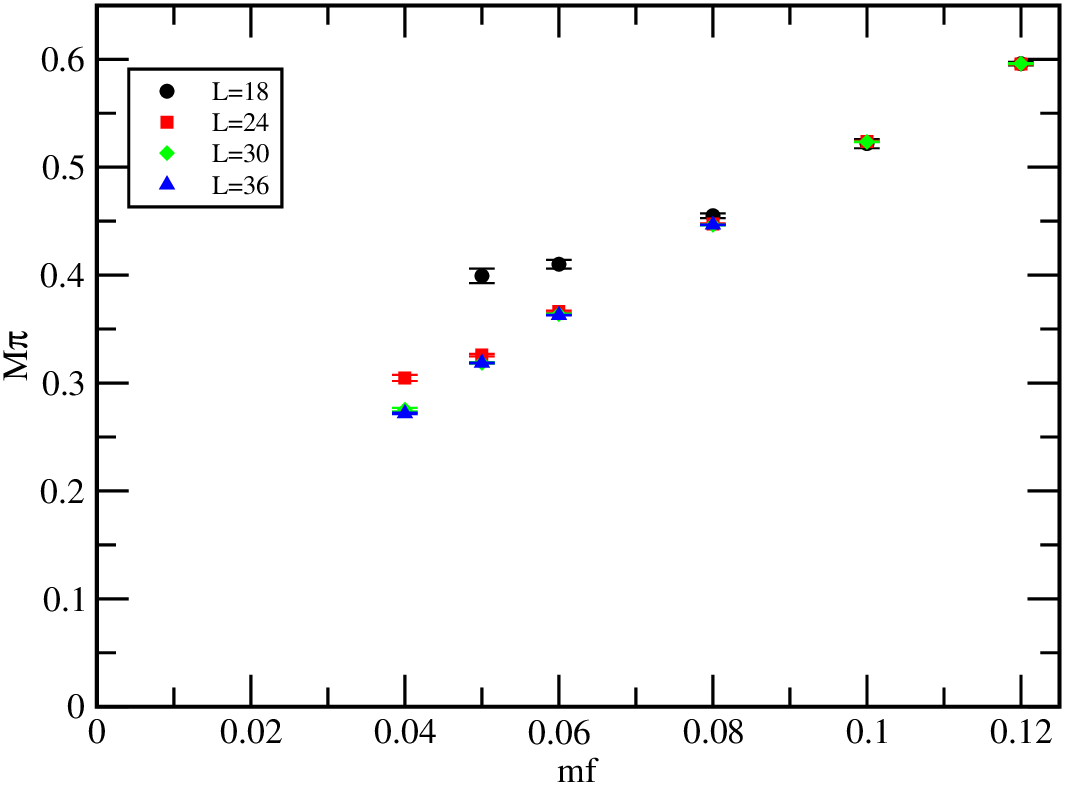}
\includegraphics[width=7.5cm,clip]{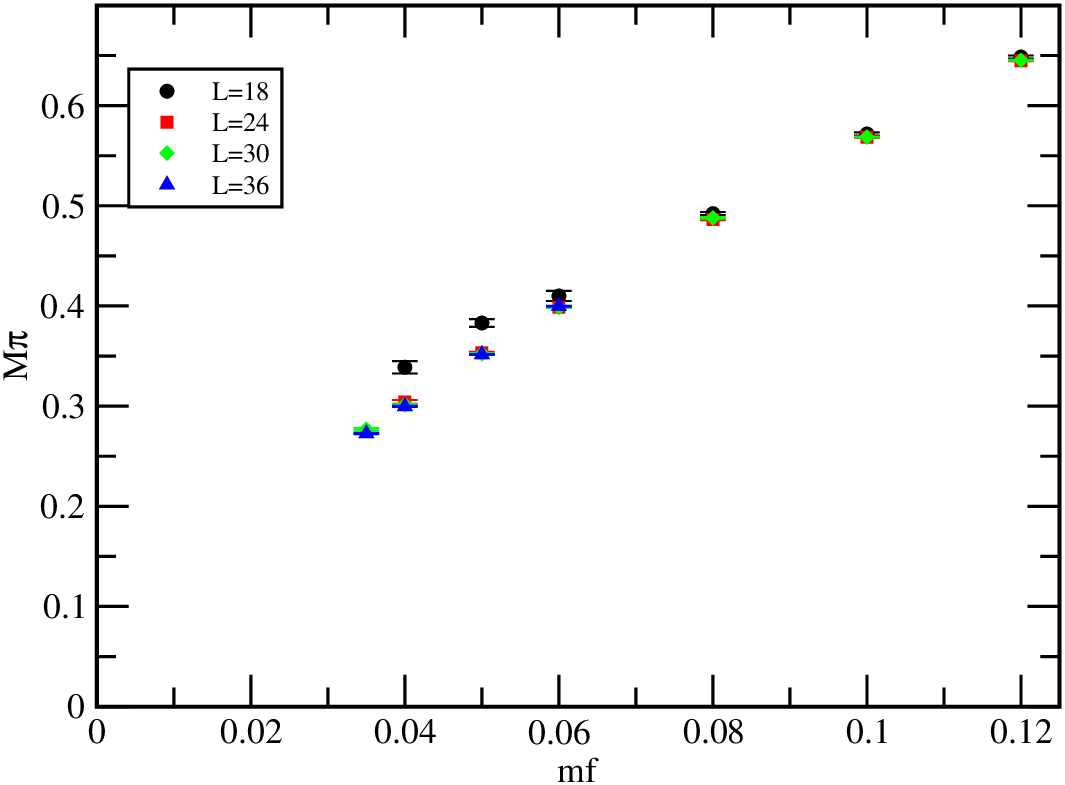}
\caption{
The mass dependence of $M_\pi$ at $\beta=4$ (left) and $\beta=3.7$ (right) in $N_f=12$.
}
\label{fig:nf12mpi}
\end{figure}

\begin{figure}[!tbp]
\includegraphics[width=7.5cm,clip]{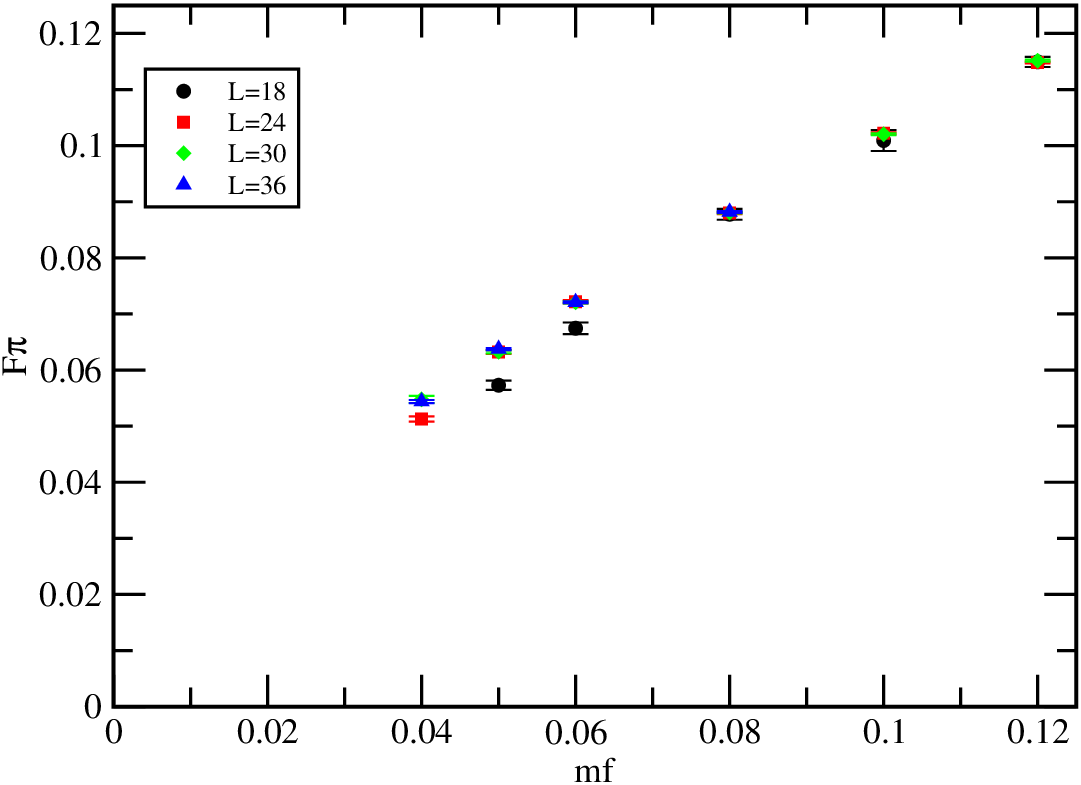}
\includegraphics[width=7.5cm,clip]{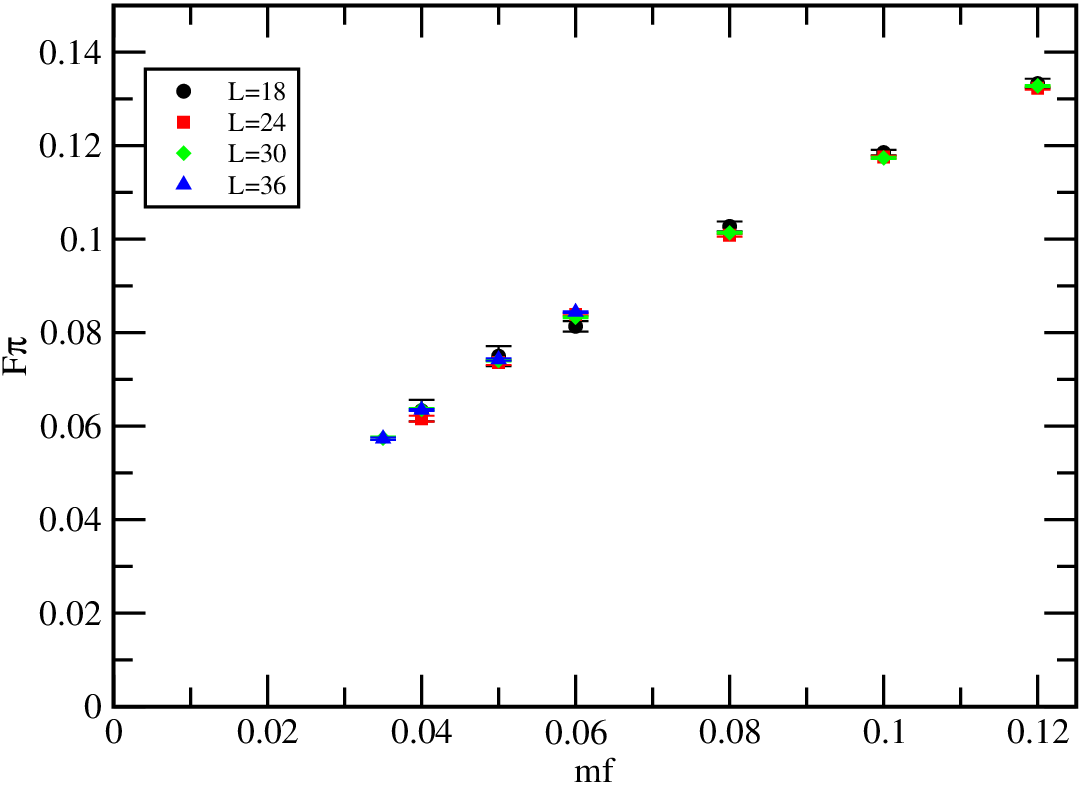}
\caption{
The mass dependence of $F_\pi$ at $\beta=4$ (left) and $\beta=3.7$ (right) in $N_f=12$.
}
\label{fig:nf12fpi}
\end{figure}

\begin{figure}[!tbp]
\includegraphics[width=7.5cm,clip]{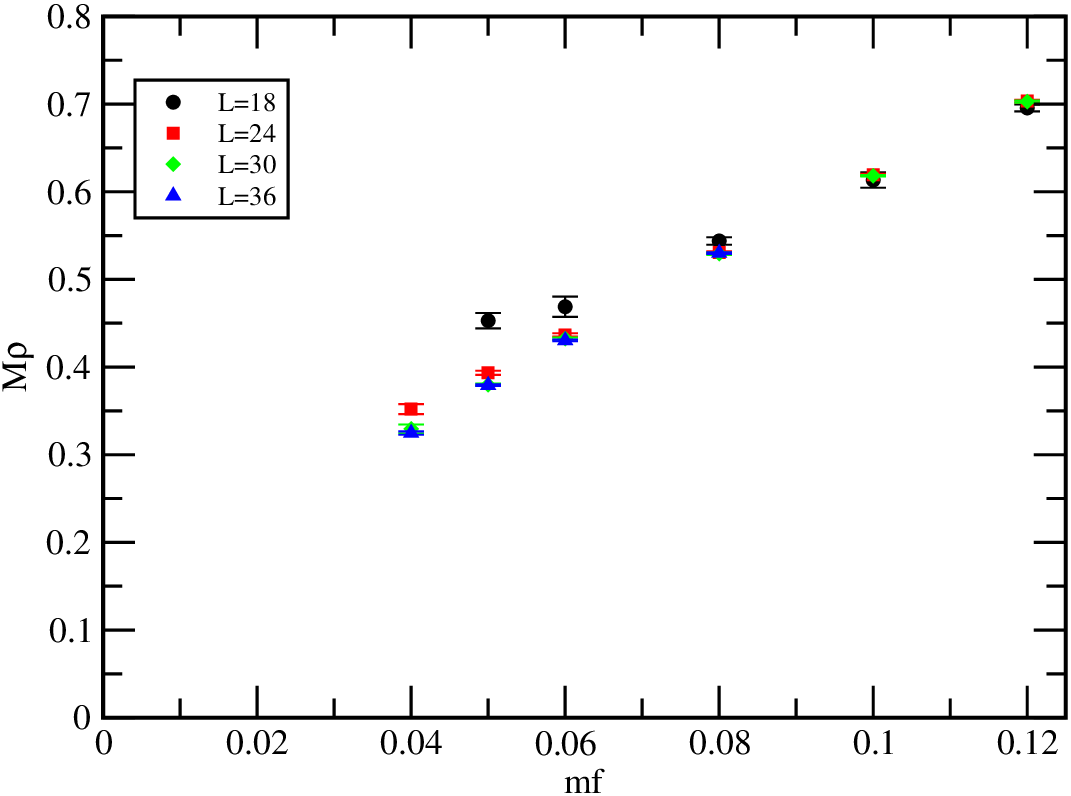}
\includegraphics[width=7.5cm,clip]{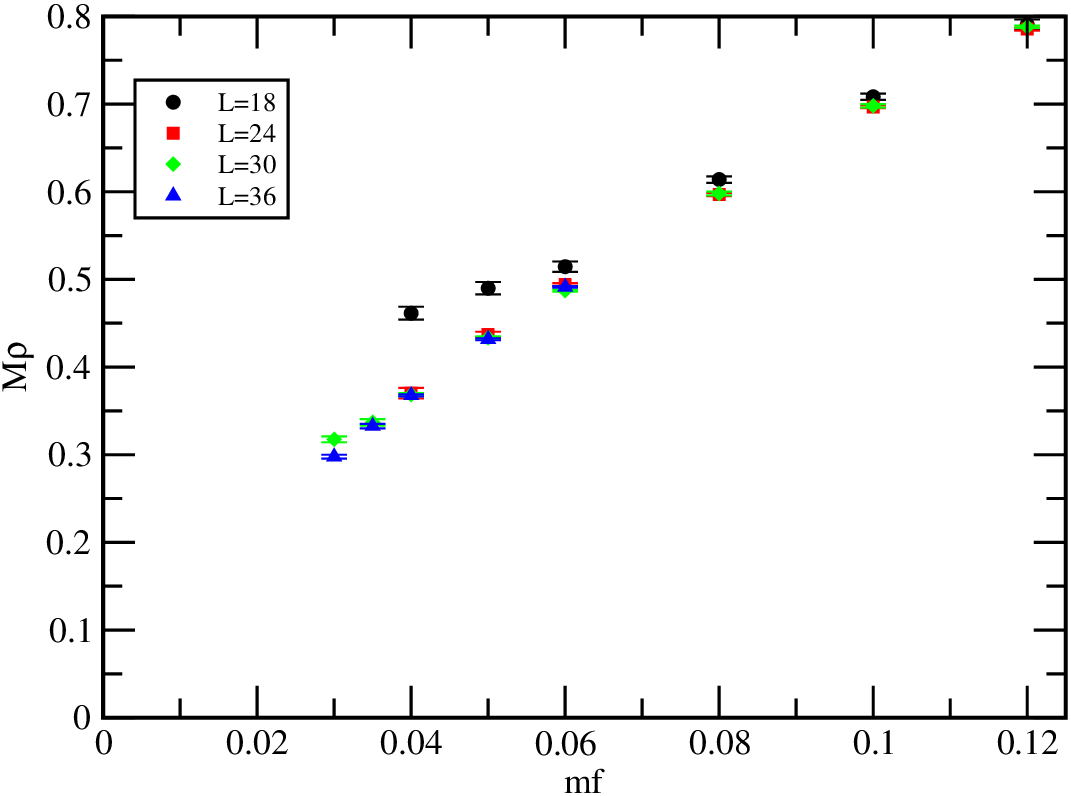}
\caption{
The mass dependence of $M_\rho$ at $\beta=4$ (left) and $\beta=3.7$ (right) in $N_f=12$.
}
\label{fig:nf12mrho}
\end{figure}

\begin{figure}[!tbp]
\includegraphics[width=7.5cm,clip]{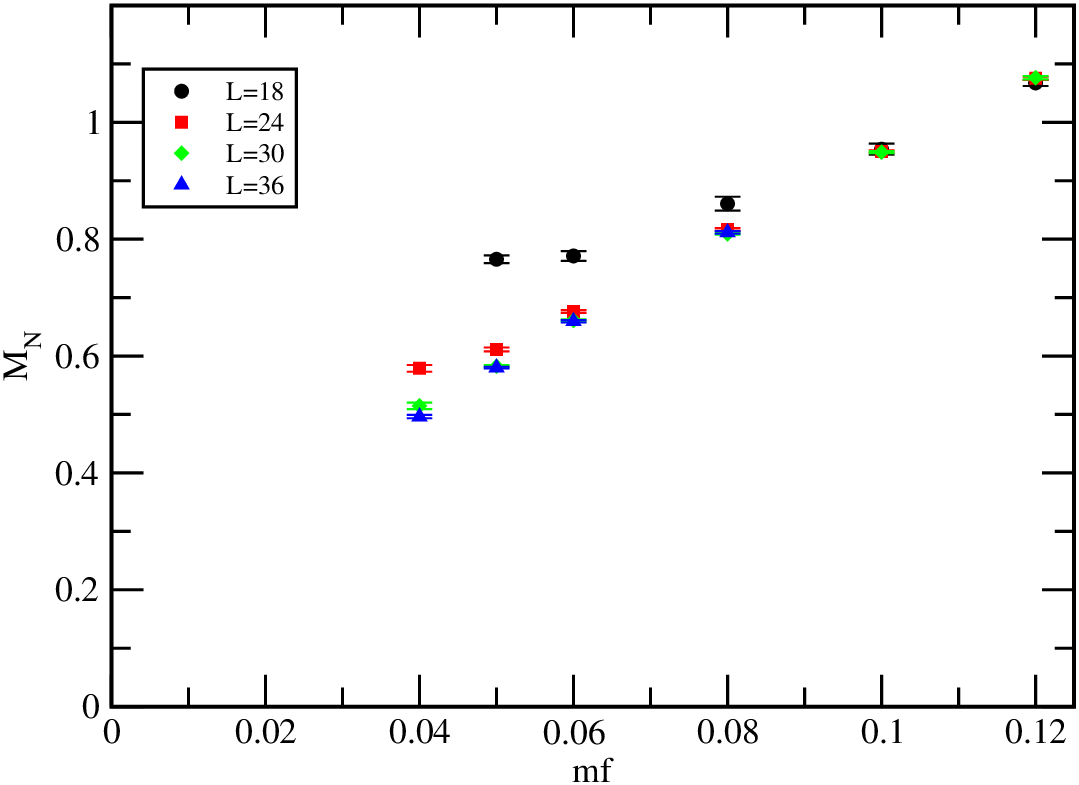}
\includegraphics[width=7.5cm,clip]{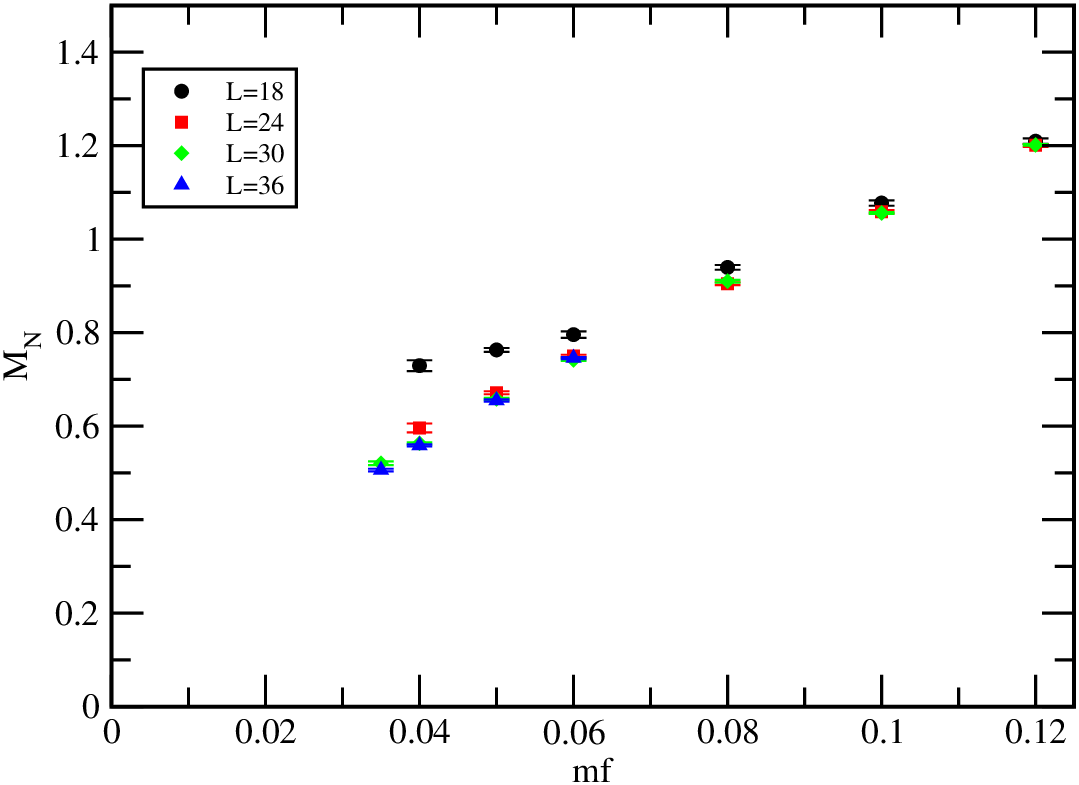}
\caption{
The mass dependence of $M_N$ at $\beta=4$ (left) and $\beta=3.7$ (right) in $N_f=12$.
}
\label{fig:nf12mN}
\end{figure}

\begin{figure}[!tbp]
\includegraphics[width=7.5cm,clip]{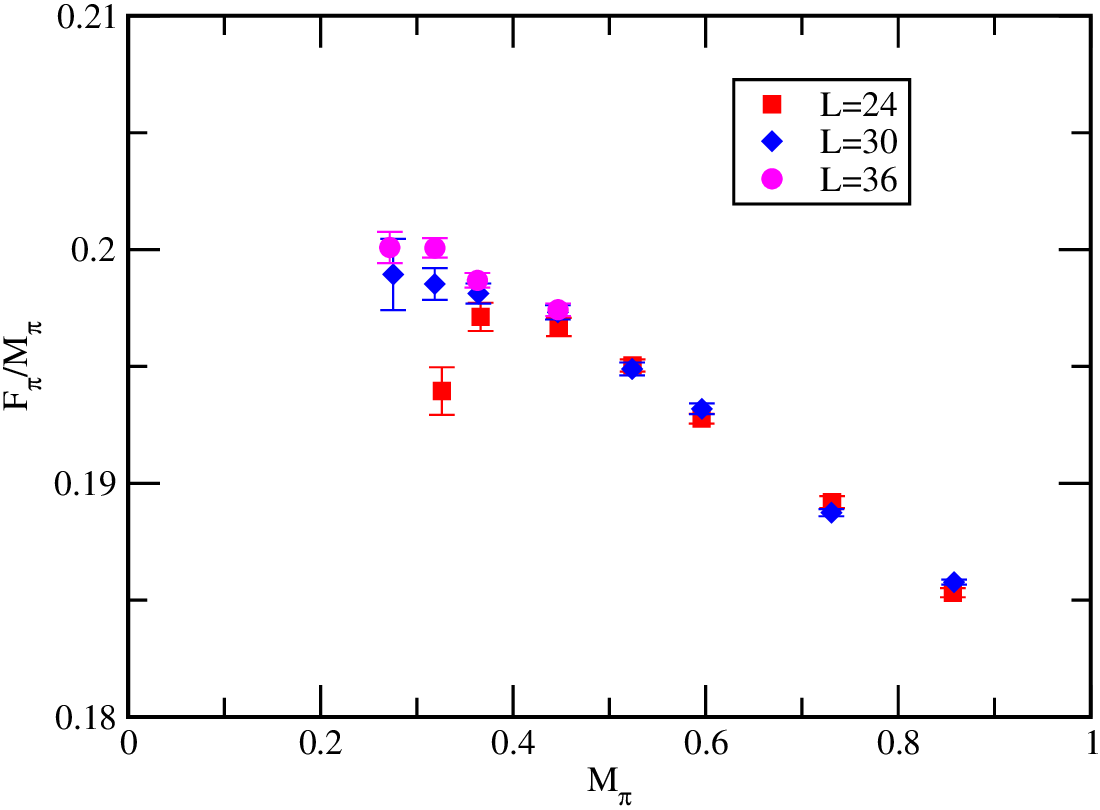}
\includegraphics[width=7.5cm,clip]{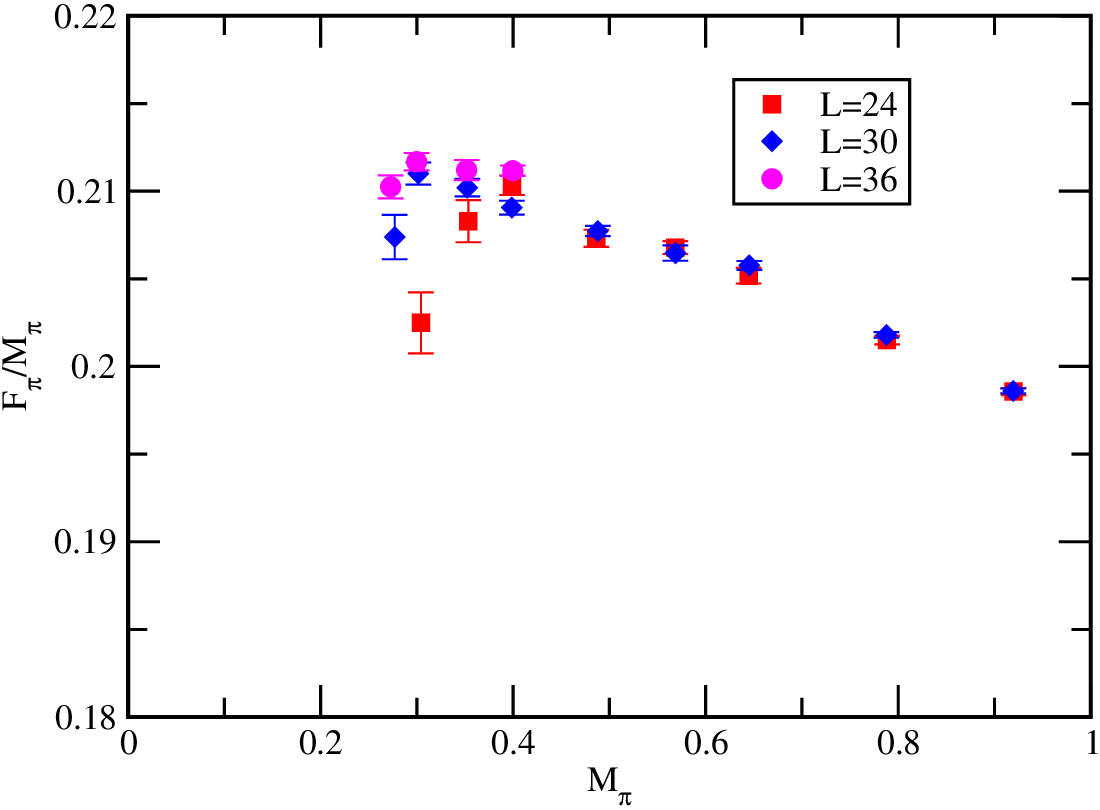}
\caption{
The dimensionless ratio of $F_\pi/M_\pi$ v.s. $M_\pi$
at $\beta=4$ (left) and $\beta=3.7$ (right) in $N_f=12$.
}
\label{fig:nf12_ratio_fpi}
\end{figure}

\begin{figure}[!tbp]
\includegraphics[width=7.5cm,clip]{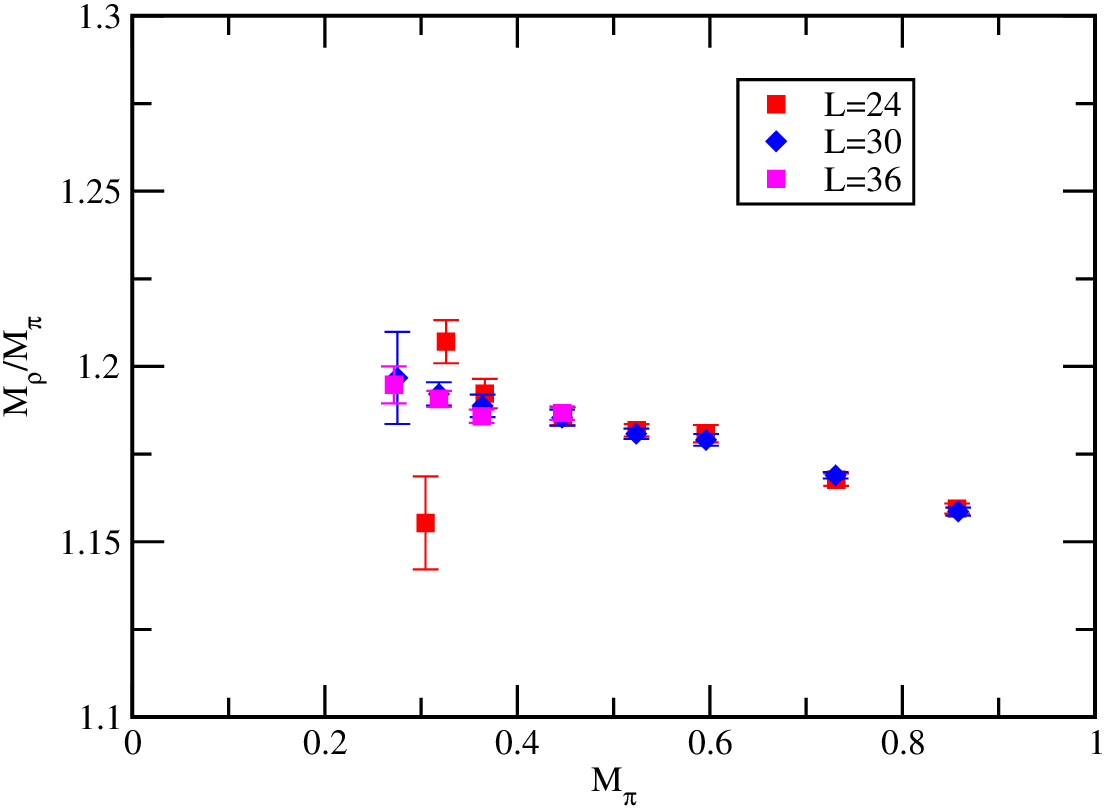}
\includegraphics[width=7.5cm,clip]{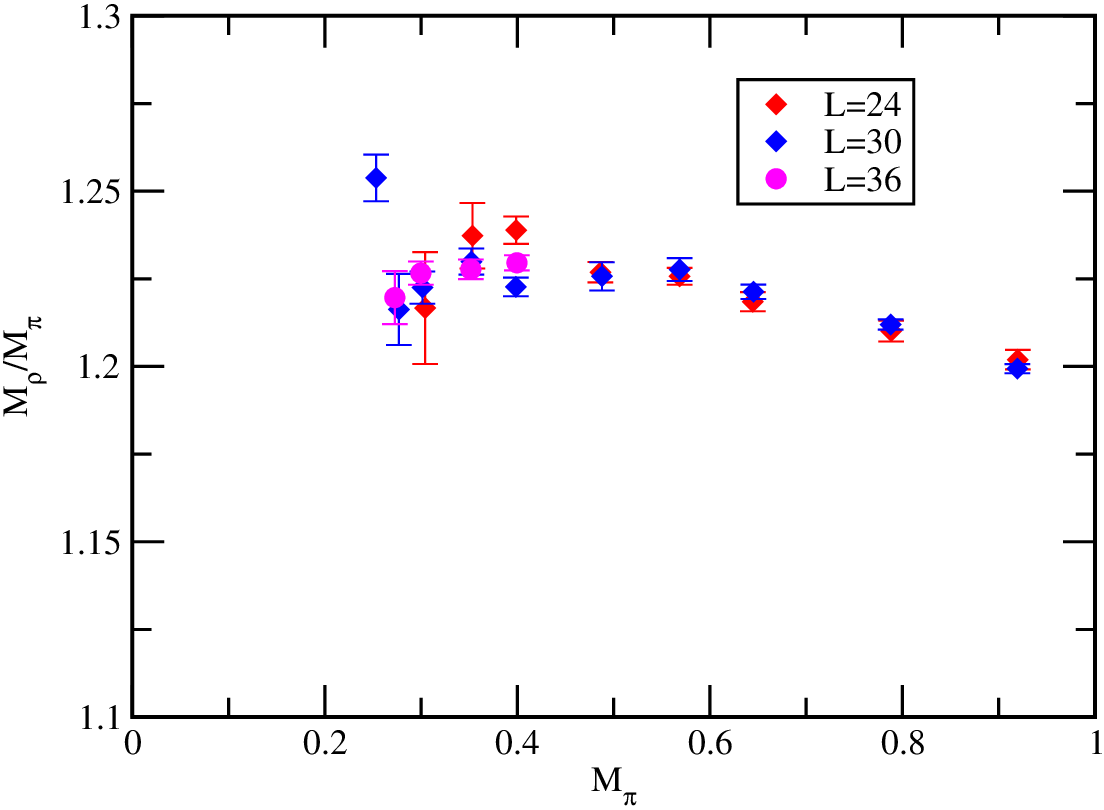}
\caption{
The dimensionless ratio of $M_\rho/M_\pi$ v.s. $M_\pi$
at $\beta=4$ (left) and $\beta=3.7$ (right) in $N_f=12$.
}
\label{fig:nf12_ratio_mrho}
\end{figure}

\begin{figure}[!tbp]
\includegraphics[width=7.5cm,clip]{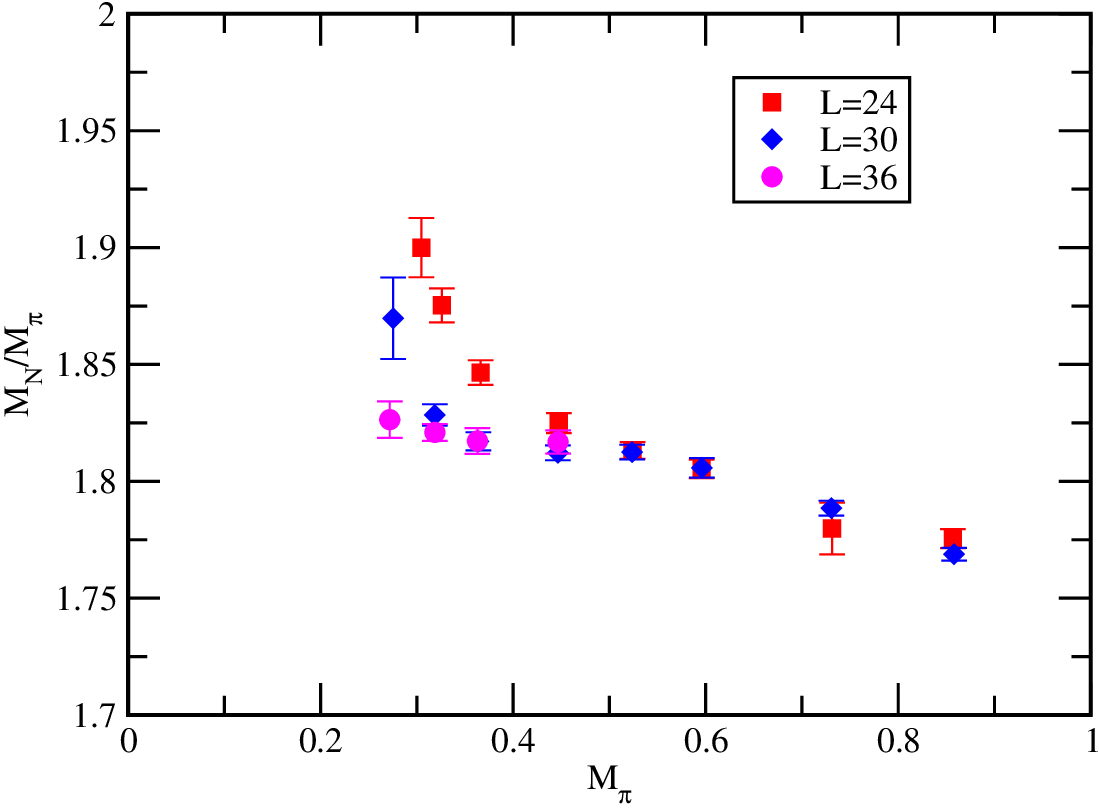}
\includegraphics[width=7.5cm,clip]{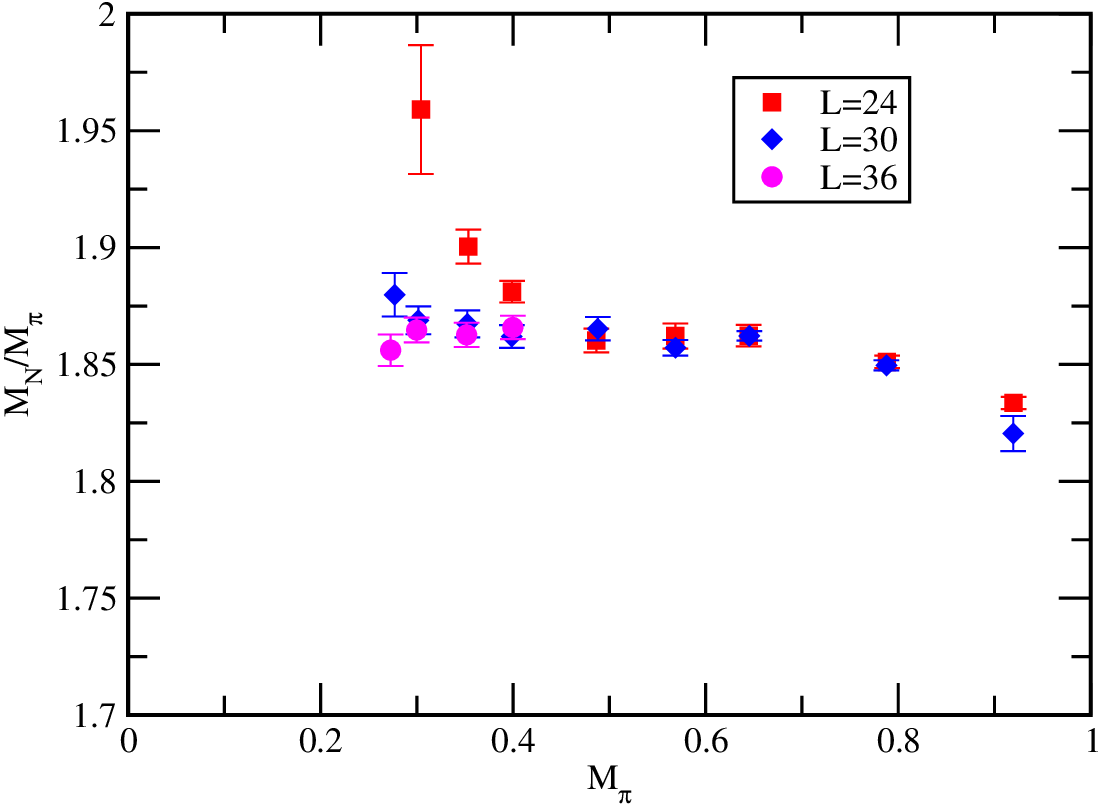}
\caption{
The dimensionless ratio of $M_N/M_\pi$ v.s. $M_\pi$
at $\beta=4$ (left) and $\beta=3.7$ (right) in $N_f=12$.
}
\label{fig:nf12_ratio_mN}
\end{figure}

The fermion mass dependence of the basic hadron spectra
of $M_\pi$, $F_\pi$, $M_{\rho {(\rm PV)}}$, and $M_N$ is shown in
Fig.~\ref{fig:nf12mpi},
Fig.~\ref{fig:nf12fpi},
Fig.~\ref{fig:nf12mrho}, and
Fig.~\ref{fig:nf12mN}.
Using those spectrum quantities,
the ratios of $F_\pi$, $M_{\rho {(\rm PV)}}$, and $M_N$
to $M_\pi$ are shown in
Fig.~\ref{fig:nf12_ratio_fpi},
Fig.~\ref{fig:nf12_ratio_mrho}, and
Fig.~\ref{fig:nf12_ratio_mN},
where all the ratios become constant in
the smaller $M_\pi$ region
for both $\beta=4$ and $3.7$,
which is consistent with being in the conformal phase in $N_f=12$.

The mild fermion mass dependence at larger $M_\pi$
could be considered as a correction to the hyperscaling due to the
large fermion mass effect as shown in the previous paper~\cite{Aoki:2012eq}.
In fact, the result of the effective $\gamma$ analysis shown in Fig.~\ref{fig:geff}
indicates universal hyperscaling in
the small mass region,
thus both the results of the effective $\gamma$ and ratio analyses
are consistent with being in the conformal phase, which is clearly different
from the result obtained in $N_f=8$ QCD.

\begin{figure}[!tbp]
\includegraphics[width=7.5cm,clip]{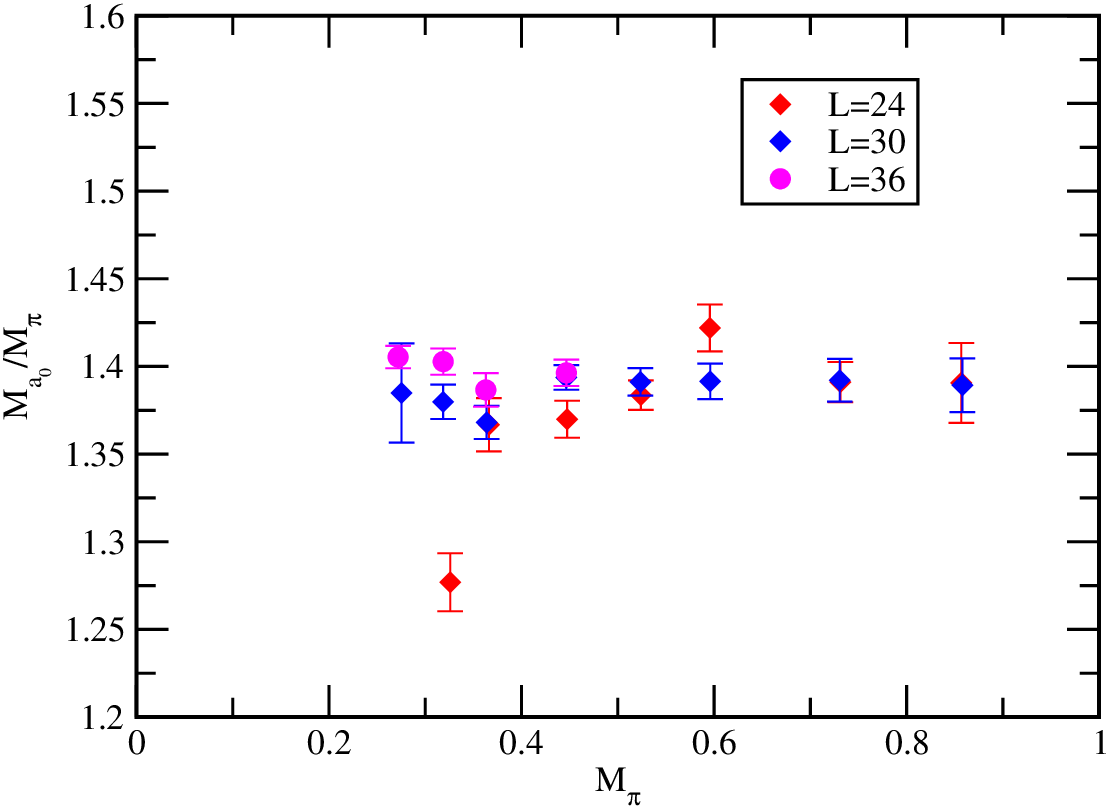}
\includegraphics[width=7.5cm,clip]{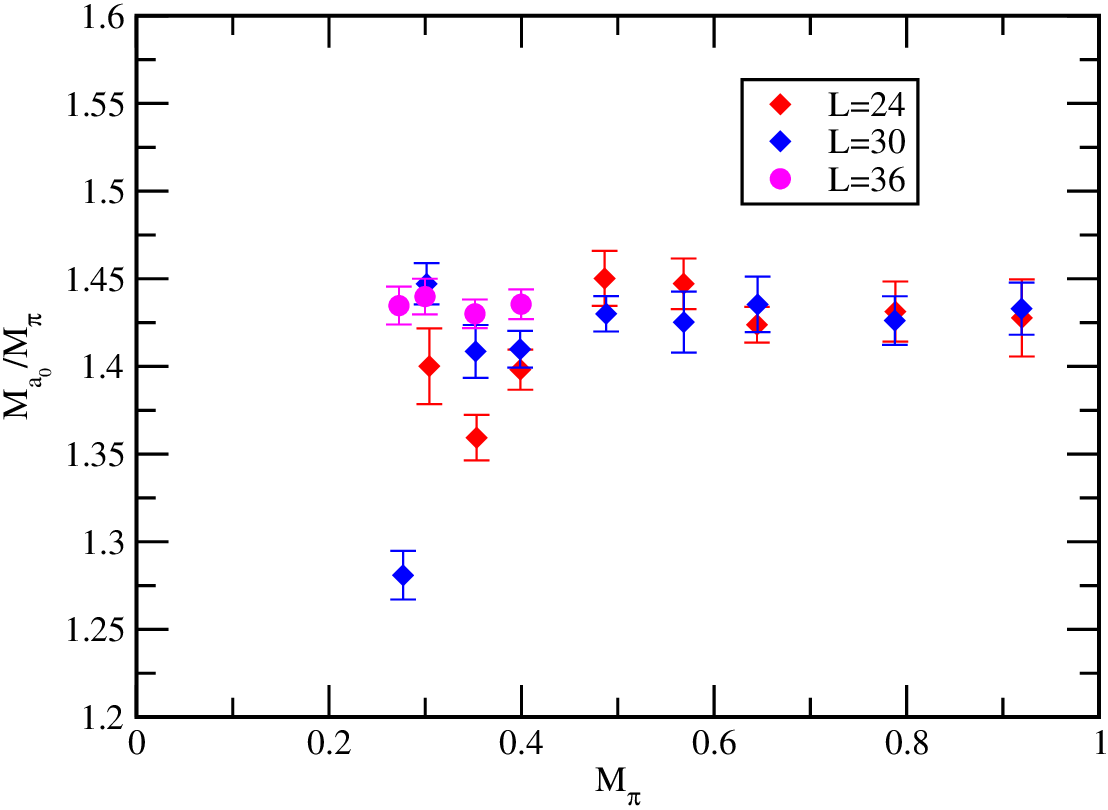}
\caption{
The dimensionless ratio of $M_{a_0}/M_\pi$ v.s. $M_\pi$
at $\beta=4$ (left) and $\beta=3.7$ (right) in $N_f=12$.
}
\label{fig:nf12_ratio_a0-pi}
\end{figure}

\begin{figure}[!tbp]
\includegraphics[width=7.5cm,clip]{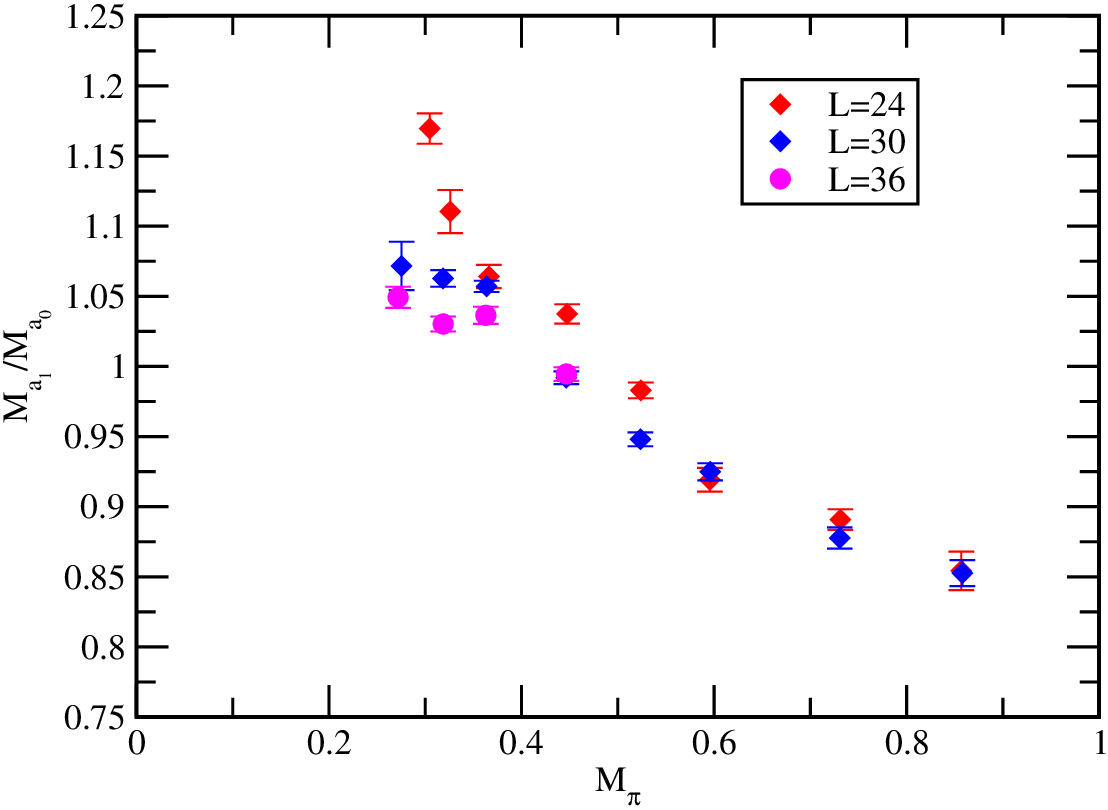}
\includegraphics[width=7.5cm,clip]{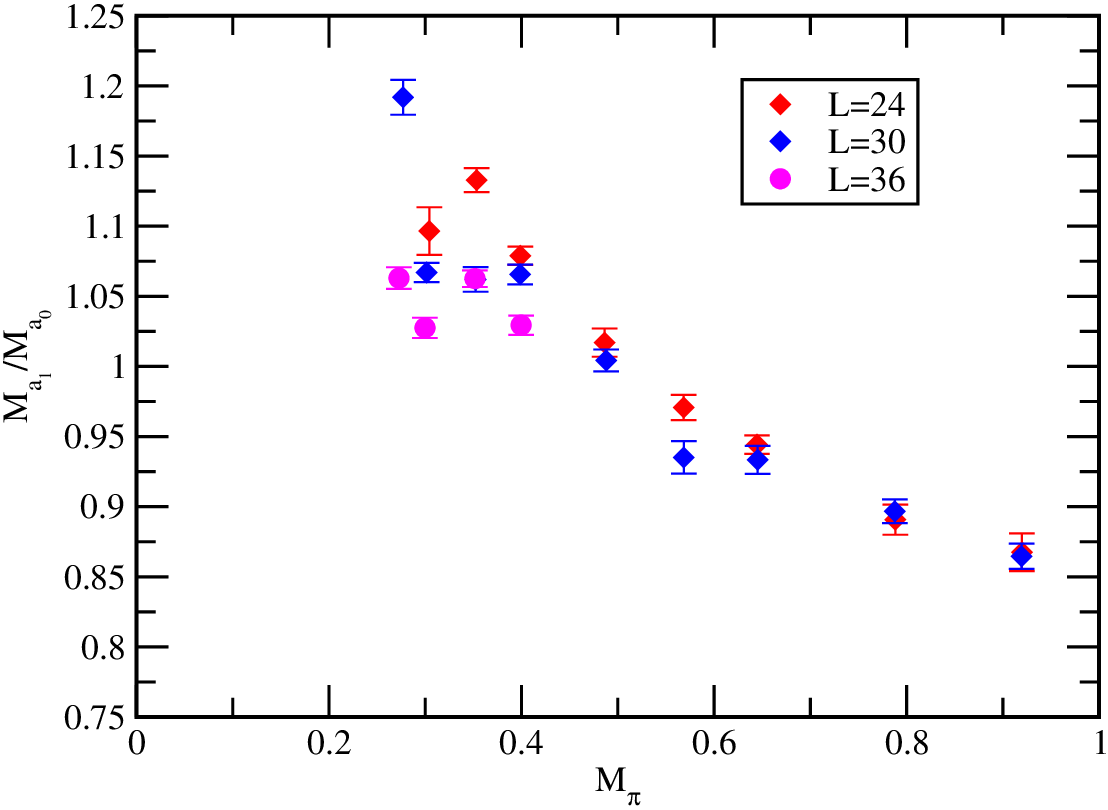}
\caption{
The dimensionless ratio of $M_{a_1}/M_{a_0}$ v.s. $M_\pi$
at $\beta=4$ (left) and $\beta=3.7$ (right) in $N_f=12$.
}
\label{fig:nf12_ratio_a1-a0}
\end{figure}

\begin{figure}[!tbp]
\includegraphics[width=7.5cm,clip]{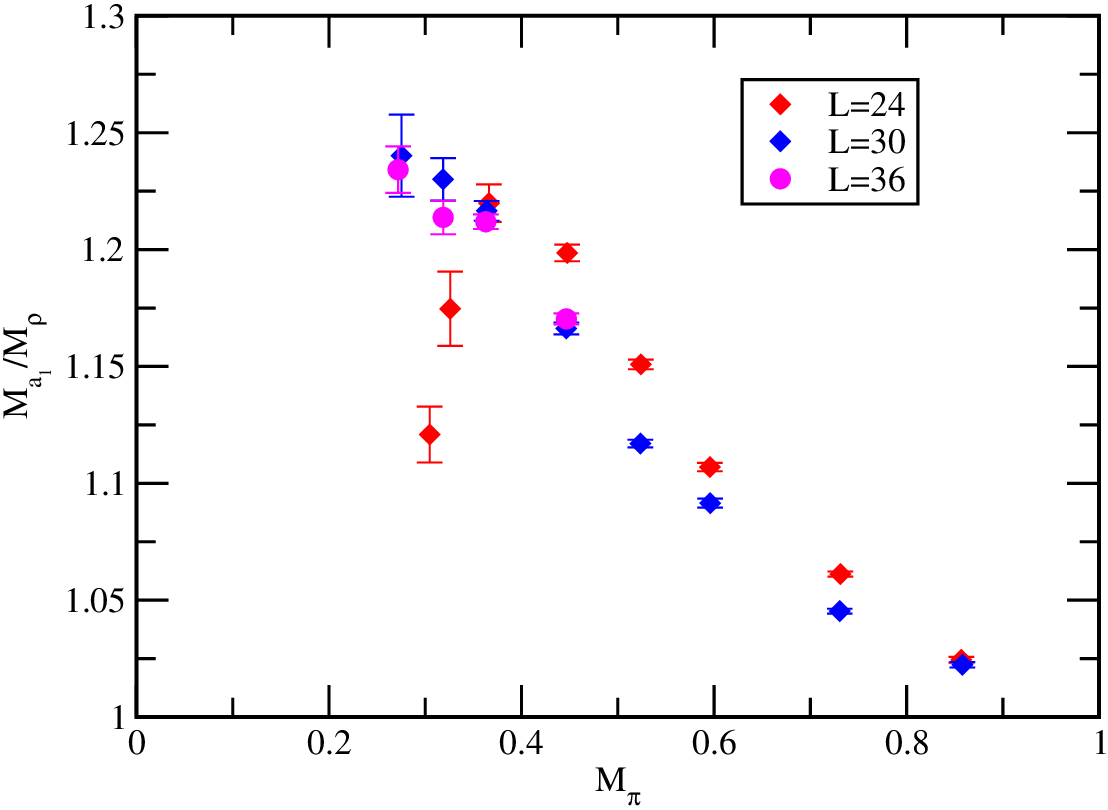}
\includegraphics[width=7.5cm,clip]{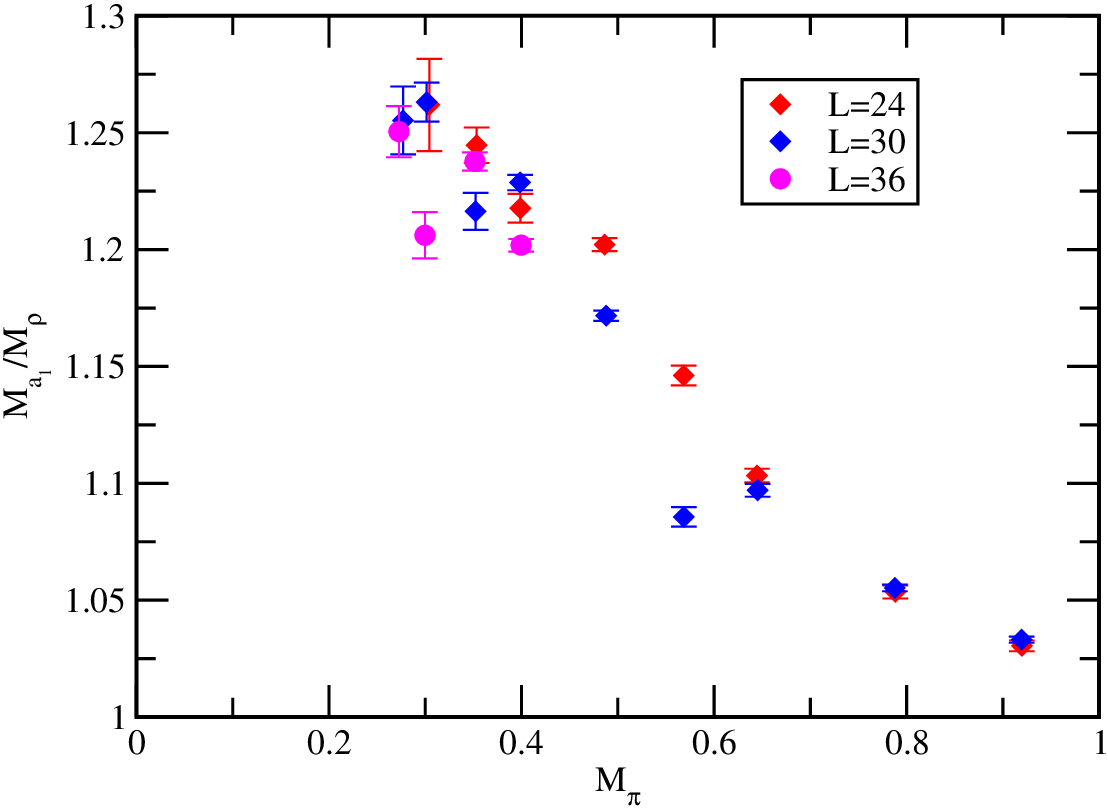}
\caption{
The dimensionless ratio of $M_{a_1}/M_\rho$ v.s. $M_\pi$
at $\beta=4$ (left) and $\beta=3.7$ (right) in $N_f=12$.
}
\label{fig:nf12_ratio_a1-rho}
\end{figure}

The ratios of $M_{a_0}/M_{\pi}$,  $M_{a_1}/M_{a_0}$,
and $M_{a_1}/M_{\rho}$ are shown in Figs.~\ref{fig:nf12_ratio_a0-pi}, \ref{fig:nf12_ratio_a1-a0}, and
~\ref{fig:nf12_ratio_a1-rho}. While those quantities have larger errors,
each ratio becomes a constant in the small mass region.
From the value of the constant for each ratio,
we may further read
$M_\rho/M_\pi \sim 1.2 < M_{a_0}/M_\pi \sim 1.4$ and
$M_{a_1}/M_{a_0} \sim 1.05 < M_{a_1}/M_\rho \sim 1.25$
in the smaller fermion mass region.
According to the ``Mended Symmetry'', our result in $N_f=12$
suggests the chiral restoration
through the ``vector manifestation'',
where $M_\rho = M_\pi$
and $M_{a_1}=M_{a_0}$. {(For details,  see Sec.~\ref{sec:summary}.)}
We also note that those mass ratios are different to those in $N_f=8$ QCD,
and hadron masses are more degenerate
in $N_f=12$ than $N_f=8$.

\begin{table}[!tbp]
\caption{Simulation parameters at $\beta=4$ in $N_f=12$.}
\label{tab:b4}
\begin{ruledtabular}
\begin{tabular}{cc l r c r r r}
\multicolumn{1}{c}{$L$} &
\multicolumn{1}{c}{$T$} &
\multicolumn{1}{c}{$m_f$} &
\multicolumn{1}{c}{$N_{\rm Traj}$} &
\multicolumn{1}{c}{$N_{\rm conf}$} &
\multicolumn{1}{c}{$N_{\rm bin}$} &
\multicolumn{1}{c}{$N_{\rm meas}$} \\
\hline
36 & 48 & 0.040  &  4000 & 250 & 160 & 4 \\
36 & 48 & 0.050  & 8000 & 250 & 320 & 4 \\
36 & 48 & 0.060  &  8000 & 250 & 320 & 4 \\
36 & 48 & 0.080  &  4000 & 250 & 160 & 4 \\
30 & 40 & 0.040  &  1000 & 250 & 100 & 4 \\
30 & 40 & 0.050  &  4000 & 250 & 160 & 4 \\
30 & 40 & 0.060  &  4000 & 250 & 160 & 4 \\
30 & 40 & 0.080  &  4000 & 250 & 320 & 4 \\
30 & 40 & 0.100  &  4000 & 250 & 160 & 4 \\
30 & 40 & 0.120  &  1000 & 250 & 40 & 4 \\
30 & 40 & 0.160  &  1000 & 250 & 40 & 4 \\
30 & 40 & 0.200  &  1000 & 250 & 40 & 4 \\
24 & 32 & 0.040  &  2000 & 250 & 80 & 4 \\
24 & 32 & 0.050  &  8000 & 250 & 160 & 4 \\
24 & 32 & 0.060  &  8000 & 250 & 160 & 4 \\
24 & 32 & 0.080  &  8000 & 250 & 160 & 4 \\
24 & 32 & 0.100  &  8000 & 250 & 160 & 4 \\
24 & 32 & 0.120  &  2000 & 250 & 80 & 4 \\
24 & 32 & 0.160  &  1000 & 250 & 100 & 4 \\
24 & 32 & 0.200  &  1000 & 250 & 100 & 4 \\
\end{tabular}
\end{ruledtabular}
\end{table}

\begin{table}[!tbp]
\caption{Simulation parameters at $\beta=3.7$ in $N_f=12$.}
\label{tab:b37}
\begin{ruledtabular}
\begin{tabular}{cc l r c r r r}
\multicolumn{1}{c}{$L$} &
\multicolumn{1}{c}{$T$} &
\multicolumn{1}{c}{$m_f$} &
\multicolumn{1}{c}{$N_{\rm Traj}$} &
\multicolumn{1}{c}{$N_{\rm conf}$} &
\multicolumn{1}{c}{$N_{\rm bin}$} &
\multicolumn{1}{c}{$N_{\rm meas}$} \\
\hline
36 & 48 & 0.035  &  4000 & 750 & 200 & 4 \\
36 & 48 & 0.040  &  2600 & 650 & 100 & 4 \\
36 & 48 & 0.050  &  2000 & 500 & 100 & 4 \\
36 & 48 & 0.060  &  2000 & 500 & 100 & 4 \\
30 & 40 & 0.035  &  2000 & 250 &   80 & 4 \\
30 & 40 & 0.040  &  3000 & 750 & 100 & 4 \\
30 & 40 & 0.050  &  1700 & 425 & 100 & 4 \\
30 & 40 & 0.060  &  2000 & 500 & 100 & 4 \\
30 & 40 & 0.080  &  1000 & 250 & 100 & 4 \\
30 & 40 & 0.100  &    500 & 125 &   20 & 4 \\
30 & 40 & 0.120  &    500 & 125 &   20 & 4 \\
30 & 40 & 0.160  &    500 & 125 &   20 & 4 \\
30 & 40 & 0.200  &    500 & 125 &   20 & 4 \\
30 & 40 & 0.200  &    400 & 100 &   20 & 4 \\
24 & 32 & 0.040  &  2000 & 250 &   80 & 4 \\
24 & 32 & 0.050  &  2000 & 250 &   80 & 4 \\
24 & 32 & 0.060  &  2000 & 250 &   80 & 4 \\
24 & 32 & 0.080  &  1000 & 125 &   40 & 4 \\
24 & 32 & 0.100  &  1000 & 125 &   40 & 4 \\
24 & 32 & 0.120  &  1000 & 125 &   40 & 4 \\
24 & 32 & 0.160  &  1000 & 125 &   40 & 4 \\
24 & 32 & 0.200  &  1000 & 125 &   40 & 4 \\
24 & 32 & 0.300  &  1000 & 125 &   40 & 4 \\
\end{tabular}
\end{ruledtabular}
\end{table}

\begin{table}[!tbp]
\caption{Spectra at $\beta=4$ in $N_f=12$.}
\label{tab:b4-2}
\begin{ruledtabular}
\begin{tabular}{cc l l l l l l l l l }
\multicolumn{1}{c}{$L$} &
\multicolumn{1}{c}{$T$} &
\multicolumn{1}{c}{$m_f$} &
\multicolumn{1}{c}{$M_\pi$} &
\multicolumn{1}{c}{$M_{\rho_{(PV)}}$} &
\multicolumn{1}{c}{$M_N$} &
\multicolumn{1}{c}{$M_{a_0}$} &
\multicolumn{1}{c}{$M_{a_1}$} &
\multicolumn{1}{c}{$M_{b_1}$} &
\multicolumn{1}{c}{$F_\pi$} \\
\hline
36 & 48 & 0.04 & 0.2718(7) & 0.3247(20)   & 0.4964(28) & 0.3820(21) & 0.4197(44) & 0.4258(61) & 0.05438(27) \\
36 & 48 & 0.05 & 0.3186(4) & 0.3794(9) & 0.5802(15) & 0.4469(25) & 0.4948(41) & 0.5012(48) & 0.06374(17) \\
36 & 48 & 0.06 & 0.3629(3) & 0.4303(9) & 0.6595(22) & 0.5032(36) & 0.5552(22) & 0.5624(27) & 0.07210(16) \\
36 & 48 & 0.08 & 0.4467(3) & 0.5301(9) & 0.8115(23) & 0.6237(35) & 0.6815(24) & 0.6890(33) & 0.08819(16) \\
30 & 40 & 0.04 & 0.2753(16)   & 0.3295(49)   & 0.5148(55) & 0.3813(88) & 0.423(12)   & 0.432(15)   & 0.05477(63) \\
30 & 40 & 0.05 & 0.3186(7) & 0.3798(14)   & 0.5825(20) & 0.4396(33) & 0.4877(49) & 0.4926(46) & 0.06325(27) \\
30 & 40 & 0.06 & 0.3638(5) & 0.4325(13)   & 0.6611(18) & 0.4977(36) & 0.5546(39) & 0.5647(39) & 0.07208(19) \\
30 & 40 & 0.08 & 0.4465(4) & 0.5292(12)   & 0.8091(16) & 0.6223(33) & 0.6778(24) & 0.6864(27) & 0.08809(17) \\
30 & 40 & 0.1   & 0.5235(3) & 0.6182(7) & 0.9489(18) & 0.7283(41) & 0.7926(28) & 0.7878(48) & 0.10203(18) \\
30 & 40 & 0.12 & 0.5960(4) & 0.7027(11)   & 1.0762(28) & 0.8293(60) & 0.8928(45) & 0.8762(68) & 0.11514(19) \\
30 & 40 & 0.16 & 0.7306(3) & 0.8541(7) & 1.3067(26) & 1.0171(90) & 1.0855(76) & 1.053(12)   & 0.13790(13) \\
30 & 40 & 0.2   & 0.8580(3) & 0.9940(11)   & 1.5176(23) & 1.192(13)   & 1.261(12)   & 1.235(25)   & 0.15938(11) \\
24 & 32 & 0.04 & 0.3047(28)   & 0.3520(57)   & 0.5789(58) & 0.3374(31) & 0.3969(60) & 0.4076(76) & 0.05128(48) \\
24 & 32 & 0.05 & 0.3259(11)   & 0.3934(24)   & 0.6112(34) & 0.4162(50)  & 0.4792(84) & 0.484(12)  & 0.06321(34) \\
24 & 32 & 0.06 & 0.3662(9) & 0.4366(19)   & 0.6761(24) & 0.5005(54) & 0.5640(46) & 0.5648(39) & 0.07218(30) \\
24 & 32 & 0.08 & 0.4474(5) & 0.5305(14)   & 0.8165(23) & 0.6129(47) & 0.6782(26) & 0.6890(31) & 0.08800(21) \\
24 & 32 & 0.1   & 0.5240(4) & 0.6193(12)   & 0.9501(20) & 0.7250(44) & 0.7895(22) & 0.7950(33) & 0.10220(18) \\
24 & 32 & 0.12 & 0.5957(5) & 0.7034(15)   & 1.0754(28) & 0.8471(83) & 0.8893(38) & 0.8920(58) & 0.11483(17) \\
24 & 32 & 0.16 & 0.7313(5) & 0.8539(15)   & 1.3015(77) & 1.0172(87) & 1.0786(68) & 1.054(12)   & 0.13835(25) \\
24 & 32 & 0.2   & 0.8568(3) & 0.9934(15)   & 1.5212(36) & 1.191(20)   & 1.264(18)   & 1.276(42)   & 0.15878(19) \\
24 & 32 & 0.3   & 1.1435(2) & 1.2996(13)   & 1.9971(28) & 1.497(24)   & 1.589(26)   & 1.488(50)   & 0.20425(14) \\
\end{tabular}
\end{ruledtabular}
\end{table}

\begin{table}[!tbp]
\caption{Spectra at $\beta=3.7$ in $N_f=12$.}
\label{tab:b37-2}
\begin{ruledtabular}
\begin{tabular}{cc l l l l l l l l l }
\multicolumn{1}{c}{$L$} &
\multicolumn{1}{c}{$T$} &
\multicolumn{1}{c}{$m_f$} &
\multicolumn{1}{c}{$M_\pi$} &
\multicolumn{1}{c}{$M_{\rho_{(PV)}}$} &
\multicolumn{1}{c}{$M_N$} &
\multicolumn{1}{c}{$M_{a_0}$} &
\multicolumn{1}{c}{$M_{a_1}$} &
\multicolumn{1}{c}{$M_{b_1}$} &
\multicolumn{1}{c}{$F_\pi$} \\
\hline
36 & 48 & 0.035 & 0.2727(7)  & 0.3326(26)  & 0.5062(28) & 0.3913(28) & 0.4364(52) & 0.4408(64) & 0.05734(24) \\
36 & 48 & 0.04  & 0.2997(6)  & 0.3677(12)  & 0.5589(20) & 0.4316(31) & 0.4707(56) & 0.4893(40) & 0.06345(21) \\
36 & 48 & 0.05  & 0.3516(4)  & 0.4317(12)  & 0.6550(22) & 0.5028(28) & 0.5658(26) & 0.5752(28) & 0.07427(26) \\
36 & 40 & 0.06  & 0.3997(3)  & 0.4914(10) & 0.7457(22) & 0.5737(34) & 0.6455(28) & 0.6546(32) & 0.08439(14) \\
30 & 40 & 0.035 & 0.2770(11)   & 0.3369(38)  & 0.5206(40) & 0.3548(4)  & 0.4314(45) & 0.4405(45) & 0.05744(36) \\
30 & 40 & 0.04  & 0.30146(7) & 0.3685(17)  & 0.5634(24) & 0.4363(33) & 0.4820(36) & 0.4971(32) & 0.06361(25) \\
30 & 40 & 0.05  & 0.3524(6)  & 0.4334(17)  & 0.6580(27) & 0.4963(55) & 0.5571(53) & 0.5650(58) & 0.07407(21) \\
30 & 40 & 0.06  & 0.3985(5)  & 0.4872(11)  & 0.7420(25) & 0.5618(43) & 0.6401(29) & 0.6458(35) & 0.08331(21) \\
30 & 40 & 0.08  & 0.4879(7)  & 0.5980(22)  & 0.9101(30) & 0.6977(56) & 0.7788(46) & 0.7820(54) & 0.10135(21) \\
30 & 40 & 0.1   & 0.5686(4)  & 0.6980(2)   & 1.0559(19) & 0.8103(97) & 0.927(14)  & 0.934(16)  & 0.11739(24) \\
30 & 40 & 0.12  & 0.6453(4)  & 0.7881(14)  & 1.2017(17) & 0.926(10)  & 1.0120(49) & 1.006(10)  & 0.13278(20) \\
30 & 40 & 0.16  & 0.7878(3)  & 0.9547(13)  & 1.4571(19) & 1.123(11)  & 1.2242(79) & 1.192(23)  & 0.15897(17) \\
30 & 40 & 0.2   & 0.9195(3)  & 1.1028(12)  & 1.6739(69) & 1.318(14)  & 1.408(10)  & 1.319(21)  & 0.18261(15) \\
30 & 40 & 0.3   & 1.2155(2)  & 1.4296(8) & 2.155(17)  & 1.726(15)  & 1.831(22)  & 1.67(19)   & 0.23492(13) \\
24 & 32 & 0.04  & 0.3043(19)   &  0.3702(61) & 0.5961(96) & 0.4260(65) & 0.4764(49) & 0.4976(64) & 0.06161(64) \\
24 & 32 & 0.05  & 0.3533(10)  & 0.4372(31)  & 0.6715(32) & 0.4803(47) & 0.5623(40) & 0.5649(45) & 0.07359(49) \\
24 & 32 & 0.06  & 0.3988(8)  & 0.4941(16)  & 0.7502(24) & 0.5576(45) & 0.6340(40) & 0.6460(45) & 0.08389(24) \\
24 & 32 & 0.08  & 0.4863(7)  & 0.5967(18)  & 0.9047(31) & 0.7053(81) & 0.7716(35) & 0.7878(48) & 0.10082(30) \\
24 & 32 & 0.1   & 0.5685(8)  & 0.6968(15)  & 1.0586(36) & 0.8227(82) & 0.9084(60) & 0.9126(75) & 0.11756(27) \\
24 & 32 & 0.12  & 0.6447(4)  & 0.7855(17)  & 1.2006(30) & 0.9179(66) & 1.0131(65) & 1.010(13)  & 0.13227(31) \\
24 & 32 & 0.16  & 0.7883(4)  & 0.9539(23)  & 1.4592(25) & 1.128(14)  & 1.222(14)  & 1.1574(15) & 0.15885(22) \\
24 & 32 & 0.2   & 0.9198(3)  & 1.1056(25)  & 1.6865(27) & 1.313(20)  & 1.430(49)  & 1.5(1.0)   & 0.18265(22) \\
24 & 32 & 0.3   & 1.2158(3)  & 1.4310(11)  & 2.167(15)  & 1.773(60)  & 1.824(25)  & 1.707(66)  & 0.23513(15) \\
\end{tabular}
\end{ruledtabular}
\end{table}

\clearpage

\bibliographystyle{apsrev4-1}
\bibliography{reference}

\begin{thebibliography}{107}%
\makeatletter
\providecommand \@ifxundefined [1]{%
 \@ifx{#1\undefined}
}%
\providecommand \@ifnum [1]{%
 \ifnum #1\expandafter \@firstoftwo
 \else \expandafter \@secondoftwo
 \fi
}%
\providecommand \@ifx [1]{%
 \ifx #1\expandafter \@firstoftwo
 \else \expandafter \@secondoftwo
 \fi
}%
\providecommand \natexlab [1]{#1}%
\providecommand \enquote  [1]{``#1''}%
\providecommand \bibnamefont  [1]{#1}%
\providecommand \bibfnamefont [1]{#1}%
\providecommand \citenamefont [1]{#1}%
\providecommand \href@noop [0]{\@secondoftwo}%
\providecommand \href [0]{\begingroup \@sanitize@url \@href}%
\providecommand \@href[1]{\@@startlink{#1}\@@href}%
\providecommand \@@href[1]{\endgroup#1\@@endlink}%
\providecommand \@sanitize@url [0]{\catcode `\\12\catcode `\$12\catcode
  `\&12\catcode `\#12\catcode `\^12\catcode `\_12\catcode `\%12\relax}%
\providecommand \@@startlink[1]{}%
\providecommand \@@endlink[0]{}%
\providecommand \url  [0]{\begingroup\@sanitize@url \@url }%
\providecommand \@url [1]{\endgroup\@href {#1}{\urlprefix }}%
\providecommand \urlprefix  [0]{URL }%
\providecommand \Eprint [0]{\href }%
\providecommand \doibase [0]{http://dx.doi.org/}%
\providecommand \selectlanguage [0]{\@gobble}%
\providecommand \bibinfo  [0]{\@secondoftwo}%
\providecommand \bibfield  [0]{\@secondoftwo}%
\providecommand \translation [1]{[#1]}%
\providecommand \BibitemOpen [0]{}%
\providecommand \bibitemStop [0]{}%
\providecommand \bibitemNoStop [0]{.\EOS\space}%
\providecommand \EOS [0]{\spacefactor3000\relax}%
\providecommand \BibitemShut  [1]{\csname bibitem#1\endcsname}%
\let\auto@bib@innerbib\@empty
%</preamble>
\bibitem [{\citenamefont {Yamawaki}\ \emph {et~al.}(1986)\citenamefont
  {Yamawaki}, \citenamefont {Bando},\ and\ \citenamefont
  {Matumoto}}]{Yamawaki:1985zg}%
  \BibitemOpen
  \bibfield  {author} {\bibinfo {author} {\bibfnamefont {K.}~\bibnamefont
  {Yamawaki}}, \bibinfo {author} {\bibfnamefont {M.}~\bibnamefont {Bando}}, \
  and\ \bibinfo {author} {\bibfnamefont {K.-i.}\ \bibnamefont {Matumoto}},\
  }\href {\doibase 10.1103/PhysRevLett.56.1335} {\bibfield  {journal} {\bibinfo
   {journal} {Phys. Rev. Lett.}\ }\textbf {\bibinfo {volume} {56}},\ \bibinfo
  {pages} {1335} (\bibinfo {year} {1986})}\BibitemShut {NoStop}%
%%CITATION = PRLTA,56,1335;%%
\bibitem [{\citenamefont {Bando}\ \emph {et~al.}(1986)\citenamefont {Bando},
  \citenamefont {Matumoto},\ and\ \citenamefont {Yamawaki}}]{Bando:1986bg}%
  \BibitemOpen
  \bibfield  {author} {\bibinfo {author} {\bibfnamefont {M.}~\bibnamefont
  {Bando}}, \bibinfo {author} {\bibfnamefont {K.-i.}\ \bibnamefont {Matumoto}},
  \ and\ \bibinfo {author} {\bibfnamefont {K.}~\bibnamefont {Yamawaki}},\
  }\href {\doibase 10.1016/0370-2693(86)91516-9} {\bibfield  {journal}
  {\bibinfo  {journal} {Phys. Lett.}\ }\textbf {\bibinfo {volume} {B178}},\
  \bibinfo {pages} {308} (\bibinfo {year} {1986})}\BibitemShut {NoStop}%
%%CITATION = PHLTA,B178,308;%%
\bibitem [{\citenamefont {Holdom}(1985)}]{Holdom:1984sk}%
  \BibitemOpen
  \bibfield  {author} {\bibinfo {author} {\bibfnamefont {B.}~\bibnamefont
  {Holdom}},\ }\href {\doibase 10.1016/0370-2693(85)91015-9} {\bibfield
  {journal} {\bibinfo  {journal} {Phys. Lett.}\ }\textbf {\bibinfo {volume}
  {B150}},\ \bibinfo {pages} {301} (\bibinfo {year} {1985})}\BibitemShut
  {NoStop}%
%%CITATION = PHLTA,B150,301;%%
\bibitem [{\citenamefont {Akiba}\ and\ \citenamefont
  {Yanagida}(1986)}]{Akiba:1985rr}%
  \BibitemOpen
  \bibfield  {author} {\bibinfo {author} {\bibfnamefont {T.}~\bibnamefont
  {Akiba}}\ and\ \bibinfo {author} {\bibfnamefont {T.}~\bibnamefont
  {Yanagida}},\ }\href {\doibase 10.1016/0370-2693(86)90385-0} {\bibfield
  {journal} {\bibinfo  {journal} {Phys. Lett.}\ }\textbf {\bibinfo {volume}
  {B169}},\ \bibinfo {pages} {432} (\bibinfo {year} {1986})}\BibitemShut
  {NoStop}%
%%CITATION = PHLTA,B169,432;%%
\bibitem [{\citenamefont {Appelquist}\ \emph {et~al.}(1986)\citenamefont
  {Appelquist}, \citenamefont {Karabali},\ and\ \citenamefont
  {Wijewardhana}}]{Appelquist:1986an}%
  \BibitemOpen
  \bibfield  {author} {\bibinfo {author} {\bibfnamefont {T.~W.}\ \bibnamefont
  {Appelquist}}, \bibinfo {author} {\bibfnamefont {D.}~\bibnamefont
  {Karabali}}, \ and\ \bibinfo {author} {\bibfnamefont {L.~C.~R.}\ \bibnamefont
  {Wijewardhana}},\ }\href {\doibase 10.1103/PhysRevLett.57.957} {\bibfield
  {journal} {\bibinfo  {journal} {Phys. Rev. Lett.}\ }\textbf {\bibinfo
  {volume} {57}},\ \bibinfo {pages} {957} (\bibinfo {year} {1986})}\BibitemShut
  {NoStop}%
%%CITATION = PRLTA,57,957;%%
\bibitem [{\citenamefont {Matsuzaki}\ and\ \citenamefont
  {Yamawaki}(2012{\natexlab{a}})}]{Matsuzaki:2012xx}%
  \BibitemOpen
  \bibfield  {author} {\bibinfo {author} {\bibfnamefont {S.}~\bibnamefont
  {Matsuzaki}}\ and\ \bibinfo {author} {\bibfnamefont {K.}~\bibnamefont
  {Yamawaki}},\ }\href {\doibase 10.1103/PhysRevD.86.115004} {\bibfield
  {journal} {\bibinfo  {journal} {Phys. Rev. D}\ }\textbf {\bibinfo {volume}
  {86}},\ \bibinfo {pages} {115004} (\bibinfo {year}
  {2012}{\natexlab{a}})}\BibitemShut {NoStop}%
%%CITATION = ARXIV:1209.2017;%%
\bibitem [{\citenamefont {Matsuzaki}\ and\ \citenamefont
  {Yamawaki}(2013)}]{Matsuzaki:2012mk}%
  \BibitemOpen
  \bibfield  {author} {\bibinfo {author} {\bibfnamefont {S.}~\bibnamefont
  {Matsuzaki}}\ and\ \bibinfo {author} {\bibfnamefont {K.}~\bibnamefont
  {Yamawaki}},\ }\href {\doibase 10.1016/j.physletb.2013.01.031} {\bibfield
  {journal} {\bibinfo  {journal} {Phys. Lett.}\ }\textbf {\bibinfo {volume}
  {B719}},\ \bibinfo {pages} {378} (\bibinfo {year} {2013})}\BibitemShut
  {NoStop}%
%%CITATION = ARXIV:1207.5911;%%
\bibitem [{\citenamefont {Yamawaki}(2017)}]{Yamawaki:2016qux}%
  \BibitemOpen
  \bibfield  {author} {\bibinfo {author} {\bibfnamefont {K.}~\bibnamefont
  {Yamawaki}},\ }\href {\doibase 10.1142/S0218301317400328} {\bibfield
  {journal} {\bibinfo  {journal} {Int. J. Mod. Phys.}\ }\textbf {\bibinfo
  {volume} {E26}},\ \bibinfo {pages} {1740032} (\bibinfo {year} {2017})},\
  \Eprint {http://arxiv.org/abs/1609.03715} {arXiv:1609.03715 [hep-ph]}
  \BibitemShut {NoStop}%
%%CITATION = ARXIV:1609.03715;%%
\bibitem [{\citenamefont {Appelquist}\ \emph {et~al.}(1996)\citenamefont
  {Appelquist}, \citenamefont {Terning},\ and\ \citenamefont
  {Wijewardhana}}]{Appelquist:1996dq}%
  \BibitemOpen
  \bibfield  {author} {\bibinfo {author} {\bibfnamefont {T.}~\bibnamefont
  {Appelquist}}, \bibinfo {author} {\bibfnamefont {J.}~\bibnamefont {Terning}},
  \ and\ \bibinfo {author} {\bibfnamefont {L.~C.~R.}\ \bibnamefont
  {Wijewardhana}},\ }\href {\doibase 10.1103/PhysRevLett.77.1214} {\bibfield
  {journal} {\bibinfo  {journal} {Phys. Rev. Lett.}\ }\textbf {\bibinfo
  {volume} {77}},\ \bibinfo {pages} {1214} (\bibinfo {year}
  {1996})}\BibitemShut {NoStop}%
%%CITATION = HEP-PH/9602385;%%
\bibitem [{\citenamefont {Miransky}\ and\ \citenamefont
  {Yamawaki}(1997)}]{Miransky:1996pd}%
  \BibitemOpen
  \bibfield  {author} {\bibinfo {author} {\bibfnamefont {V.~A.}\ \bibnamefont
  {Miransky}}\ and\ \bibinfo {author} {\bibfnamefont {K.}~\bibnamefont
  {Yamawaki}},\ }\href {\doibase 10.1103/PhysRevD.55.5051} {\bibfield
  {journal} {\bibinfo  {journal} {Phys. Rev. D}\ }\textbf {\bibinfo {volume}
  {55}},\ \bibinfo {pages} {5051} (\bibinfo {year} {1997})}\BibitemShut
  {NoStop}%
%%CITATION = HEP-TH/9611142;%%
\bibitem [{\citenamefont {Caswell}(1974)}]{Caswell:1974gg}%
  \BibitemOpen
  \bibfield  {author} {\bibinfo {author} {\bibfnamefont {W.~E.}\ \bibnamefont
  {Caswell}},\ }\href {\doibase 10.1103/PhysRevLett.33.244} {\bibfield
  {journal} {\bibinfo  {journal} {Phys. Rev. Lett.}\ }\textbf {\bibinfo
  {volume} {33}},\ \bibinfo {pages} {244} (\bibinfo {year} {1974})}\BibitemShut
  {NoStop}%
%%CITATION = PRLTA,33,244;%%
\bibitem [{\citenamefont {Banks}\ and\ \citenamefont
  {Zaks}(1982)}]{Banks:1981nn}%
  \BibitemOpen
  \bibfield  {author} {\bibinfo {author} {\bibfnamefont {T.}~\bibnamefont
  {Banks}}\ and\ \bibinfo {author} {\bibfnamefont {A.}~\bibnamefont {Zaks}},\
  }\href {\doibase 10.1016/0550-3213(82)90035-9} {\bibfield  {journal}
  {\bibinfo  {journal} {Nucl. Phys.}\ }\textbf {\bibinfo {volume} {B196}},\
  \bibinfo {pages} {189} (\bibinfo {year} {1982})}\BibitemShut {NoStop}%
%%CITATION = NUPHA,B196,189;%%
\bibitem [{\citenamefont {Aoki}\ \emph
  {et~al.}(2012{\natexlab{a}})\citenamefont {Aoki}, \citenamefont {Aoyama},
  \citenamefont {Kurachi}, \citenamefont {Maskawa}, \citenamefont {Nagai},
  \citenamefont {Ohki}, \citenamefont {Shibata}, \citenamefont {Yamawaki},\
  and\ \citenamefont {Yamazaki}}]{Aoki:2012ve}%
  \BibitemOpen
  \bibfield  {author} {\bibinfo {author} {\bibfnamefont {Y.}~\bibnamefont
  {Aoki}}, \bibinfo {author} {\bibfnamefont {T.}~\bibnamefont {Aoyama}},
  \bibinfo {author} {\bibfnamefont {M.}~\bibnamefont {Kurachi}}, \bibinfo
  {author} {\bibfnamefont {T.}~\bibnamefont {Maskawa}}, \bibinfo {author}
  {\bibfnamefont {K.-i.}\ \bibnamefont {Nagai}}, \bibinfo {author}
  {\bibfnamefont {H.}~\bibnamefont {Ohki}}, \bibinfo {author} {\bibfnamefont
  {A.}~\bibnamefont {Shibata}}, \bibinfo {author} {\bibfnamefont
  {K.}~\bibnamefont {Yamawaki}}, \ and\ \bibinfo {author} {\bibfnamefont
  {T.}~\bibnamefont {Yamazaki}} (\bibinfo {collaboration} {LatKMI
  Collaboration}),\ }\href {\doibase 10.1103/PhysRevD.85.074502} {\bibfield
  {journal} {\bibinfo  {journal} {Phys. Rev. D}\ }\textbf {\bibinfo {volume}
  {85}},\ \bibinfo {pages} {074502} (\bibinfo {year}
  {2012}{\natexlab{a}})}\BibitemShut {NoStop}%
%%CITATION = ARXIV:1201.4157;%%
\bibitem [{\citenamefont {Matsuzaki}\ and\ \citenamefont
  {Yamawaki}(2015)}]{Matsuzaki:2015sya}%
  \BibitemOpen
  \bibfield  {author} {\bibinfo {author} {\bibfnamefont {S.}~\bibnamefont
  {Matsuzaki}}\ and\ \bibinfo {author} {\bibfnamefont {K.}~\bibnamefont
  {Yamawaki}},\ }\href {\doibase 10.1007/JHEP12(2015)053} {\bibfield  {journal}
  {\bibinfo  {journal} {JHEP}\ }\textbf {\bibinfo {volume} {12}},\ \bibinfo
  {pages} {053} (\bibinfo {year} {2015})}\BibitemShut {NoStop}%
%%CITATION = ARXIV:1508.07688;%%
\bibitem [{\citenamefont {Aoki}\ \emph
  {et~al.}(2013{\natexlab{a}})\citenamefont {Aoki}, \citenamefont {Aoyama},
  \citenamefont {Kurachi}, \citenamefont {Maskawa}, \citenamefont {Nagai},
  \citenamefont {Ohki}, \citenamefont {Shibata}, \citenamefont {Yamawaki},\
  and\ \citenamefont {Yamazaki}}]{Aoki:2013xza}%
  \BibitemOpen
  \bibfield  {author} {\bibinfo {author} {\bibfnamefont {Y.}~\bibnamefont
  {Aoki}}, \bibinfo {author} {\bibfnamefont {T.}~\bibnamefont {Aoyama}},
  \bibinfo {author} {\bibfnamefont {M.}~\bibnamefont {Kurachi}}, \bibinfo
  {author} {\bibfnamefont {T.}~\bibnamefont {Maskawa}}, \bibinfo {author}
  {\bibfnamefont {K.-i.}\ \bibnamefont {Nagai}}, \bibinfo {author}
  {\bibfnamefont {H.}~\bibnamefont {Ohki}}, \bibinfo {author} {\bibfnamefont
  {A.}~\bibnamefont {Shibata}}, \bibinfo {author} {\bibfnamefont
  {K.}~\bibnamefont {Yamawaki}}, \ and\ \bibinfo {author} {\bibfnamefont
  {T.}~\bibnamefont {Yamazaki}} (\bibinfo {collaboration} {LatKMI
  Collaboration}),\ }\href {\doibase 10.1103/PhysRevD.87.094511} {\bibfield
  {journal} {\bibinfo  {journal} {Phys. Rev. D}\ }\textbf {\bibinfo {volume}
  {87}},\ \bibinfo {pages} {094511} (\bibinfo {year}
  {2013}{\natexlab{a}})}\BibitemShut {NoStop}%
%%CITATION = ARXIV:1302.6859;%%
\bibitem [{\citenamefont {Kuti}(2014)}]{Kuti:2014epa}%
  \BibitemOpen
  \bibfield  {author} {\bibinfo {author} {\bibfnamefont {J.}~\bibnamefont
  {Kuti}},\ }\href@noop {} {\bibfield  {journal} {\bibinfo  {journal} {PoS}\
  }\textbf {\bibinfo {volume} {LATTICE2013}},\ \bibinfo {pages} {004} (\bibinfo
  {year} {2014})}\BibitemShut {NoStop}%
%%CITATION = POSCI,LATTICE2013,004;%%
\bibitem [{\citenamefont {DeGrand}(2016)}]{DeGrand:2015zxa}%
  \BibitemOpen
  \bibfield  {author} {\bibinfo {author} {\bibfnamefont {T.}~\bibnamefont
  {DeGrand}},\ }\href {\doibase 10.1103/RevModPhys.88.015001} {\bibfield
  {journal} {\bibinfo  {journal} {Rev. Mod. Phys.}\ }\textbf {\bibinfo {volume}
  {88}},\ \bibinfo {pages} {015001} (\bibinfo {year} {2016})}\BibitemShut
  {NoStop}%
%%CITATION = ARXIV:1510.05018;%%
\bibitem [{\citenamefont {Aoki}(2014)}]{Aoki:lat2014}%
  \BibitemOpen
  \bibfield  {author} {\bibinfo {author} {\bibfnamefont {Y.}~\bibnamefont
  {Aoki}},\ }\href@noop {} {\bibfield  {journal} {\bibinfo  {journal} {Plenary
  talk presented at the 32nd International Symposium on Lattice Field Theory
  (Lattice 2014)}\ } (\bibinfo {year} {2014})}\BibitemShut {NoStop}%
\bibitem [{\citenamefont {Hasenfratz}(2015)}]{Hasenfratz:lat2015}%
  \BibitemOpen
  \bibfield  {author} {\bibinfo {author} {\bibfnamefont {A.}~\bibnamefont
  {Hasenfratz}},\ }\href@noop {} {\bibfield  {journal} {\bibinfo  {journal}
  {Plenary talk presented at 33rd International Symposium on Lattice Field
  Theory (Lattice 2015)}\ } (\bibinfo {year} {2015})}\BibitemShut {NoStop}%
\bibitem [{\citenamefont {Iwasaki}\ \emph {et~al.}(1992)\citenamefont
  {Iwasaki}, \citenamefont {Kanaya}, \citenamefont {Sakai},\ and\ \citenamefont
  {Yoshie}}]{Iwasaki:1991mr}%
  \BibitemOpen
  \bibfield  {author} {\bibinfo {author} {\bibfnamefont {Y.}~\bibnamefont
  {Iwasaki}}, \bibinfo {author} {\bibfnamefont {K.}~\bibnamefont {Kanaya}},
  \bibinfo {author} {\bibfnamefont {S.}~\bibnamefont {Sakai}}, \ and\ \bibinfo
  {author} {\bibfnamefont {T.}~\bibnamefont {Yoshie}},\ }\href {\doibase
  10.1103/PhysRevLett.69.21} {\bibfield  {journal} {\bibinfo  {journal} {Phys.
  Rev. Lett.}\ }\textbf {\bibinfo {volume} {69}},\ \bibinfo {pages} {21}
  (\bibinfo {year} {1992})}\BibitemShut {NoStop}%
%%CITATION = PRLTA,69,21;%%
\bibitem [{\citenamefont {Aoki}\ \emph
  {et~al.}(2012{\natexlab{b}})\citenamefont {Aoki}, \citenamefont {Aoyama},
  \citenamefont {Kurachi}, \citenamefont {Maskawa}, \citenamefont {Nagai},
  \citenamefont {Ohki}, \citenamefont {Shibata}, \citenamefont {Yamawaki},\
  and\ \citenamefont {Yamazaki}}]{Aoki:2012eq}%
  \BibitemOpen
  \bibfield  {author} {\bibinfo {author} {\bibfnamefont {Y.}~\bibnamefont
  {Aoki}}, \bibinfo {author} {\bibfnamefont {T.}~\bibnamefont {Aoyama}},
  \bibinfo {author} {\bibfnamefont {M.}~\bibnamefont {Kurachi}}, \bibinfo
  {author} {\bibfnamefont {T.}~\bibnamefont {Maskawa}}, \bibinfo {author}
  {\bibfnamefont {K.-i.}\ \bibnamefont {Nagai}}, \bibinfo {author}
  {\bibfnamefont {H.}~\bibnamefont {Ohki}}, \bibinfo {author} {\bibfnamefont
  {A.}~\bibnamefont {Shibata}}, \bibinfo {author} {\bibfnamefont
  {K.}~\bibnamefont {Yamawaki}}, \ and\ \bibinfo {author} {\bibfnamefont
  {T.}~\bibnamefont {Yamazaki}} (\bibinfo {collaboration} {LatKMI
  Collaboration}),\ }\href {\doibase 10.1103/PhysRevD.86.059903,
  10.1103/PhysRevD.86.054506} {\bibfield  {journal} {\bibinfo  {journal} {Phys.
  Rev. D}\ }\textbf {\bibinfo {volume} {86}},\ \bibinfo {pages} {054506}
  (\bibinfo {year} {2012}{\natexlab{b}})}\BibitemShut {NoStop}%
%%CITATION = ARXIV:1207.3060;%%
\bibitem [{\citenamefont {Fodor}\ \emph {et~al.}(2011)\citenamefont {Fodor},
  \citenamefont {Holland}, \citenamefont {Kuti}, \citenamefont {Nogradi},
  \citenamefont {Schroeder}, \citenamefont {Holland}, \citenamefont {Kuti},
  \citenamefont {Nogradi},\ and\ \citenamefont {Schroeder}}]{Fodor:2011tu}%
  \BibitemOpen
  \bibfield  {author} {\bibinfo {author} {\bibfnamefont {Z.}~\bibnamefont
  {Fodor}}, \bibinfo {author} {\bibfnamefont {K.}~\bibnamefont {Holland}},
  \bibinfo {author} {\bibfnamefont {J.}~\bibnamefont {Kuti}}, \bibinfo {author}
  {\bibfnamefont {D.}~\bibnamefont {Nogradi}}, \bibinfo {author} {\bibfnamefont
  {C.}~\bibnamefont {Schroeder}}, \bibinfo {author} {\bibfnamefont
  {K.}~\bibnamefont {Holland}}, \bibinfo {author} {\bibfnamefont
  {J.}~\bibnamefont {Kuti}}, \bibinfo {author} {\bibfnamefont {D.}~\bibnamefont
  {Nogradi}}, \ and\ \bibinfo {author} {\bibfnamefont {C.}~\bibnamefont
  {Schroeder}},\ }\href {\doibase 10.1016/j.physletb.2011.07.037} {\bibfield
  {journal} {\bibinfo  {journal} {Phys. Lett.}\ }\textbf {\bibinfo {volume}
  {B703}},\ \bibinfo {pages} {348} (\bibinfo {year} {2011})}\BibitemShut
  {NoStop}%
%%CITATION = ARXIV:1104.3124;%%
\bibitem [{\citenamefont {Hayakawa}\ \emph {et~al.}(2011)\citenamefont
  {Hayakawa}, \citenamefont {Ishikawa}, \citenamefont {Osaki}, \citenamefont
  {Takeda}, \citenamefont {Uno},\ and\ \citenamefont
  {Yamada}}]{Hayakawa:2010yn}%
  \BibitemOpen
  \bibfield  {author} {\bibinfo {author} {\bibfnamefont {M.}~\bibnamefont
  {Hayakawa}}, \bibinfo {author} {\bibfnamefont {K.~I.}\ \bibnamefont
  {Ishikawa}}, \bibinfo {author} {\bibfnamefont {Y.}~\bibnamefont {Osaki}},
  \bibinfo {author} {\bibfnamefont {S.}~\bibnamefont {Takeda}}, \bibinfo
  {author} {\bibfnamefont {S.}~\bibnamefont {Uno}}, \ and\ \bibinfo {author}
  {\bibfnamefont {N.}~\bibnamefont {Yamada}},\ }\href {\doibase
  10.1103/PhysRevD.83.074509} {\bibfield  {journal} {\bibinfo  {journal} {Phys.
  Rev. D}\ }\textbf {\bibinfo {volume} {83}},\ \bibinfo {pages} {074509}
  (\bibinfo {year} {2011})}\BibitemShut {NoStop}%
%%CITATION = ARXIV:1011.2577;%%
\bibitem [{\citenamefont {Dimopoulos}(1980)}]{Dimopoulos:1979sp}%
  \BibitemOpen
  \bibfield  {author} {\bibinfo {author} {\bibfnamefont {S.}~\bibnamefont
  {Dimopoulos}},\ }\href {\doibase 10.1016/0550-3213(80)90277-1} {\bibfield
  {journal} {\bibinfo  {journal} {Nucl. Phys.}\ }\textbf {\bibinfo {volume}
  {B168}},\ \bibinfo {pages} {69} (\bibinfo {year} {1980})}\BibitemShut
  {NoStop}%
%%CITATION = NUPHA,B168,69;%%
\bibitem [{\citenamefont {Farhi}\ and\ \citenamefont
  {Susskind}(1981)}]{Farhi:1980xs}%
  \BibitemOpen
  \bibfield  {author} {\bibinfo {author} {\bibfnamefont {E.}~\bibnamefont
  {Farhi}}\ and\ \bibinfo {author} {\bibfnamefont {L.}~\bibnamefont
  {Susskind}},\ }\href {\doibase 10.1016/0370-1573(81)90173-3} {\bibfield
  {journal} {\bibinfo  {journal} {Phys. Rept.}\ }\textbf {\bibinfo {volume}
  {74}},\ \bibinfo {pages} {277} (\bibinfo {year} {1981})}\BibitemShut
  {NoStop}%
%%CITATION = PRPLC,74,277;%%
\bibitem [{\citenamefont {Dimopoulos}\ and\ \citenamefont
  {Susskind}(1979)}]{Dimopoulos:1979es}%
  \BibitemOpen
  \bibfield  {author} {\bibinfo {author} {\bibfnamefont {S.}~\bibnamefont
  {Dimopoulos}}\ and\ \bibinfo {author} {\bibfnamefont {L.}~\bibnamefont
  {Susskind}},\ }\href {\doibase 10.1016/0550-3213(79)90364-X} {\bibfield
  {journal} {\bibinfo  {journal} {Nucl. Phys.}\ }\textbf {\bibinfo {volume}
  {B155}},\ \bibinfo {pages} {237} (\bibinfo {year} {1979})}\BibitemShut
  {NoStop}%
%%CITATION = NUPHA,B155,237;%%
\bibitem [{\citenamefont {Eichten}\ and\ \citenamefont
  {Lane}(1980)}]{Eichten:1979ah}%
  \BibitemOpen
  \bibfield  {author} {\bibinfo {author} {\bibfnamefont {E.}~\bibnamefont
  {Eichten}}\ and\ \bibinfo {author} {\bibfnamefont {K.~D.}\ \bibnamefont
  {Lane}},\ }\href {\doibase 10.1016/0370-2693(80)90065-9} {\bibfield
  {journal} {\bibinfo  {journal} {Phys. Lett.}\ }\textbf {\bibinfo {volume}
  {B90}},\ \bibinfo {pages} {125} (\bibinfo {year} {1980})}\BibitemShut
  {NoStop}%
%%CITATION = PHLTA,B90,125;%%
\bibitem [{\citenamefont {{Aoki}}\ \emph {et~al.}(2014)\citenamefont {{Aoki}},
  \citenamefont {{Aoyama}}, \citenamefont {{Kurachi}}, \citenamefont
  {{Maskawa}}, \citenamefont {{Miura}}, \citenamefont {{Nagai}}, \citenamefont
  {{Ohki}}, \citenamefont {{Rinaldi}}, \citenamefont {{Shibata}}, \citenamefont
  {{Yamawaki}},\ and\ \citenamefont {{Yamazaki}}}]{Aoki:2013qxa}%
  \BibitemOpen
  \bibfield  {author} {\bibinfo {author} {\bibfnamefont {Y.}~\bibnamefont
  {{Aoki}}}, \bibinfo {author} {\bibfnamefont {T.}~\bibnamefont {{Aoyama}}},
  \bibinfo {author} {\bibfnamefont {M.}~\bibnamefont {{Kurachi}}}, \bibinfo
  {author} {\bibfnamefont {T.}~\bibnamefont {{Maskawa}}}, \bibinfo {author}
  {\bibfnamefont {K.}~\bibnamefont {{Miura}}}, \bibinfo {author} {\bibfnamefont
  {K.-i.}\ \bibnamefont {{Nagai}}}, \bibinfo {author} {\bibfnamefont
  {H.}~\bibnamefont {{Ohki}}}, \bibinfo {author} {\bibfnamefont
  {E.}~\bibnamefont {{Rinaldi}}}, \bibinfo {author} {\bibfnamefont
  {A.}~\bibnamefont {{Shibata}}}, \bibinfo {author} {\bibfnamefont
  {K.}~\bibnamefont {{Yamawaki}}}, \ and\ \bibinfo {author} {\bibfnamefont
  {T.}~\bibnamefont {{Yamazaki}}} (\bibinfo {collaboration} {LatKMI
  Collaboration}),\ }\href@noop {} {\bibfield  {journal} {\bibinfo  {journal}
  {PoS}\ }\textbf {\bibinfo {volume} {LATTICE2013}},\ \bibinfo {pages} {070}
  (\bibinfo {year} {2014})}\BibitemShut {NoStop}%
%%CITATION = ARXIV:1309.0711;%%
\bibitem [{\citenamefont {Aoki}\ \emph {et~al.}(2014)\citenamefont {Aoki},
  \citenamefont {Aoyama}, \citenamefont {Kurachi}, \citenamefont {Maskawa},
  \citenamefont {Miura}, \citenamefont {Nagai}, \citenamefont {Ohki},
  \citenamefont {Rinaldi}, \citenamefont {Shibata}, \citenamefont {Yamawaki},\
  and\ \citenamefont {Yamazaki}}]{Aoki:2014oha}%
  \BibitemOpen
  \bibfield  {author} {\bibinfo {author} {\bibfnamefont {Y.}~\bibnamefont
  {Aoki}}, \bibinfo {author} {\bibfnamefont {T.}~\bibnamefont {Aoyama}},
  \bibinfo {author} {\bibfnamefont {M.}~\bibnamefont {Kurachi}}, \bibinfo
  {author} {\bibfnamefont {T.}~\bibnamefont {Maskawa}}, \bibinfo {author}
  {\bibfnamefont {K.}~\bibnamefont {Miura}}, \bibinfo {author} {\bibfnamefont
  {K.-i.}\ \bibnamefont {Nagai}}, \bibinfo {author} {\bibfnamefont
  {H.}~\bibnamefont {Ohki}}, \bibinfo {author} {\bibfnamefont {E.}~\bibnamefont
  {Rinaldi}}, \bibinfo {author} {\bibfnamefont {A.}~\bibnamefont {Shibata}},
  \bibinfo {author} {\bibfnamefont {K.}~\bibnamefont {Yamawaki}}, \ and\
  \bibinfo {author} {\bibfnamefont {T.}~\bibnamefont {Yamazaki}} (\bibinfo
  {collaboration} {LatKMI Collaboration}),\ }\href {\doibase
  10.1103/PhysRevD.89.111502} {\bibfield  {journal} {\bibinfo  {journal} {Phys.
  Rev. D}\ }\textbf {\bibinfo {volume} {89}},\ \bibinfo {pages} {111502}
  (\bibinfo {year} {2014})}\BibitemShut {NoStop}%
\bibitem [{\citenamefont {Appelquist}\ \emph {et~al.}(2008)\citenamefont
  {Appelquist}, \citenamefont {Fleming},\ and\ \citenamefont
  {Neil}}]{Appelquist:2007hu}%
  \BibitemOpen
  \bibfield  {author} {\bibinfo {author} {\bibfnamefont {T.}~\bibnamefont
  {Appelquist}}, \bibinfo {author} {\bibfnamefont {G.~T.}\ \bibnamefont
  {Fleming}}, \ and\ \bibinfo {author} {\bibfnamefont {E.~T.}\ \bibnamefont
  {Neil}},\ }\href {\doibase 10.1103/PhysRevLett.100.171607} {\bibfield
  {journal} {\bibinfo  {journal} {Phys. Rev. Lett.}\ }\textbf {\bibinfo
  {volume} {100}},\ \bibinfo {pages} {171607} (\bibinfo {year}
  {2008})}\BibitemShut {NoStop}%
%%CITATION = 0712.0609;%%
\bibitem [{\citenamefont {Hasenfratz}\ \emph {et~al.}(2015)\citenamefont
  {Hasenfratz}, \citenamefont {Schaich},\ and\ \citenamefont
  {Veernala}}]{Hasenfratz:2014rna}%
  \BibitemOpen
  \bibfield  {author} {\bibinfo {author} {\bibfnamefont {A.}~\bibnamefont
  {Hasenfratz}}, \bibinfo {author} {\bibfnamefont {D.}~\bibnamefont {Schaich}},
  \ and\ \bibinfo {author} {\bibfnamefont {A.}~\bibnamefont {Veernala}},\
  }\href {\doibase 10.1007/JHEP06(2015)143} {\bibfield  {journal} {\bibinfo
  {journal} {JHEP}\ }\textbf {\bibinfo {volume} {06}},\ \bibinfo {pages} {143}
  (\bibinfo {year} {2015})}\BibitemShut {NoStop}%
%%CITATION = ARXIV:1410.5886;%%
\bibitem [{\citenamefont {Ishikawa}\ \emph {et~al.}(2013)\citenamefont
  {Ishikawa}, \citenamefont {Iwasaki}, \citenamefont {Nakayama},\ and\
  \citenamefont {Yoshie}}]{Ishikawa:2013wf}%
  \BibitemOpen
  \bibfield  {author} {\bibinfo {author} {\bibfnamefont {K.~I.}\ \bibnamefont
  {Ishikawa}}, \bibinfo {author} {\bibfnamefont {Y.}~\bibnamefont {Iwasaki}},
  \bibinfo {author} {\bibfnamefont {Y.}~\bibnamefont {Nakayama}}, \ and\
  \bibinfo {author} {\bibfnamefont {T.}~\bibnamefont {Yoshie}},\ }\href
  {\doibase 10.1103/PhysRevD.87.071503} {\bibfield  {journal} {\bibinfo
  {journal} {Phys. Rev.}\ }\textbf {\bibinfo {volume} {D87}},\ \bibinfo {pages}
  {071503} (\bibinfo {year} {2013})}\BibitemShut {NoStop}%
%%CITATION = ARXIV:1301.4785;%%
\bibitem [{\citenamefont {Appelquist}\ \emph
  {et~al.}(2014{\natexlab{a}})\citenamefont {Appelquist}, \citenamefont
  {Brower}, \citenamefont {Fleming}, \citenamefont {Kiskis}, \citenamefont
  {Lin}, \citenamefont {Neil}, \citenamefont {Osborn}, \citenamefont {Rebbi},
  \citenamefont {Rinaldi}, \citenamefont {Schaich}, \citenamefont {Schroeder},
  \citenamefont {Syritsyn}, \citenamefont {Voronov}, \citenamefont {Vranas},
  \citenamefont {Weinberg},\ and\ \citenamefont {Witzel}}]{Appelquist:2014zsa}%
  \BibitemOpen
  \bibfield  {author} {\bibinfo {author} {\bibfnamefont {T.}~\bibnamefont
  {Appelquist}}, \bibinfo {author} {\bibfnamefont {R.~C.}\ \bibnamefont
  {Brower}}, \bibinfo {author} {\bibfnamefont {G.~T.}\ \bibnamefont {Fleming}},
  \bibinfo {author} {\bibfnamefont {J.}~\bibnamefont {Kiskis}}, \bibinfo
  {author} {\bibfnamefont {M.~F.}\ \bibnamefont {Lin}}, \bibinfo {author}
  {\bibfnamefont {E.~T.}\ \bibnamefont {Neil}}, \bibinfo {author}
  {\bibfnamefont {J.~C.}\ \bibnamefont {Osborn}}, \bibinfo {author}
  {\bibfnamefont {C.}~\bibnamefont {Rebbi}}, \bibinfo {author} {\bibfnamefont
  {E.}~\bibnamefont {Rinaldi}}, \bibinfo {author} {\bibfnamefont
  {D.}~\bibnamefont {Schaich}}, \bibinfo {author} {\bibfnamefont
  {C.}~\bibnamefont {Schroeder}}, \bibinfo {author} {\bibfnamefont
  {S.}~\bibnamefont {Syritsyn}}, \bibinfo {author} {\bibfnamefont
  {G.}~\bibnamefont {Voronov}}, \bibinfo {author} {\bibfnamefont
  {P.}~\bibnamefont {Vranas}}, \bibinfo {author} {\bibfnamefont
  {E.}~\bibnamefont {Weinberg}}, \ and\ \bibinfo {author} {\bibfnamefont
  {O.}~\bibnamefont {Witzel}} (\bibinfo {collaboration} {Lattice Strong
  Dynamics (LSD) Collaboration}),\ }\href {\doibase 10.1103/PhysRevD.90.114502}
  {\bibfield  {journal} {\bibinfo  {journal} {Phys. Rev. D}\ }\textbf {\bibinfo
  {volume} {90}},\ \bibinfo {pages} {114502} (\bibinfo {year}
  {2014}{\natexlab{a}})}\BibitemShut {NoStop}%
\bibitem [{\citenamefont {Appelquist}\ \emph {et~al.}(2016)\citenamefont
  {Appelquist}, \citenamefont {Brower}, \citenamefont {Fleming}, \citenamefont
  {Hasenfratz}, \citenamefont {Jin}, \citenamefont {Kiskis}, \citenamefont
  {Neil}, \citenamefont {Osborn}, \citenamefont {Rebbi}, \citenamefont
  {Rinaldi}, \citenamefont {Schaich}, \citenamefont {Vranas}, \citenamefont
  {Weinberg},\ and\ \citenamefont {Witzel}}]{Appelquist:2016viq}%
  \BibitemOpen
  \bibfield  {author} {\bibinfo {author} {\bibfnamefont {T.}~\bibnamefont
  {Appelquist}}, \bibinfo {author} {\bibfnamefont {R.~C.}\ \bibnamefont
  {Brower}}, \bibinfo {author} {\bibfnamefont {G.~T.}\ \bibnamefont {Fleming}},
  \bibinfo {author} {\bibfnamefont {A.}~\bibnamefont {Hasenfratz}}, \bibinfo
  {author} {\bibfnamefont {X.~Y.}\ \bibnamefont {Jin}}, \bibinfo {author}
  {\bibfnamefont {J.}~\bibnamefont {Kiskis}}, \bibinfo {author} {\bibfnamefont
  {E.~T.}\ \bibnamefont {Neil}}, \bibinfo {author} {\bibfnamefont {J.~C.}\
  \bibnamefont {Osborn}}, \bibinfo {author} {\bibfnamefont {C.}~\bibnamefont
  {Rebbi}}, \bibinfo {author} {\bibfnamefont {E.}~\bibnamefont {Rinaldi}},
  \bibinfo {author} {\bibfnamefont {D.}~\bibnamefont {Schaich}}, \bibinfo
  {author} {\bibfnamefont {P.}~\bibnamefont {Vranas}}, \bibinfo {author}
  {\bibfnamefont {E.}~\bibnamefont {Weinberg}}, \ and\ \bibinfo {author}
  {\bibfnamefont {O.}~\bibnamefont {Witzel}} (\bibinfo {collaboration} {Lattice
  Strong Dynamics (LSD) Collaboration}),\ }\href {\doibase
  10.1103/PhysRevD.93.114514} {\bibfield  {journal} {\bibinfo  {journal} {Phys.
  Rev. D}\ }\textbf {\bibinfo {volume} {93}},\ \bibinfo {pages} {114514}
  (\bibinfo {year} {2016})}\BibitemShut {NoStop}%
\bibitem [{\citenamefont {Aoki}\ \emph
  {et~al.}(2013{\natexlab{b}})\citenamefont {Aoki}, \citenamefont {Aoyama},
  \citenamefont {Kurachi}, \citenamefont {Maskawa}, \citenamefont {Nagai},
  \citenamefont {Ohki}, \citenamefont {Rinaldi}, \citenamefont {Shibata},
  \citenamefont {Yamawaki},\ and\ \citenamefont {Yamazaki}}]{Aoki:2013zsa}%
  \BibitemOpen
  \bibfield  {author} {\bibinfo {author} {\bibfnamefont {Y.}~\bibnamefont
  {Aoki}}, \bibinfo {author} {\bibfnamefont {T.}~\bibnamefont {Aoyama}},
  \bibinfo {author} {\bibfnamefont {M.}~\bibnamefont {Kurachi}}, \bibinfo
  {author} {\bibfnamefont {T.}~\bibnamefont {Maskawa}}, \bibinfo {author}
  {\bibfnamefont {K.-i.}\ \bibnamefont {Nagai}}, \bibinfo {author}
  {\bibfnamefont {H.}~\bibnamefont {Ohki}}, \bibinfo {author} {\bibfnamefont
  {E.}~\bibnamefont {Rinaldi}}, \bibinfo {author} {\bibfnamefont
  {A.}~\bibnamefont {Shibata}}, \bibinfo {author} {\bibfnamefont
  {K.}~\bibnamefont {Yamawaki}}, \ and\ \bibinfo {author} {\bibfnamefont
  {T.}~\bibnamefont {Yamazaki}} (\bibinfo {collaboration} {LatKMI
  Collaboration}),\ }\href {\doibase 10.1103/PhysRevLett.111.162001} {\bibfield
   {journal} {\bibinfo  {journal} {Phys. Rev. Lett.}\ }\textbf {\bibinfo
  {volume} {111}},\ \bibinfo {pages} {162001} (\bibinfo {year}
  {2013}{\natexlab{b}})}\BibitemShut {NoStop}%
%%CITATION = ARXIV:1305.6006;%%
\bibitem [{\citenamefont {Fodor}\ \emph {et~al.}(2014)\citenamefont {Fodor},
  \citenamefont {Holland}, \citenamefont {Kuti}, \citenamefont {Nogradi},\ and\
  \citenamefont {Wong}}]{Fodor:2014pqa}%
  \BibitemOpen
  \bibfield  {author} {\bibinfo {author} {\bibfnamefont {Z.}~\bibnamefont
  {Fodor}}, \bibinfo {author} {\bibfnamefont {K.}~\bibnamefont {Holland}},
  \bibinfo {author} {\bibfnamefont {J.}~\bibnamefont {Kuti}}, \bibinfo {author}
  {\bibfnamefont {D.}~\bibnamefont {Nogradi}}, \ and\ \bibinfo {author}
  {\bibfnamefont {C.~H.}\ \bibnamefont {Wong}},\ }\href@noop {} {\bibfield
  {journal} {\bibinfo  {journal} {PoS}\ }\textbf {\bibinfo {volume}
  {LATTICE2013}},\ \bibinfo {pages} {062} (\bibinfo {year} {2014})}\BibitemShut
  {NoStop}%
%%CITATION = ARXIV:1401.2176;%%
\bibitem [{\citenamefont {Kunihiro}\ \emph {et~al.}(2004)\citenamefont
  {Kunihiro}, \citenamefont {Muroya}, \citenamefont {Nakamura}, \citenamefont
  {Nonaka}, \citenamefont {Sekiguchi},\ and\ \citenamefont
  {Wada}}]{Kunihiro:2003yj}%
  \BibitemOpen
  \bibfield  {author} {\bibinfo {author} {\bibfnamefont {T.}~\bibnamefont
  {Kunihiro}}, \bibinfo {author} {\bibfnamefont {S.}~\bibnamefont {Muroya}},
  \bibinfo {author} {\bibfnamefont {A.}~\bibnamefont {Nakamura}}, \bibinfo
  {author} {\bibfnamefont {C.}~\bibnamefont {Nonaka}}, \bibinfo {author}
  {\bibfnamefont {M.}~\bibnamefont {Sekiguchi}}, \ and\ \bibinfo {author}
  {\bibfnamefont {H.}~\bibnamefont {Wada}} (\bibinfo {collaboration} {SCALAR
  Collaboration}),\ }\href {\doibase 10.1103/PhysRevD.70.034504} {\bibfield
  {journal} {\bibinfo  {journal} {Phys. Rev. D}\ }\textbf {\bibinfo {volume}
  {70}},\ \bibinfo {pages} {034504} (\bibinfo {year} {2004})}\BibitemShut
  {NoStop}%
%%CITATION = HEP-PH/0310312;%%
\bibitem [{\citenamefont {{Aoki}}\ \emph
  {et~al.}(2015{\natexlab{a}})\citenamefont {{Aoki}}, \citenamefont {{Aoyama}},
  \citenamefont {{Bennett}}, \citenamefont {{Kurachi}}, \citenamefont
  {{Maskawa}}, \citenamefont {{Miura}}, \citenamefont {{Nagai}}, \citenamefont
  {{Ohki}}, \citenamefont {{Rinaldi}}, \citenamefont {{Shibata}}, \citenamefont
  {{Yamawaki}},\ and\ \citenamefont {{Yamazaki}}}]{Aoki:2015gea}%
  \BibitemOpen
  \bibfield  {author} {\bibinfo {author} {\bibfnamefont {Y.}~\bibnamefont
  {{Aoki}}}, \bibinfo {author} {\bibfnamefont {T.}~\bibnamefont {{Aoyama}}},
  \bibinfo {author} {\bibfnamefont {E.}~\bibnamefont {{Bennett}}}, \bibinfo
  {author} {\bibfnamefont {M.}~\bibnamefont {{Kurachi}}}, \bibinfo {author}
  {\bibfnamefont {T.}~\bibnamefont {{Maskawa}}}, \bibinfo {author}
  {\bibfnamefont {K.}~\bibnamefont {{Miura}}}, \bibinfo {author} {\bibfnamefont
  {K.-i.}\ \bibnamefont {{Nagai}}}, \bibinfo {author} {\bibfnamefont
  {H.}~\bibnamefont {{Ohki}}}, \bibinfo {author} {\bibfnamefont
  {E.}~\bibnamefont {{Rinaldi}}}, \bibinfo {author} {\bibfnamefont
  {A.}~\bibnamefont {{Shibata}}}, \bibinfo {author} {\bibfnamefont
  {K.}~\bibnamefont {{Yamawaki}}}, \ and\ \bibinfo {author} {\bibfnamefont
  {T.}~\bibnamefont {{Yamazaki}}} (\bibinfo {collaboration} {LatKMI
  Collaboration}),\ }\href@noop {} {\bibfield  {journal} {\bibinfo  {journal}
  {PoS}\ }\textbf {\bibinfo {volume} {LATTICE2014}},\ \bibinfo {pages} {256}
  (\bibinfo {year} {2015}{\natexlab{a}})}\BibitemShut {NoStop}%
%%CITATION = ARXIV:1501.06660;%%
\bibitem [{\citenamefont {{Aoki}}\ \emph
  {et~al.}(2015{\natexlab{b}})\citenamefont {{Aoki}}, \citenamefont {{Aoyama}},
  \citenamefont {{Bennett}}, \citenamefont {{Kurachi}}, \citenamefont
  {{Maskawa}}, \citenamefont {{Miura}}, \citenamefont {{Nagai}}, \citenamefont
  {{Ohki}}, \citenamefont {{Rinaldi}}, \citenamefont {{Shibata}}, \citenamefont
  {{Yamawaki}},\ and\ \citenamefont {{Yamazaki}}}]{Aoki:2015jfa}%
  \BibitemOpen
  \bibfield  {author} {\bibinfo {author} {\bibfnamefont {Y.}~\bibnamefont
  {{Aoki}}}, \bibinfo {author} {\bibfnamefont {T.}~\bibnamefont {{Aoyama}}},
  \bibinfo {author} {\bibfnamefont {E.}~\bibnamefont {{Bennett}}}, \bibinfo
  {author} {\bibfnamefont {M.}~\bibnamefont {{Kurachi}}}, \bibinfo {author}
  {\bibfnamefont {T.}~\bibnamefont {{Maskawa}}}, \bibinfo {author}
  {\bibfnamefont {K.}~\bibnamefont {{Miura}}}, \bibinfo {author} {\bibfnamefont
  {K.-i.}\ \bibnamefont {{Nagai}}}, \bibinfo {author} {\bibfnamefont
  {H.}~\bibnamefont {{Ohki}}}, \bibinfo {author} {\bibfnamefont
  {E.}~\bibnamefont {{Rinaldi}}}, \bibinfo {author} {\bibfnamefont
  {A.}~\bibnamefont {{Shibata}}}, \bibinfo {author} {\bibfnamefont
  {K.}~\bibnamefont {{Yamawaki}}}, \ and\ \bibinfo {author} {\bibfnamefont
  {T.}~\bibnamefont {{Yamazaki}}} (\bibinfo {collaboration} {LatKMI
  Collaboration}),\ }\href
  {http://inspirehep.net/record/1400811/files/arXiv:1510.07373.pdf} {\
  (\bibinfo {year} {2015}{\natexlab{b}})},\ \Eprint
  {http://arxiv.org/abs/1510.07373} {arXiv:1510.07373 [hep-lat]} \BibitemShut
  {NoStop}%
%%CITATION = ARXIV:1510.07373;%%
\bibitem [{\citenamefont {{Aoki}}\ \emph
  {et~al.}(2016{\natexlab{a}})\citenamefont {{Aoki}}, \citenamefont {{Aoyama}},
  \citenamefont {{Bennett}}, \citenamefont {{Kurachi}}, \citenamefont
  {{Maskawa}}, \citenamefont {{Miura}}, \citenamefont {{Nagai}}, \citenamefont
  {{Ohki}}, \citenamefont {{Rinaldi}}, \citenamefont {{Shibata}}, \citenamefont
  {{Yamawaki}},\ and\ \citenamefont {{Yamazaki}}}]{Aoki:2016fxd}%
  \BibitemOpen
  \bibfield  {author} {\bibinfo {author} {\bibfnamefont {Y.}~\bibnamefont
  {{Aoki}}}, \bibinfo {author} {\bibfnamefont {T.}~\bibnamefont {{Aoyama}}},
  \bibinfo {author} {\bibfnamefont {E.}~\bibnamefont {{Bennett}}}, \bibinfo
  {author} {\bibfnamefont {M.}~\bibnamefont {{Kurachi}}}, \bibinfo {author}
  {\bibfnamefont {T.}~\bibnamefont {{Maskawa}}}, \bibinfo {author}
  {\bibfnamefont {K.}~\bibnamefont {{Miura}}}, \bibinfo {author} {\bibfnamefont
  {K.-i.}\ \bibnamefont {{Nagai}}}, \bibinfo {author} {\bibfnamefont
  {H.}~\bibnamefont {{Ohki}}}, \bibinfo {author} {\bibfnamefont
  {E.}~\bibnamefont {{Rinaldi}}}, \bibinfo {author} {\bibfnamefont
  {A.}~\bibnamefont {{Shibata}}}, \bibinfo {author} {\bibfnamefont
  {K.}~\bibnamefont {{Yamawaki}}}, \ and\ \bibinfo {author} {\bibfnamefont
  {T.}~\bibnamefont {{Yamazaki}}} (\bibinfo {collaboration} {LatKMI
  Collaboration}),\ }\href@noop {} {\bibfield  {journal} {\bibinfo  {journal}
  {PoS}\ }\textbf {\bibinfo {volume} {LATTICE2015}},\ \bibinfo {pages} {213}
  (\bibinfo {year} {2016}{\natexlab{a}})}\BibitemShut {NoStop}%
%%CITATION = ARXIV:1601.02287;%%
\bibitem [{\citenamefont {{Aoki}}\ \emph
  {et~al.}(2016{\natexlab{b}})\citenamefont {{Aoki}}, \citenamefont {{Aoyama}},
  \citenamefont {{Bennett}}, \citenamefont {{Kurachi}}, \citenamefont
  {{Maskawa}}, \citenamefont {{Miura}}, \citenamefont {{Nagai}}, \citenamefont
  {{Ohki}}, \citenamefont {{Rinaldi}}, \citenamefont {{Shibata}}, \citenamefont
  {{Yamawaki}},\ and\ \citenamefont {{Yamazaki}}}]{Aoki:2015zny}%
  \BibitemOpen
  \bibfield  {author} {\bibinfo {author} {\bibfnamefont {Y.}~\bibnamefont
  {{Aoki}}}, \bibinfo {author} {\bibfnamefont {T.}~\bibnamefont {{Aoyama}}},
  \bibinfo {author} {\bibfnamefont {E.}~\bibnamefont {{Bennett}}}, \bibinfo
  {author} {\bibfnamefont {M.}~\bibnamefont {{Kurachi}}}, \bibinfo {author}
  {\bibfnamefont {T.}~\bibnamefont {{Maskawa}}}, \bibinfo {author}
  {\bibfnamefont {K.}~\bibnamefont {{Miura}}}, \bibinfo {author} {\bibfnamefont
  {K.-i.}\ \bibnamefont {{Nagai}}}, \bibinfo {author} {\bibfnamefont
  {H.}~\bibnamefont {{Ohki}}}, \bibinfo {author} {\bibfnamefont
  {E.}~\bibnamefont {{Rinaldi}}}, \bibinfo {author} {\bibfnamefont
  {A.}~\bibnamefont {{Shibata}}}, \bibinfo {author} {\bibfnamefont
  {K.}~\bibnamefont {{Yamawaki}}}, \ and\ \bibinfo {author} {\bibfnamefont
  {T.}~\bibnamefont {{Yamazaki}}} (\bibinfo {collaboration} {LatKMI
  Collaboration}),\ }\href@noop {} {\bibfield  {journal} {\bibinfo  {journal}
  {PoS}\ }\textbf {\bibinfo {volume} {LATTICE2015}},\ \bibinfo {pages} {215}
  (\bibinfo {year} {2016}{\natexlab{b}})}\BibitemShut {NoStop}%
%%CITATION = ARXIV:1512.00957;%%
\bibitem [{\citenamefont {Cheng}\ \emph {et~al.}(2014)\citenamefont {Cheng},
  \citenamefont {Hasenfratz}, \citenamefont {Liu}, \citenamefont
  {Petropoulos},\ and\ \citenamefont {Schaich}}]{Cheng:2013xha}%
  \BibitemOpen
  \bibfield  {author} {\bibinfo {author} {\bibfnamefont {A.}~\bibnamefont
  {Cheng}}, \bibinfo {author} {\bibfnamefont {A.}~\bibnamefont {Hasenfratz}},
  \bibinfo {author} {\bibfnamefont {Y.}~\bibnamefont {Liu}}, \bibinfo {author}
  {\bibfnamefont {G.}~\bibnamefont {Petropoulos}}, \ and\ \bibinfo {author}
  {\bibfnamefont {D.}~\bibnamefont {Schaich}},\ }\href {\doibase
  10.1103/PhysRevD.90.014509} {\bibfield  {journal} {\bibinfo  {journal} {Phys.
  Rev. D}\ }\textbf {\bibinfo {volume} {90}},\ \bibinfo {pages} {014509}
  (\bibinfo {year} {2014})}\BibitemShut {NoStop}%
%%CITATION = ARXIV:1401.0195;%%
\bibitem [{\citenamefont {Matsuzaki}\ and\ \citenamefont
  {Yamawaki}(2014)}]{Matsuzaki:2013eva}%
  \BibitemOpen
  \bibfield  {author} {\bibinfo {author} {\bibfnamefont {S.}~\bibnamefont
  {Matsuzaki}}\ and\ \bibinfo {author} {\bibfnamefont {K.}~\bibnamefont
  {Yamawaki}},\ }\href {\doibase 10.1103/PhysRevLett.113.082002} {\bibfield
  {journal} {\bibinfo  {journal} {Phys. Rev. Lett.}\ }\textbf {\bibinfo
  {volume} {113}},\ \bibinfo {pages} {082002} (\bibinfo {year}
  {2014})}\BibitemShut {NoStop}%
%%CITATION = ARXIV:1311.3784;%%
\bibitem [{\citenamefont {Follana}\ \emph {et~al.}(2007)\citenamefont
  {Follana}, \citenamefont {Mason}, \citenamefont {Davies}, \citenamefont
  {Hornbostel}, \citenamefont {Lepage}, \citenamefont {Shigemitsu},
  \citenamefont {Trottier},\ and\ \citenamefont {Wong}}]{Follana:2006rc}%
  \BibitemOpen
  \bibfield  {author} {\bibinfo {author} {\bibfnamefont {E.}~\bibnamefont
  {Follana}}, \bibinfo {author} {\bibfnamefont {Q.}~\bibnamefont {Mason}},
  \bibinfo {author} {\bibfnamefont {C.}~\bibnamefont {Davies}}, \bibinfo
  {author} {\bibfnamefont {K.}~\bibnamefont {Hornbostel}}, \bibinfo {author}
  {\bibfnamefont {G.~P.}\ \bibnamefont {Lepage}}, \bibinfo {author}
  {\bibfnamefont {J.}~\bibnamefont {Shigemitsu}}, \bibinfo {author}
  {\bibfnamefont {H.}~\bibnamefont {Trottier}}, \ and\ \bibinfo {author}
  {\bibfnamefont {K.}~\bibnamefont {Wong}},\ }\href {\doibase
  10.1103/PhysRevD.75.054502} {\bibfield  {journal} {\bibinfo  {journal} {Phys.
  Rev. D}\ }\textbf {\bibinfo {volume} {75}},\ \bibinfo {pages} {054502}
  (\bibinfo {year} {2007})}\BibitemShut {NoStop}%
\bibitem [{\citenamefont {Bazavov}\ \emph {et~al.}(2010)\citenamefont
  {Bazavov}, \citenamefont {Bernard}, \citenamefont {DeTar}, \citenamefont
  {Freeman}, \citenamefont {Gottlieb}, \citenamefont {Heller}, \citenamefont
  {Hetrick}, \citenamefont {Laiho}, \citenamefont {Levkova}, \citenamefont
  {Oktay}, \citenamefont {Osborn}, \citenamefont {Sugar}, \citenamefont
  {Toussaint},\ and\ \citenamefont {Van~de Water}}]{Bazavov:2010ru}%
  \BibitemOpen
  \bibfield  {author} {\bibinfo {author} {\bibfnamefont {A.}~\bibnamefont
  {Bazavov}}, \bibinfo {author} {\bibfnamefont {C.}~\bibnamefont {Bernard}},
  \bibinfo {author} {\bibfnamefont {C.}~\bibnamefont {DeTar}}, \bibinfo
  {author} {\bibfnamefont {W.}~\bibnamefont {Freeman}}, \bibinfo {author}
  {\bibfnamefont {S.}~\bibnamefont {Gottlieb}}, \bibinfo {author}
  {\bibfnamefont {U.~M.}\ \bibnamefont {Heller}}, \bibinfo {author}
  {\bibfnamefont {J.~E.}\ \bibnamefont {Hetrick}}, \bibinfo {author}
  {\bibfnamefont {J.}~\bibnamefont {Laiho}}, \bibinfo {author} {\bibfnamefont
  {L.}~\bibnamefont {Levkova}}, \bibinfo {author} {\bibfnamefont
  {M.}~\bibnamefont {Oktay}}, \bibinfo {author} {\bibfnamefont
  {J.}~\bibnamefont {Osborn}}, \bibinfo {author} {\bibfnamefont {R.~L.}\
  \bibnamefont {Sugar}}, \bibinfo {author} {\bibfnamefont {D.}~\bibnamefont
  {Toussaint}}, \ and\ \bibinfo {author} {\bibfnamefont {R.~S.}\ \bibnamefont
  {Van~de Water}} (\bibinfo {collaboration} {MILC Collaboration}),\ }\href
  {\doibase 10.1103/PhysRevD.82.074501} {\bibfield  {journal} {\bibinfo
  {journal} {Phys. Rev. D}\ }\textbf {\bibinfo {volume} {82}},\ \bibinfo
  {pages} {074501} (\bibinfo {year} {2010})}\BibitemShut {NoStop}%
\bibitem [{\citenamefont {Bazavov}\ \emph {et~al.}(2012)\citenamefont
  {Bazavov}, \citenamefont {Bhattacharya}, \citenamefont {Cheng}, \citenamefont
  {DeTar}, \citenamefont {Ding}, \citenamefont {Gottlieb}, \citenamefont
  {Gupta}, \citenamefont {Hegde}, \citenamefont {Heller}, \citenamefont
  {Karsch}, \citenamefont {Laermann}, \citenamefont {Levkova}, \citenamefont
  {Mukherjee}, \citenamefont {Petreczky}, \citenamefont {Schmidt},
  \citenamefont {Soltz}, \citenamefont {Soeldner}, \citenamefont {Sugar},
  \citenamefont {Toussaint}, \citenamefont {Unger},\ and\ \citenamefont
  {Vranas}}]{Bazavov:2011nk}%
  \BibitemOpen
  \bibfield  {author} {\bibinfo {author} {\bibfnamefont {A.}~\bibnamefont
  {Bazavov}}, \bibinfo {author} {\bibfnamefont {T.}~\bibnamefont
  {Bhattacharya}}, \bibinfo {author} {\bibfnamefont {M.}~\bibnamefont {Cheng}},
  \bibinfo {author} {\bibfnamefont {C.}~\bibnamefont {DeTar}}, \bibinfo
  {author} {\bibfnamefont {H.-T.}\ \bibnamefont {Ding}}, \bibinfo {author}
  {\bibfnamefont {S.}~\bibnamefont {Gottlieb}}, \bibinfo {author}
  {\bibfnamefont {R.}~\bibnamefont {Gupta}}, \bibinfo {author} {\bibfnamefont
  {P.}~\bibnamefont {Hegde}}, \bibinfo {author} {\bibfnamefont {U.~M.}\
  \bibnamefont {Heller}}, \bibinfo {author} {\bibfnamefont {F.}~\bibnamefont
  {Karsch}}, \bibinfo {author} {\bibfnamefont {E.}~\bibnamefont {Laermann}},
  \bibinfo {author} {\bibfnamefont {L.}~\bibnamefont {Levkova}}, \bibinfo
  {author} {\bibfnamefont {S.}~\bibnamefont {Mukherjee}}, \bibinfo {author}
  {\bibfnamefont {P.}~\bibnamefont {Petreczky}}, \bibinfo {author}
  {\bibfnamefont {C.}~\bibnamefont {Schmidt}}, \bibinfo {author} {\bibfnamefont
  {R.~A.}\ \bibnamefont {Soltz}}, \bibinfo {author} {\bibfnamefont
  {W.}~\bibnamefont {Soeldner}}, \bibinfo {author} {\bibfnamefont
  {R.}~\bibnamefont {Sugar}}, \bibinfo {author} {\bibfnamefont
  {D.}~\bibnamefont {Toussaint}}, \bibinfo {author} {\bibfnamefont
  {W.}~\bibnamefont {Unger}}, \ and\ \bibinfo {author} {\bibfnamefont
  {P.}~\bibnamefont {Vranas}} (\bibinfo {collaboration} {HotQCD
  Collaboration}),\ }\href {\doibase 10.1103/PhysRevD.85.054503} {\bibfield
  {journal} {\bibinfo  {journal} {Phys. Rev. D}\ }\textbf {\bibinfo {volume}
  {85}},\ \bibinfo {pages} {054503} (\bibinfo {year} {2012})}\BibitemShut
  {NoStop}%
\bibitem [{\citenamefont {Peskin}\ and\ \citenamefont
  {Takeuchi}(1990)}]{Peskin:1990zt}%
  \BibitemOpen
  \bibfield  {author} {\bibinfo {author} {\bibfnamefont {M.~E.}\ \bibnamefont
  {Peskin}}\ and\ \bibinfo {author} {\bibfnamefont {T.}~\bibnamefont
  {Takeuchi}},\ }\href@noop {} {\bibfield  {journal} {\bibinfo  {journal}
  {Phys. Rev. Lett.}\ }\textbf {\bibinfo {volume} {65}},\ \bibinfo {pages}
  {964} (\bibinfo {year} {1990})}\BibitemShut {NoStop}%
%%CITATION = PRLTA,65,964;%%
\bibitem [{\citenamefont {{Aoki}}\ \emph
  {et~al.}(2016{\natexlab{c}})\citenamefont {{Aoki}}, \citenamefont {{Aoyama}},
  \citenamefont {{Bennett}}, \citenamefont {{Kurachi}}, \citenamefont
  {{Maskawa}}, \citenamefont {{Miura}}, \citenamefont {{Nagai}}, \citenamefont
  {{Ohki}}, \citenamefont {{Rinaldi}}, \citenamefont {{Shibata}}, \citenamefont
  {{Yamawaki}},\ and\ \citenamefont {{Yamazaki}}}]{Aoki:2016bfp}%
  \BibitemOpen
  \bibfield  {author} {\bibinfo {author} {\bibfnamefont {Y.}~\bibnamefont
  {{Aoki}}}, \bibinfo {author} {\bibfnamefont {T.}~\bibnamefont {{Aoyama}}},
  \bibinfo {author} {\bibfnamefont {E.}~\bibnamefont {{Bennett}}}, \bibinfo
  {author} {\bibfnamefont {M.}~\bibnamefont {{Kurachi}}}, \bibinfo {author}
  {\bibfnamefont {T.}~\bibnamefont {{Maskawa}}}, \bibinfo {author}
  {\bibfnamefont {K.}~\bibnamefont {{Miura}}}, \bibinfo {author} {\bibfnamefont
  {K.-i.}\ \bibnamefont {{Nagai}}}, \bibinfo {author} {\bibfnamefont
  {H.}~\bibnamefont {{Ohki}}}, \bibinfo {author} {\bibfnamefont
  {E.}~\bibnamefont {{Rinaldi}}}, \bibinfo {author} {\bibfnamefont
  {A.}~\bibnamefont {{Shibata}}}, \bibinfo {author} {\bibfnamefont
  {K.}~\bibnamefont {{Yamawaki}}}, \ and\ \bibinfo {author} {\bibfnamefont
  {T.}~\bibnamefont {{Yamazaki}}} (\bibinfo {collaboration} {LatKMI
  Collaboration}),\ }\href@noop {} {\bibfield  {journal} {\bibinfo  {journal}
  {PoS}\ }\textbf {\bibinfo {volume} {LATTICE2015}},\ \bibinfo {pages} {245}
  (\bibinfo {year} {2016}{\natexlab{c}})}\BibitemShut {NoStop}%
%%CITATION = ARXIV:1602.00796;%%
\bibitem [{\citenamefont {Duane}\ \emph {et~al.}(1987)\citenamefont {Duane},
  \citenamefont {Kennedy}, \citenamefont {Pendleton},\ and\ \citenamefont
  {Roweth}}]{Duane:1987de}%
  \BibitemOpen
  \bibfield  {author} {\bibinfo {author} {\bibfnamefont {S.}~\bibnamefont
  {Duane}}, \bibinfo {author} {\bibfnamefont {A.~D.}\ \bibnamefont {Kennedy}},
  \bibinfo {author} {\bibfnamefont {B.~J.}\ \bibnamefont {Pendleton}}, \ and\
  \bibinfo {author} {\bibfnamefont {D.}~\bibnamefont {Roweth}},\ }\href@noop {}
  {\bibfield  {journal} {\bibinfo  {journal} {Phys. Lett.}\ }\textbf {\bibinfo
  {volume} {B195}},\ \bibinfo {pages} {216} (\bibinfo {year}
  {1987})}\BibitemShut {NoStop}%
%%CITATION = PHLTA,B195,216;%%
\bibitem [{\citenamefont {Hasenbusch}(2001)}]{Hasenbusch:2001ne}%
  \BibitemOpen
  \bibfield  {author} {\bibinfo {author} {\bibfnamefont {M.}~\bibnamefont
  {Hasenbusch}},\ }\href {\doibase 10.1016/S0370-2693(01)01102-9} {\bibfield
  {journal} {\bibinfo  {journal} {Phys. Lett.}\ }\textbf {\bibinfo {volume}
  {B519}},\ \bibinfo {pages} {177} (\bibinfo {year} {2001})}\BibitemShut
  {NoStop}%
%%CITATION = HEP-LAT/0107019;%%
\bibitem [{Mil()}]{MilcCodeV7}%
  \BibitemOpen
  \href@noop {} {\bibinfo  {journal} {See {\texttt
  http://www.physics.utah.edu\//\~{}detar\//milc/milc\_qcd.html}}\
  }\BibitemShut {NoStop}%
\bibitem [{\citenamefont {Gupta}(1997)}]{Gupta:1997nd}%
  \BibitemOpen
\bibfield  {journal} {  }\bibfield  {author} {\bibinfo {author} {\bibfnamefont
  {R.}~\bibnamefont {Gupta}}\ }(\bibinfo {year} {1997})\ \Eprint
  {http://arxiv.org/abs/hep-lat/9807028} {hep-lat/9807028} \BibitemShut
  {NoStop}%
%%CITATION = HEP-LAT/9807028;%%
\bibitem [{\citenamefont {Kilcup}\ and\ \citenamefont
  {Sharpe}(1987)}]{Kilcup:1986dg}%
  \BibitemOpen
  \bibfield  {author} {\bibinfo {author} {\bibfnamefont {G.~W.}\ \bibnamefont
  {Kilcup}}\ and\ \bibinfo {author} {\bibfnamefont {S.~R.}\ \bibnamefont
  {Sharpe}},\ }\href@noop {} {\bibfield  {journal} {\bibinfo  {journal} {Nucl.
  Phys.}\ }\textbf {\bibinfo {volume} {B283}},\ \bibinfo {pages} {493}
  (\bibinfo {year} {1987})}\BibitemShut {NoStop}%
%%CITATION = NUPHA,B283,493;%%
\bibitem [{\citenamefont {Blum}\ \emph {et~al.}(2003)\citenamefont {Blum} \emph
  {et~al.}}]{Blum:2001xb}%
  \BibitemOpen
  \bibfield  {author} {\bibinfo {author} {\bibfnamefont {T.}~\bibnamefont
  {Blum}} \emph {et~al.} (\bibinfo {collaboration} {RBC}),\ }\href@noop {}
  {\bibfield  {journal} {\bibinfo  {journal} {Phys. Rev.}\ }\textbf {\bibinfo
  {volume} {D68}},\ \bibinfo {pages} {114506} (\bibinfo {year} {2003})},\
  \Eprint {http://arxiv.org/abs/hep-lat/0110075} {hep-lat/0110075} \BibitemShut
  {NoStop}%
%%CITATION = HEP-LAT 0110075;%%
\bibitem [{\citenamefont {Umeda}(2007)}]{Umeda:2007hy}%
  \BibitemOpen
  \bibfield  {author} {\bibinfo {author} {\bibfnamefont {T.}~\bibnamefont
  {Umeda}},\ }\href {\doibase 10.1103/PhysRevD.75.094502} {\bibfield  {journal}
  {\bibinfo  {journal} {Phys. Rev.}\ }\textbf {\bibinfo {volume} {D75}},\
  \bibinfo {pages} {094502} (\bibinfo {year} {2007})},\ \Eprint
  {http://arxiv.org/abs/hep-lat/0701005} {arXiv:hep-lat/0701005 [hep-lat]}
  \BibitemShut {NoStop}%
%%CITATION = HEP-LAT/0701005;%%
\bibitem [{\citenamefont {Golterman}\ and\ \citenamefont
  {Smit}(1984)}]{Golterman:1984cy}%
  \BibitemOpen
  \bibfield  {author} {\bibinfo {author} {\bibfnamefont {M.~F.}\ \bibnamefont
  {Golterman}}\ and\ \bibinfo {author} {\bibfnamefont {J.}~\bibnamefont
  {Smit}},\ }\href {\doibase 10.1016/0550-3213(84)90424-3} {\bibfield
  {journal} {\bibinfo  {journal} {Nucl.Phys.}\ }\textbf {\bibinfo {volume}
  {B245}},\ \bibinfo {pages} {61} (\bibinfo {year} {1984})}\BibitemShut
  {NoStop}%
%%CITATION = NUPHA,B245,61;%%
\bibitem [{\citenamefont {Golterman}(1986)}]{Golterman:1985dz}%
  \BibitemOpen
  \bibfield  {author} {\bibinfo {author} {\bibfnamefont {M.~F.~L.}\
  \bibnamefont {Golterman}},\ }\href {\doibase 10.1016/0550-3213(86)90383-4}
  {\bibfield  {journal} {\bibinfo  {journal} {Nucl. Phys.}\ }\textbf {\bibinfo
  {volume} {B273}},\ \bibinfo {pages} {663} (\bibinfo {year}
  {1986})}\BibitemShut {NoStop}%
%%CITATION = NUPHA,B273,663;%%
\bibitem [{\citenamefont {Ishizuka}\ \emph {et~al.}(1994)\citenamefont
  {Ishizuka}, \citenamefont {Fukugita}, \citenamefont {Mino}, \citenamefont
  {Okawa},\ and\ \citenamefont {Ukawa}}]{Ishizuka:1993mt}%
  \BibitemOpen
  \bibfield  {author} {\bibinfo {author} {\bibfnamefont {N.}~\bibnamefont
  {Ishizuka}}, \bibinfo {author} {\bibfnamefont {M.}~\bibnamefont {Fukugita}},
  \bibinfo {author} {\bibfnamefont {H.}~\bibnamefont {Mino}}, \bibinfo {author}
  {\bibfnamefont {M.}~\bibnamefont {Okawa}}, \ and\ \bibinfo {author}
  {\bibfnamefont {A.}~\bibnamefont {Ukawa}},\ }\href@noop {} {\bibfield
  {journal} {\bibinfo  {journal} {Nucl. Phys.}\ }\textbf {\bibinfo {volume}
  {B411}},\ \bibinfo {pages} {875} (\bibinfo {year} {1994})}\BibitemShut
  {NoStop}%
%%CITATION = NUPHA,B411,875;%%
\bibitem [{\citenamefont {Golterman}\ and\ \citenamefont
  {Smit}(1985)}]{Golterman:1984dn}%
  \BibitemOpen
  \bibfield  {author} {\bibinfo {author} {\bibfnamefont {M.~F.~L.}\
  \bibnamefont {Golterman}}\ and\ \bibinfo {author} {\bibfnamefont
  {J.}~\bibnamefont {Smit}},\ }\href@noop {} {\bibfield  {journal} {\bibinfo
  {journal} {Nucl. Phys.}\ }\textbf {\bibinfo {volume} {B255}},\ \bibinfo
  {pages} {328} (\bibinfo {year} {1985})}\BibitemShut {NoStop}%
%%CITATION = NUPHA,B255,328;%%
\bibitem [{\citenamefont {Gasser}\ and\ \citenamefont
  {Leutwyler}(1988)}]{Gasser:1987zq}%
  \BibitemOpen
  \bibfield  {author} {\bibinfo {author} {\bibfnamefont {J.}~\bibnamefont
  {Gasser}}\ and\ \bibinfo {author} {\bibfnamefont {H.}~\bibnamefont
  {Leutwyler}},\ }\href {\doibase 10.1016/0550-3213(88)90107-1} {\bibfield
  {journal} {\bibinfo  {journal} {Nucl. Phys.}\ }\textbf {\bibinfo {volume}
  {B307}},\ \bibinfo {pages} {763} (\bibinfo {year} {1988})}\BibitemShut
  {NoStop}%
%%CITATION = NUPHA,B307,763;%%
\bibitem [{\citenamefont {Luscher}(1986)}]{Luscher:1985dn}%
  \BibitemOpen
  \bibfield  {author} {\bibinfo {author} {\bibfnamefont {M.}~\bibnamefont
  {Luscher}},\ }\href {\doibase 10.1007/BF01211589} {\bibfield  {journal}
  {\bibinfo  {journal} {Commun. Math. Phys.}\ }\textbf {\bibinfo {volume}
  {104}},\ \bibinfo {pages} {177} (\bibinfo {year} {1986})}\BibitemShut
  {NoStop}%
%%CITATION = CMPHA,104,177;%%
\bibitem [{\citenamefont {Aubin}\ \emph {et~al.}(2004)\citenamefont {Aubin},
  \citenamefont {Bernard}, \citenamefont {DeTar}, \citenamefont {Osborn},
  \citenamefont {Gottlieb}, \citenamefont {Gregory}, \citenamefont {Toussaint},
  \citenamefont {Heller}, \citenamefont {Hetrick},\ and\ \citenamefont
  {Sugar}}]{Aubin:2004fs}%
  \BibitemOpen
  \bibfield  {author} {\bibinfo {author} {\bibfnamefont {C.}~\bibnamefont
  {Aubin}}, \bibinfo {author} {\bibfnamefont {C.}~\bibnamefont {Bernard}},
  \bibinfo {author} {\bibfnamefont {C.}~\bibnamefont {DeTar}}, \bibinfo
  {author} {\bibfnamefont {J.}~\bibnamefont {Osborn}}, \bibinfo {author}
  {\bibfnamefont {S.}~\bibnamefont {Gottlieb}}, \bibinfo {author}
  {\bibfnamefont {E.~B.}\ \bibnamefont {Gregory}}, \bibinfo {author}
  {\bibfnamefont {D.}~\bibnamefont {Toussaint}}, \bibinfo {author}
  {\bibfnamefont {U.~M.}\ \bibnamefont {Heller}}, \bibinfo {author}
  {\bibfnamefont {J.~E.}\ \bibnamefont {Hetrick}}, \ and\ \bibinfo {author}
  {\bibfnamefont {R.}~\bibnamefont {Sugar}} (\bibinfo {collaboration} {MILC
  Collaboration}),\ }\href {\doibase 10.1103/PhysRevD.70.114501} {\bibfield
  {journal} {\bibinfo  {journal} {Phys. Rev. D}\ }\textbf {\bibinfo {volume}
  {70}},\ \bibinfo {pages} {114501} (\bibinfo {year} {2004})}\BibitemShut
  {NoStop}%
\bibitem [{\citenamefont {Fodor}\ \emph {et~al.}(2009)\citenamefont {Fodor},
  \citenamefont {Holland}, \citenamefont {Kuti}, \citenamefont {Nogradi},\ and\
  \citenamefont {Schroeder}}]{Fodor:2009wk}%
  \BibitemOpen
  \bibfield  {author} {\bibinfo {author} {\bibfnamefont {Z.}~\bibnamefont
  {Fodor}}, \bibinfo {author} {\bibfnamefont {K.}~\bibnamefont {Holland}},
  \bibinfo {author} {\bibfnamefont {J.}~\bibnamefont {Kuti}}, \bibinfo {author}
  {\bibfnamefont {D.}~\bibnamefont {Nogradi}}, \ and\ \bibinfo {author}
  {\bibfnamefont {C.}~\bibnamefont {Schroeder}},\ }\href {\doibase
  10.1016/j.physletb.2009.10.040} {\bibfield  {journal} {\bibinfo  {journal}
  {Phys. Lett.}\ }\textbf {\bibinfo {volume} {B681}},\ \bibinfo {pages} {353}
  (\bibinfo {year} {2009})}\BibitemShut {NoStop}%
%%CITATION = 0907.4562;%%
\bibitem [{\citenamefont {Soldate}\ and\ \citenamefont
  {Sundrum}(1990)}]{Soldate:1989fh}%
  \BibitemOpen
  \bibfield  {author} {\bibinfo {author} {\bibfnamefont {M.}~\bibnamefont
  {Soldate}}\ and\ \bibinfo {author} {\bibfnamefont {R.}~\bibnamefont
  {Sundrum}},\ }\href {\doibase 10.1016/0550-3213(90)90156-8} {\bibfield
  {journal} {\bibinfo  {journal} {Nucl. Phys.}\ }\textbf {\bibinfo {volume}
  {B340}},\ \bibinfo {pages} {1} (\bibinfo {year} {1990})}\BibitemShut
  {NoStop}%
%%CITATION = NUPHA,B340,1;%%
\bibitem [{\citenamefont {Chivukula}\ \emph {et~al.}(1993)\citenamefont
  {Chivukula}, \citenamefont {Dugan},\ and\ \citenamefont
  {Golden}}]{Chivukula:1992gi}%
  \BibitemOpen
  \bibfield  {author} {\bibinfo {author} {\bibfnamefont {R.~S.}\ \bibnamefont
  {Chivukula}}, \bibinfo {author} {\bibfnamefont {M.~J.}\ \bibnamefont
  {Dugan}}, \ and\ \bibinfo {author} {\bibfnamefont {M.}~\bibnamefont
  {Golden}},\ }\href {\doibase 10.1103/PhysRevD.47.2930} {\bibfield  {journal}
  {\bibinfo  {journal} {Phys. Rev. D}\ }\textbf {\bibinfo {volume} {47}},\
  \bibinfo {pages} {2930} (\bibinfo {year} {1993})}\BibitemShut {NoStop}%
%%CITATION = HEP-PH/9206222;%%
\bibitem [{\citenamefont {Harada}\ and\ \citenamefont
  {Yamawaki}(2003)}]{Harada:2003jx}%
  \BibitemOpen
  \bibfield  {author} {\bibinfo {author} {\bibfnamefont {M.}~\bibnamefont
  {Harada}}\ and\ \bibinfo {author} {\bibfnamefont {K.}~\bibnamefont
  {Yamawaki}},\ }\href {\doibase 10.1016/S0370-1573(03)00139-X} {\bibfield
  {journal} {\bibinfo  {journal} {Phys. Rept.}\ }\textbf {\bibinfo {volume}
  {381}},\ \bibinfo {pages} {1} (\bibinfo {year} {2003})}\BibitemShut {NoStop}%
%%CITATION = HEP-PH/0302103;%%
\bibitem [{\citenamefont {Gasser}\ and\ \citenamefont
  {Leutwyler}(1984)}]{Gasser:1983yg}%
  \BibitemOpen
  \bibfield  {author} {\bibinfo {author} {\bibfnamefont {J.}~\bibnamefont
  {Gasser}}\ and\ \bibinfo {author} {\bibfnamefont {H.}~\bibnamefont
  {Leutwyler}},\ }\href {\doibase 10.1016/0003-4916(84)90242-2} {\bibfield
  {journal} {\bibinfo  {journal} {Ann. Phys.}\ }\textbf {\bibinfo {volume}
  {158}},\ \bibinfo {pages} {142} (\bibinfo {year} {1984})}\BibitemShut
  {NoStop}%
%%CITATION = APNYA,158,142;%%
\bibitem [{\citenamefont {Weinberg}(1967)}]{Weinberg:1967kj}%
  \BibitemOpen
  \bibfield  {author} {\bibinfo {author} {\bibfnamefont {S.}~\bibnamefont
  {Weinberg}},\ }\href {\doibase 10.1103/PhysRevLett.18.507} {\bibfield
  {journal} {\bibinfo  {journal} {Phys. Rev. Lett.}\ }\textbf {\bibinfo
  {volume} {18}},\ \bibinfo {pages} {507} (\bibinfo {year} {1967})}\BibitemShut
  {NoStop}%
%%CITATION = PRLTA,18,507;%%
\bibitem [{\citenamefont {Gilman}\ and\ \citenamefont
  {Harari}(1968)}]{Gilman:1967qs}%
  \BibitemOpen
  \bibfield  {author} {\bibinfo {author} {\bibfnamefont {F.~J.}\ \bibnamefont
  {Gilman}}\ and\ \bibinfo {author} {\bibfnamefont {H.}~\bibnamefont
  {Harari}},\ }\href {\doibase 10.1103/PhysRev.165.1803} {\bibfield  {journal}
  {\bibinfo  {journal} {Phys. Rev.}\ }\textbf {\bibinfo {volume} {165}},\
  \bibinfo {pages} {1803} (\bibinfo {year} {1968})}\BibitemShut {NoStop}%
%%CITATION = PHRVA,165,1803;%%
\bibitem [{\citenamefont {Weinberg}(1969)}]{Weinberg:1969hw}%
  \BibitemOpen
  \bibfield  {author} {\bibinfo {author} {\bibfnamefont {S.}~\bibnamefont
  {Weinberg}},\ }\href {\doibase 10.1103/PhysRev.177.2604} {\bibfield
  {journal} {\bibinfo  {journal} {Phys. Rev.}\ }\textbf {\bibinfo {volume}
  {177}},\ \bibinfo {pages} {2604} (\bibinfo {year} {1969})}\BibitemShut
  {NoStop}%
%%CITATION = PHRVA,177,2604;%%
\bibitem [{\citenamefont {Weinberg}(1990)}]{Weinberg:1990xn}%
  \BibitemOpen
  \bibfield  {author} {\bibinfo {author} {\bibfnamefont {S.}~\bibnamefont
  {Weinberg}},\ }\href {\doibase 10.1103/PhysRevLett.65.1177} {\bibfield
  {journal} {\bibinfo  {journal} {Phys. Rev. Lett.}\ }\textbf {\bibinfo
  {volume} {65}},\ \bibinfo {pages} {1177} (\bibinfo {year}
  {1990})}\BibitemShut {NoStop}%
%%CITATION = PRLTA,65,1177;%%
\bibitem [{\citenamefont {Maskawa}\ and\ \citenamefont
  {Yamawaki}(1976)}]{Maskawa:1975ky}%
  \BibitemOpen
  \bibfield  {author} {\bibinfo {author} {\bibfnamefont {T.}~\bibnamefont
  {Maskawa}}\ and\ \bibinfo {author} {\bibfnamefont {K.}~\bibnamefont
  {Yamawaki}},\ }\href {\doibase 10.1143/PTP.56.270} {\bibfield  {journal}
  {\bibinfo  {journal} {Prog. Theor. Phys.}\ }\textbf {\bibinfo {volume}
  {56}},\ \bibinfo {pages} {270} (\bibinfo {year} {1976})}\BibitemShut
  {NoStop}%
%%CITATION = PTPKA,56,270;%%
\bibitem [{\citenamefont {Yamawaki}(1998)}]{Yamawaki:1998cy}%
  \BibitemOpen
  \bibfield  {author} {\bibinfo {author} {\bibfnamefont {K.}~\bibnamefont
  {Yamawaki}},\ }in\ \href {http://alice.cern.ch/format/showfull?sysnb=0269152}
  {\emph {\bibinfo {booktitle} {{QCD, light cone physics and hadron
  phenomenology. Proceedings, 10th Nuclear Summer School and Symposium,
  NuSS'97, Seoul, Korea, June 23-28, 1997}}}}\ (\bibinfo {year} {1998})\ pp.\
  \bibinfo {pages} {116--199},\ \Eprint {http://arxiv.org/abs/hep-th/9802037}
  {arXiv:hep-th/9802037 [hep-th]} \BibitemShut {NoStop}%
%%CITATION = HEP-TH/9802037;%%
\bibitem [{\citenamefont {Ida}(1974)}]{Ida:1973ec}%
  \BibitemOpen
  \bibfield  {author} {\bibinfo {author} {\bibfnamefont {M.}~\bibnamefont
  {Ida}},\ }\href {\doibase 10.1143/PTP.51.1521} {\bibfield  {journal}
  {\bibinfo  {journal} {Prog. Theor. Phys.}\ }\textbf {\bibinfo {volume}
  {51}},\ \bibinfo {pages} {1521} (\bibinfo {year} {1974})},\ \bibinfo {note}
  {[Erratum: Prog. Theor. Phys.52,1978(1974)]}\BibitemShut {NoStop}%
%%CITATION = PTPKA,51,1521;%%
\bibitem [{\citenamefont {Del~Debbio}\ and\ \citenamefont
  {Zwicky}(2010)}]{DelDebbio:2010ze}%
  \BibitemOpen
  \bibfield  {author} {\bibinfo {author} {\bibfnamefont {L.}~\bibnamefont
  {Del~Debbio}}\ and\ \bibinfo {author} {\bibfnamefont {R.}~\bibnamefont
  {Zwicky}},\ }\href {\doibase 10.1103/PhysRevD.82.014502} {\bibfield
  {journal} {\bibinfo  {journal} {Phys. Rev. D}\ }\textbf {\bibinfo {volume}
  {82}},\ \bibinfo {pages} {014502} (\bibinfo {year} {2010})}\BibitemShut
  {NoStop}%
%%CITATION = 1005.2371;%%
\bibitem [{\citenamefont {Albanese}\ \emph {et~al.}(1987)\citenamefont
  {Albanese}, \citenamefont {Costantini}, \citenamefont {Fiorentini},
  \citenamefont {Flore}, \citenamefont {Lombardo}, \citenamefont {Tripiccione},
  \citenamefont {Bacilieri}, \citenamefont {Fonti}, \citenamefont {Giacomelli},
  \citenamefont {Remiddi}, \citenamefont {Bernaschi}, \citenamefont {Cabibbo},
  \citenamefont {Marinari}, \citenamefont {Parisi}, \citenamefont {Salina},
  \citenamefont {Cabasino}, \citenamefont {Marzano}, \citenamefont {Paolucci},
  \citenamefont {Petrarca}, \citenamefont {Rapuano}, \citenamefont
  {Marchesini},\ and\ \citenamefont {Rusack}}]{Albanese:1987ds}%
  \BibitemOpen
  \bibfield  {author} {\bibinfo {author} {\bibfnamefont {M.}~\bibnamefont
  {Albanese}}, \bibinfo {author} {\bibfnamefont {F.}~\bibnamefont
  {Costantini}}, \bibinfo {author} {\bibfnamefont {G.}~\bibnamefont
  {Fiorentini}}, \bibinfo {author} {\bibfnamefont {F.}~\bibnamefont {Flore}},
  \bibinfo {author} {\bibfnamefont {M.}~\bibnamefont {Lombardo}}, \bibinfo
  {author} {\bibfnamefont {R.}~\bibnamefont {Tripiccione}}, \bibinfo {author}
  {\bibfnamefont {P.}~\bibnamefont {Bacilieri}}, \bibinfo {author}
  {\bibfnamefont {L.}~\bibnamefont {Fonti}}, \bibinfo {author} {\bibfnamefont
  {P.}~\bibnamefont {Giacomelli}}, \bibinfo {author} {\bibfnamefont
  {E.}~\bibnamefont {Remiddi}}, \bibinfo {author} {\bibfnamefont
  {M.}~\bibnamefont {Bernaschi}}, \bibinfo {author} {\bibfnamefont
  {N.}~\bibnamefont {Cabibbo}}, \bibinfo {author} {\bibfnamefont
  {E.}~\bibnamefont {Marinari}}, \bibinfo {author} {\bibfnamefont
  {G.}~\bibnamefont {Parisi}}, \bibinfo {author} {\bibfnamefont
  {G.}~\bibnamefont {Salina}}, \bibinfo {author} {\bibfnamefont
  {S.}~\bibnamefont {Cabasino}}, \bibinfo {author} {\bibfnamefont
  {F.}~\bibnamefont {Marzano}}, \bibinfo {author} {\bibfnamefont
  {P.}~\bibnamefont {Paolucci}}, \bibinfo {author} {\bibfnamefont
  {S.}~\bibnamefont {Petrarca}}, \bibinfo {author} {\bibfnamefont
  {F.}~\bibnamefont {Rapuano}}, \bibinfo {author} {\bibfnamefont
  {P.}~\bibnamefont {Marchesini}}, \ and\ \bibinfo {author} {\bibfnamefont
  {R.}~\bibnamefont {Rusack}} (\bibinfo {collaboration} {APE Collaboration}),\
  }\href {\doibase http://dx.doi.org/10.1016/0370-2693(87)91160-9} {\bibfield
  {journal} {\bibinfo  {journal} {Phys. Lett. B}\ }\textbf {\bibinfo {volume}
  {192}},\ \bibinfo {pages} {163 } (\bibinfo {year} {1987})}\BibitemShut
  {NoStop}%
\bibitem [{\citenamefont {Hasenfratz}\ and\ \citenamefont
  {Knechtli}(2001)}]{Hasenfratz:2001hp}%
  \BibitemOpen
  \bibfield  {author} {\bibinfo {author} {\bibfnamefont {A.}~\bibnamefont
  {Hasenfratz}}\ and\ \bibinfo {author} {\bibfnamefont {F.}~\bibnamefont
  {Knechtli}},\ }\href {\doibase 10.1103/PhysRevD.64.034504} {\bibfield
  {journal} {\bibinfo  {journal} {Phys. Rev. D}\ }\textbf {\bibinfo {volume}
  {64}},\ \bibinfo {pages} {034504} (\bibinfo {year} {2001})}\BibitemShut
  {NoStop}%
%%CITATION = HEP-LAT/0103029;%%
\bibitem [{\citenamefont {Della~Morte}\ \emph {et~al.}(2004)\citenamefont
  {Della~Morte}, \citenamefont {Durr}, \citenamefont {Heitger}, \citenamefont
  {Molke}, \citenamefont {Rolf}, \citenamefont {Shindler},\ and\ \citenamefont
  {Sommer}}]{DellaMorte:2003mn}%
  \BibitemOpen
  \bibfield  {author} {\bibinfo {author} {\bibfnamefont {M.}~\bibnamefont
  {Della~Morte}}, \bibinfo {author} {\bibfnamefont {S.}~\bibnamefont {Durr}},
  \bibinfo {author} {\bibfnamefont {J.}~\bibnamefont {Heitger}}, \bibinfo
  {author} {\bibfnamefont {H.}~\bibnamefont {Molke}}, \bibinfo {author}
  {\bibfnamefont {J.}~\bibnamefont {Rolf}}, \bibinfo {author} {\bibfnamefont
  {A.}~\bibnamefont {Shindler}}, \ and\ \bibinfo {author} {\bibfnamefont
  {R.}~\bibnamefont {Sommer}} (\bibinfo {collaboration} {ALPHA
  Collaboration}),\ }\href {\doibase 10.1016/j.physletb.2005.03.017,
  10.1016/j.physletb.2003.11.064} {\bibfield  {journal} {\bibinfo  {journal}
  {Phys. Lett.}\ }\textbf {\bibinfo {volume} {B581}},\ \bibinfo {pages} {93}
  (\bibinfo {year} {2004})},\ \bibinfo {note} {[Erratum: Phys.
  Lett.B612,313(2005)]}\BibitemShut {NoStop}%
%%CITATION = HEP-LAT/0307021;%%
\bibitem [{\citenamefont {Della~Morte}\ \emph {et~al.}(2005)\citenamefont
  {Della~Morte}, \citenamefont {Shindler},\ and\ \citenamefont
  {Sommer}}]{DellaMorte:2005nwx}%
  \BibitemOpen
  \bibfield  {author} {\bibinfo {author} {\bibfnamefont {M.}~\bibnamefont
  {Della~Morte}}, \bibinfo {author} {\bibfnamefont {A.}~\bibnamefont
  {Shindler}}, \ and\ \bibinfo {author} {\bibfnamefont {R.}~\bibnamefont
  {Sommer}},\ }\href {\doibase 10.1088/1126-6708/2005/08/051} {\bibfield
  {journal} {\bibinfo  {journal} {JHEP}\ }\textbf {\bibinfo {volume} {08}},\
  \bibinfo {pages} {051} (\bibinfo {year} {2005})}\BibitemShut {NoStop}%
%%CITATION = HEP-LAT/0506008;%%
\bibitem [{\citenamefont {Hasenfratz}(2010)}]{Hasenfratz:2010fi}%
  \BibitemOpen
  \bibfield  {author} {\bibinfo {author} {\bibfnamefont {A.}~\bibnamefont
  {Hasenfratz}},\ }\href {\doibase 10.1103/PhysRevD.82.014506} {\bibfield
  {journal} {\bibinfo  {journal} {Phys. Rev. D}\ }\textbf {\bibinfo {volume}
  {82}},\ \bibinfo {pages} {014506} (\bibinfo {year} {2010})}\BibitemShut
  {NoStop}%
%%CITATION = ARXIV:1004.1004;%%
\bibitem [{\citenamefont {Brower}\ \emph {et~al.}(2016)\citenamefont {Brower},
  \citenamefont {Hasenfratz}, \citenamefont {Rebbi}, \citenamefont {Weinberg},\
  and\ \citenamefont {Witzel}}]{Brower:2015owo}%
  \BibitemOpen
  \bibfield  {author} {\bibinfo {author} {\bibfnamefont {R.~C.}\ \bibnamefont
  {Brower}}, \bibinfo {author} {\bibfnamefont {A.}~\bibnamefont {Hasenfratz}},
  \bibinfo {author} {\bibfnamefont {C.}~\bibnamefont {Rebbi}}, \bibinfo
  {author} {\bibfnamefont {E.}~\bibnamefont {Weinberg}}, \ and\ \bibinfo
  {author} {\bibfnamefont {O.}~\bibnamefont {Witzel}},\ }\href {\doibase
  10.1103/PhysRevD.93.075028} {\bibfield  {journal} {\bibinfo  {journal} {Phys.
  Rev. D}\ }\textbf {\bibinfo {volume} {93}},\ \bibinfo {pages} {075028}
  (\bibinfo {year} {2016})}\BibitemShut {NoStop}%
%%CITATION = ARXIV:1512.02576;%%
\bibitem [{\citenamefont {Venkataraman}\ and\ \citenamefont
  {Kilcup}(1997)}]{Venkataraman:1997xi}%
  \BibitemOpen
  \bibfield  {author} {\bibinfo {author} {\bibfnamefont {L.}~\bibnamefont
  {Venkataraman}}\ and\ \bibinfo {author} {\bibfnamefont {G.}~\bibnamefont
  {Kilcup}},\ }\href@noop {} {\bibfield  {journal} {\bibinfo  {journal}
  {Submitted to: Phys. Rev. D}\ } (\bibinfo {year} {1997})}\BibitemShut
  {NoStop}%
%%CITATION = HEP-LAT/9711006;%%
\bibitem [{\citenamefont {Gregory}\ \emph {et~al.}(2008)\citenamefont
  {Gregory}, \citenamefont {Irving}, \citenamefont {Richards},\ and\
  \citenamefont {McNeile}}]{Gregory:2007ev}%
  \BibitemOpen
  \bibfield  {author} {\bibinfo {author} {\bibfnamefont {E.~B.}\ \bibnamefont
  {Gregory}}, \bibinfo {author} {\bibfnamefont {A.~C.}\ \bibnamefont {Irving}},
  \bibinfo {author} {\bibfnamefont {C.~M.}\ \bibnamefont {Richards}}, \ and\
  \bibinfo {author} {\bibfnamefont {C.}~\bibnamefont {McNeile}},\ }\href
  {\doibase 10.1103/PhysRevD.77.065019} {\bibfield  {journal} {\bibinfo
  {journal} {Phys. Rev. D}\ }\textbf {\bibinfo {volume} {77}},\ \bibinfo
  {pages} {065019} (\bibinfo {year} {2008})}\BibitemShut {NoStop}%
%%CITATION = ARXIV:0709.4224;%%
\bibitem [{\citenamefont {McNeile}\ \emph {et~al.}(2013)\citenamefont
  {McNeile}, \citenamefont {Bazavov}, \citenamefont {Davies}, \citenamefont
  {Dowdall}, \citenamefont {Hornbostel}, \citenamefont {Lepage},\ and\
  \citenamefont {Trottier}}]{McNeile:2012xh}%
  \BibitemOpen
  \bibfield  {author} {\bibinfo {author} {\bibfnamefont {C.}~\bibnamefont
  {McNeile}}, \bibinfo {author} {\bibfnamefont {A.}~\bibnamefont {Bazavov}},
  \bibinfo {author} {\bibfnamefont {C.~T.~H.}\ \bibnamefont {Davies}}, \bibinfo
  {author} {\bibfnamefont {R.~J.}\ \bibnamefont {Dowdall}}, \bibinfo {author}
  {\bibfnamefont {K.}~\bibnamefont {Hornbostel}}, \bibinfo {author}
  {\bibfnamefont {G.~P.}\ \bibnamefont {Lepage}}, \ and\ \bibinfo {author}
  {\bibfnamefont {H.~D.}\ \bibnamefont {Trottier}},\ }\href {\doibase
  10.1103/PhysRevD.87.034503} {\bibfield  {journal} {\bibinfo  {journal} {Phys.
  Rev. D}\ }\textbf {\bibinfo {volume} {87}},\ \bibinfo {pages} {034503}
  (\bibinfo {year} {2013})}\BibitemShut {NoStop}%
%%CITATION = ARXIV:1211.6577;%%
\bibitem [{\citenamefont {Jin}\ and\ \citenamefont
  {Mawhinney}(2011)}]{Jin:2012dw}%
  \BibitemOpen
  \bibfield  {author} {\bibinfo {author} {\bibfnamefont {X.-Y.}\ \bibnamefont
  {Jin}}\ and\ \bibinfo {author} {\bibfnamefont {R.~D.}\ \bibnamefont
  {Mawhinney}},\ }\href@noop {} {\bibfield  {journal} {\bibinfo  {journal}
  {PoS}\ }\textbf {\bibinfo {volume} {LATTICE2011}},\ \bibinfo {pages} {066}
  (\bibinfo {year} {2011})}\BibitemShut {NoStop}%
%%CITATION = ARXIV:1203.5855;%%
\bibitem [{\citenamefont {Kosower}(1984)}]{Kosower:1984aw}%
  \BibitemOpen
  \bibfield  {author} {\bibinfo {author} {\bibfnamefont {D.~A.}\ \bibnamefont
  {Kosower}},\ }\href {\doibase 10.1016/0370-2693(84)91806-9} {\bibfield
  {journal} {\bibinfo  {journal} {Phys. Lett.}\ }\textbf {\bibinfo {volume}
  {B144}},\ \bibinfo {pages} {215} (\bibinfo {year} {1984})}\BibitemShut
  {NoStop}%
%%CITATION = PHLTA,B144,215;%%
\bibitem [{\citenamefont {Peskin}(1980)}]{Peskin:1980gc}%
  \BibitemOpen
  \bibfield  {author} {\bibinfo {author} {\bibfnamefont {M.~E.}\ \bibnamefont
  {Peskin}},\ }\href {\doibase 10.1016/0550-3213(80)90051-6} {\bibfield
  {journal} {\bibinfo  {journal} {Nucl. Phys.}\ }\textbf {\bibinfo {volume}
  {B175}},\ \bibinfo {pages} {197} (\bibinfo {year} {1980})}\BibitemShut
  {NoStop}%
%%CITATION = NUPHA,B175,197;%%
\bibitem [{\citenamefont {Bando}\ \emph {et~al.}(1988)\citenamefont {Bando},
  \citenamefont {Kugo},\ and\ \citenamefont {Yamawaki}}]{Bando:1987br}%
  \BibitemOpen
  \bibfield  {author} {\bibinfo {author} {\bibfnamefont {M.}~\bibnamefont
  {Bando}}, \bibinfo {author} {\bibfnamefont {T.}~\bibnamefont {Kugo}}, \ and\
  \bibinfo {author} {\bibfnamefont {K.}~\bibnamefont {Yamawaki}},\ }\href
  {\doibase 10.1016/0370-1573(88)90019-1} {\bibfield  {journal} {\bibinfo
  {journal} {Phys. Rept.}\ }\textbf {\bibinfo {volume} {164}},\ \bibinfo
  {pages} {217} (\bibinfo {year} {1988})}\BibitemShut {NoStop}%
%%CITATION = PRPLC,164,217;%%
\bibitem [{\citenamefont {Aad}\ \emph {et~al.}(2015)\citenamefont {Aad} \emph
  {et~al.}}]{Aad:2015mxa}%
  \BibitemOpen
  \bibfield  {author} {\bibinfo {author} {\bibfnamefont {G.}~\bibnamefont
  {Aad}} \emph {et~al.} (\bibinfo {collaboration} {ATLAS}),\ }\href {\doibase
  10.1140/epjc/s10052-015-3685-1, 10.1140/epjc/s10052-016-3934-y} {\bibfield
  {journal} {\bibinfo  {journal} {Eur. Phys. J.}\ }\textbf {\bibinfo {volume}
  {C75}},\ \bibinfo {pages} {476} (\bibinfo {year} {2015})},\ \bibinfo {note}
  {[Erratum: Eur. Phys. J.C76,no.3,152(2016)]},\ \Eprint
  {http://arxiv.org/abs/1506.05669} {arXiv:1506.05669 [hep-ex]} \BibitemShut
  {NoStop}%
%%CITATION = ARXIV:1506.05669;%%
\bibitem [{\citenamefont {Khachatryan}\ \emph {et~al.}(2015)\citenamefont
  {Khachatryan} \emph {et~al.}}]{Khachatryan:2014kca}%
  \BibitemOpen
  \bibfield  {author} {\bibinfo {author} {\bibfnamefont {V.}~\bibnamefont
  {Khachatryan}} \emph {et~al.} (\bibinfo {collaboration} {CMS}),\ }\href
  {\doibase 10.1103/PhysRevD.92.012004} {\bibfield  {journal} {\bibinfo
  {journal} {Phys. Rev.}\ }\textbf {\bibinfo {volume} {D92}},\ \bibinfo {pages}
  {012004} (\bibinfo {year} {2015})},\ \Eprint {http://arxiv.org/abs/1411.3441}
  {arXiv:1411.3441 [hep-ex]} \BibitemShut {NoStop}%
%%CITATION = ARXIV:1411.3441;%%
\bibitem [{\citenamefont {Matsuzaki}\ and\ \citenamefont
  {Yamawaki}(2012{\natexlab{b}})}]{Matsuzaki:2012vc}%
  \BibitemOpen
  \bibfield  {author} {\bibinfo {author} {\bibfnamefont {S.}~\bibnamefont
  {Matsuzaki}}\ and\ \bibinfo {author} {\bibfnamefont {K.}~\bibnamefont
  {Yamawaki}},\ }\href {\doibase 10.1103/PhysRevD.86.035025} {\bibfield
  {journal} {\bibinfo  {journal} {Phys. Rev. D}\ }\textbf {\bibinfo {volume}
  {86}},\ \bibinfo {pages} {035025} (\bibinfo {year}
  {2012}{\natexlab{b}})}\BibitemShut {NoStop}%
%%CITATION = ARXIV:1206.6703;%%
\bibitem [{\citenamefont {Appelquist}\ \emph
  {et~al.}(2014{\natexlab{b}})\citenamefont {Appelquist}, \citenamefont
  {Berkowitz}, \citenamefont {Brower}, \citenamefont {Buchoff}, \citenamefont
  {Fleming}, \citenamefont {Kiskis}, \citenamefont {Lin}, \citenamefont {Neil},
  \citenamefont {Osborn}, \citenamefont {Rebbi}, \citenamefont {Rinaldi},
  \citenamefont {Schaich}, \citenamefont {Schroeder}, \citenamefont {Syritsyn},
  \citenamefont {Voronov}, \citenamefont {Vranas}, \citenamefont {Weinberg},
  \citenamefont {Witzel},\ and\ \citenamefont {Kribs}}]{Appelquist:2014jch}%
  \BibitemOpen
  \bibfield  {author} {\bibinfo {author} {\bibfnamefont {T.}~\bibnamefont
  {Appelquist}}, \bibinfo {author} {\bibfnamefont {E.}~\bibnamefont
  {Berkowitz}}, \bibinfo {author} {\bibfnamefont {R.~C.}\ \bibnamefont
  {Brower}}, \bibinfo {author} {\bibfnamefont {M.~I.}\ \bibnamefont {Buchoff}},
  \bibinfo {author} {\bibfnamefont {G.~T.}\ \bibnamefont {Fleming}}, \bibinfo
  {author} {\bibfnamefont {J.}~\bibnamefont {Kiskis}}, \bibinfo {author}
  {\bibfnamefont {M.~F.}\ \bibnamefont {Lin}}, \bibinfo {author} {\bibfnamefont
  {E.~T.}\ \bibnamefont {Neil}}, \bibinfo {author} {\bibfnamefont {J.~C.}\
  \bibnamefont {Osborn}}, \bibinfo {author} {\bibfnamefont {C.}~\bibnamefont
  {Rebbi}}, \bibinfo {author} {\bibfnamefont {E.}~\bibnamefont {Rinaldi}},
  \bibinfo {author} {\bibfnamefont {D.}~\bibnamefont {Schaich}}, \bibinfo
  {author} {\bibfnamefont {C.}~\bibnamefont {Schroeder}}, \bibinfo {author}
  {\bibfnamefont {S.}~\bibnamefont {Syritsyn}}, \bibinfo {author}
  {\bibfnamefont {G.}~\bibnamefont {Voronov}}, \bibinfo {author} {\bibfnamefont
  {P.}~\bibnamefont {Vranas}}, \bibinfo {author} {\bibfnamefont
  {E.}~\bibnamefont {Weinberg}}, \bibinfo {author} {\bibfnamefont
  {O.}~\bibnamefont {Witzel}}, \ and\ \bibinfo {author} {\bibfnamefont {G.~D.}\
  \bibnamefont {Kribs}} (\bibinfo {collaboration} {Lattice Strong Dynamics
  (LSD) Collaboration}),\ }\href {\doibase 10.1103/PhysRevD.89.094508}
  {\bibfield  {journal} {\bibinfo  {journal} {Phys. Rev. D}\ }\textbf {\bibinfo
  {volume} {89}},\ \bibinfo {pages} {094508} (\bibinfo {year}
  {2014}{\natexlab{b}})}\BibitemShut {NoStop}%
\bibitem [{\citenamefont {Akerib}\ \emph {et~al.}(2014)\citenamefont {Akerib}
  \emph {et~al.}}]{Akerib:2013tjd}%
  \BibitemOpen
  \bibfield  {author} {\bibinfo {author} {\bibfnamefont {D.~S.}\ \bibnamefont
  {Akerib}} \emph {et~al.} (\bibinfo {collaboration} {LUX Collaboration}),\
  }\href {\doibase 10.1103/PhysRevLett.112.091303} {\bibfield  {journal}
  {\bibinfo  {journal} {Phys. Rev. Lett.}\ }\textbf {\bibinfo {volume} {112}},\
  \bibinfo {pages} {091303} (\bibinfo {year} {2014})}\BibitemShut {NoStop}%
%%CITATION = ARXIV:1310.8214;%%
\bibitem [{\citenamefont {Akerib}\ \emph {et~al.}(2016)\citenamefont {Akerib}
  \emph {et~al.}}]{Akerib:2016vxi}%
  \BibitemOpen
  \bibfield  {author} {\bibinfo {author} {\bibfnamefont {D.~S.}\ \bibnamefont
  {Akerib}} \emph {et~al.},\ }\href@noop {} {\  (\bibinfo {year} {2016})},\
  \Eprint {http://arxiv.org/abs/1608.07648} {arXiv:1608.07648 [astro-ph.CO]}
  \BibitemShut {NoStop}%
%%CITATION = ARXIV:1608.07648;%%
\bibitem [{\citenamefont {Tan}\ \emph {et~al.}(2016)\citenamefont {Tan} \emph
  {et~al.}}]{Tan:2016zwf}%
  \BibitemOpen
  \bibfield  {author} {\bibinfo {author} {\bibfnamefont {A.}~\bibnamefont
  {Tan}} \emph {et~al.} (\bibinfo {collaboration} {PandaX-II Collaboration}),\
  }\href {\doibase 10.1103/PhysRevLett.117.121303} {\bibfield  {journal}
  {\bibinfo  {journal} {Phys. Rev. Lett.}\ }\textbf {\bibinfo {volume} {117}},\
  \bibinfo {pages} {121303} (\bibinfo {year} {2016})}\BibitemShut {NoStop}%
%%CITATION = ARXIV:1607.07400;%%
\bibitem [{\citenamefont {Fodor}\ \emph {et~al.}(2016)\citenamefont {Fodor},
  \citenamefont {Holland}, \citenamefont {Kuti}, \citenamefont {Mondal},
  \citenamefont {Nogradi},\ and\ \citenamefont {Wong}}]{Fodor:2016wal}%
  \BibitemOpen
  \bibfield  {author} {\bibinfo {author} {\bibfnamefont {Z.}~\bibnamefont
  {Fodor}}, \bibinfo {author} {\bibfnamefont {K.}~\bibnamefont {Holland}},
  \bibinfo {author} {\bibfnamefont {J.}~\bibnamefont {Kuti}}, \bibinfo {author}
  {\bibfnamefont {S.}~\bibnamefont {Mondal}}, \bibinfo {author} {\bibfnamefont
  {D.}~\bibnamefont {Nogradi}}, \ and\ \bibinfo {author} {\bibfnamefont
  {C.~H.}\ \bibnamefont {Wong}},\ }\href {\doibase 10.1103/PhysRevD.94.014503}
  {\bibfield  {journal} {\bibinfo  {journal} {Phys. Rev. D}\ }\textbf {\bibinfo
  {volume} {94}},\ \bibinfo {pages} {014503} (\bibinfo {year}
  {2016})}\BibitemShut {NoStop}%
%%CITATION = ARXIV:1601.03302;%%
\bibitem [{\citenamefont {Holdom}\ and\ \citenamefont
  {Terning}(1990)}]{Holdom:1990tc}%
  \BibitemOpen
  \bibfield  {author} {\bibinfo {author} {\bibfnamefont {B.}~\bibnamefont
  {Holdom}}\ and\ \bibinfo {author} {\bibfnamefont {J.}~\bibnamefont
  {Terning}},\ }\href {\doibase 10.1016/0370-2693(90)91054-F} {\bibfield
  {journal} {\bibinfo  {journal} {Phys. Lett.}\ }\textbf {\bibinfo {volume}
  {B247}},\ \bibinfo {pages} {88} (\bibinfo {year} {1990})}\BibitemShut
  {NoStop}%
%%CITATION = PHLTA,B247,88;%%
\bibitem [{\citenamefont {Golden}\ and\ \citenamefont
  {Randall}(1991)}]{Golden:1990ig}%
  \BibitemOpen
  \bibfield  {author} {\bibinfo {author} {\bibfnamefont {M.}~\bibnamefont
  {Golden}}\ and\ \bibinfo {author} {\bibfnamefont {L.}~\bibnamefont
  {Randall}},\ }\href {\doibase 10.1016/0550-3213(91)90614-4} {\bibfield
  {journal} {\bibinfo  {journal} {Nucl. Phys.}\ }\textbf {\bibinfo {volume}
  {B361}},\ \bibinfo {pages} {3} (\bibinfo {year} {1991})}\BibitemShut
  {NoStop}%
%%CITATION = NUPHA,B361,3;%%
\bibitem [{\citenamefont {Cacciapaglia}\ \emph {et~al.}(2005)\citenamefont
  {Cacciapaglia}, \citenamefont {Csaki}, \citenamefont {Grojean},\ and\
  \citenamefont {Terning}}]{Cacciapaglia:2004rb}%
  \BibitemOpen
  \bibfield  {author} {\bibinfo {author} {\bibfnamefont {G.}~\bibnamefont
  {Cacciapaglia}}, \bibinfo {author} {\bibfnamefont {C.}~\bibnamefont {Csaki}},
  \bibinfo {author} {\bibfnamefont {C.}~\bibnamefont {Grojean}}, \ and\
  \bibinfo {author} {\bibfnamefont {J.}~\bibnamefont {Terning}},\ }\href
  {\doibase 10.1103/PhysRevD.71.035015} {\bibfield  {journal} {\bibinfo
  {journal} {Phys. Rev. D}\ }\textbf {\bibinfo {volume} {71}},\ \bibinfo
  {pages} {035015} (\bibinfo {year} {2005})}\BibitemShut {NoStop}%
%%CITATION = HEP-PH/0409126;%%
\bibitem [{\citenamefont {Foadi}\ \emph {et~al.}(2005)\citenamefont {Foadi},
  \citenamefont {Gopalakrishna},\ and\ \citenamefont {Schmidt}}]{Foadi:2004ps}%
  \BibitemOpen
  \bibfield  {author} {\bibinfo {author} {\bibfnamefont {R.}~\bibnamefont
  {Foadi}}, \bibinfo {author} {\bibfnamefont {S.}~\bibnamefont
  {Gopalakrishna}}, \ and\ \bibinfo {author} {\bibfnamefont {C.}~\bibnamefont
  {Schmidt}},\ }\href {\doibase 10.1016/j.physletb.2004.11.055} {\bibfield
  {journal} {\bibinfo  {journal} {Phys. Lett.}\ }\textbf {\bibinfo {volume}
  {B606}},\ \bibinfo {pages} {157} (\bibinfo {year} {2005})}\BibitemShut
  {NoStop}%
%%CITATION = HEP-PH/0409266;%%
\bibitem [{\citenamefont {Chivukula}\ \emph {et~al.}(2005)\citenamefont
  {Chivukula}, \citenamefont {Simmons}, \citenamefont {He}, \citenamefont
  {Kurachi},\ and\ \citenamefont {Tanabashi}}]{Chivukula:2005xm}%
  \BibitemOpen
  \bibfield  {author} {\bibinfo {author} {\bibfnamefont {R.~S.}\ \bibnamefont
  {Chivukula}}, \bibinfo {author} {\bibfnamefont {E.~H.}\ \bibnamefont
  {Simmons}}, \bibinfo {author} {\bibfnamefont {H.-J.}\ \bibnamefont {He}},
  \bibinfo {author} {\bibfnamefont {M.}~\bibnamefont {Kurachi}}, \ and\
  \bibinfo {author} {\bibfnamefont {M.}~\bibnamefont {Tanabashi}},\ }\href
  {\doibase 10.1103/PhysRevD.72.015008} {\bibfield  {journal} {\bibinfo
  {journal} {Phys. Rev. D}\ }\textbf {\bibinfo {volume} {72}},\ \bibinfo
  {pages} {015008} (\bibinfo {year} {2005})}\BibitemShut {NoStop}%
%%CITATION = HEP-PH/0504114;%%
\bibitem [{\citenamefont {Aoki}(2013)}]{Aoki:lat2013}%
  \BibitemOpen
  \bibfield  {author} {\bibinfo {author} {\bibfnamefont {Y.}~\bibnamefont
  {Aoki}} (\bibinfo {collaboration} {LatKMI Collaboration}),\ }\href@noop {}
  {\bibfield  {journal} {\bibinfo  {journal} {Talk presented at the 31st
  International Symposium on Lattice Field Theory (Lattice 2013)}\ } (\bibinfo
  {year} {2013})}\BibitemShut {NoStop}%
\bibitem [{\citenamefont {Luscher}(2010)}]{Luscher:2010iy}%
  \BibitemOpen
  \bibfield  {author} {\bibinfo {author} {\bibfnamefont {M.}~\bibnamefont
  {Luscher}},\ }\href {\doibase 10.1007/JHEP08(2010)071} {\bibfield  {journal}
  {\bibinfo  {journal} {JHEP}\ }\textbf {\bibinfo {volume} {1008}},\ \bibinfo
  {pages} {071} (\bibinfo {year} {2010})}\BibitemShut {NoStop}%
%%CITATION = ARXIV:1006.4518;%%
\bibitem [{\citenamefont {{Aoki}}\ \emph
  {et~al.}(2016{\natexlab{d}})\citenamefont {{Aoki}}, \citenamefont {{Aoyama}},
  \citenamefont {{Bennett}}, \citenamefont {{Kurachi}}, \citenamefont
  {{Maskawa}}, \citenamefont {{Miura}}, \citenamefont {{Nagai}}, \citenamefont
  {{Ohki}}, \citenamefont {{Rinaldi}}, \citenamefont {{Shibata}}, \citenamefont
  {{Yamawaki}},\ and\ \citenamefont {{Yamazaki}}}]{Aoki:2016yrm}%
  \BibitemOpen
  \bibfield  {author} {\bibinfo {author} {\bibfnamefont {Y.}~\bibnamefont
  {{Aoki}}}, \bibinfo {author} {\bibfnamefont {T.}~\bibnamefont {{Aoyama}}},
  \bibinfo {author} {\bibfnamefont {E.}~\bibnamefont {{Bennett}}}, \bibinfo
  {author} {\bibfnamefont {M.}~\bibnamefont {{Kurachi}}}, \bibinfo {author}
  {\bibfnamefont {T.}~\bibnamefont {{Maskawa}}}, \bibinfo {author}
  {\bibfnamefont {K.}~\bibnamefont {{Miura}}}, \bibinfo {author} {\bibfnamefont
  {K.-i.}\ \bibnamefont {{Nagai}}}, \bibinfo {author} {\bibfnamefont
  {H.}~\bibnamefont {{Ohki}}}, \bibinfo {author} {\bibfnamefont
  {E.}~\bibnamefont {{Rinaldi}}}, \bibinfo {author} {\bibfnamefont
  {A.}~\bibnamefont {{Shibata}}}, \bibinfo {author} {\bibfnamefont
  {K.}~\bibnamefont {{Yamawaki}}}, \ and\ \bibinfo {author} {\bibfnamefont
  {T.}~\bibnamefont {{Yamazaki}}} (\bibinfo {collaboration} {LatKMI
  Collaboration}),\ }\href@noop {} {\bibfield  {journal} {\bibinfo  {journal}
  {PoS}\ }\textbf {\bibinfo {volume} {LATTICE2015}},\ \bibinfo {pages} {214}
  (\bibinfo {year} {2016}{\natexlab{d}})}\BibitemShut {NoStop}%
%%CITATION = ARXIV:1601.04687;%%
\bibitem [{\citenamefont {Teper}(1985)}]{Teper:1985rb}%
  \BibitemOpen
  \bibfield  {author} {\bibinfo {author} {\bibfnamefont {M.}~\bibnamefont
  {Teper}},\ }\href {\doibase 10.1016/0370-2693(85)90939-6} {\bibfield
  {journal} {\bibinfo  {journal} {Phys. Lett.}\ }\textbf {\bibinfo {volume}
  {B162}},\ \bibinfo {pages} {357} (\bibinfo {year} {1985})}\BibitemShut
  {NoStop}%
%%CITATION = PHLTA,B162,357;%%
\bibitem [{\citenamefont {Bonnet}\ \emph {et~al.}(2002)\citenamefont {Bonnet},
  \citenamefont {Leinweber}, \citenamefont {Williams},\ and\ \citenamefont
  {Zanotti}}]{Bonnet:2001rc}%
  \BibitemOpen
  \bibfield  {author} {\bibinfo {author} {\bibfnamefont {F.~D.~R.}\
  \bibnamefont {Bonnet}}, \bibinfo {author} {\bibfnamefont {D.~B.}\
  \bibnamefont {Leinweber}}, \bibinfo {author} {\bibfnamefont {A.~G.}\
  \bibnamefont {Williams}}, \ and\ \bibinfo {author} {\bibfnamefont {J.~M.}\
  \bibnamefont {Zanotti}},\ }\href {\doibase 10.1103/PhysRevD.65.114510}
  {\bibfield  {journal} {\bibinfo  {journal} {Phys. Rev. D}\ }\textbf {\bibinfo
  {volume} {65}},\ \bibinfo {pages} {114510} (\bibinfo {year}
  {2002})}\BibitemShut {NoStop}%
%%CITATION = HEP-LAT/0106023;%%
\bibitem [{\citenamefont {Bors{\'a}nyi}\ \emph {et~al.}(2012)\citenamefont
  {Bors{\'a}nyi}, \citenamefont {D{\"u}rr}, \citenamefont {Fodor},
  \citenamefont {Hoelbling}, \citenamefont {Katz}, \citenamefont {Krieg},
  \citenamefont {Kurth}, \citenamefont {Lellouch}, \citenamefont {Lippert},
  \citenamefont {McNeile},\ and\ \citenamefont {Szab{\'o}}}]{Borsanyi:2012zs}%
  \BibitemOpen
  \bibfield  {author} {\bibinfo {author} {\bibfnamefont {S.}~\bibnamefont
  {Bors{\'a}nyi}}, \bibinfo {author} {\bibfnamefont {S.}~\bibnamefont
  {D{\"u}rr}}, \bibinfo {author} {\bibfnamefont {Z.}~\bibnamefont {Fodor}},
  \bibinfo {author} {\bibfnamefont {C.}~\bibnamefont {Hoelbling}}, \bibinfo
  {author} {\bibfnamefont {S.~D.}\ \bibnamefont {Katz}}, \bibinfo {author}
  {\bibfnamefont {S.}~\bibnamefont {Krieg}}, \bibinfo {author} {\bibfnamefont
  {T.}~\bibnamefont {Kurth}}, \bibinfo {author} {\bibfnamefont
  {L.}~\bibnamefont {Lellouch}}, \bibinfo {author} {\bibfnamefont
  {T.}~\bibnamefont {Lippert}}, \bibinfo {author} {\bibfnamefont
  {C.}~\bibnamefont {McNeile}}, \ and\ \bibinfo {author} {\bibfnamefont
  {K.~K.}\ \bibnamefont {Szab{\'o}}},\ }\href {\doibase
  10.1007/JHEP09(2012)010} {\bibfield  {journal} {\bibinfo  {journal} {JHEP}\
  }\textbf {\bibinfo {volume} {1209}},\ \bibinfo {pages} {010} (\bibinfo {year}
  {2012})}\BibitemShut {NoStop}%
%%CITATION = ARXIV:1203.4469;%%
\end{thebibliography}%

\end{document}